\documentclass[apj]{emulateapj}
\usepackage{amssymb,amsmath}
\usepackage{color,hyperref}
\definecolor{linkcolor}{rgb}{0,0,0.25}
\hypersetup{
  colorlinks=true,        
  linkcolor=linkcolor,    
  citecolor=linkcolor,    
  filecolor=linkcolor,    
  urlcolor=linkcolor      
}
\newcounter{address}
\newcommand{\ie}{i.e.}
\newcommand{\etal}{et al.}
\newcommand{\dd}{\mathrm{d}}
\newcommand{\eg}{e.g.}
\newcommand{\eqnname}{equation}
\newcommand{\equationname}{\eqnname}
\newcommand{\figurenames}{\figurename s}
\newcommand{\sectionname}{$\mathsection$}
\newcommand{\segue}{\emph{SEGUE}}
\newcommand{\sdss}{\emph{SDSS}}

\newcommand{\apoor}{\ensuremath{\alpha}-young}
\newcommand{\aenhanced}{\ensuremath{\alpha}-old}
\newcommand{\flag}[1]{\texttt{\lowercase{#1}}}
\newcommand{\platesn}{\texttt{plateSN\_r}}
\newcommand{\sn}{SN}
\newcommand{\feh}{\ensuremath{[\mathrm{Fe/H}]}}
\newcommand{\afe}{\ensuremath{[\alpha\mathrm{/Fe}]}}
\newcommand{\logg}{log g}
\newcommand{\Ro}{\ensuremath{R_0}}
\newcommand{\hz}{\ensuremath{h_z}}
\newcommand{\hR}{\ensuremath{h_R}}

\newcommand{\dens}{\ensuremath{\nu_*}}

\begin{document}

\submitted{Astrophys.~J., in press}

\title{The spatial structure of mono-abundance sub-populations of the
  Milky Way disk}
\author{Jo~Bovy\altaffilmark{1,2,3},
  Hans-Walter~Rix\altaffilmark{4}, 
  Chao~Liu\altaffilmark{4}, 
  David~W.~Hogg\altaffilmark{4,5},
  Timothy~C.~Beers\altaffilmark{6,7},
  and Young Sun Lee\altaffilmark{7}}
\altaffiltext{\theaddress}{\stepcounter{address} Institute for Advanced Study, Einstein Drive, Princeton, NJ 08540, USA}
\altaffiltext{\theaddress}{\stepcounter{address} Hubble fellow}
\altaffiltext{\theaddress}{\stepcounter{address}
  Correspondence should be addressed to bovy@ias.edu~.}
\altaffiltext{\theaddress}{\stepcounter{address}
  Max-Planck-Institut f\"ur Astronomie, K\"onigstuhl 17, D-69117
  Heidelberg, Germany}
\altaffiltext{\theaddress}{\stepcounter{address} Center for
  Cosmology and Particle Physics, Department of Physics, New York
  University, 4 Washington Place, New York, NY 10003, USA}
\altaffiltext{\theaddress}{\stepcounter{address} National Optical Astronomy Observatory, Tucson, AZ 85719, USA}
\altaffiltext{\theaddress}{\stepcounter{address}
Department of Physics \& Astronomy and JINA (Joint Institute for
Nuclear Astrophysics), Michigan State University, East Lansing, MI
48824, USA}

\begin{abstract} The spatial, kinematic, and elemental-abundance
structure of the Milky Way's stellar disk is complex, and has been
difficult to dissect with local spectroscopic or global photometric
data. Here, we develop and apply a rigorous density modeling approach
for Galactic spectroscopic surveys that enables investigation of the
global spatial structure of stellar sub-populations in narrow bins of
\afe\ and \feh, using 23,767 G-type dwarfs from \sdss/\segue, which
effectively sample $5 < R_{GC} < 12$ kpc and 0.3 $\lesssim |Z|
\lesssim 3$ kpc.  We fit models for the number density of each such
([$\alpha$/Fe] \& [Fe/H]) mono-abundance component, properly
accounting for the complex spectroscopic \segue\ sampling of the
underlying stellar population, as well as for the metallicity and
color distributions of the samples. We find that each mono-abundance
sub-population has a simple spatial structure that can be described by
a single exponential in both the vertical and radial direction, with
continuously increasing scale heights ($\approx$200 pc to 1 kpc) and
decreasing scale lengths ($>$4.5 kpc to 2 kpc) for increasingly older
sub-populations, as indicated by their lower metallicities and \afe\
enhancements.  That the abundance-selected sub-components with the
largest scale heights have the shortest scale lengths is in sharp
contrast with purely geometric `thick--thin disk' decompositions. To
the extent that [$\alpha$/Fe] is an adequate proxy for age, our
results directly show that older disk sub-populations are more
centrally concentrated, which implies inside-out formation of galactic
disks. The fact that the largest scale-height sub-components are most
centrally concentrated in the Milky Way is an almost inevitable
consequence of explaining the vertical structure of the disk through
internal evolution. Whether the simple spatial structure of the
mono-abundance sub-components, and the striking correlations between
age, scale length, and scale height can be plausibly explained by
satellite accretion or other external heating remains to be seen.
\end{abstract}


\section{Introduction}

The formation of galactic disks is a long-standing problem in galaxy
formation. In numerical simulations, disks form through gas dissipation
\citep{Sandage70a,Larson76a}, and the formation of the outer regions
of the disk happens on longer time scales than the inner disk
\citep{Gott76a,Larson76a,Katz91a}. The disks that form have
exponential density profiles \citep{Lake88a}, possibly due to the
detailed conservation of angular momentum of an initially spherical
cloud in solid-body rotation \citep{Fall80a,Gunn82a}. Yet, direct
observational evidence for this picture is scant
\citep[\eg,][]{Dalcanton97a,Somerville08a,Wang11a}, and forming
realistic disks within the $\Lambda$CDM paradigm remains challenging
\citep[\eg,][]{Abadi03a,Scannapieco09a,Guedes11a}.

Central to the question as to how galactic disks form and evolve is
the existence of ``thick'' disk components. First discovered in
external galaxies \citep{Tsikoudi79a,Burstein79a,vanderkruit81a},
thick-disk components represent excess light or stars beyond the
canonical thin disk's exponential vertical profile. The Milky Way's
thick-disk component
\citep{Yoshii82a,Gilmore83a,Reid93a,Majewski93a,Juric08a} provides us
with a detailed look at this common galactic component. Generally,
thick-disk components are found to be old
\citep{Bensby05a,Yoachim08a}, kinematically hot
\citep{Chiba00a,Soubiran03a,Gilmore02a,Yoachim05a}, and metal-poor
(compared to the thin-disk components), as well as enhanced in
$\alpha$-elements
\citep{Fuhrmann98a,Prochaska00a,Taut01a,Bensby03a,Feltzing03a,Mishenina04a,Bensby05a,Reddy06a,Haywood08a}. Density
decompositions of the stellar disk into thinner and thicker components
of external galaxies and the Milky Way have found that thicker-disk
components have larger scale heights and longer scale lengths than
their corresponding thin disk
\citep{Robin96a,Buser99a,Chen01a,Ojha01a,Neeser02a,Larsen03a,Yoachim06a,Pohlen07a,Juric08a}. However,
these decompositions are purely geometric, and do not take kinematics
or abundance information into account when assigning thinner- or
thicker-disk membership.

A number of qualitatively very different models have been proposed for
the formation of thick-disk components. External mechanisms, such as
the direct accretion of stars from a disrupted satellite galaxy
\citep{Abadi03a} or the heating of a pre-existing thin disk through a
minor merger
\citep{Quinn93a,Wyse06a,Kazantzidis08a,Villalobos08a,Moster10a}, can
explain many of the observed properties of thick-disk
components. Thick-disk components can also be formed internally
through star formation following a gas-rich merger
\citep[\eg,][]{Brook04a} or by quiescent internal dynamical evolution
\citep{Schoenrich08a,Schoenrich09a,Loebman11a}.

The idea that the thick-disk component could arise in good part through
internal evolution is an intriguing possibility, as it explains a
range of other observations. Significant redistribution of angular
momentum without radially heating the disk (``radial migration'')
happens naturally if spiral structure is transient
\citep{sellwood02a,roskar08a,roskar11a}, and has also been shown to
occur through bar--spiral structure interactions
\citep{Minchev10a,Minchev11a}. It can also be induced by an orbiting
satellite \citep{Quillen09a,Bird11a}. The transient nature of spiral
structure is favored both theoretically
\citep[\eg,][]{Sellwood84a,Carlberg85a,Sellwood89a} and
observationally from surveys of the Solar neighborhood
\citep[\eg,][]{Dehnen98b,deSimone04a,Bovy09a,Bovy10a,Sellwood11a} and
of external galaxies \citep[\eg,][]{Meidt09a,Foyle11a}. Radial
migration naturally explains the flatness and spread in the
age-metallicity relation in the Solar neighborhood, as the large-scale
changes in the guiding radii of stars tend to flatten radial-abundance
gradients. A thicker-disk component arises through radial migration
when stars from the inner Galaxy migrate outward, where the
gravitational attraction toward the mid-plane is smaller, such that
they reach larger heights above the plane.  However, to date,
radial-migration models have essentially only been confronted with
data at the Solar radius, and observational tests to discriminate
formation scenarios for thicker-disk components have not been
conclusive \citep[\eg,][]{Dierickx10a}.

Because radial migration is effectively a diffusion process, it
complicates, if not erases, the link between the present-day
chemo-orbital distribution and the orbital characteristics and
abundance distribution at the time of a given stars' birth. Without
detailed modeling of the episodes of transient spiral structure (or
the equivalent in other radial-migration scenarios), reconstructing
the radial and azimuthal actions is problematic as well. However, the
vertical action is an adiabatic invariant during the slow change in
the vertical potential that ensues from this migration. The overall
(mass-weighted) radial structure of the disk is left relatively
unchanged as radial migration proceeds
\citep{sellwood02a,Minchev11a}---essentially because any
redistribution of the surface-mass density would provide energy to
heat the disk, and to avoid this heating the overall surface-mass
density profile needs to be conserved---but if different (age- or
abundance-) components of the disk have a different initial structure,
radial migration will work to bring the spatial distribution of
different populations closer to the mean.

In this paper we implement for the first time an alternative approach
to globally `dissecting' the Milky Way's stellar disk: we study the
overall (vertical and radial) spatial structure of large samples of
stars selected to be sub-populations in the elemental-abundance space
spanned by metallicity [Fe/H] and $\alpha$-abundance \afe\footnote{The
[$\alpha$/Fe] ratio in this paper is an average of the [Mg/Fe],
[Si/Fe], [Ca/Fe], and [Ti/Fe] ratios \citep{Lee11a}.}, as it is
becoming increasingly clear that a characterization of the thicker
disk components based only on stellar abundances is superior to
kinematic definitions \citep{Navarro11a,Lee11b}. The \afe\ ratio in
particular is a crucial parameter, as it can be used as a relative age
indicator \citep{Wyse88a}. At early times, the low-metallicity
interstellar medium is enriched by type II supernovae (SNeII). After
about 2 to 3 Gyr, type Ia SNe occur \citep[\eg][]{Maoz11a}, and the
stellar yields shift toward Fe, leading to a decreasing \afe\ with
increasing age. Therefore, populations of stars with enhanced \afe\
ratios are chemically older than those with \afe\ closer to the solar
ratio. By using the \sdss/\segue\ G-dwarf sample, we observe stars
globally across the Milky Way, constraining their vertical
distributions from 300 pc to 4 kpc from the mid-plane, and their
radial densities from Galactocentric radii ranging from 5 to 12
kpc. We show that the scale length of the $\alpha$-enhanced---and thus
probably oldest---population is much shorter than that of the
chemically more-evolved stars with solar \afe.  This is opposite to
previous disk decompositions into thicker and thinner components that
make use of geometric information alone \citep[\eg,][and see
above]{Juric08a}. Also, we do not detect any discontinuity in the
vertical scale height as a function of \afe\ that might be expected if
the thick-disk component was formed through a singular external or
internal event, but instead observe a continuous increase in
scale-height with \afe. This casts doubt on how sensible or useful it
is to think of distinct thin- and thick-disk components in the Milky
Way.

The outline of this paper is as follows. In
\sectionname~\ref{sec:data} we present the details of our data
sample. Our density-fit methodology, accounting for the various
aspects of the \segue\ selection function, is given in
\sectionname~\ref{sec:density}. We give the results of the density
fits to the various abundance-selected samples in
\sectionname~\ref{sec:results}, and discuss these results in terms of
disk formation and evolution models in
\sectionname~\ref{sec:discussion}. We summarize the main conclusions
of the paper in \sectionname~\ref{sec:conclusion}. The appendices
describe our model for the \segue\ selection function, some details of
our fitting methodology, and detailed comparisons between our fits and
the data. Modeling the spectroscopic \segue\ selection function is
central to our analysis. It is described in an appendix to
aid the readability of the paper, as its implementation requires a
detailed and hence extensive description that may not be of interest
to all readers. Throughout this paper, we assume that the Sun's
displacement from the mid-plane is 25 pc toward the North Galactic Pole
\citep{Chen01a,Juric08a}, and that the Sun is located at 8 kpc from
the Galactic center \citep[\eg,][]{Bovy09b}.

\section{Data}\label{sec:data}

The \emph{Sloan Digital Sky Survey} (\sdss; \citealt{York:2000gk}) has
obtained \emph{u,g,r,i} and \emph{z} CCD imaging of $\approx$ 10$^4$
deg$^2$ of the northern and southern Galactic sky
\citep{Gunn:1998vh,Stoughton:2002ae,Gunn06a}. All the data processing,
including astrometry \citep{Pier:2002iq}, source identification,
deblending, and photometry \citep{Lupton:2001zb}, calibration
\citep{Fukugita:1996qt, Hogg01a,Smith:2002pca,
Ivezic:2004bf,Padmanabhan08a}, and spectroscopic fiber placement
\citep{Blanton:2001yk}, are performed with automated \sdss\
software. The \sdss\ spectroscopic survey uses two fiber-fed
spectrographs that have 320 fibers each.

The \emph{Sloan Extension for Galactic Understanding and Exploration}
(\segue; \citealt{Yanny09a}) is a low-resolution ($R\approx2,000$)
spectroscopic sub-survey of the \sdss\ focused on Galactic science. We
select a sample of G-type dwarfs from the \sdss/\segue\ Data Release 7
(DR7; \citealt{Abazajian09a}). G-type dwarfs are the most luminous
tracers whose main-sequence lifetime is larger than the expected disk
age at basically all metallicities. G-type stars are selected from the
full DR7 \segue\ sample using a simple color--magnitude-cut that
corresponds to the \segue\ G-star target type: 0.48 $\leq g-r \leq
0.55$ and $r < 20.2$. All magnitudes here and in what follows are
absorption-corrected and dereddened, respectively, using the reddening
maps of \citet{Schlegel98a}; as we only use lines of sight with
relatively small extinction and we do not use the \sdss\ $u$ band,
using the improved reddening maps of \citet{Schlafly11a} leads to
insignificant differences for the purpose of our analysis. We further
limit the spectroscopic sample to those lines of sight with $E(B-V) <
0.3$, to minimize effects due to uncertainty in extinction, to objects
having spectra with signal-to-noise ratio \sn\ $> 15$, and to objects
with valid metallicities, heliocentric line-of-sight velocities, and
proper motions (even though the latter two are not used in the
analysis). All of the selected objects have valid values for their
stellar atmospheric parameters as determined by the \segue\ Stellar
Parameter Pipeline
\citep{Lee08a,Lee08b,AllendePrieto08a,Lee11a,Smolinski11a}. Typical
uncertainties in these parameters are 0.2 dex for the spectroscopic
metallicity \feh, 0.1 dex for \afe, 0.25 dex for the surface gravity
\logg, and 180 K for the effective temperature
\citep{Schlesinger10a,Smolinski11a}. In what follows we are primarily
interested in the relative rankings of stars based on \afe\ and \feh,
such that random uncertainties are all that matter. Note that our
signal-to-noise-ratio cut of \sn\ $>15$ is more inclusive than
recommended by \citet{Lee11a} (who recommend \sn\ $> 20$), but this
does not increase the uncertainties in \afe\ by much. For dwarfs with
\feh\ $> -2$, there is no significant correlation between the \feh\
and \afe\ estimates. We use this sample of G-type stars to determine
the \segue\ G-star selection function in
\appendixname~\ref{sec:selection} below.

\begin{figure}[tp]
\includegraphics[width=0.5\textwidth,clip=]{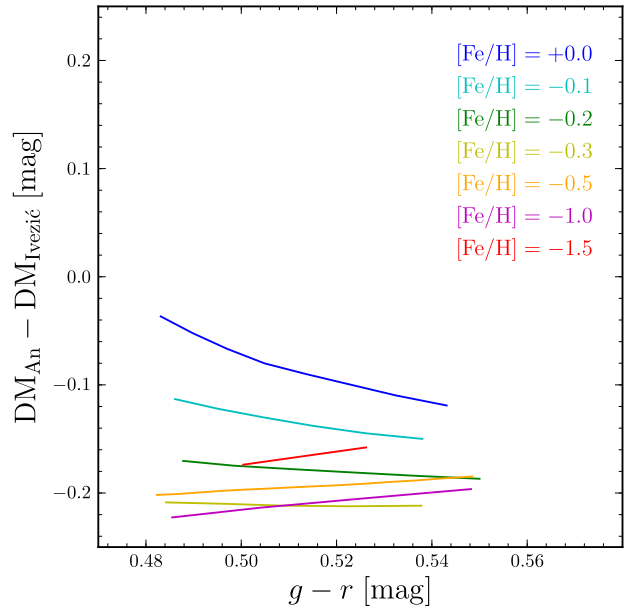}
\caption{Comparison between distance moduli derived from the
\citet{Ivezic08a} photometric distance relation (their eqn. A7), and those derived from
the \citet{An09a} theoretical
isochrones.}\label{fig:dm_anjuric_ivezic}
\end{figure}

We select G-type dwarfs by selecting stars with \logg\ $> 4.2$, to
eliminate giant stars (we have verified that more conservative \logg\
cuts give the same results, see \appendixname~\ref{sec:datamodel}). We
perform no other cuts (\eg, other color cuts or distance cuts) beyond
these basic cuts in order to preserve a relatively simple spatial
selection function. This sample contains about 28,000 stars, 23,767 of
which lie within the well-populated bins in the (\feh,\afe) plane that
we analyze below. Distances to individual stars are obtained from the
\citet{Ivezic08a} photometric color--metallicity-absolute-magnitude
relation (their eqn. A7) applied to the $g-r$ color, rather than the
$g-i$ color, using \begin{equation}\label{eq:ri_gr} r-i =
\frac{(g-r-0.12)}{2.34}\, , \end{equation} and employing the
spectroscopic metallicity. These distances are about 10\,percent
larger than the distances obtained from the \citet{An09a} stellar
isochrones, with little to no color or metallicity dependence over the
color and metallicity ranges considered here (see
\figurename~\ref{fig:dm_anjuric_ivezic} and further discussion in
\sectionname~\ref{sec:discussion}). Individual distance uncertainties
are typically $\lesssim$ 10\,percent, and thus do not greatly smooth
the underlying Galactic density, whose scales are much larger than
this (for an illustration of this see \citealt{Juric08a}, where much
larger distance uncertainties of around 20\,percent were shown to
influence the inferred scale heights by less than 5\,percent).

\begin{figure}[tp]
\includegraphics[width=0.5\textwidth]{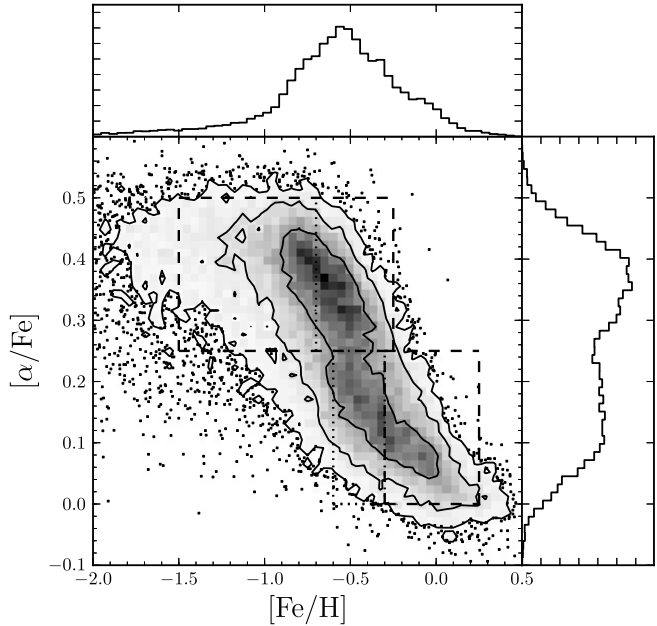}
\caption{Distribution of the spectroscopic sample of G dwarfs in
  elemental-abundance space. The density is linear, and the contours
  contain 68, 95, and 99\,percent of the distribution. Outliers beyond
  99\,percent are individually shown. Our cuts to select \aenhanced\
  (top, left) and \apoor\ (bottom, right) samples are shown as dashed
  boxes. The dotted lines indicate the median [Fe/H] for the
  \aenhanced\ sample (rounded to the nearest 0.05 dex), used to split
  the \aenhanced\ sample in [Fe/H], and for \afe\ $<$ 0.25 the dotted
  box indicates the metal-poor \apoor\ sample, used in \sectionname\sectionname~\ref{sec:aenhancedresults} and \ref{sec:apoorresults}.}\label{fig:afeh_g}
\end{figure}

The distribution of the G-dwarf sample in the elemental abundance
space, made up of [Fe/H] and [$\alpha$/Fe], is shown in
\figurename~\ref{fig:afeh_g}. This distribution is characterized by
two modes, one a metal-poor, $\alpha$-enhanced population that must
represent the oldest part of the Galactic disk, and another that is
metal-rich and has a solar [$\alpha$/Fe] ratio. The two boxes
delineated by dashed lines constitute our broad separation of these two
populations, which we will refer to as \aenhanced\ and \apoor,
respectively
\begin{align}\label{eq:sampledef}
\alpha\text{-}\mathrm{old\ sample}:\qquad \qquad \qquad &\nonumber\\
\qquad -1.5 < \feh < -0.25,\, & 0.25 < \afe < 0.50\,,\\
\nonumber\\
\alpha\text{-}\mathrm{young\ sample}:\qquad \qquad \ \ &\nonumber\\
\qquad -0.3 < \feh < \phantom{-}0.25,\,& 0.00 < \afe < 0.25\,.
\end{align}

\begin{figure}[tp]
\includegraphics[width=0.5\textwidth]{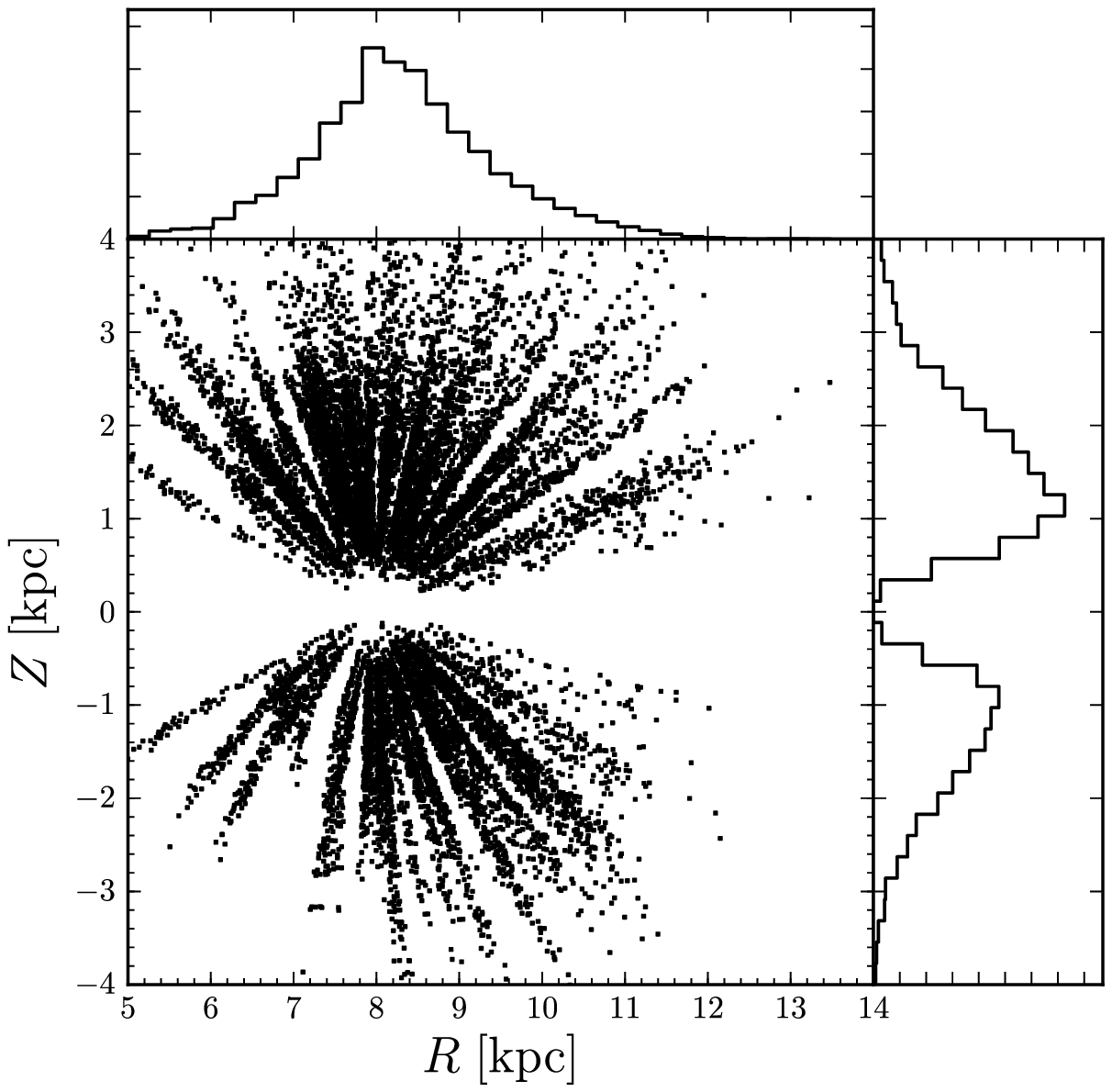}\\
\includegraphics[width=0.5\textwidth]{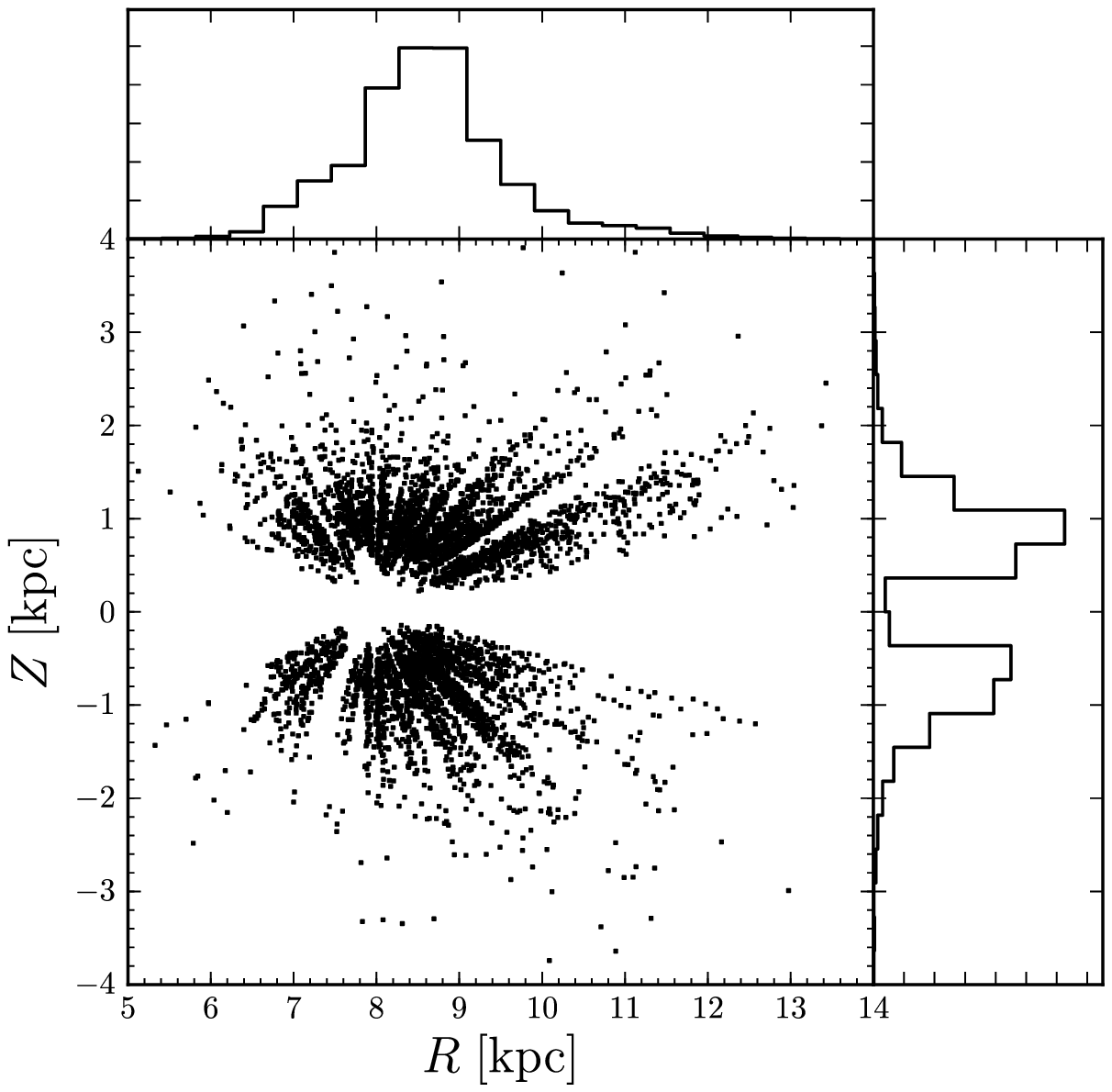}
\caption{Spatial distribution of the spectroscopic sample of \aenhanced\
  (\emph{top panel}) and \apoor\ (\emph{bottom panel}) G dwarfs in the
  $R,Z$ plane.}\label{fig:dataxyrz_g_poor}
\end{figure}

The spatial distributions of the \aenhanced\ and \apoor\ G-dwarf
samples are shown in \figurename~\ref{fig:dataxyrz_g_poor}, without
accounting for the selection function. It is clear from this figure
that the bright limit of the G-dwarf sample ($r > 14.5$, see below) is
such that the effective minimum distance is approximately 600 pc. This
means that for the thinner disk components, stars within one scale
height are not sampled by \sdss/\segue\ (this is also apparent in
\figurename~\ref{fig:afeh_g}, where most stars have sub-solar
metallicities). However, because the thinner components also contain
the most stars (see below and \citealt{Bovy12a}), there are still
sufficient stars above one scale height of these components such that
the G-dwarf data set contains a large number of them.

Understanding and modeling the \segue\ selection function, \ie, the
relation between the stars with successfully determined spectral
parameters that enter our sample, and their photometric or
volume-complete parent population, is central to any analysis that
involves the spatial structure of spectroscopically-selected
samples. It has not been worked out previously, and while in principle
straightforward, it requires attention to a number of details. We
describe our model for the \segue\ selection function in
\appendixname~\ref{sec:selection}. The \segue\ G-star sample was
obtained by uniformly sampling the dereddened color--magnitude boxes
with color range 0.48 $\leq g-r \leq 0.55$ and a ``bright'' (14.5
$\leq r \leq 17.8$) and ``faint'' (17.8 $\leq r \leq 20.2$) apparent
magnitude range along a set of $\approx 150$ lines of sight. Due to
our signal-to-noise ratio cut, this uniform sampling is truncated at a
brighter magnitude, where the cut-off is different for each line of
sight. We determine the cut-off for each \segue\ plug-plate (which we
refer to simply as ``plates'' in what follows) as the faintest star in
the color--magnitude box, and model the $r$-dependence of the
selection function using a hyperbolic-tangent step around the
cut-off. We obtain the overall selection fraction for each line of
sight by comparing the size of the spectroscopic sample to that of the
photometric sample in the targeted color--magnitude box for each
individual line of sight. This model is described in more detail in
\appendixname~\ref{sec:selection}.

\section{Density fitting methodology}\label{sec:density}

\subsection{Generalities}

Fitting the spatial-density profiles of various G-dwarf sub-samples
must account for the fact that the observed star counts do not reflect
the underlying stellar distribution, but are strongly shaped by (a)
the strongly position-dependent selection fraction of stars with
spectra (see \figurename~\ref{fig:sfxyrz_g}), (b) the need to use
photometric distances that in turn depend on the color and metallicity
distribution of the sample (as the magnitude-limited \segue\ sample
corresponds to a color- and metallicity-dependent distance-limited
sample), and (c) the pencil-beam nature of the \segue\ survey. To
properly take all of these effects into account, we need to use
forward modeling: in what follows we fit stellar-density models to the
data by generating the expected observed distribution of stars in the
spectroscopic sample, based on our model for the \segue\ selection
function and the photometric distance relation; this predicted
distribution is then compared to the observed star counts. We show
below how this can be expressed as a maximum likelihood problem. This
general density-fitting methodology applies to any spectroscopic
survey, with minor modifications, and needs to be applied to obtain
selection-corrected distributions from spectroscopically selected
stellar samples. In particular, this methodology needs to be applied
to constrain the structural parameters of abundance-selected samples in
the Milky Way.

As the photometric distance estimates depend on the $g-r$ color, metallicity [Fe/H], and
apparent $r$-band magnitude, and because the selection function is a
function of position, $r$, and $g-r$, we need to model the observed
density of stars in color--magnitude--metallicity--position
space, $\lambda(l,b,d,r,g-r,\feh)$. This density of stars can be written as
\begin{equation}\label{eq:rate}
\begin{split}
\lambda(l,b,d,r,&g-r,\feh)= \\
& \rho(r,g-r,\feh|R,Z,\phi)\times\dens(R,Z,\phi)
\\ &  \times |J(R,Z,\phi;l,b,d)| \times S(\mathrm{plate},r,g-r)\,.
\end{split}
\end{equation}
Here, $(R,Z,\phi)$ are Galactocentric cylindrical coordinates
corresponding to rectangular coordinates $(X,Y,Z)$, which can be
calculated from $(l,b,d)$. The factor $\rho(r,g-r,\feh|R,Z,\phi)$ is the number density
in magnitude--color--metallicity space as a function of
position (see further discussion in \appendixname~\ref{sec:colorFeh}). The $|J(R,z;l,b,d)|$ is a Jacobian term because of the $(X,Y,Z)
\rightarrow (l,b,d)$ coordinate transformation;
the crucial factor $S(\mathrm{plate},r,g-r)$ is the selection function as given in
\equationname~(\ref{eq:seguesf}). Finally, $\dens(R,Z,\phi)$ is the underlying
spatial number density of the sample; we stress that this is a density as a
function of rectangular coordinates $(X,Y,Z)$ that we evaluate through $(R,Z,\phi)$ (as we will assume later that the density is axisymmetric), \ie, its dimension is 1
/ (spatial unit)$^3$. In what follows we will assume that our models
for this density (\eg, exponentials in the vertical and radial direction) are characterized by a set of parameters denoted as
$\theta$ and that the density is axisymmetric, such that
$\dens \equiv \dens(R,Z|\theta)$.

The likelihood of a given model for the density $\dens(R,z|\theta)$
is given by that of a Poisson
process with rate parameter $\lambda$
\begin{equation}\label{eq:likelike}
\begin{split}
  \ln & \mathcal{L} = \sum_i \left[ \ln \lambda(\{l,b,d,r,\right.\left.g-r,\feh\}_i|\theta)
  \right]\\
& -\int \dd l \,\dd b\, \dd d\, \dd r\, \dd (g-r) \,\dd \feh
  \,\lambda(l,b,d,r,g-r,\feh|\theta)\, ,
\end{split}
\end{equation}
where the integral is over the domain surveyed and $i$ indexes the
observed objects. Because the Jacobian, the selection function, and
the density in magnitude--color--metallicity space only enter $\lambda$
multiplicatively (\eqnname~(\ref{eq:rate})) their contribution to the first term ($\ln \lambda$) in \eqnname~(\ref{eq:likelike}) is a constant that does not depend on the density parameters. Thus, up to a term that does not depend on $\theta$, the log likelihood
is equivalent to
\begin{equation}\label{eq:densitylike}
\begin{split}
  \ln \mathcal{L} &= \sum_i \left[ \ln \dens(R,z|\{l,b,d\}_i,\theta)
  \right]\\
& \ -\int \dd l \,\dd b\, \dd d\, \dd r\, \dd (g-r) \,\dd \feh
  \,\lambda(l,b,d,r,g-r,\feh|\theta) \, .
\end{split}
\end{equation}
Note that the Jacobian, the density in the
magnitude--color--metallicity space, and the selection function only enter
through the second term, and do not need to be evaluated on a
star--by--star basis. The second term in
\eqnname~(\ref{eq:densitylike})---the normalization integral---can be
written as (assuming that the density does not depend on $(l,b)$ over
the area of a plate, although this can easily be relaxed)
\begin{equation}\label{eq:normint}
\begin{split}
\int & \dd l\, \dd b\,\dd d\, \dd r\, \dd (g-r) \,\dd \feh \,\lambda(l,b,d,r,g-r,\feh|\theta) \\
&= A_p\,\sum_{\mathrm{plates}\ p}
\int \dd (g-r)\,\dd\feh\,\dd r \, S(p,r,g-r) \\
& \qquad \int \dd d\,
\rho(r,g-r,\feh|R,Z,\phi)\,d^2\,\dens(R,z|l,b,d,\theta)\,,
\end{split}
\end{equation}
where $A_p$ is the area of a \segue\ plate (approximately 7 deg$^2$). 

In the following, we analytically marginalize over the amplitude of the
rate $\lambda$ with a logarithmically flat prior. In that case, the log
likelihood becomes
\begin{equation}\label{eq:densitylike2}
\begin{split}
  \ln& \mathcal{L} =\\& \sum_i \left[ \ln \dens(R,z|\{l,b,d\}_i,\theta)\right.\\ 
  &\left.-\ln \int \dd l \,\dd b\, \dd d\, \dd r\, \dd (g-r) \,\dd\feh
  \,\lambda(l,b,d,r,g-r,\feh|\theta)\right]\,.
\end{split}
\end{equation}
Note that the normalization integral is now  moved inside of the logarithm.

In \appendixname~\ref{sec:colorFeh}, we discuss how we include the
magnitude--color--metallicity factor $\rho(r,g-r,\feh|R,Z,\phi)$ in
the likelihood.

\subsection{Stellar number density models}

We fit number-density models for the various abundance sub-populations, consisting of a disk with an exponential profile in both the vertical and radial direction, plus a constant density
\begin{equation}\label{eq:densmodel} \begin{split} \dens(&R,Z) =\\
& N(\Ro)\left[\frac{1}{2\,\hz}\,\exp\left(-\frac{R-\Ro}{\hR}\right)\,\exp\left(-\frac{|Z|}{\hz}\right) + \frac{\beta_c}{24}\right]\,,
\end{split} \end{equation} where $N(\Ro)$ is the vertically-integrated number density at
\Ro. We refer to this model below as a single-exponential disk fit, as in all cases the data imply $\beta_c \ll 1$. We also fit combinations of exponential disks as
\begin{equation} \begin{split}\dens & (R,Z) =\\
& N(\Ro)\,\left[\frac{1-\beta_2}{2\,\hz}\,\exp\left(-\frac{R-\Ro}{\hR}\right)\,\exp\left(-\frac{|Z|}{\hz}\right)
\right. \\ & \left. +
\frac{\beta_2}{2\,h_{z,2}}\,\exp\left(-\frac{R-\Ro}{h_{R,2}}\right)\,\exp\left(-\frac{|Z|}{h_{z,2}}\right)\right]\, . \end{split}\end{equation}
In particular in the $Z$ direction, this is analogous to traditional density fits based on photometric data, which require (at least) two exponential components. We do not fit for the overall normalization, $N(\Ro)$, as we are interested primarily in the shape of the stellar-density profile.

To determine the best-fit parameters and their uncertainties we use
Powell's method for minimization \citep{Press07a}, and then MCMC-sample
the posterior distribution function, obtained by multiplying the
likelihood in \eqnname~(\ref{eq:densitylike2}) with flat logarithmic
priors for the scale parameters (\hz, \hR, $h_{z,2}$, $h_{R,2}$) and
flat priors on the contamination-fraction parameters ($\beta_c$,
$\beta_2$), using an ensemble MCMC sampler (\citealt{Goodman10a};
Foreman-Mackey \etal, 2011, in preparation).

\subsection{Tests on mock data}

In \appendixname~\ref{sec:fakedata}, we discuss tests of the fitting
methodology on mock data sets made up of single-exponential disk
components observed using the \segue\ sampling. These tests show that
we can recover the input density structure to within the
MCMC-determined uncertainties over the range of inferred scale
heights, scale lengths, and sample sizes found below.

\section{Density structure}\label{sec:results}

First, we briefly discuss the result of fitting the broad bins in
abundance as defined in \eqnname~(\ref{eq:sampledef}), in order to
explore the broad trends in spatial structure with elemental
abundance. In \sectionname~\ref{sec:monoresults}, we then split the
sample finely in elemental-abundance space and map the structure of
mono-abundance populations.

\subsection{The \aenhanced\ stars}\label{sec:aenhancedresults}

\begin{deluxetable*}{cr@{}lr@{}lr@{}lr@{}lr@{}lr@{}l}
\tablecaption{}
\tablecolumns{11}
\tablewidth{0pt}
\tabletypesize{\footnotesize}
\tablecaption{Results for the \aenhanced\ G-dwarf sample ( $-1.5 <$ [Fe/H] $< -0.25$, $0.25 <$ \afe\ $< 0.50$)}
\tablehead{\colhead{} & \multicolumn{2}{c}{$h_z$} & \multicolumn{2}{c}{$h_R$} & \multicolumn{2}{c}{$h_{z,2}$} & \multicolumn{2}{c}{$h_{R,2}$}  &\multicolumn{2}{c}{$\beta_2$} & \multicolumn{2}{c}{$\beta_c$}\\
\colhead{} & \multicolumn{2}{c}{(pc)} & \multicolumn{2}{c}{(kpc)} & \multicolumn{2}{c}{(pc)} & \multicolumn{2}{c}{(kpc)} & \multicolumn{2}{c}{} & \multicolumn{2}{c}{}}
\startdata
all plates & 701&$\pm$5& 2.06&$\pm$0.03& \ldots & & \ldots & & \ldots & & 0.0000&$\pm$0.0009\\
bright plates  & 769&$\pm$14& 1.79&$\pm$0.05& \ldots & & \ldots & & \ldots & & 0.004&$\pm$0.009\\
faint plates  & 714&$\pm$11& 2.25&$\pm$0.05& \ldots & & \ldots & & \ldots & & 0.001&$\pm$0.001\\
$b < 0^\circ$  & 694&$\pm$9& 2.02&$\pm$0.05& \ldots & & \ldots & & \ldots & & 0.0000&$\pm$0.0010\\
$b > 0^\circ$  & 699&$\pm$8& 2.10&$\pm$0.04& \ldots & & \ldots & & \ldots & & 0.000&$\pm$0.001\\
$|b| > 45^\circ$ & 696&$\pm$6& 2.23&$\pm$0.06& \ldots & & \ldots & & \ldots & & 0.0000&$\pm$0.0009\\
$|b| < 45^\circ$  & 640&$\pm$10& 2.05&$\pm$0.04& \ldots & & \ldots & & \ldots & & 0.002&$\pm$0.002\\
\\
all plates & 686&$\pm$11& 2.01&$\pm$0.05& 933&$\pm$49& 3.0&$\pm$0.4& 0.04&$\pm$0.03& \ldots & \\
bright plates  & 764&$\pm$20& 1.78&$\pm$0.04& 3126&$\pm$271& $>$64 & & 0.01&$\pm$0.02& \ldots & \\
faint plates  & 688&$\pm$40& 2.2&$\pm$0.1& 1311&$\pm$189& $>$3.0 (5&$\pm$1)& 0.03&$\pm$0.04& \ldots & \\
$b < 0^\circ$  & 671&$\pm$22& 1.97&$\pm$0.08& 993&$\pm$169& 3.7&$\pm$0.4& 0.05&$\pm$0.05& \ldots & \\
$b > 0^\circ$  & 687&$\pm$11& 2.06&$\pm$0.07& 886&$^{+350}_{-708}$& 3&$\pm$1& 0.04&$\pm$0.04& \ldots & \\
$|b| > 45^\circ$ & 692&$\pm$11& 2.2&$\pm$0.1& 800&$\pm$88& 4.3&$\pm$0.4& 0.01&$\pm$0.07& \ldots & \\
$|b| < 45^\circ$  & 639&$\pm$17& 2.03&$\pm$0.07& 1142&$\pm$99& $>$5 & & 0.01&$\pm$0.02& \ldots & \\
\\
\protect{[}Fe/H] $<$ -0.7  & 856&$\pm$20& 2.06&$\pm$0.08& 865&$\pm$108& 2.1&$\pm$0.3& 0.07&$\pm$0.08& \ldots & \\
\protect{[}Fe/H] $>$ -0.7  & 583&$\pm$16& 1.97&$\pm$0.08& 873&$\pm$62& 4.0&$\pm$0.5& 0.03&$\pm$0.04& \ldots & \\
\\
0.25 $\leq$ [$\alpha$/Fe] $<$ 0.35  & 627&$\pm$18& 2.23&$\pm$0.10& 802&$\pm$104& 3.5&$\pm$0.3& 0.03&$\pm$0.06& \ldots & \\
0.35 $\leq$ [$\alpha$/Fe] $<$ 0.5\phantom{0}  & 765&$\pm$15& 1.89&$\pm$0.04& 826&$\pm$45& 2.0&$\pm$0.1& 0.03&$\pm$0.06& \ldots & 
\enddata
\tablecomments{\protect{L}ower limits are at 99\,percent posterior confidence. Lower limits are given when the best-fit value is larger than 4.5 kpc. The best-fit value is not given if the best-fit value is larger than 6 kpc.\label{table:poor_results}}
\end{deluxetable*}

For the \aenhanced\ sample, the fit results for single exponential profiles in $R$ and $Z$, and for
a combination of two exponential profiles for both $R$ and $Z$, are given in
\tablename~\ref{table:poor_results}. The model with two exponentials
in both $R$ and $Z$ is preferred, but the parameters of the dominant
double-exponential disk are similar for both fits. That is, \emph{even
when we give the model the additional freedom of two vertical scale
heights, the data lead us to employ only a single exponential scale
height}. There is no evidence for a thinner component in the
\aenhanced\ abundance range. We see that the \aenhanced\ sample is
dominated by a population of stars with a scale height of 686$\pm$11
pc, and a short scale length of 2.01$\pm$0.05 kpc (consistent with the
rough estimate of 2 kpc based on a handful of stars by
\citealt{Bensby11a} and the indirect dynamical estimate of
2.2$\pm$0.35 kpc of \citealt{Carollo10a}).

We have split the \aenhanced\ sample into more metal-poor and more
metal-rich sub-samples by cutting the sample at [Fe/H] = $-0.7$. This
is close to the median [Fe/H] of the \aenhanced\ sample. The
metal-poor sub-sample may be identified with the metal-weak thick disk
(MWTD) population discussed by \citet{Carollo10a}, which they argue
covers the metallicity range $-1.8 \leq \feh\ \leq -0.8$. The
resulting fits for the spatial structure of these sub-samples are given
in \tablename~\ref{table:poor_results}. The inferred scale lengths for
these sub-samples are equal to within the uncertainties. However, the
scale height of the more metal-poor sample is 856$\pm$20 pc while that
of the more metal-rich sample is 583$\pm$16 pc. The radial scale
length of the MWTD determined from the indirect dynamical analysis of
\citet{Carollo10a} is roughly 2 kpc, while the scale height is
1.36$\pm$0.13 kpc.

We have also split the \aenhanced\ sample into two bins in \afe\ by
splitting the sample at \afe\ = 0.35. The best-fit density profiles,
given at the bottom of \tablename~\ref{table:poor_results}, again have
similar scale lengths, around 2 kpc, and different scale heights. The
stars that are most enhanced in $\alpha$-elements have the largest
scale height (765$\pm$15 pc) and the shortest scale length
(1.89$\pm$0.04 kpc), while the less $\alpha$-enhanced stars have a
smaller scale height (627$\pm$18 pc) and longer scale length
(2.23$\pm$0.1 kpc). As the latter dominate the full \aenhanced\
sample, their scale height is very similar to that inferred for the
full sample. We explore the dependence of the disk parameters on \feh\
and \afe\ in more detail in \sectionname~\ref{sec:monoresults} below.

\subsection{The \apoor\ sample}\label{sec:apoorresults}

The results for single exponential disk fits and double exponential
disk fits for the \apoor\ sample are given in
\tablename~\ref{table:rich_results}. The double exponential
disk fit model is formally preferred, but the parameters of the dominant
double-exponential disk are again similar for both fits. We see that the
\apoor\ sample is dominated by a population of stars with a low scale
height of 256$\pm$4 pc and a long scale length of 3.6$\pm$0.2 kpc.

The second double-exponential disk in the best-fit model for the
\apoor\ sample has a scale height of 664$\pm$132 pc, which is
consistent with the scale-height measurement of the \aenhanced\ sample
above. However, the fraction of stars in this secondary component is
too small to constrain its scale length, and is conceivably simply a
result of `abundance contamination' of the sample.

Density fits for \apoor\ samples with the same \afe\ limits as the
nominal \apoor\ sample shown in the top panel, but that are more
metal-poor, are also given in \tablename~\ref{table:rich_results}. We
do not measure any radial density decline for these more metal-poor
\apoor\ samples, and short scale lengths for these samples are ruled
out by the data. We consider this further in
\sectionname~\ref{sec:monoresults} and in the discussion section
below. 

When we split the \apoor\ sample into two pieces, by cutting at \afe\
= 0.15, we find that the more $\alpha$-enhanced sample has the
shortest scale length (2.3$\pm$0.2 kpc) and the largest scale height
(348$\pm$13 pc). The sample with \afe\ closer to solar has a longer
scale length of 4.3$\pm$0.2 kpc and a smaller scale height of
239$\pm$4 pc.

\begin{deluxetable*}{cr@{}lr@{}lr@{}lr@{}lr@{}lr@{}l}
\tablecaption{}
\tablecolumns{11}
\tabletypesize{\footnotesize}
\tablewidth{0pt}
\tablecaption{Results for the \apoor\ G-dwarf sample ( $-0.3 <$ [Fe/H] $< 0.25$, $0.00 <$ \afe\ $< 0.25$)}
\tablehead{\colhead{} & \multicolumn{2}{c}{$h_z$} & \multicolumn{2}{c}{$h_R$} & \multicolumn{2}{c}{$h_{z,2}$} & \multicolumn{2}{c}{$h_{R,2}$}  &\multicolumn{2}{c}{$\beta_2$} & \multicolumn{2}{c}{$\beta_c$}\\
\colhead{} & \multicolumn{2}{c}{(pc)} & \multicolumn{2}{c}{(kpc)} & \multicolumn{2}{c}{(pc)} & \multicolumn{2}{c}{(kpc)} & \multicolumn{2}{c}{} & \multicolumn{2}{c}{}}
\startdata
all plates & 270&$\pm$3& 3.8&$\pm$0.2& \ldots & & \ldots & & \ldots & & 0.0005&$\pm$0.0010\\
bright plates  & 267&$\pm$3& 3.6&$\pm$0.2& \ldots & & \ldots & & \ldots & & 0.0009&$\pm$0.0003\\
faint plates  & 329&$\pm$14& $>$3.8 (5.1&$\pm$1.0)& \ldots & & \ldots & & \ldots & & 0.0010&$\pm$0.0003\\
$b < 0^\circ$  & 264&$\pm$4& 3.6&$\pm$0.2& \ldots & & \ldots & & \ldots & & 0.0008&$\pm$0.0009\\
$b > 0^\circ$  & 271&$\pm$4& 3.80&$\pm$0.10& \ldots & & \ldots & & \ldots & & 0.000&$\pm$0.001\\
$|b| > 45^\circ$ & 270&$\pm$5& 4.2&$\pm$0.8& \ldots & & \ldots & & \ldots & & 0.0004&$\pm$0.0008\\
$|b| < 45^\circ$  & 264&$\pm$3& 4.0&$\pm$0.2& \ldots & & \ldots & & \ldots & & 0.0006&$\pm$0.0007\\
\\
all plates & 256&$\pm$4& 3.6&$\pm$0.2& 664&$\pm$132& $>$5 & & 0.012&$\pm$0.004& \ldots & \\
bright plates  & 260&$\pm$5& 3.5&$\pm$0.3& 491&$\pm$83& $>$2 & & 0.02&$\pm$0.02& \ldots & \\
faint plates  & 268&$\pm$23& $>$3.8 (5.0&$\pm$0.8)& 910&$\pm$152& $>$2.9 (6&$\pm$2)& 0.014&$\pm$0.008& \ldots & \\
$b < 0^\circ$  & 242&$\pm$8& 3.2&$\pm$0.2& 639&$\pm$81& $>$5 & & 0.017&$\pm$0.010& \ldots & \\
$b > 0^\circ$  & 263&$\pm$6& 3.7&$\pm$0.2& 834&$\pm$70& $>$4 & & 0.004&$\pm$0.002& \ldots & \\
$|b| > 45^\circ$ & 249&$\pm$6& 3.8&$\pm$0.8& 631&$\pm$142& $>$6 & & 0.015&$\pm$0.005& \ldots & \\
$|b| < 45^\circ$  & 252&$\pm$5& 3.9&$\pm$0.3& 656&$\pm$65& $>$5 & & 0.012&$\pm$0.005& \ldots & \\
\\
-1.5 $<$ \protect{[}Fe/H] $<$ -0.6\tablenotemark{1}  & 689&$\pm$25& $>$37 & & 1431&$^{+704}_{-1916}$& 1.1&$^{+0.6}_{-1.0}$& 0.03&$\pm$0.07& \ldots & \\
-0.6 $<$ [Fe/H] $<$ -0.3\tablenotemark{1}  & 360&$\pm$9& $>$16 & & 946&$\pm$92& $>$14 & & 0.018&$\pm$0.009& \ldots & \\
\\
0.00 $<$ [$\alpha$/Fe] $<$ 0.15  & 239&$\pm$4& 4.3&$\pm$0.2& 647&$\pm$53& $>$7 & & 0.010&$\pm$0.003& \ldots & \\
0.15 $\leq$ [$\alpha$/Fe] $<$ 0.25  & 348&$\pm$13& 2.3&$\pm$0.2& 959&$\pm$335& $>$2.0 (5&$\pm$2)& 0.018&$\pm$0.009& \ldots & 
\enddata
\tablecomments{\protect{L}ower limits are at 99\,percent posterior confidence. Lower limits are given when the best-fit value is larger than 4.5 kpc. The best-fit value is not given if the best-fit value is larger than 6 kpc.\label{table:rich_results}}
\tablenotetext{1}{\label{fehrange} These samples have the same \afe\ range as the nominal \apoor\ sample.}
\end{deluxetable*}

\subsection{The spatial structure of mono-abundance
sub-populations}\label{sec:monoresults}

In the previous two sections we found that sub-samples of stars
defined by their element abundances appear to have a simple spatial
structure, approximated by a single exponential in the radial and
vertical direction. The scale lengths and heights of these sub-sets
seem to vary systematically with the abundances: the \aenhanced\
sample has a shorter scale length than the \apoor\ sample, and if
we split those two samples further in \afe, the part of the \apoor\
sample that has the closest to the solar \afe\ ratio has the longest scale
length and the smallest scale height. We also noticed that populations
with \afe\ $< 0.25$ have longer scale lengths and scale heights with
decreasing \feh. 

\begin{figure}[tp]
\includegraphics[width=0.5\textwidth,clip=]{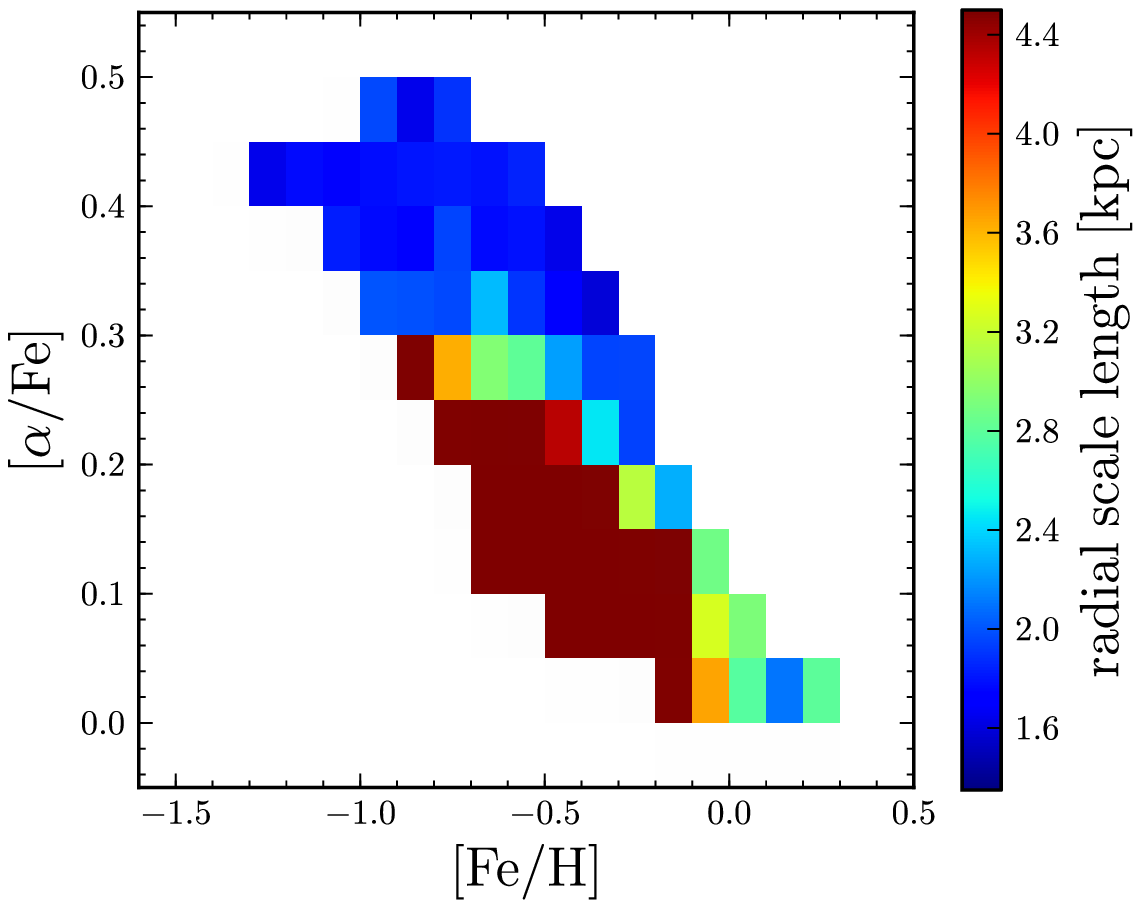}
\includegraphics[width=0.5\textwidth,clip=]{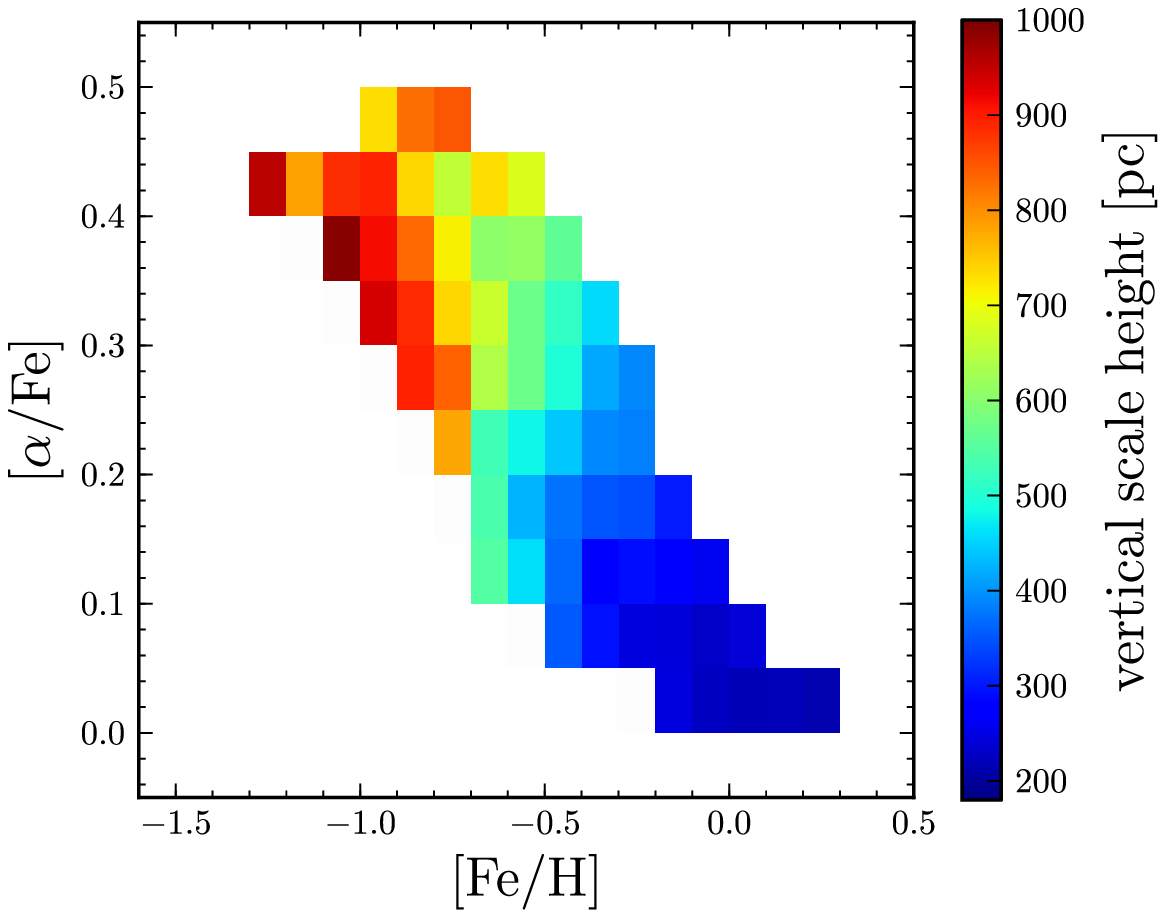}\\
\caption{Radial scale length (\emph{left panel}) and vertical
scale height (\emph{right panel}) of the best-fitting density models with a single exponential in $R$ and $Z$, shown for different mono-abundance sub-populations as a function of their position in
the \feh\ and \afe\ plane. Pixels span 0.1 dex in \feh\ and 0.05 dex in \afe. Only pixels with more than 100 stars are shown. The striking correlations between the spatial structure and the elemental abundances of the sub-populations is apparent.}\label{fig:pixelFit_g}
\end{figure}

To further investigate these trends, we have fit disk models with
single exponential profiles in $R$ and $Z$ to sub-populations of stars
with narrow bins in \feh\ and \afe. We divide stars into bins of width
0.1 dex in \feh\ and 0.05 dex in \afe, and only fit those bins with more than
100 stars. The results from these fits are shown in
\figurename~\ref{fig:pixelFit_g}. The populations in the lower left
part of the \afe--\feh\ diagram all have best-fit scale lengths in
excess of 4.5 kpc.

We also fit two-component, \ie, two exponential disks to each of the
bins, but found that these led to overfitting, and only marginal
improvements in the likelihood for the best fit. Thus, for narrow bins
in elemental-abundance space, the sub-populations are very-well
described by single exponential profiles in the $R$ and $Z$ directions.

\begin{figure}[tp]
\includegraphics[width=0.5\textwidth,clip=]{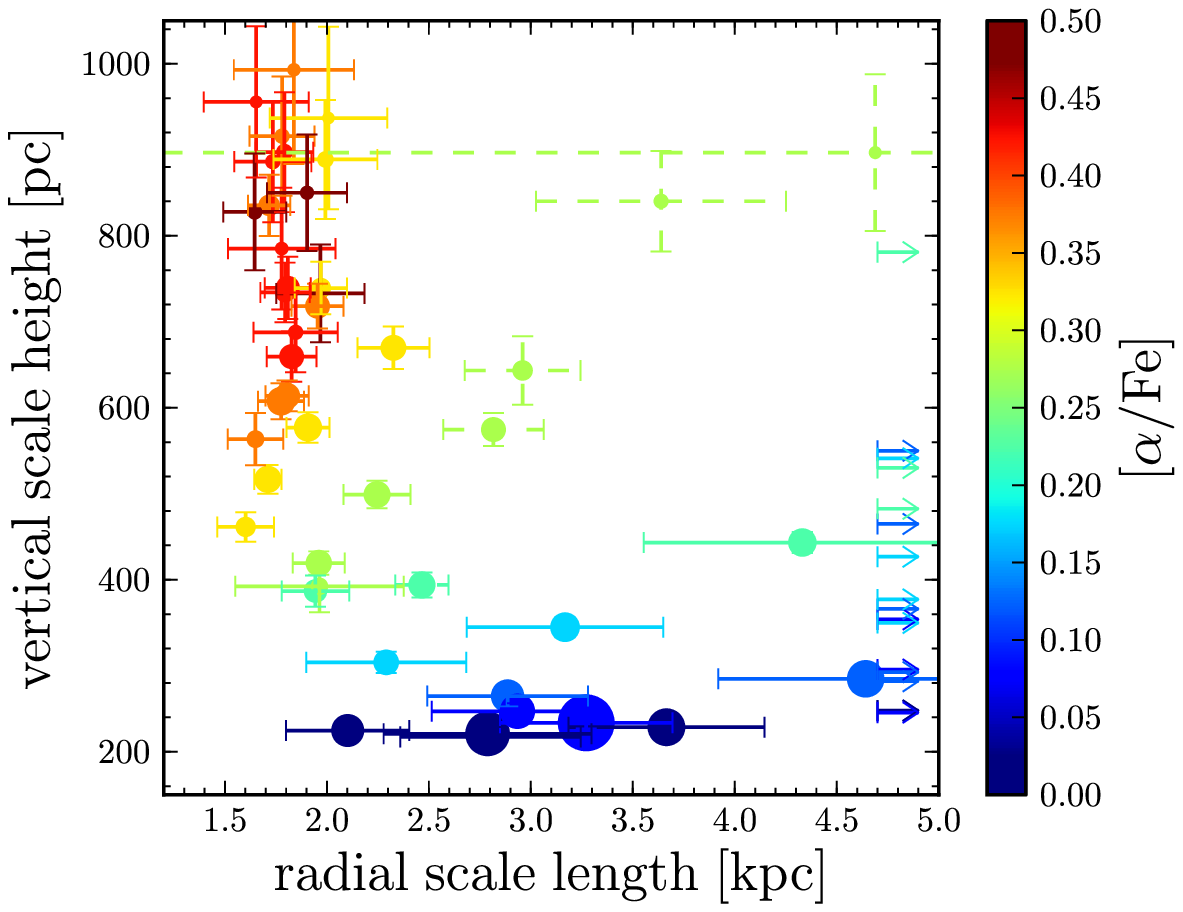}\\
\includegraphics[width=0.5\textwidth,clip=]{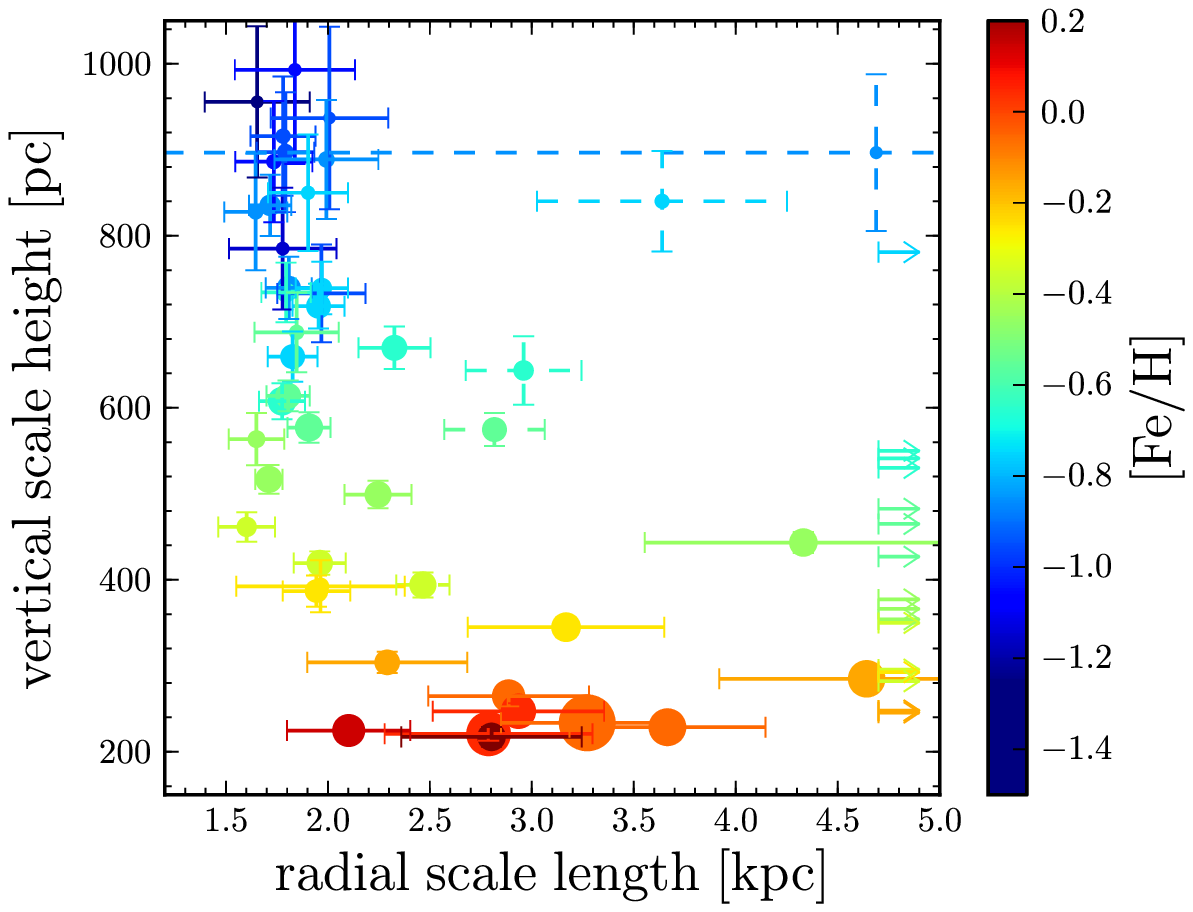}
\caption{Correlation between the radial scale length and vertical scale height of mono-abundance sub-populations, essentially a different projection of the results in
\figurename~\ref{fig:pixelFit_g}. The top panel shows the vertical scale height versus the radial scale length, color-coded by \afe. The bottom panel shows the same as the top panel,
but with \feh\ color-coding. Bins with best-fit scale lengths in
excess of 5 kpc are indicated with lower limits in the bottom row. The
area of the disks are proportional to the total surface-mass density contained in each point's \feh--\afe\  bin, corrected for mass and sample selection effects (see \citealt{Bovy12a}). Some of the uncertainties on the vertical scale heights are smaller than the points. The bins with dashed errorbars lie in a part of the abundance plane where abundance contamination is likely to be most severe, where the \afe-based age ranking is least reliable, and where the spatial properties change most rapidly. They contain $<$ 5\,percent of the disk surface mass.}\label{fig:pixelFit_g_alt}
\end{figure}

A different view of the results in \figurename~\ref{fig:pixelFit_g} is
given in \figurename~\ref{fig:pixelFit_g_alt}. The results in the
different \feh--\afe\ bins are shown as a function of scale length and
scale height; the points are color-coded by their \afe\ or \feh\
dependence, and the size of the points corresponds to the total
stellar surface-mass density---corrected for mass and sample selection
effects--in each population (calculated in
\citealt{Bovy12a}). \figurename~\ref{fig:pixelFit_g_alt} also shows
the uncertainties in the inferred parameters; the formal uncertainty
in the scale height for some points is so small that it cannot be
seen. The bins with dashed error bars lie in a part of the abundance
plane where abundance contamination is likely to be the most severe, where
the \afe-based age ranking is least reliable, and where the spatial
properties change most rapidly. They contain $<$ 5\,percent of the
disk surface mass.

We see that these fits for mono-abundance sub-components flesh out the
main trends we noted in the broader \feh\ and \afe\ ranges above. At
any given metallicity \feh, the scale length increases and the scale
height decreases when moving from \aenhanced\ to \apoor\
populations. At any given $\alpha$-age, the scale length and the scale
height increase for the more metal-poor components, implying an
outward metallicity gradient. And, as \figurename~\ref{fig:pixelFit_g_alt}
shows most clearly, increasing scale lengths are correlated with
decreasing scale heights (except for a few bins on the boundary
between the very long scale lengths at low metallicity and
solar $\alpha$-enhancement and the shorter scale lengths of the
\aenhanced\ populations; see further discussion in
\sectionname~\ref{sec:discuss_evolve}). From
\figurename~\ref{fig:pixelFit_g} it is clear that neither \afe\ or
\feh, on its own, accounts for the trends in scale height and scale
length. We discuss what this implies for disk formation and evolution
in \sectionname\sectionname~\ref{sec:discuss_form} and
\ref{sec:discuss_evolve}, respectively.

\begin{figure}[tp]
\includegraphics[width=0.5\textwidth,clip=]{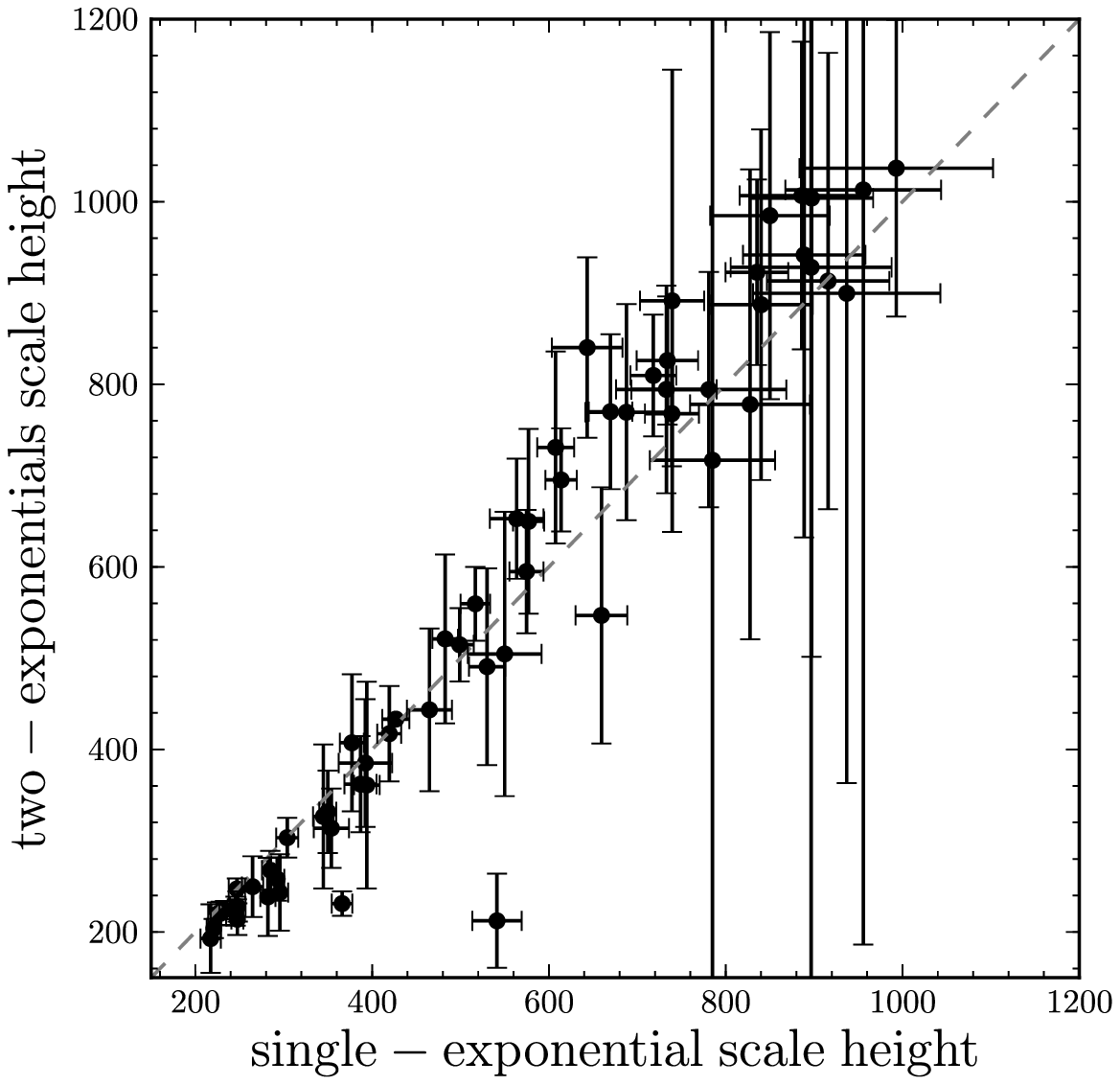}\\
\caption{Scale height of the dominant component, when fitting two exponential components in $R$ and $Z$, versus the scale height for a single exponential component in $R$ and $Z$. The dashed line gives the one--to--one correspondence. For most mono-abundance populations a single exponential description in $Z$ is a good model.}\label{fig:one_vs_two}
\end{figure}

\figurename~\ref{fig:one_vs_two} shows the results of fitting two
components with exponential profiles in both $R$ and $Z$ to each
abundance bin. The scale height of the dominant component is shown
against the best-fit scale height, when fitting a single exponential
profile in $R$ and $Z$. We see that these scale heights are strongly
clustered around the one--to--one correspondence line. Thus, for each
bin, a single vertical exponential suffices to explain the observed
number counts. The fact that the two measurements agree better than
would be expected, given the uncertainties shown, is due to the fact
that the scale heights for each bin are strongly correlated when
fitting a single or a double exponential profile in $R$ and
$Z$. Overall, \figurename~\ref{fig:one_vs_two} confirms that a single
exponential model in $Z$ and $R$ is a good model for the spatial
structure of mono-abundance sub-populations. 

In \appendixname~\ref{sec:fakedata}, we perform a test to determine
whether abundance uncertainties can plausibly lead us to find spurious
disk components between a ``thin'' and a ``thick'' component. That is,
we ask whether it is plausible that an underlying density dominated by
distinct thin- and  thick-disk components can be smoothed by abundance
errors into the density structure we inferred in \figurename
s~\ref{fig:pixelFit_g} to \ref{fig:one_vs_two}. This test shows that
if this were the case, every bin is preferentially fit with two
components, corresponding to the input thin and thick components. The
equivalent of \figurename~\ref{fig:one_vs_two}, shown in the bottom
right panel of \figurename~\ref{fig:fakeBimodalResults}, is
qualitatively different, with a distinct difference between the
single-component scale height and the scale height of the dominant
component in the two-component fit.

To test whether the analysis in this section is influenced by our
signal-to-noise ratio cut of \sn\ $> 15$, we have repeated the
analysis with a cut of \sn\ $> 30$, as also used by
\citet{Lee11b}. The equivalents of \figurename s~\ref{fig:pixelFit_g}
to \ref{fig:one_vs_two} look qualitatively the same, albeit with
larger uncertainties for each bin, and the dependence of \hz\ and \hR\
on elemental abundance is the same as that inferred from the sample
with the \sn\ $> 15$ cut. The number of (\feh,\afe) bins with more
than 100 stars is smaller, but the inferred (\hz,\hR) for those bins
with more than 100 stars when using \sn\ $> 30$ cut are consistent
within the uncertainties with those found with the less restrictive
signal-to-noise ratio cut. We stress that even when selecting stars
with \sn\ $>30$, the equivalent of \figurename~\ref{fig:one_vs_two}
does not show any sign of a second component in the mono-abundance
bins.

To perform the binning in this section, we used narrow bins of 0.1 dex
in \feh\ and 0.05 dex in \afe. These bins are somewhat narrower than
the total typical uncertainty ($\approx 0.15$ dex in \feh, $\approx
0.07$ dex in \afe; \citealt{Bovy12b}), but we prefer to oversample,
rather than undersample, to avoid smoothing out underlying
structure. The analysis in each bin holds irrespective of the bin
size. What matters for the analysis is that the data in each bin are
disjoint, such that the bins are statistically independent.

\section{Discussion}\label{sec:discussion}

Our basic result is that various stellar disk sub-components, when
defined purely through stellar abundances, are \emph{simple}, \ie, can
be described by a single exponential in $R$ and $Z$, and exhibit
distinctive trends of the scale height and scale length with chemical
abundance. This suggests that dissecting the Milky Way's disk on the
basis of chemical abundances alone is a useful approach. In this
section we go through a number of practical issues pertaining to
these estimates, before discussing possible implications for galactic
disk formation and evolution.

\subsection{Distance systematics}

The absolute values of the distance scales measured in this paper are
subject to distance systematics, which we discuss in this
subsection. We have used the data-driven photometric-distance relation
from \citet{Ivezic08a} to infer the spatial structure of the various
samples of stars, but an alternative photometric-distance relation can
be obtained by using the \citet{An09a} stellar isochrones in the
\sdss\ passbands. These isochrones depend on \feh\ as well as on \afe,
although in practice a linear relation between \afe\ and \feh\ is
assumed, and the spectroscopically measured \afe\ is not used directly
to estimate the photometric distance. In the top panel of
\figurename~\ref{fig:dm_anjuric_ivezic} we compared the distance
moduli derived using the \citet{An09a} stellar isochrones with those
obtained using the \citet{Ivezic08a} relation for a few values of
\feh. We see that, for the values of \feh\ that span most of our sample,
the distance modulus difference is -0.12 mag, corresponding to a
systematic difference in the inferred distances of about 6\,percent,
nearly independent of color. Thus, if we had used the \citet{An09a}
photometric distances we would have obtained scale lengths and scale
heights that were 6\,percent shorter.

A second distance systematic that could influence our results is the
Malmquist bias \citep{Malmquist20a,Malmquist22a}---the fact that
brighter stars are over-represented in a magnitude-limited survey. For
our relatively bright sample, this is dominated by the finite width of
the photometric distance relation. The Malmquist bias in absolute
magnitude is apparent-magnitude dependent and approximately equal to
$-\sigma^2 \dd\ln A(r) / \dd r$, where $A(r)$ is the differential
number count as a function of apparent magnitude and $\sigma$ is the
dispersion in the absolute magnitudes (either due to photometric
uncertainties or due to intrinsic scatter in the photometric distance
relation). Conservatively assuming that the combination of the finite
width of the photometric distance relation and the photometric
uncertainties is 0.2 mag, and that the underlying density is constant,
the Malmquist bias would be of order 2.5\,percent. However, due to the
exponential fall-off of the density in both the $R$ and $Z$
directions, the differential number counts are (a) flat near the peak
induced by the vertical exponential, and (b) for most apparent
magnitudes $|\dd \ln A(r) / \dd r|$ is less than 1. Therefore, the
Malmquist bias is at most about 2\,percent, and will not strongly
affect the measurement of the vertical scale height in particular.

We have assumed throughout our analysis that all of the stars in our
sample are single. The presence of unresolved binaries will lead us to
underestimate scales, as these binaries will appear to us as brighter,
and thus closer, single stars. The binary fraction and companion-mass
distribution for G-type dwarfs remains controversial, but it appears that the
overall binary fraction for G dwarfs is approximately 40\,percent
\citep{Abt76a,Duquennoy91a,Raghavan10a}, similar to but slightly
larger than that of M dwarfs \citep{Fischer92a,Raghavan10a}. The
distribution of companion masses is poorly known, and could range from
being peaked around 20\,percent of the primary's mass
\citep{Duquennoy91a}, to being relatively flat between 20 and
100\,percent of the primary's mass \citep{Raghavan10a}, with numerical
simulations indicating that multiple-star systems form preferentially
with approximately equal-mass members \citep{Bate05a}, and an overall
multiplicity fraction of around 40\,percent
\citep{Bate03a}. Lower-metallicity stars most likely have a higher
binary fraction \citep{Machida09a}, and could reach 100\,percent for
\feh\ $< -0.8$ \citep{Raghavan10a}.

For a likely scenario where 40\,percent of our \apoor\ sample is made
up of binary stars (ignoring higher-order multiplicities) with a flat
distribution of companion masses between 20\, and 100\,percent of the
primary's mass, the magnitude would be overestimated on average by
0.12 mag, such that the scale height and scale length would be
underestimated by about 6\,percent.  If 70\,percent of the \aenhanced\
sample would consist of binary systems (taking into account the rising
binary fraction with decreasing metallicity), the magnitudes would be
overestimated by approximately 0.21 mag, and the \aenhanced\ scale
heights and scale lengths would be underestimated by
10\,percent. These biases are somewhat larger than the statistical
uncertainties on our results, but they are similar to the overall
distance-scale uncertainty (see above), and they do not change the
conclusion that the \aenhanced\ scale length is much shorter than that
of the \apoor\ sample. Even in a worst-case scenario, where all binary
systems have equal-mass companions and where 100\,percent of the
\aenhanced\ stars are in binaries, the \aenhanced\ scale-length would
still be $\lesssim 2.8$ kpc (40\,percent up from 2 kpc), which is
shorter than the scale length measured for the \apoor\ sample in
\tablename~\ref{table:rich_results} and
\figurename~\ref{fig:pixelFit_g_alt}, and the \apoor\ scale lengths
themselves would also increase by about 15\,percent in this
scenario. In principle, a careful spectral analysis of the \segue\
spectra itself could provide direct constraints on the (unresolved)
binary contamination in this sample.

\subsection{Halo contamination}

In our density fits we have mostly fit disk components to the data,
except for the single exponential disk model where we added a uniform
density (\eqnname~(\ref{eq:densmodel})). We thus assumed that the
stellar halo does not influence our disk fits, beyond what can be
described by a uniform density across our survey volume. We can
estimate the expected number of halo stars in our sample using the
\citet{Bell08a} density fits to the smooth stellar halo.  We run the
\citet{Bell08a} stellar-halo density through the G-star \segue\
selection function, and marginalize over $g-r$ color using a flat
distribution over $0.48 \leq g-r \leq 0.55$, and over \feh\ using the
\citet{Ivezic08a} halo metallicity distribution (mean \feh\ = -1.52,
width = 0.32). We then find that for $\approx 10^8$ G-type stars
between 1 and 40 Galactocentric kpc in the stellar halo, there should
be about 100 halo stars in our sample, compared to the total sample
size of 30,353 G-type dwarfs. Hence, the halo contamination is very
small and does not influence the fits. Additionally, halo
contamination will be most severe for the \aenhanced\ sub-populations,
and this contamination should work to \emph{increase} the inferred
scales (length and height). Therefore, the result that the radial
scale length of \aenhanced\ sub-populations is shorter than that of
\apoor\ sub-populations is robust against any halo contamination.

\subsection{Comparison to traditional geometric disk decompositions}

The density fits in this paper are the first to constrain the vertical
scale height and radial scale length of numerous disk sub-components,
defined using elemental abundances alone, from a large sample of
stars. Our results show that the vertically thicker disk
sub-components---when chemically defined---have a much shorter scale
length than the thinner-disk sub-components, which is opposite to
traditional purely geometric disk decompositions
\citep[\eg,][]{Robin96a,Ojha01a,Larsen03a}, which typically find that
the thick-disk component has a \emph{longer} scale length than the
thin disk, and that this scale length is $\gtrsim 3.5$ kpc
\citep[\eg,][]{Juric08a}.

When we fit the spatial structure in our approach, taking stars of all
metallicities (specifically, the combination of our \aenhanced\
(``thick'') and \apoor\ (``thin'') samples), we can recover the result
of purely geometric decompositions: the thin-disk component---\ie, the
component with the lowest scale height, $\approx$ 300 pc---gets paired
with the shortest scale length ($\approx$ 2 kpc), while the
thicker-disk component gets assigned both the largest scale height and
scale length (for our particular sample fit with a combination of
three double-exponential disks these are $\approx$ 600 pc and
$\approx$ 2.4 kpc, with a small component with an even larger scale
height and scale length). Thus, it seems that purely geometric
decompositions naturally associate the longest scale length with the
largest scale height. That \emph{both} geometrically determined scale
lengths are shorter than the scale length of the \apoor\ sample is due
to the fact that the metallicity distribution for the entire sample
extends down to \feh\ = -1.5, such that the model `expects' many
low-metallicity stars in the ``thin'' component at large distances (as
the model does not contain the information that the thin component has
higher metallicities), which are not observed. Therefore, metallicity
and $\alpha$-element abundances, which are manifestly quantities that
can identify sub-samples of stars independent of their spatial
structure and kinematics, lead to a qualitatively different
decomposition into two (or more) sub-components than the purely
geometrical approach, with its inherent risk of circular reasoning.

\subsection{Implications for disk formation}\label{sec:discuss_form}

The distinctive changes of the global disk structure with abundance,
especially with the age proxy \afe, should provide valuable clues to
the formation of the Milky Way's disk. While a concrete and
quantitative model comparison is beyond the scope of this paper, we
discuss some of the qualitative implications here. As mentioned in
Section 1, the overall radial-density profile of the stellar disk is
expected to be conserved even in the face of large-scale radial
migration, but the radial profile of sub-components will tend to relax
to the mass-weighted mean radial profile. Thus, a difference in the
radial distribution of various populations of stars today is a
less-pronounced version of more different initial radial distributions
(at formation). Assuming that the \afe\ ratio is an adequate proxy for
age \citep[\eg,][]{Schoenrich09a}, our results then imply that the
$\alpha$-enhanced, hence oldest, populations are more centrally
concentrated---have a shorter scale length---than populations with
$\alpha$-abundances that are closer to Solar, and therefore
younger. This is direct observational evidence for inside-out
formation of galactic disks across the presumed age range of our
sample, 1 -- 10 Gyr, where the inner parts of the disk form before the
outer part of the disk. A similar age-dependence of the exponential
scale length has been found in several external galaxies
\citep{deJong07a,Radburn12a}.

\begin{figure}[tp]
\includegraphics[width=0.5\textwidth,clip=]{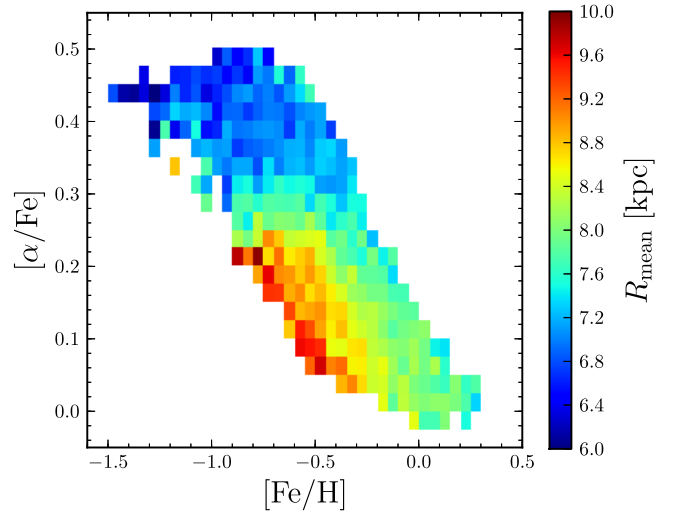}
\caption{Mean orbital radii of the G-dwarf sample. Median
values of the mean orbital radii are shown, in bins of width 0.05 dex in \feh\
and 0.025 dex in \afe. Only bins with at least 20 stars are
shown.}\label{fig:gOrbits}
\end{figure}

Second, our analysis shows that our Milky Way has not only a
metallicity gradient among its youngest stars, but that it has always
had one \citep{Cheng11a}: at a given \afe, standing in for age,
sub-populations with lower \feh\ have a longer scale length than more
metal-rich stars. This picture is confirmed by looking at the orbital
properties of the stars when integrating our sample of G-type dwarfs
in a simple model for the Milky Way's potential, made up of a
Miyamoto-Nagai disk with a radial scale of 4 kpc and vertical scale of
300 pc contributing 60\,percent of the radial force at the Solar
radius, a Hernquist bulge with a scale radius of 600 pc contributing
5\,percent of the radial force, and a Navarro-Frenk-White halo with a
scale radius of 36 kpc that contributes 35\,percent of the rotational
support at the Sun's position. The median of the mean orbital radii as
a function of elemental abundance is shown in
\figurename~\ref{fig:gOrbits} (see also \citealt{Lee11b} and Liu \&
van de Ven, 2011, in prep. for similar figures of the eccentricity and
rotational velocity). We see that stars with \afe\ $< 0.25$ and lower
\feh\ are thin-disk stars that live, on average, farther out than more
metal-rich stars. Thus, the longer scale length for outer-disk stars,
combined with the fact that, for solar \afe\, decreasing \feh\ is
correlated with decreasing age \citep[\eg,][]{Schoenrich09a}, implies
that the outer part of the disk formed later than the inner part.

We have assumed that \afe\ is an adequate proxy for age, such that the
mono-abundance populations that are more \afe-enhanced are older than
the populations with solar \afe. This is typically the case in
standard scenarios for the star formation history of the Milky Way
disk, in which \afe\ steeply drops around 2 to 3 Gyr due to the onset
of type Ia supernovae \citep{Dahlen08a,Maoz11a}, and then stays
roughly constant, although the value of \feh\ at which the \afe\
downturn happens depends on the star formation history
\citep{Matteucci01a}. Only if the local star formation was
characterized by bursts of star formation can younger populations of
stars have similar levels of \afe\ as older stars
\citep{Gilmore91a}. Most current fits of the local star formation
history prefer a smooth history \citep[\eg,][]{Aumer09a}, although it
is difficult to rule out epochs of enhanced star formation
\citep[\eg,][]{RochaPinto00a}.

\subsection{Implications for disk evolution}\label{sec:discuss_evolve}

The spatial structure inferred for mono-abundance sub-populations
(\figurename s~\ref{fig:pixelFit_g} and \ref{fig:pixelFit_g_alt}) show
two important results: first, there is a tight anti-correlation
between the scale heights and scale lengths of the
sub-components. Secondly, there is a continuous distribution in scale
height when moving from $\alpha$-enhanced, metal-poor populations to
stars with solar \afe\ and \feh. This suggests that the
\aenhanced---``thick''---and the $\alpha$-younger, ``thin'', regime of
the stellar disk are not two separate entities, but merely opposite
ends of the disk evolution spectrum \citep[suggested before in][but
never directly measured as we do here]{Norris87a}. This issue, which
requires proper stellar-mass weighting of the sub-components, is
worked out in a separate paper \citep{Bovy12a}. Taken together, these
findings suggest a continuous evolutionary mechanism created the
observed scale-height distribution, rather than a discrete external
heating or accretion event.  Radial migration is an obvious candidate
for this internal evolution mechanism. That the most centrally
concentrated component of the disk is not only the ($\alpha$-)oldest
part, but also has the largest scale height, is a nearly inevitable
condition, and hence a natural prediction, of any scenario where much
of the disk scale-height distribution is created through radial
migration. The $\alpha$-old sub-population not only had the most time
to evolve, but its centrally concentrated parent population implies
that stars at $6 < R < 12$ kpc have migrated out by the largest
factor.

A different internal explanation for the thicker disk components in
the Milky Way is that, rather than being thickened over the history of
the Galactic disk, thick components were created thick during an
early, turbulent phase in the formation of the disk
\citep[\eg,][]{Bournaud09a,ForsterSchreiber09a}. If such a scenario is
combined with a inside-out growth of the disk, and the disk remains
turbulent over a significant fraction of its history, this formation
scenario could plausibly explain the continuous dependence of disk
structure on elemental abundance found in this paper.

Our result that the transition between the \apoor, ``thin'',
components and the \aenhanced, ``thick'', components is smooth, rather
than showing a clear separation between thin and thick components, may
appear to be in conflict with local, high-resolution spectroscopic
samples of stars \citep[\eg,][]{Reddy06a,Fuhrmann11a,Navarro11a} or
other analyses of the \segue\ data \citep[\eg,][]{Lee11b}. A detailed
comparison between these and our results requires careful accounting
for the spectroscopic volume sampling, which has not been done in the
\citet{Lee11b} analysis or for the high-resolution samples, except for
the sample of \citet{Fuhrmann11a}, which is volume complete out to 25
pc. Without taking the volume selection into account, the sample used
here also displays a bi-modality in the \feh--\afe\ plane (see
\figurename~\ref{fig:afeh_g}). We discuss this issue in more detail in
\citet{Bovy12a}, but we note here that the apparent bi-modality in the
observed number density of stars disappears when properly correcting
for the spectroscopic sampling. Furthermore, the local,
high-resolution analyses cannot directly measure the spatial
distribution of stars of different elemental abundances (\eg,
\citealt{Fuhrmann11a}, which only has 15 high-\afe\ stars out to 25
pc; \citealt{Reddy06a}) and therefore rely on kinematics to argue that
the vertical distribution of stars in the Solar neighborhood is
characterized by a bi-modal ``thin''--``thick''-disk dichotomy. This
interpretation is driven by the selection of stars that are disjoint
in \afe, which leads to disjoint kinematics because the kinematics is
a strong---and smooth---function of abundance as well
\citep{Bovy12b}. While the stellar content of different survey volumes
can (and should) be connected by dynamics, we note that the effective
volumes sampled by, \eg, the \citet{Fuhrmann11a} survey and by our
analysis differ by a factor of about $10^5$; hence the extrapolation
from one to the other is enormous. The analysis of the vertical
kinematics of stars in our sample confirms the existence of the
intermediate populations with scale heights between 400 and 600 pc and
vertical-velocity dispersions of 30 to 35 km s$^{-1}$ \citep{Bovy12b}.

Our finding that the scale length does not behave as smoothly as the
scale height, as a function of \afe, is presumably a consequence of
the disk's formation history: here the increasing metallicity as a
function of time (\ie, youth) and the radial metallicity gradient
compete. As the mapping between \afe\ and age is not linear, but
rather, \afe\ steeply drops around 2 to 3 Gyr due to the onset of type
Ia supernovae \citep{Matteucci01a,Dahlen08a,Maoz11a}, and then stays
roughly constant, the scale length should change similarly rapidly
with \afe. The scale height, however, is determined by subsequent
evolution, where radial migration transports stars to larger
Galactocentric radii, where the lower disk density allows them to
travel farther from the plane. Since this evolution is continuous,
rather than sudden, and includes additional contributions from
heating, trends in scale height versus elemental abundance should be
expected to be smoother, even if radial migration is not the disk's
dominant evolutionary mode. Our results are therefore consistent with
a scenario where the thick-disk component is the inner part of the
disk that formed at the earliest time, and either by having formed
thick or through the effect of radial migration, has a large scale
height at the present time.

A gas-rich merger, followed by intense star formation at an early
time, could have affected the formation of the early disk
\citep{Brook04a}, as seems consistent with the observed distribution
of eccentricities of the thick-disk component
\citep{Sales09a,Dierickx10a,Wilson11a}.  However, it would lead to a
scale length for the thicker component that is larger than that of the
thinner component \citep{Qu11a}. It is clear that internal mechanisms
must have played an important role during the evolution of the
disk. However, we caution that the radial and vertical consequences of
neither radial migration, nor turbulent disk evolution, nor of
satellite thickening, have been worked out in quantitative detail,
and, in particular, resonant coupling between satellites and the disk
might induce some similar observational signatures to radial
migration.

The rapid change in the mean stellar population in an \afe -- \feh\
abundance bin at the onset of type Ia supernovae is also likely the
explanation for the presence of the few points of intermediate \afe\
and \feh\ in \figurename~\ref{fig:pixelFit_g_alt} that do not follow
the anti-correlation between scale height and scale length; these
bins, which do not contribute significantly to the total stellar mass
(indicated by the size of the symbols in
\figurename~\ref{fig:pixelFit_g_alt}), are also the bins that fall
short of the one-to-one correlation between single- and two-disk fits
in \figurename~\ref{fig:one_vs_two}. This provides further evidence of
the fact that at the rapid \afe\ (age) transition our bins do not
adequately resolve single components.

\section{Conclusions}\label{sec:conclusion}

The main conclusions of this paper are as follows

$\bullet$ An assessment of the global ($R,Z$) structure of the Milky
Way's stellar disk for sub-components selected solely by their
elemental abundances is now feasible, \eg, with spectroscopic surveys
such as \segue, but requires a thorough accounting for the effective
selection function of the spectroscopic sample.

$\bullet$ A decomposition of the Galactic disk, based on \sdss/\segue\
data for G-type dwarfs, into mono-abundance sub-populations in the
\feh--\afe\ plane, reveals that each such component has a simple
spatial structure that can be described by \emph{single exponential profiles}
in \emph{both} the vertical and the radial direction.

$\bullet$ Adopting increasing levels of \afe\ enhancement as a proxy
for the increasing age of the stellar population, the disk dissection
into narrow mono-abundance populations in the space of \feh\ and \afe\
exhibits a continuous trend of increasing scale height and decreasing
scale length, when moving from younger to older populations of
stars. 

$\bullet$ We find that the oldest---most $\alpha$-enhanced---part of
the disk is both the thickest and the most centrally concentrated. If
we split the sample in only two broad abundance regimes we can make a
precise determination of the \aenhanced\ scale length, 2.01$\pm$0.05
kpc, and scale height, 686$\pm$11 pc. The scale length of the
$\alpha$-younger disk is around 3.5 kpc (3.6$\pm$0.2 kpc for our
nominal \apoor\ sample) and is far thinner, with a vertical scale
height of 256$\pm$4 pc.

$\bullet$ These observations show quite directly that the bulk of the
Galactic disk has formed from the inside out. 

$\bullet$ The tight (anti-) correlations between population age,
vertical scale height, and radial scale length strongly suggest that
the disk's subsequent evolution must have been heavily influenced by
internal mechanisms, such as radial migration or turbulent,
gravitationally-unstable disk evolution, as this naturally explains
the continuous increase of scale height with decreasing scale
length. At first sight, external mechanisms to form the Milky Way's
thick disk component through external heating or accretion appear to
be inconsistent with our results, but a thorough model comparison is
warranted.

While, at face value, our results emphasize the importance of
evolutionary processes that could be purely internal to the Milky Way
(radial migration, turbulent disk formation), the overall $\Lambda$CDM
cosmogony makes it likely that external processes must also have
played some role. In the end, it is likely that the Milky Way disk's
formation history may be more complex than inferred here, especially
once not only the spatial distribution but also the orbital
distribution of the mono-abundance sub-populations is fully analyzed.

\acknowledgements It is a pleasure to thank the anonymous referee,
James Binney, Doug Finkbeiner, Dan Foreman-Mackey, Patrick Hall, Juna
Kollmeier, George Lake, Rok Ro{\v s}kar, Scott Tremaine, Glenn van de
Ven, and Lan Zhang for helpful comments and assistance. We thank the
\segue\ team for their efforts in producing the \segue\ data set, and
Connie Rockosi and Katie Schlesinger in particular for help with the
\segue\ selection function. Support for Program number
HST-HF-51285.01-A was provided by NASA through a Hubble Fellowship
grant from the Space Telescope Science Institute, which is operated by
the Association of Universities for Research in Astronomy,
Incorporated, under NASA contract NAS5-26555.  J.B. and D.W.H. were
partially supported by NASA (grant NNX08AJ48G) and the NSF (grant
AST-0908357). D.W.H. is a research fellow of the Alexander von
Humboldt Foundation of Germany. J.B. and H.W.R acknowledge partial
support from SFB 881 funded by the German Research Foundation
DFG. Y.S.L. and T.C.B. acknowledge partial funding of this work from
grants PHY 02-16783 and PHY 08-22648: Physics Frontier Center / Joint
Institute for Nuclear Physics (JINA), awarded by the National Science
Foundation.

Funding for the SDSS and SDSS-II has been provided
by the Alfred P. Sloan Foundation, the Participating Institutions, the
National Science Foundation, the U.S. Department of Energy, the
National Aeronautics and Space Administration, the Japanese
Monbukagakusho, the Max Planck Society, and the Higher Education
Funding Council for England. The SDSS Web Site is
http://www.sdss.org/.

\appendix

\section{The \segue\ G-star selection function}\label{sec:selection}

To determine the spatial distribution of the G dwarfs, we require a good
understanding of the \segue\ G-star selection function, \ie, the
fraction of stars that has been targeted by \segue\ and produced good
enough spectra to derive the parameters needed in the present (or any
other) analysis (\eg, \sn\ $> 15$), and we need this selection
fraction as a function of position, color, and apparent magnitude. The
observed density of G-type stars is simply the product of the underlying
density with the sampling selection function, suggesting that one
constrains this underlying density by forward modeling of the
observations.

The spectroscopic G-star target type was selected uniformly from the
set of objects in the G-star color--magnitude box in the area and
apparent magnitude range of the spectroscopic plug-plates (simply
``plates'' hereafter), thus the selection function can be
reconstructed. The \segue\ survey implementation distinguishes between
``bright'' and ``faint'' plates, with bright plates containing stars
with $r \leq 17.8$ mag and faint plates containing stars with $r >
17.8$ mag. For the purposes of the selection function, we assume that
this separation at 17.8 mag is a hard cut, even though in reality some
stars were observed on both bright and faint plates for calibration
purposes, and some ``bright'' stars are part of faint plates, and vice
versa, because of changes between the photometry used for target
selection and that released as part of the \sdss\ DR7, which we employ
here. Duplicates are resolved in favor of the higher signal-to-noise
ratio observation (typically on the faint plate as this has a longer
integration time). We retain stars with $r \geq 17.8$ mag when they
were observed on a bright plate, and we keep objects with $r < 17.8$
mag when they were observed on a faint plate, even though this should
not happen in our model for the \segue\ selection function below. A
total of 586 stars in the \aenhanced\ sample and 47 stars in the
\apoor\ sample fall into this category; they do not influence any of
the fits or conclusions in this paper.

We select the superset of targets by querying the \sdss\ DR7 imaging
CAS\footnote{http://cas.sdss.org/dr7/en/~.} for all potential targets
in the color--magnitude box of the G-star target type in the area of a
\segue\ plate \citep{Yanny09a}. These objects are
\flag{primary}\footnote{See
\url{http://sdss3.org/dr8/algorithms/bitmask\_flags1.php} and\\
\url{http://sdss3.org/dr8/algorithms/bitmask\_flags2.php} for a
description of these flags.} detections (removing duplicates and
objects from overlapping imaging scans) with stellar PSFs (\flag{type}
equal to 6).  Objects must not be \flag{saturated}, nor be close to
the \flag{edge}, nor have an interpolated PSF (\flag{interp\_psf}),
and must not have an inconsistent flux count (\flag{badcounts}).
Furthermore, if the center is interpolated (\flag{interp\_center}),
there should not be a cosmic ray indicated (\flag{CR}). See
\citet{Stoughton:2002ae} for a description of the \sdss\ photometric
\flag{flags}. Using the superset of targets we determine for each
plate the fraction of stars that were observed spectroscopically of
all available targets.

\begin{figure}[tp]
\includegraphics[width=0.5\textwidth]{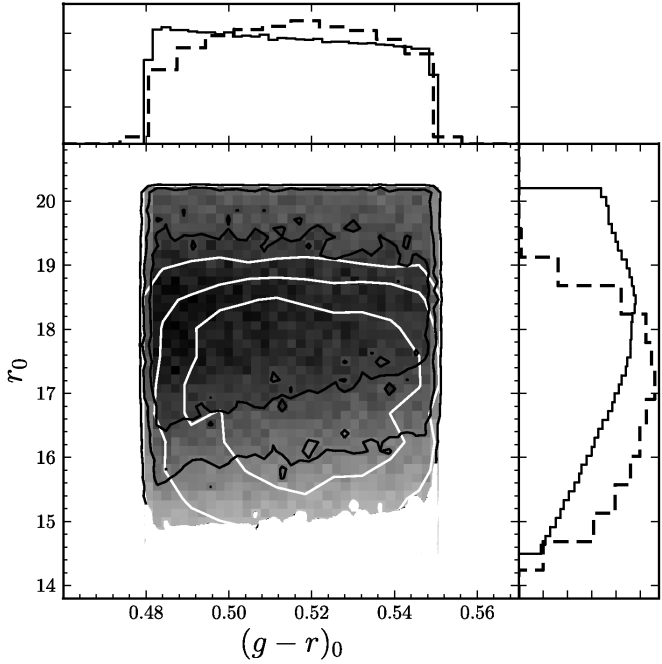}
\caption{Distribution of the photometric sample of G-type stars (linear
  density grayscale; black curves) and the spectroscopic sample (white
  contours, dashed histograms) after the signal-to-noise ratio cut of \sn\ $>
  15$. The contours contain 68, 95, and 99\,percent of the
  distribution.}\label{fig:colormag_g}
\end{figure}

\begin{figure}[tp]
\includegraphics[width=0.5\textwidth,clip=]{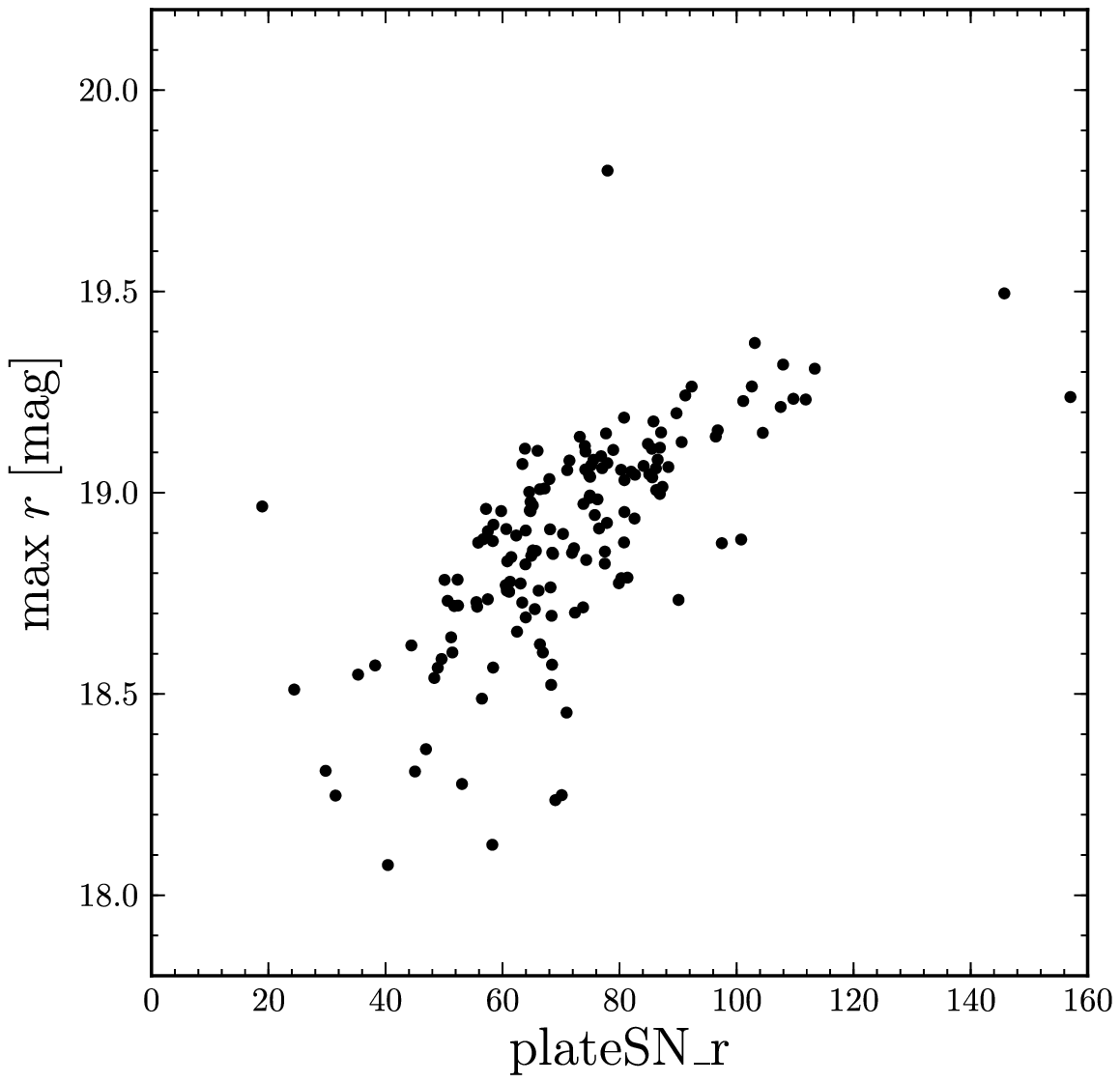}
\includegraphics[width=0.5\textwidth,clip=]{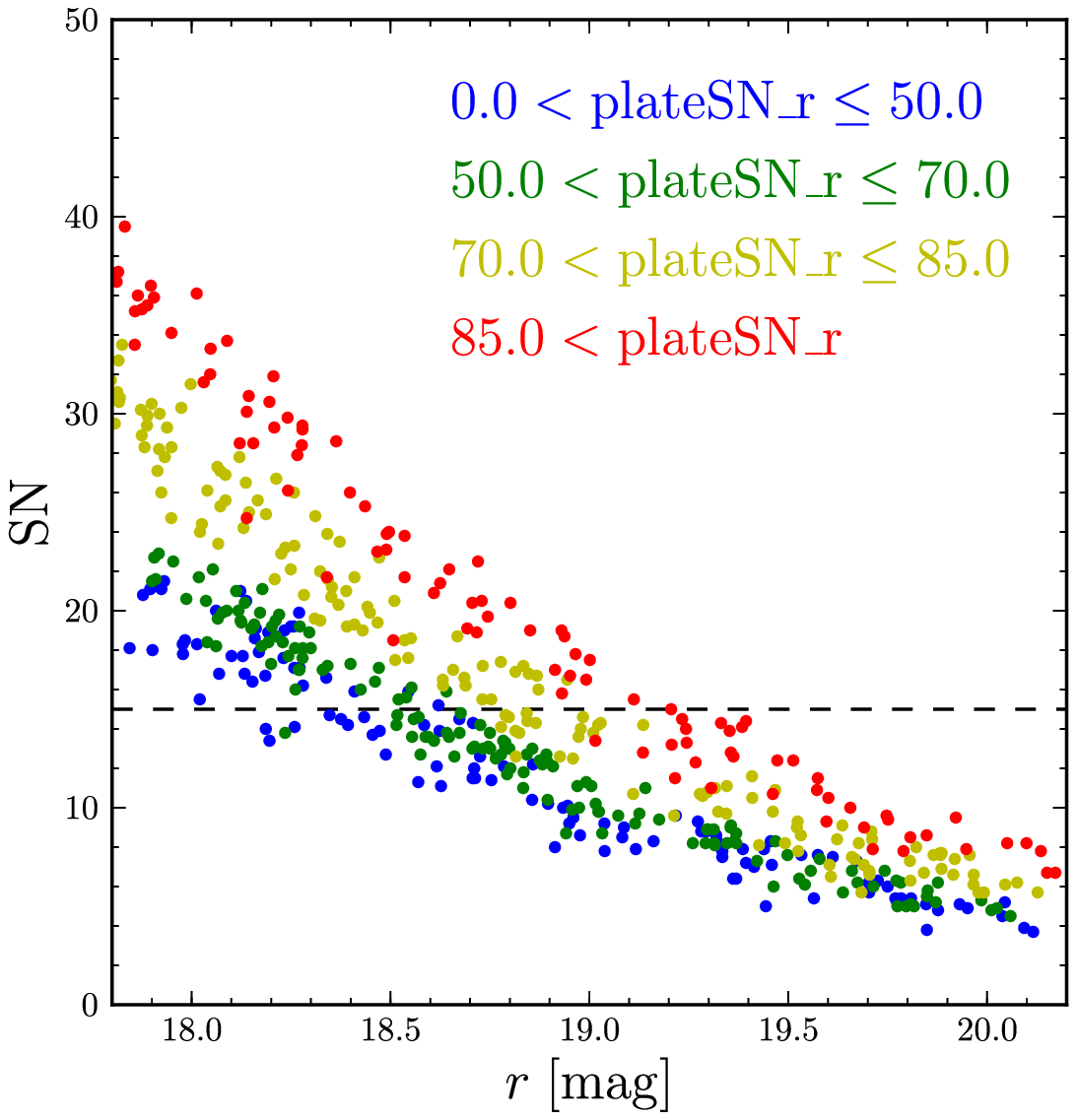}
\caption{Maximum apparent $r$-band magnitude per plate versus overall
plate signal-to-noise ratio \platesn\ for the G-star sample with \sn\
$> 15$ for faint plates (\emph{top panel}). Signal-to-noise ratio \sn\ for stars on
four typical \segue\ plates as a function of the apparent $r$-band magnitude
(\emph{bottom panel}). The four plates have been chosen to show a range in overall
plate signal-to-noise ratio \platesn.}
\label{fig:platesn_maxr_g}
\end{figure}

To infer the dependence on color and apparent magnitude of the
selection function, we look at the distribution of the potential G-star
targets in color--magnitude space. This is shown in
\figurename~\ref{fig:colormag_g}. The distribution of the
spectroscopic sample is overlaid. This shows that the spectroscopic
sampling is relatively fair in $g-r$ color, with some frayed edges
because of changes between target and current photometry, and that the
selection as a function of $r$-band magnitude tapers at the faint end, as should be
expected when using a signal-to-noise ratio cut. If all \segue\ plates
were integrated to the same depth, the signal-to-noise ratio cut should
be a clean cut in $r$, but it is clear from
\figurename~\ref{fig:colormag_g} that this is not the case. To
distinguish between relatively shallow and relatively deep plates, we
introduce the overall plate signal-to-noise ratio \platesn
\begin{equation}\label{eq:platesn} \mathrm{plateSN\_r} =
(\mathrm{sn1\_1}+\mathrm{sn2\_1})/2\,, \end{equation} where sn1\_1 and
sn2\_1 are the $r$-band plate signal-to-noise ratio for the two \sdss\
spectrographs \citep[see \tablename~17 in][]{Stoughton:2002ae}. The
faintest spectroscopic G-type star per plate as a function of \platesn\ for
the faint plates is shown in
\figurename~\ref{fig:platesn_maxr_g} for the faint plates. This figure shows that
there is a clear difference in the faintest object that could have
been successfully observed at $\sn > 15$ between relatively shallow
and relatively deep plates. The bottom panel of
\figurename~\ref{fig:platesn_maxr_g} shows the signal-to-noise ratio
of stars on four plates chosen to cover a range in the overall plate
signal-to-noise ratio. This shows that the $\sn > 15$ cut for the
entire sample translates into a fairly sharp $r$-band cut for each
individual plate.

Our model for the \segue\ G-star selection function is then the
following: For each plate we find the faintest targeted object in $r$-band magnitude with \sn\ larger than our signal-to-noise-ratio cut,
with apparent magnitude $r_{\mathrm{cut}}$ (if this object is fainter than the nominal limit
$r_{\mathrm{max}}$ for bright or faint plates, we set
$r_{\mathrm{cut}}$ equal to this limit; $r_{\mathrm{max}} = 17.8$ mag
for bright plates and 20.2 mag for faint plates), and then assume
that the selection function for that plate is given by a hyperbolic
tangent cut-off, centered on $r_{\mathrm{cut}}-0.1$ mag, and with a
width-parameter whose natural logarithm is -3 ($\approx 0.05$ mag),
such that the total width of the cut-off is about 0.2 mag and the
faintest object on the plate is about 2 widths from the center of the
cut-off. The function value at the bright end is equal to the number
of spectroscopic objects brighter than $r_{\mathrm{cut}}$ divided by
the total number of targets brighter than $r_{\mathrm{cut}}$. Thus,
the plate-dependent selection function is given by
\begin{equation}\label{eq:seguesf}
\begin{split}
S(\mathrm{plate},&r,g-r)= \\
& \frac{\# \mathrm{spectroscopic\ objects\ with}\ r_{\mathrm{min}} \leq r \leq r_{\mathrm{cut}}}{\# \mathrm{targets\ with}\ r_{\mathrm{min}} \leq r \leq r_{\mathrm{cut}}}\times
\left[1-\tanh\left(\frac{r-r_{\mathrm{cut}}+0.1}{\exp\left(-3\right)}\right)\right]\Big/ 2\,,
\end{split}
\end{equation}
where the numbers of objects are evaluated within the $\approx 7$
deg$^2$ area of the plate in question and in the $0.48 \leq g-r \leq
0.55$ G-star color range; $r_{\mathrm{min}}$ is 14.5 mag for bright
plates and 17.8 mag for faint plates. The selection function is zero
outside of the apparent $r$-band magnitude range of the plate
($[14.5,17.8]$ for bright plates and $[17.8,20.2]$ for faint plates).

\begin{figure}[tp]
\includegraphics[width=0.5\textwidth,clip=]{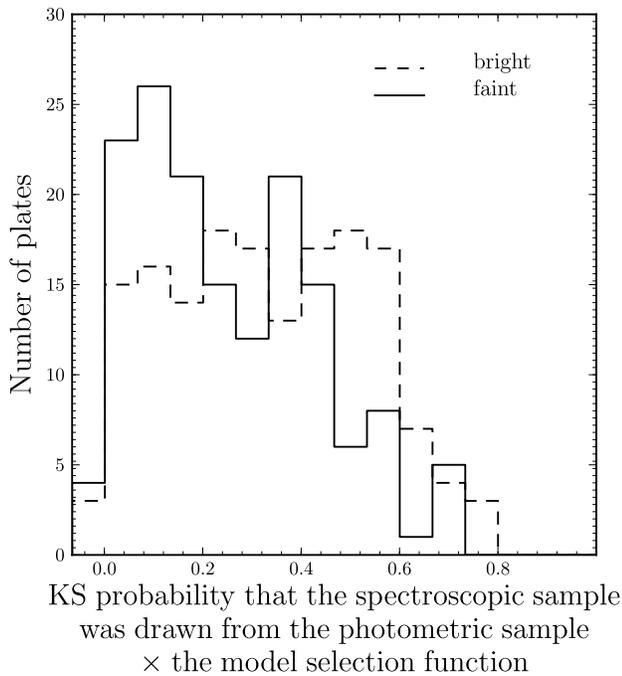}
\caption{Distribution of the probability that a plate's spectroscopic
sample was drawn from the photometric sample combined with the model
selection function. The leftmost bin contains the plates with
probability $< 0.001$. For the correct model of the \segue\ selection function, the distribution should be approximately flat between zero and one. }\label{fig:ks_tanhr} 
\end{figure}

We use this model both for the bright plates and the faint plates,
although most bright plates are in fact consistent with being complete
up to 17.8 mag. \figurename~\ref{fig:ks_tanhr} shows the distribution
of Kolmogorov-Smirnov (KS) probabilities that the spectroscopic sample
for any given plate was selected from the target sample with this
model for the selection function. All but 7 plates have probabilities
larger than 0.001 and the distribution of probabilities is relatively
flat, as expected. 

Rather than using a smooth hyperbolic tangent cut-off, we also tried a
sharp cut at $r_{\mathrm{cut}}$. With this model for the selection
function, 79 plates have a KS probability $< 0.05$
($\approx$25\,percent of the number of plates), as opposed to 30
plates in the hyperbolic-tangent-cut-off model ($\approx$9\,percent of
the sample). Therefore, the smooth cut-off is necessary to fully model
the selection function. The fact that the distribution of KS
probabilities in \figurename~\ref{fig:ks_tanhr} is not entirely flat
is due to remaining details in the faint cut-off of the selection
function, as we know that the selection function is flat at brighter
magnitudes. This does not impact our analsis greatly, as most stars
are much brighter than the cut-off (as compared to the scale over
which the selection function changes near the cut-off).

\begin{figure}[tp]
\includegraphics[width=0.5\textwidth,clip=]{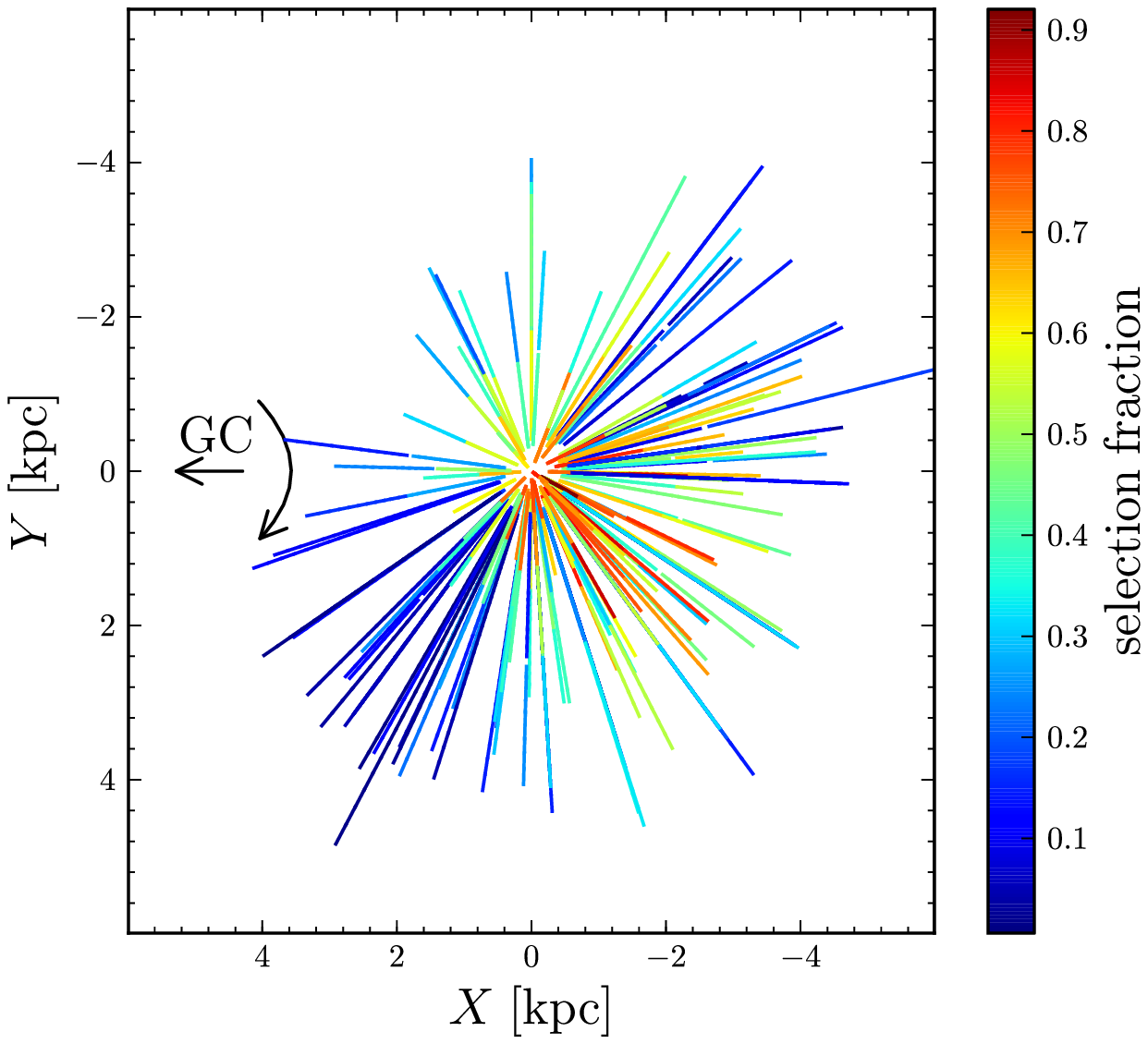}
\includegraphics[width=0.5\textwidth,clip=]{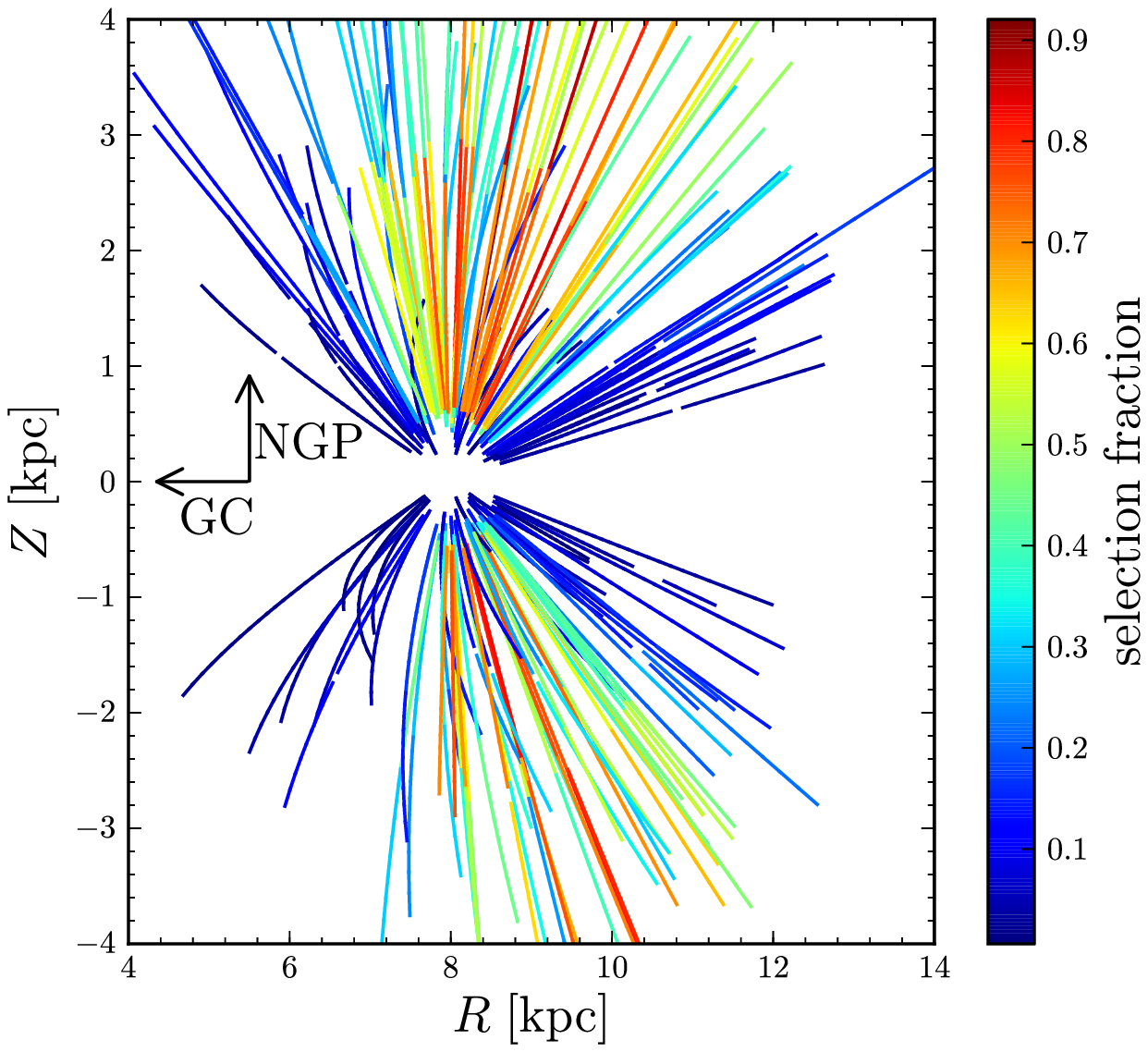}
\caption{The \segue\ selection function---the fraction of objects
  successfully observed spectroscopically with \sn\ $> 15$ ---for the
  G-star sample, as a function of Galactic coordinates $X$ and $Y$
  (\emph{left panel}), and of Galactocentric radius $R$ and vertical
  height $Z$ (\emph{right panel}). The $r$-dependent \segue\ selection
  function is here transformed into spatial coordinates using the
  photometric distance relation applied to a color $g-r = 0.515$ mag
  and [Fe/H] = -0.5.}\label{fig:sfxyrz_g}
\end{figure}

The selection function is simplest in its native coordinates, survey
plate, and $r$-band magnitude. For each value of $g-r$ and [Fe/H], the
$r$-dependent selection function above translates into a (different)
spatial selection function through the use of the photometric distance
relation. The selection function projected into spatial coordinates
for a typical value of $g-r$ and [Fe/H] is shown in
\figurename~\ref{fig:sfxyrz_g}. Near $|b| = 90^\circ$ the
spectroscopic sample is relatively complete, whereas near the Galactic
plane the selection is much less complete.

We have posted Python code that implements this model for the \segue\
selection function. It is publicly available at

\url{https://github.com/jobovy/segueSelect}\,.

\section{The Magnitude--color--metallicity density and estimates of the effective survey volume}\label{sec:colorFeh}

The density in magnitude--color--metallicity space needs to be
included in the likelihood in \eqnname~(\ref{eq:densitylike}), because
it forms the basis of the photometric distance relation used to
translate observed colors, metallicities, and apparent magnitudes into
distances, which ultimately relate to the effective search volume. We assume here for simplicity that stars of a given
$g-r$ and \feh\ follow a single stellar isochrone given by the
\citet{Ivezic08a} photometric distance relation in terms of $g-r$
using \eqnname~(\ref{eq:ri_gr}) to translate $g-r$ into the $g-i$
color used by the \citet{Ivezic08a} relation. The reason for
expressing the \citet{Ivezic08a} $g-i$--metallicity--magnitude
relation into $g-r$ is to keep the integration in
\eqnname~(\ref{eq:normint}) simple; if we had chosen to use the $g-i$
relation we would have to include the $r-i$ color as well, and model
and integrate over the full $g-r$,$r-i$ plane. As the stellar locus
is very narrow ($\lesssim 0.1$ mag), this adds less (random) scatter
than is intrinsic to the photometric distance relation.

In the single-isochrone model, $\rho(r,g-r,\feh|R,Z,\phi)$ becomes the
product of a delta function with the density in the color--metallicity
plane
\begin{equation}
\rho(r,g-r,\feh|R,Z,\phi) = \delta(r-r[g-r,\feh,d]|R,Z,\phi)\,\rho(g-r,\feh|R,Z)\,,
\end{equation}
where $r[g-r,\feh,d]$ is the apparent magnitude derived from the
photometric distance relation combined with the distance, and by a
slight abuse of notation we have used the same symbol to denote the
density in the color--metallicity plane. We assume that this density
is independent of Galactocentric azimuth $\phi$, but for now allow it to depend on $R$ and $Z$. Using this, the
normalization integral in \eqnname~(\ref{eq:normint}) simplifies to
\begin{equation}\label{eq:normint2}
\begin{split}
\int & \dd l\,\dd b\,\dd d\, \dd r\, \dd (g-r) \,\dd \feh \,\lambda(l,b,d,r,g-r,\feh|\theta) \\
&= A_p\,\sum_{\mathrm{plates}\ p}
\int \dd (g-r)\,\dd\feh\, \\
& \,\times 
\int_{d[r_{\mathrm{min}},g-r,\feh]}^{d[r_{\mathrm{max}},g-r,\feh]} \dd d\,
S(p,r[g-r,\feh,d],g-r)\,\rho(g-r,\feh|R,Z)\,d^2\,\dens(R,z|l,b,d,\theta)\,,
\end{split}
\end{equation}
where $r_{\mathrm{min}}$ and $r_{\mathrm{max}}$ are the minimum and
maximum apparent magnitude of plate $p$, and the functions $d[\cdot]$
and $r[\cdot]$ use the photometric distance relation.

\begin{figure}[tp]
\includegraphics[width=0.26\textwidth,clip=]{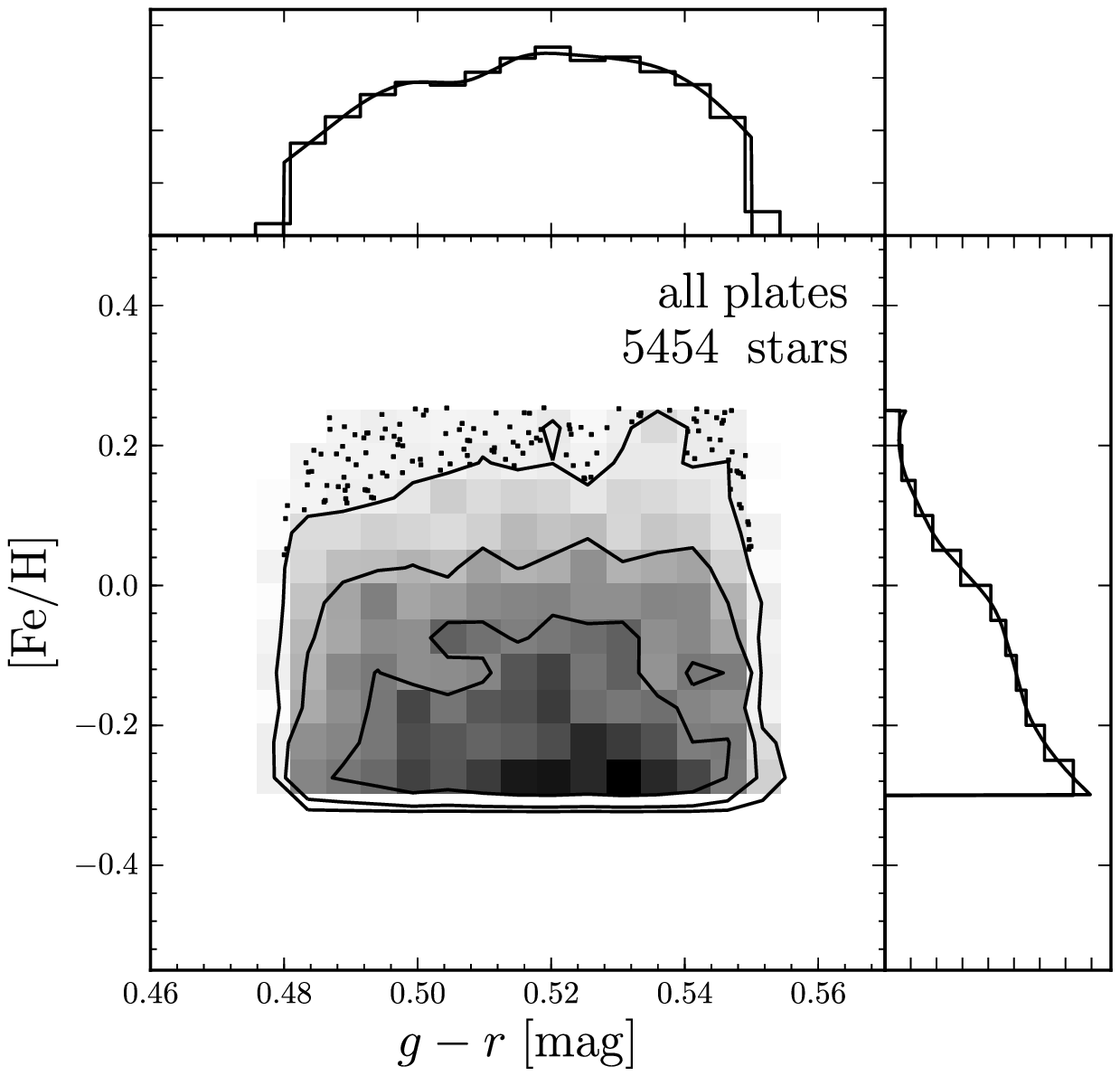}
\includegraphics[width=0.228\textwidth,clip=]{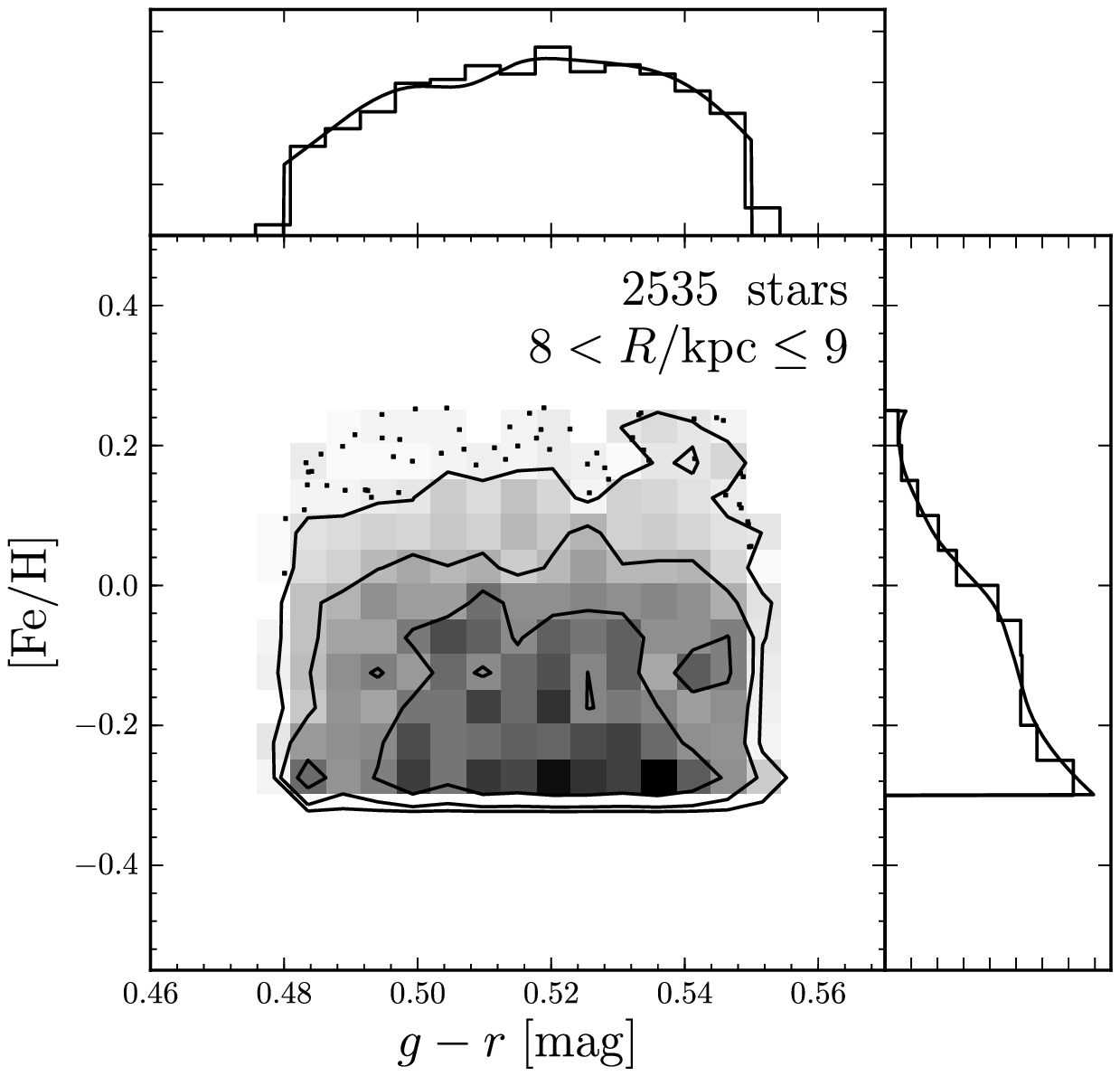}
\includegraphics[width=0.228\textwidth,clip=]{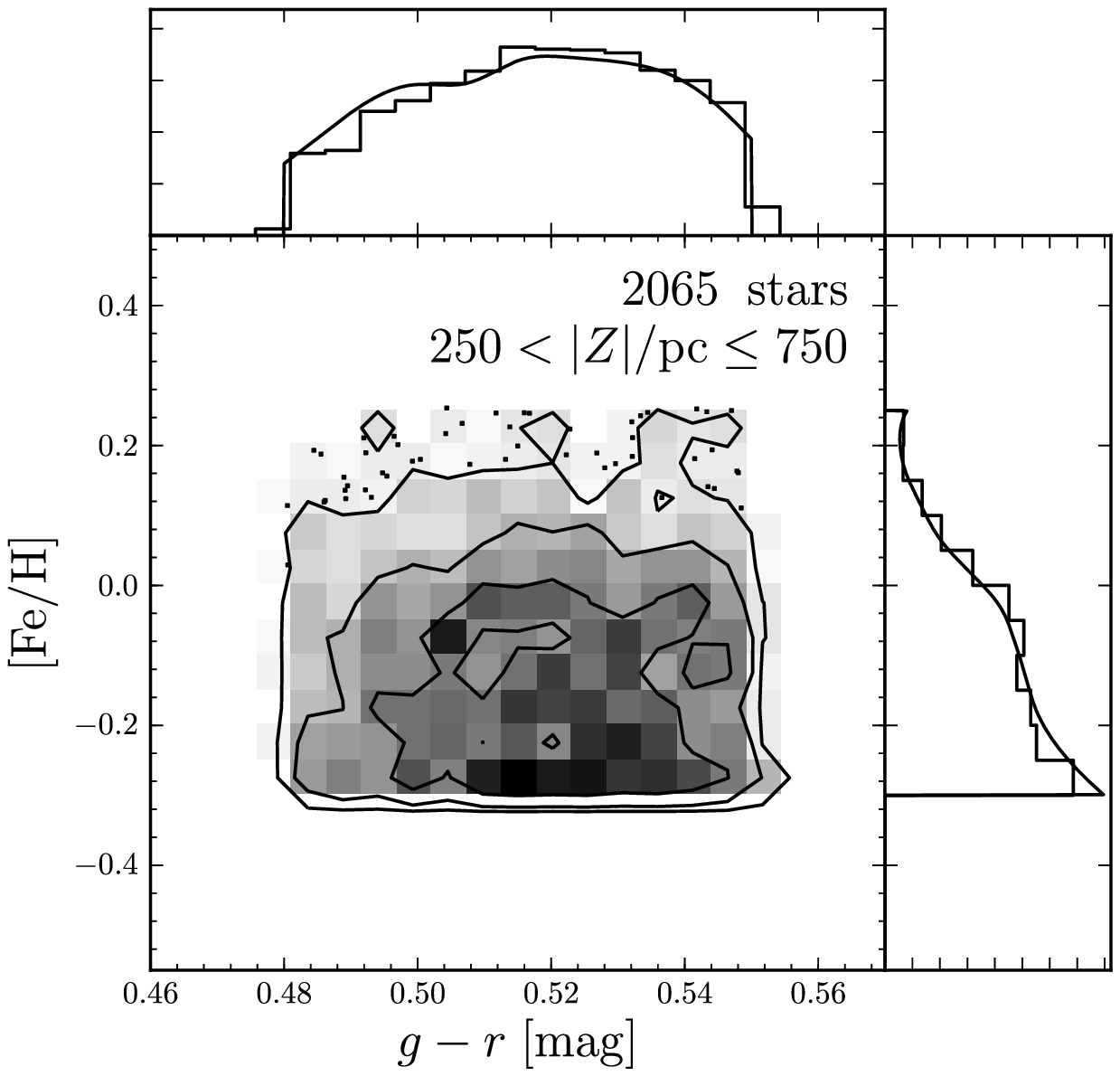}
\includegraphics[width=0.228\textwidth,clip=]{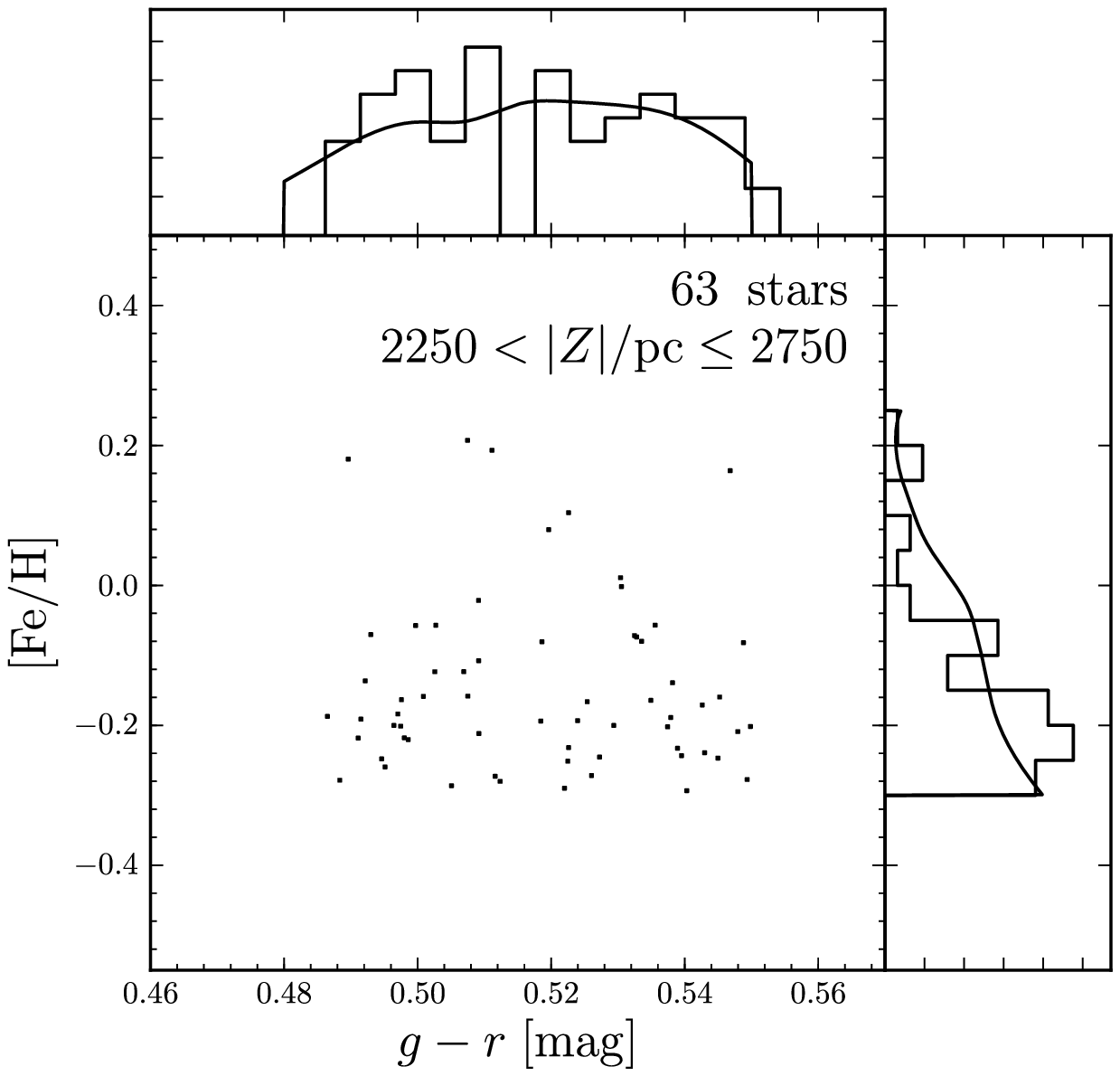}\\
\includegraphics[width=0.26\textwidth,clip=]{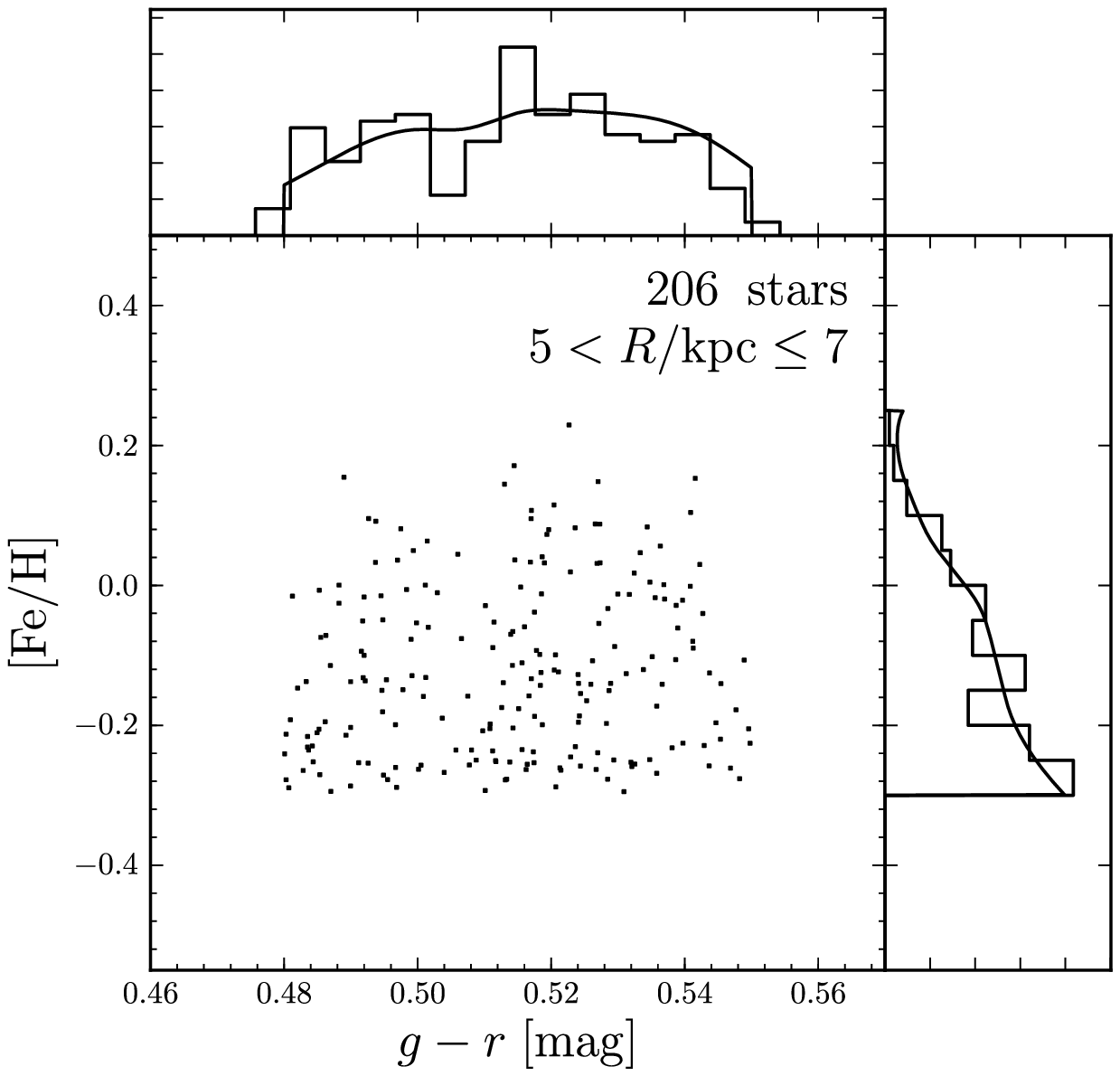}
\includegraphics[width=0.228\textwidth,clip=]{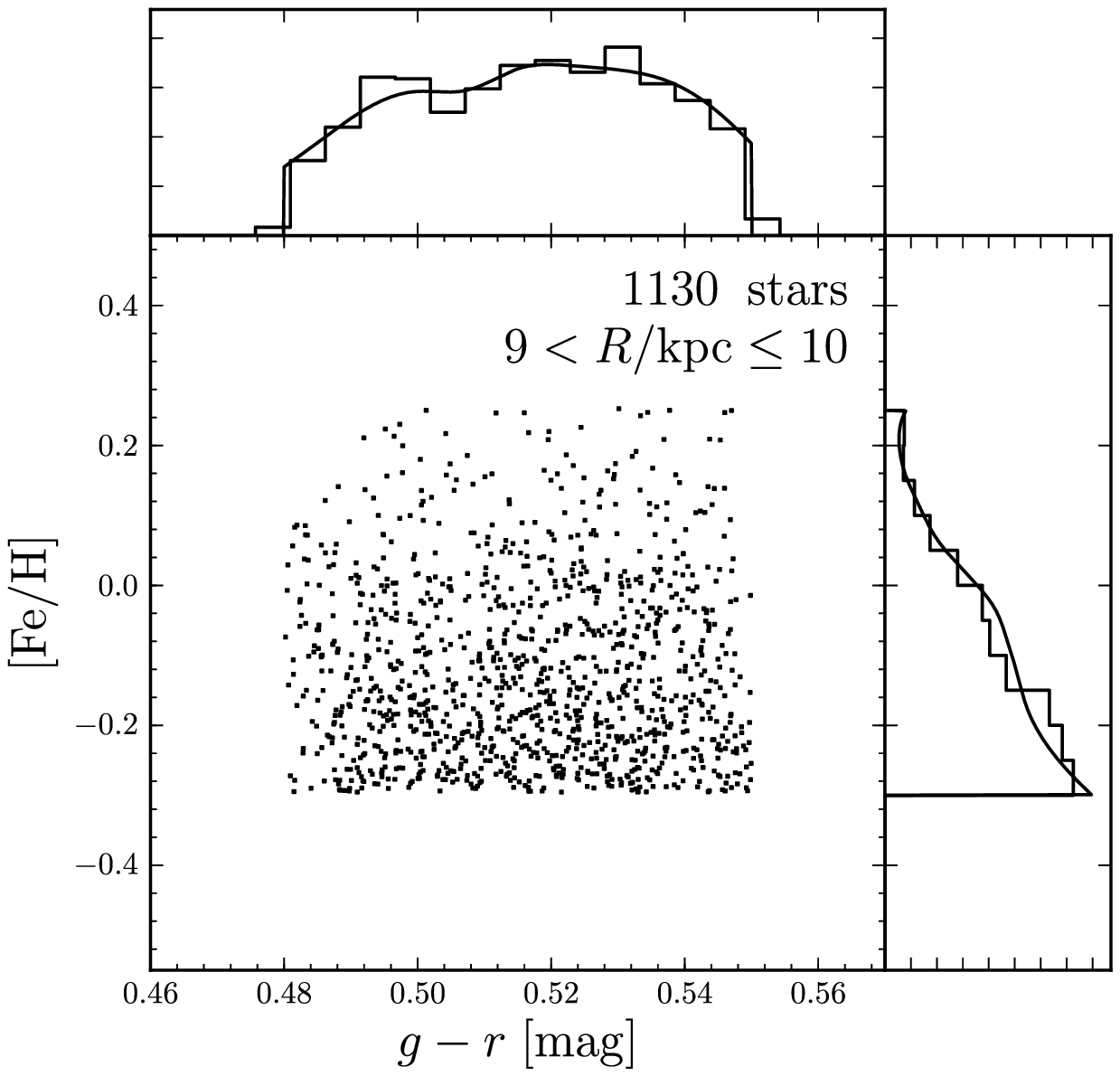}
\includegraphics[width=0.228\textwidth,clip=]{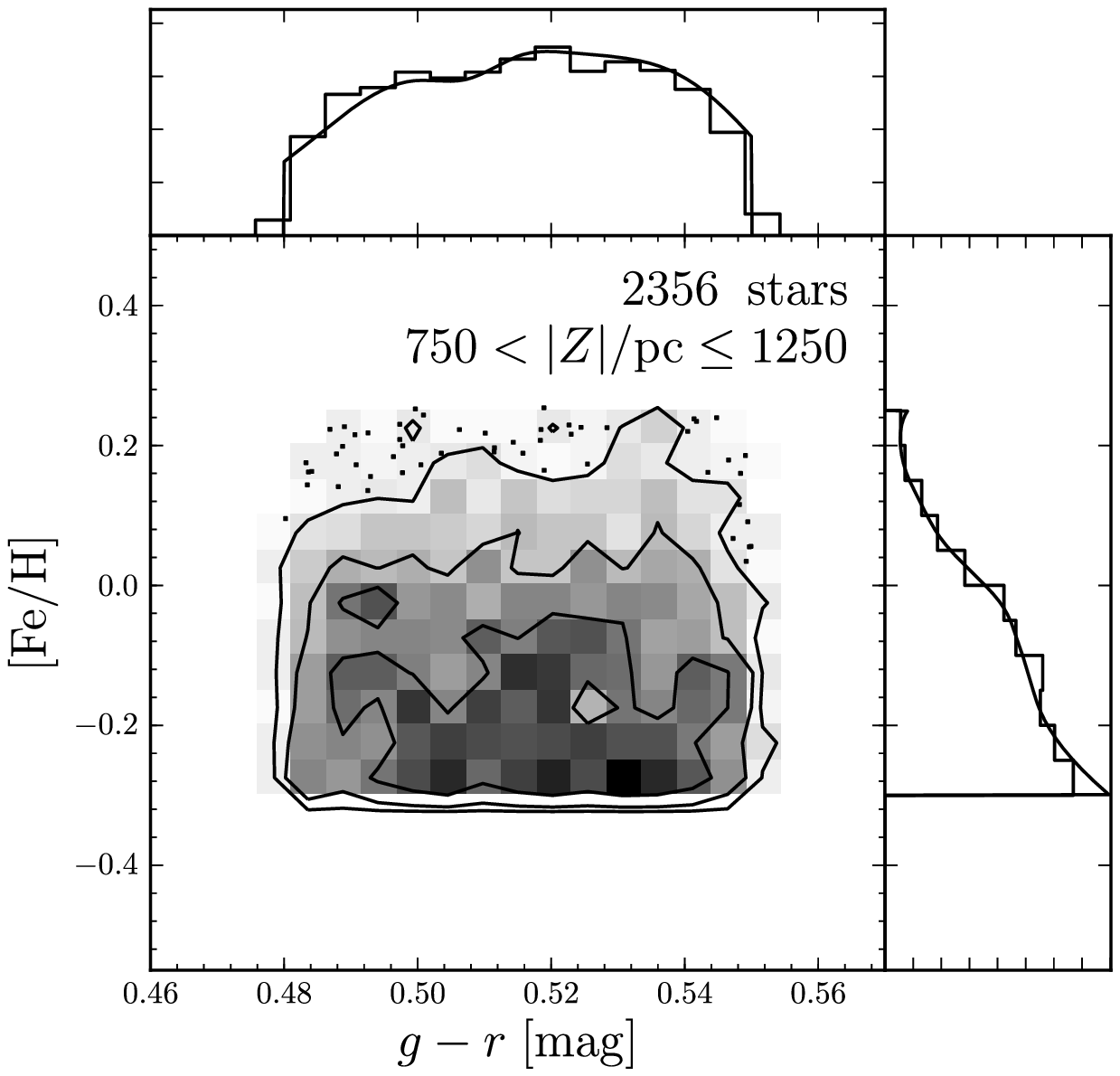}
\includegraphics[width=0.228\textwidth,clip=]{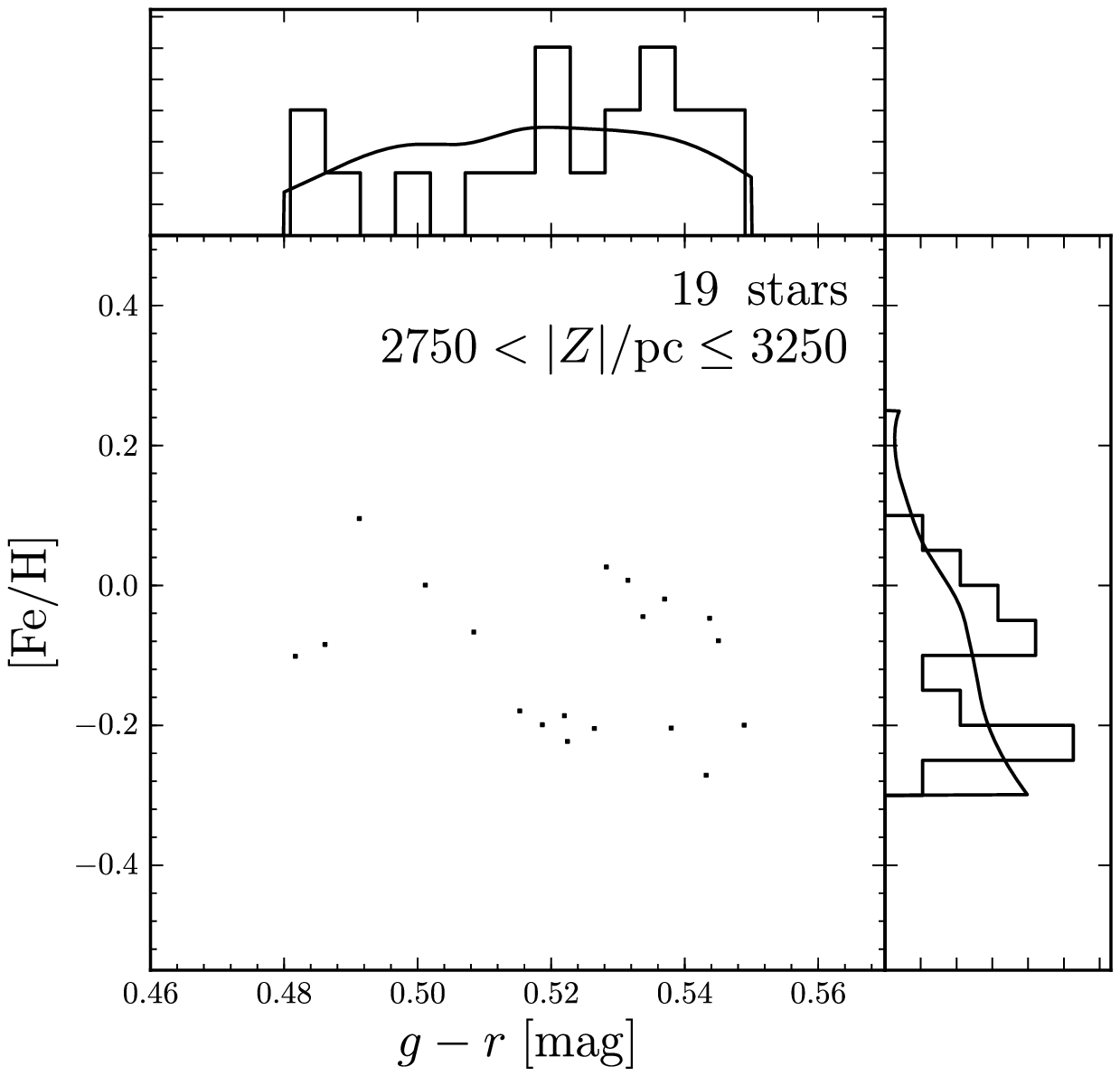}\\
\includegraphics[width=0.26\textwidth,clip=]{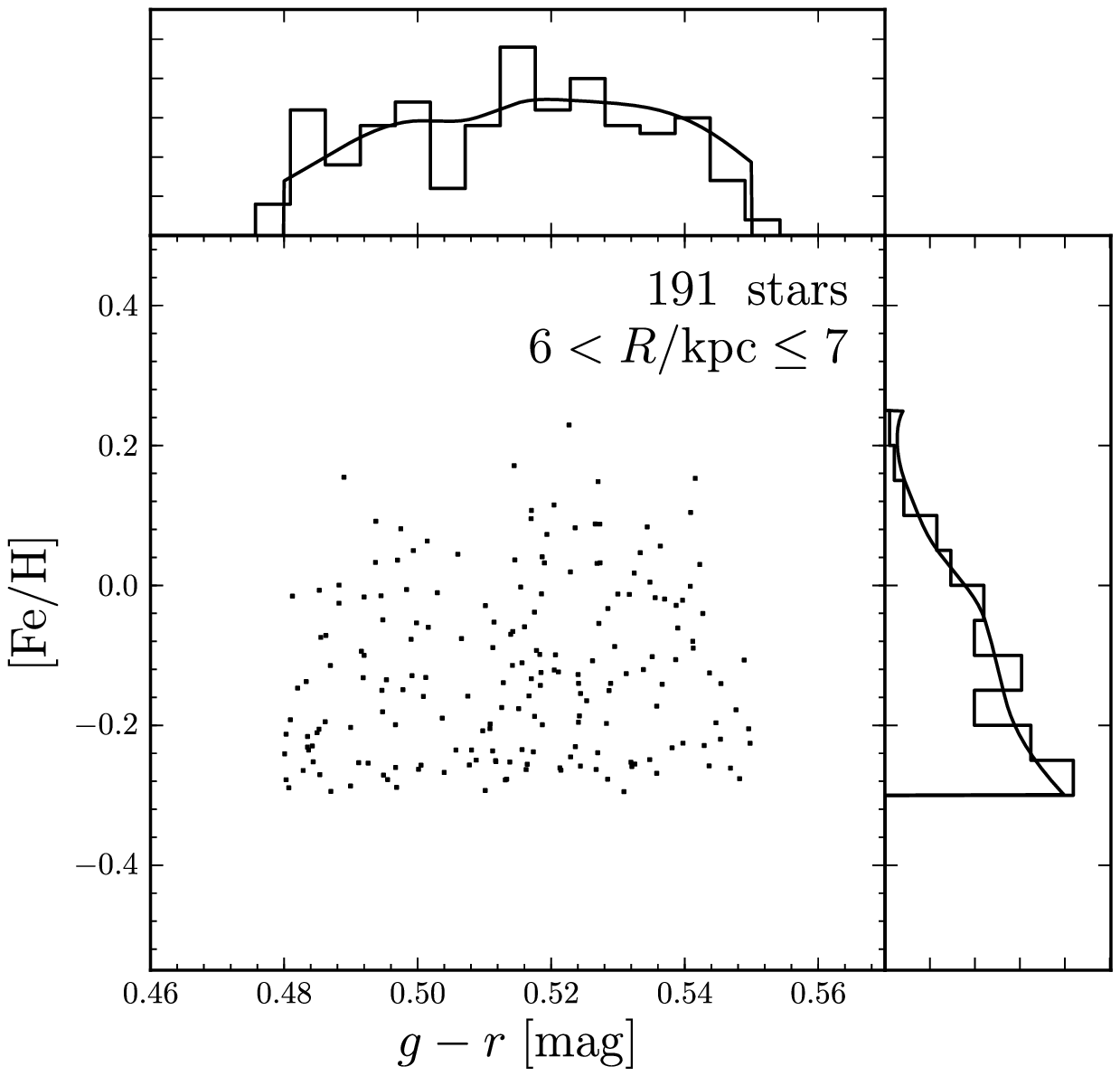}
\includegraphics[width=0.228\textwidth,clip=]{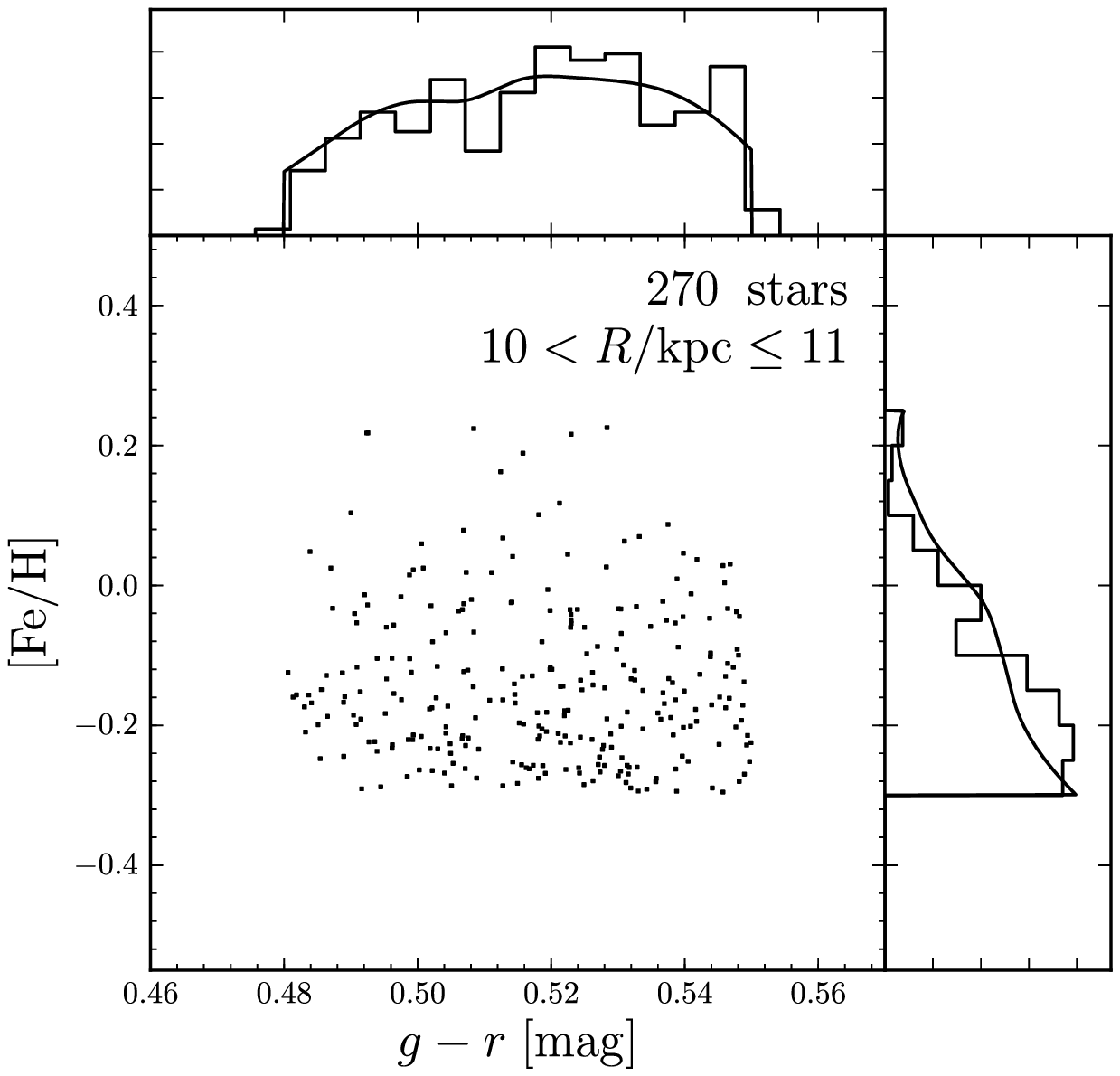}
\includegraphics[width=0.228\textwidth,clip=]{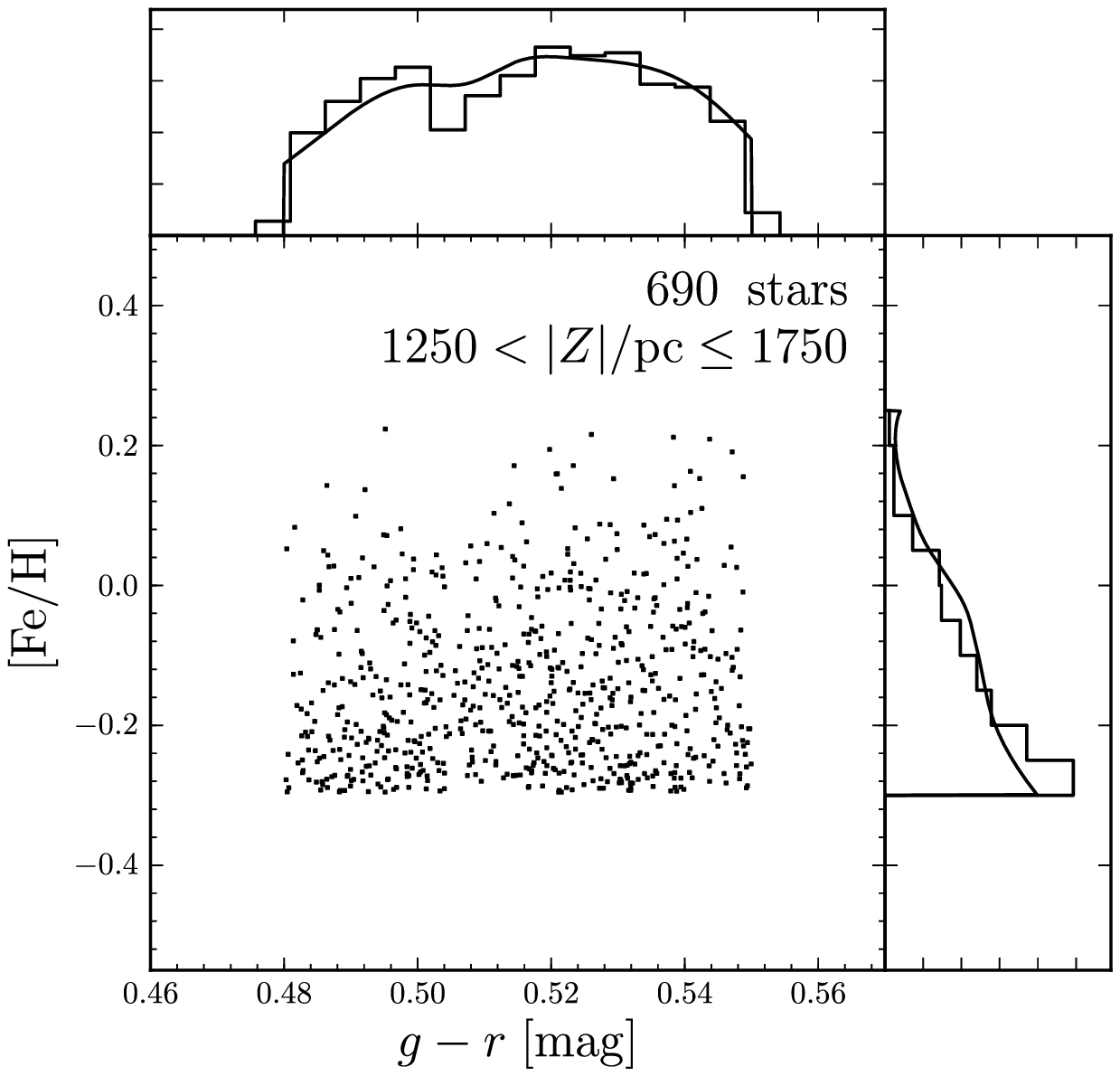}
\includegraphics[width=0.228\textwidth,clip=]{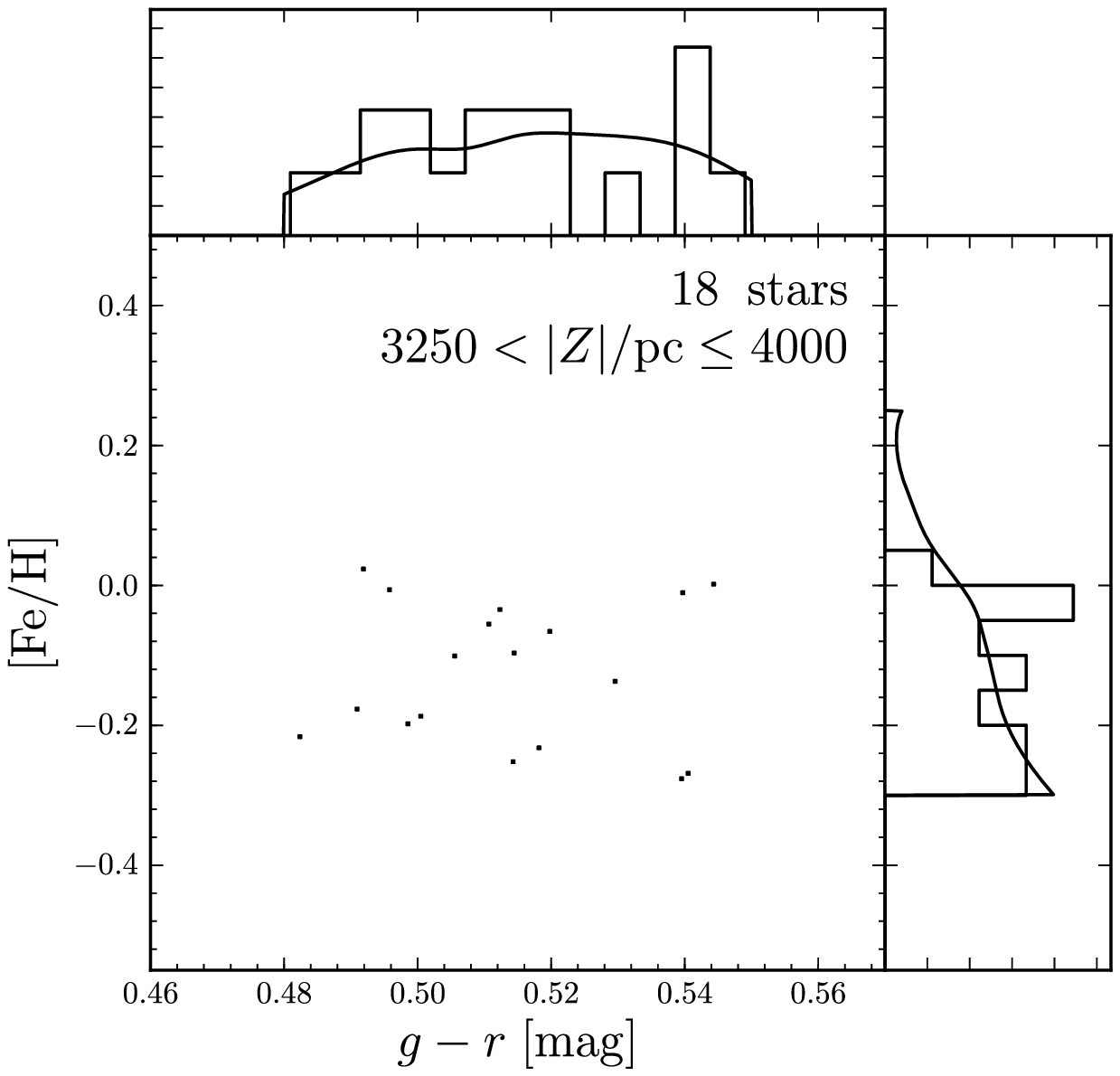}\\
\includegraphics[width=0.26\textwidth,clip=]{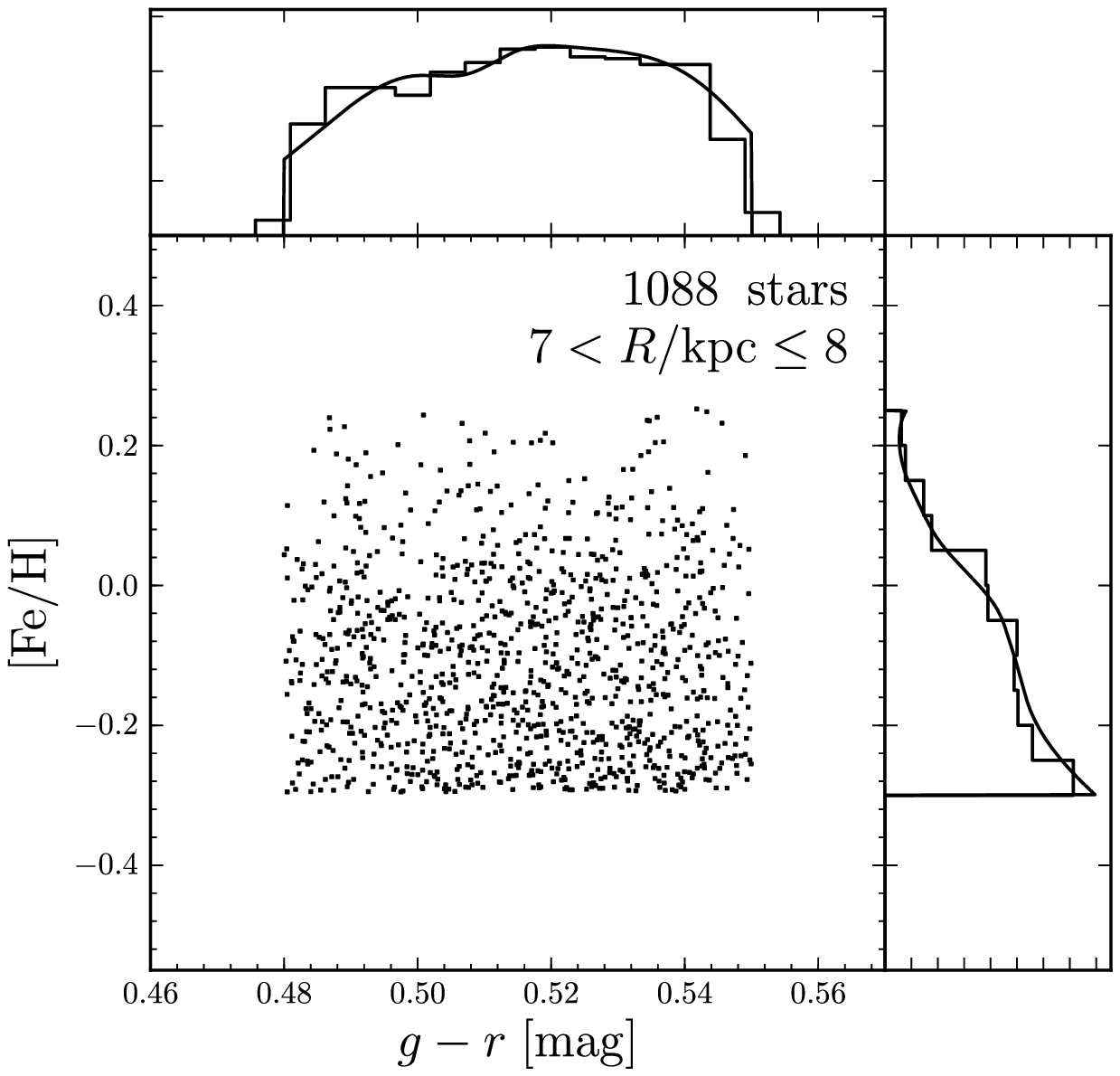}
\includegraphics[width=0.228\textwidth,clip=]{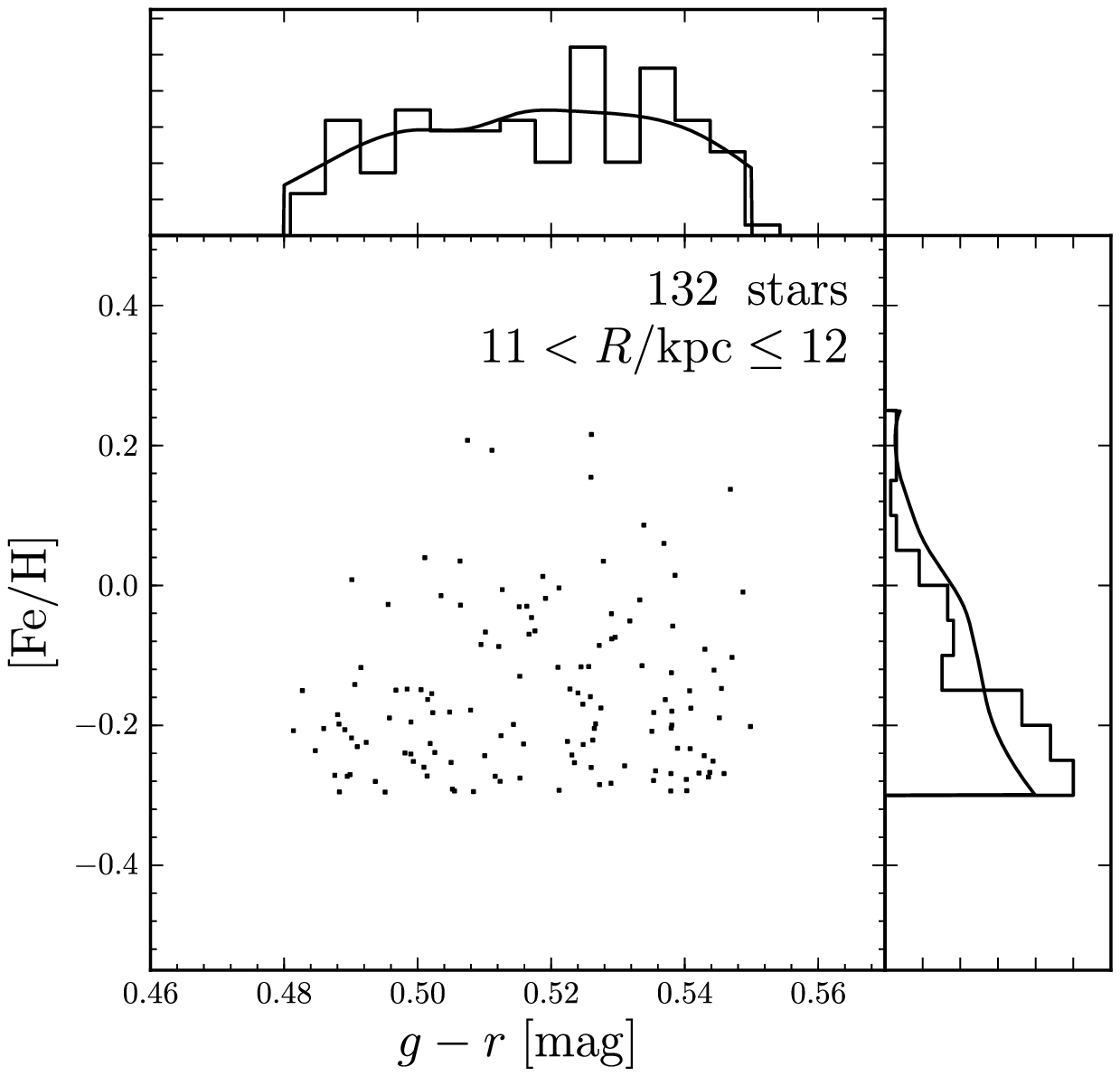}
\includegraphics[width=0.228\textwidth,clip=]{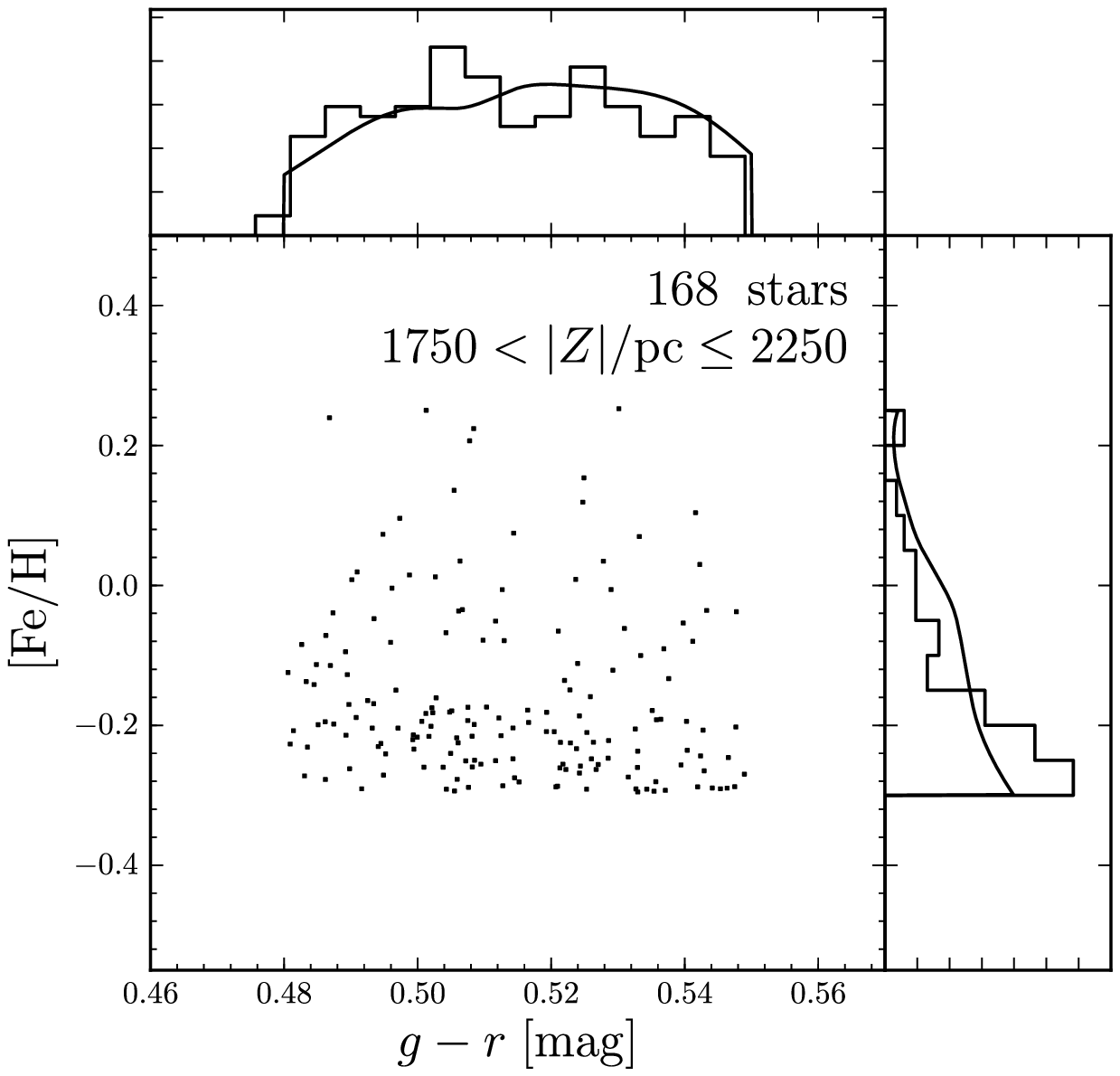}
\includegraphics[width=0.2335\textwidth,clip=]{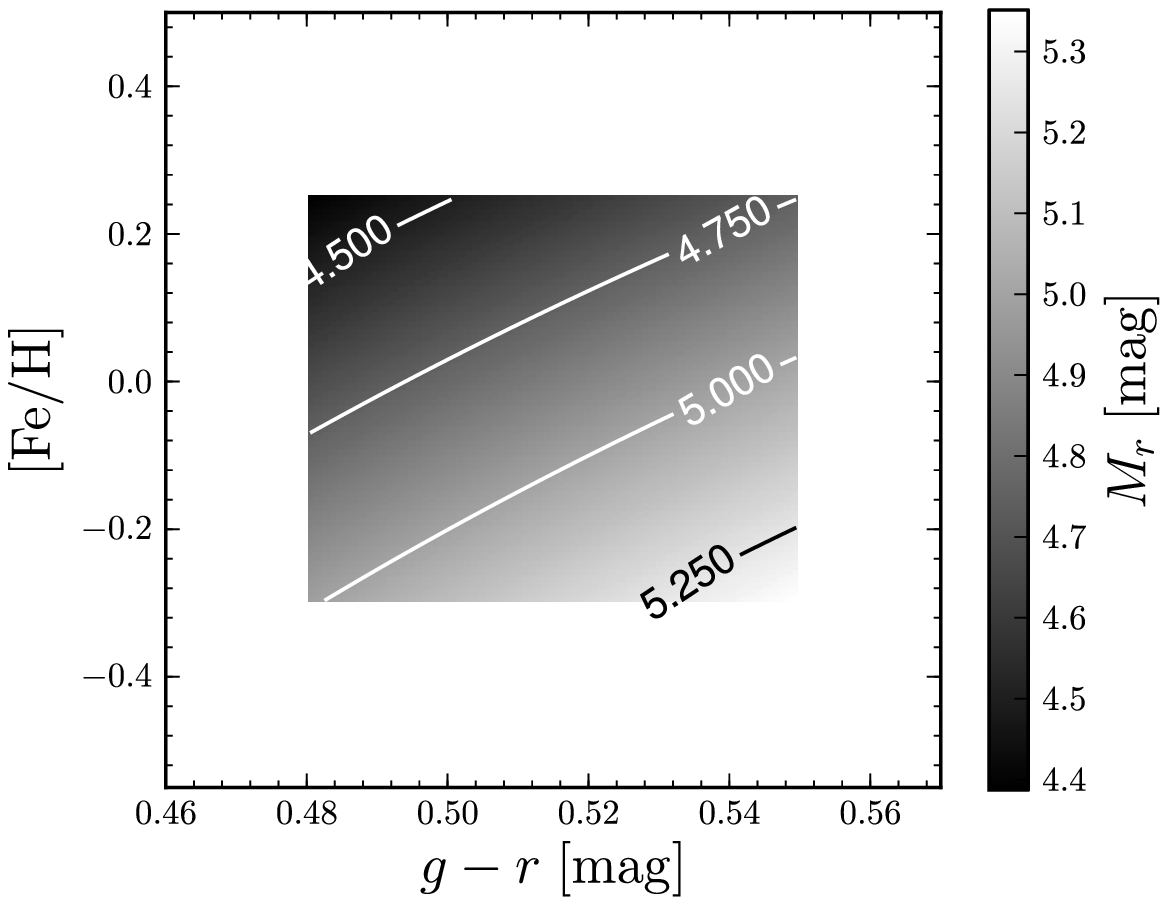}
\caption{Distribution of [Fe/H] and $g-r$ for the full \apoor\ G-dwarf
sample (\emph{top, left panel}) and split into ranges in
Galactocentric radius $R$ and vertical height $|Z|$ (\emph{other
panels}). Linear binned densities with contours containing 68, 95, and
99\,percent of the distribution and individual outliers beyond
99\,percent are shown in spatial bins with more than 1,500 stars. A smooth
interpolation of the one-dimensional distributions in the top left
panel is shown in all panels. The bottom right panel shows the
absolute $r$-band magnitude as a function of $g-r$ and [Fe/H] from the
\citet{Ivezic08a} color--metallicity--magnitude
relation (their eqn. A7).}\label{fig:FeH_RZ_rich_g}
 \end{figure}
\begin{figure}[tp]
\includegraphics[width=0.26\textwidth,clip=]{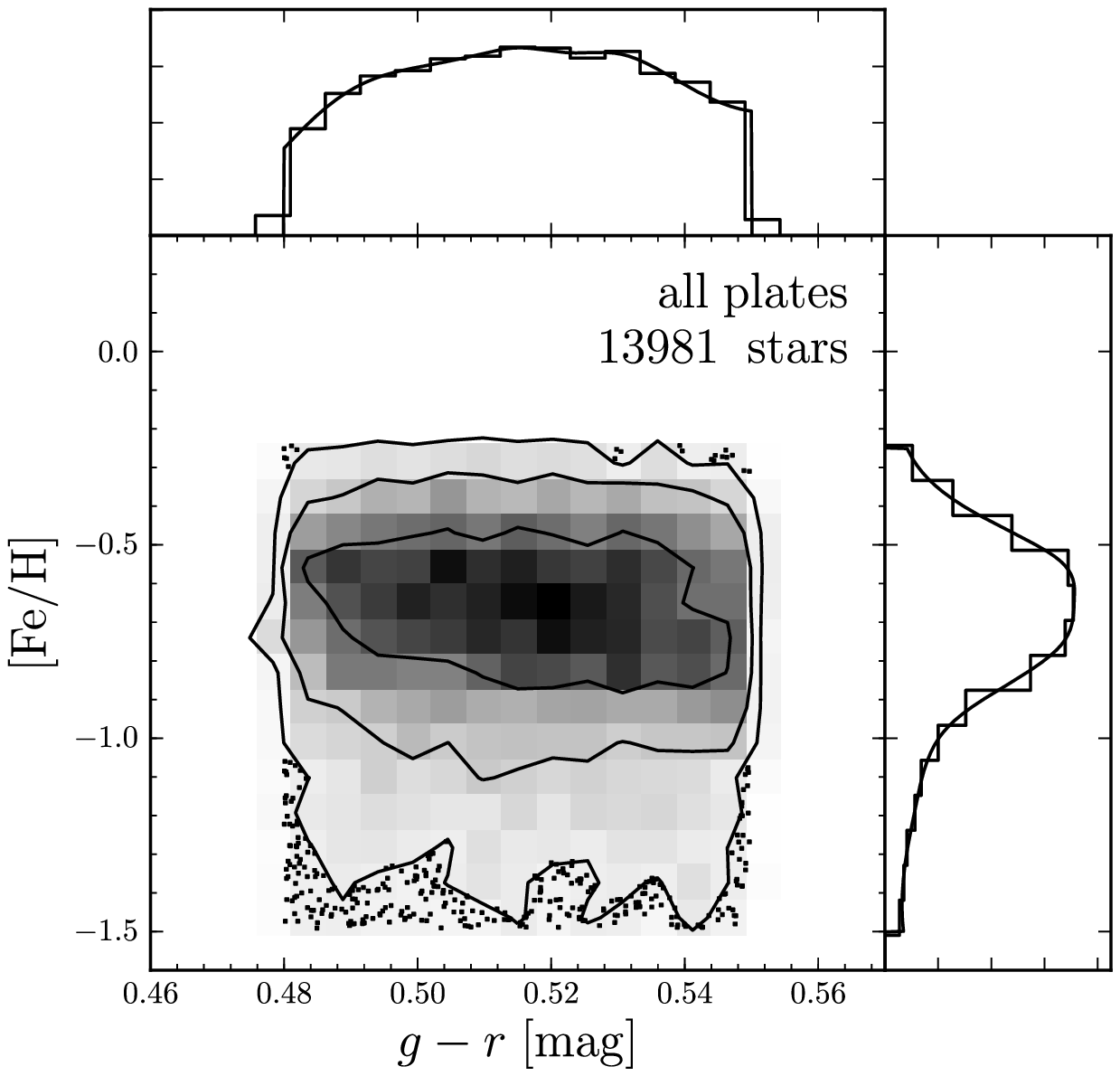}
\includegraphics[width=0.228\textwidth,clip=]{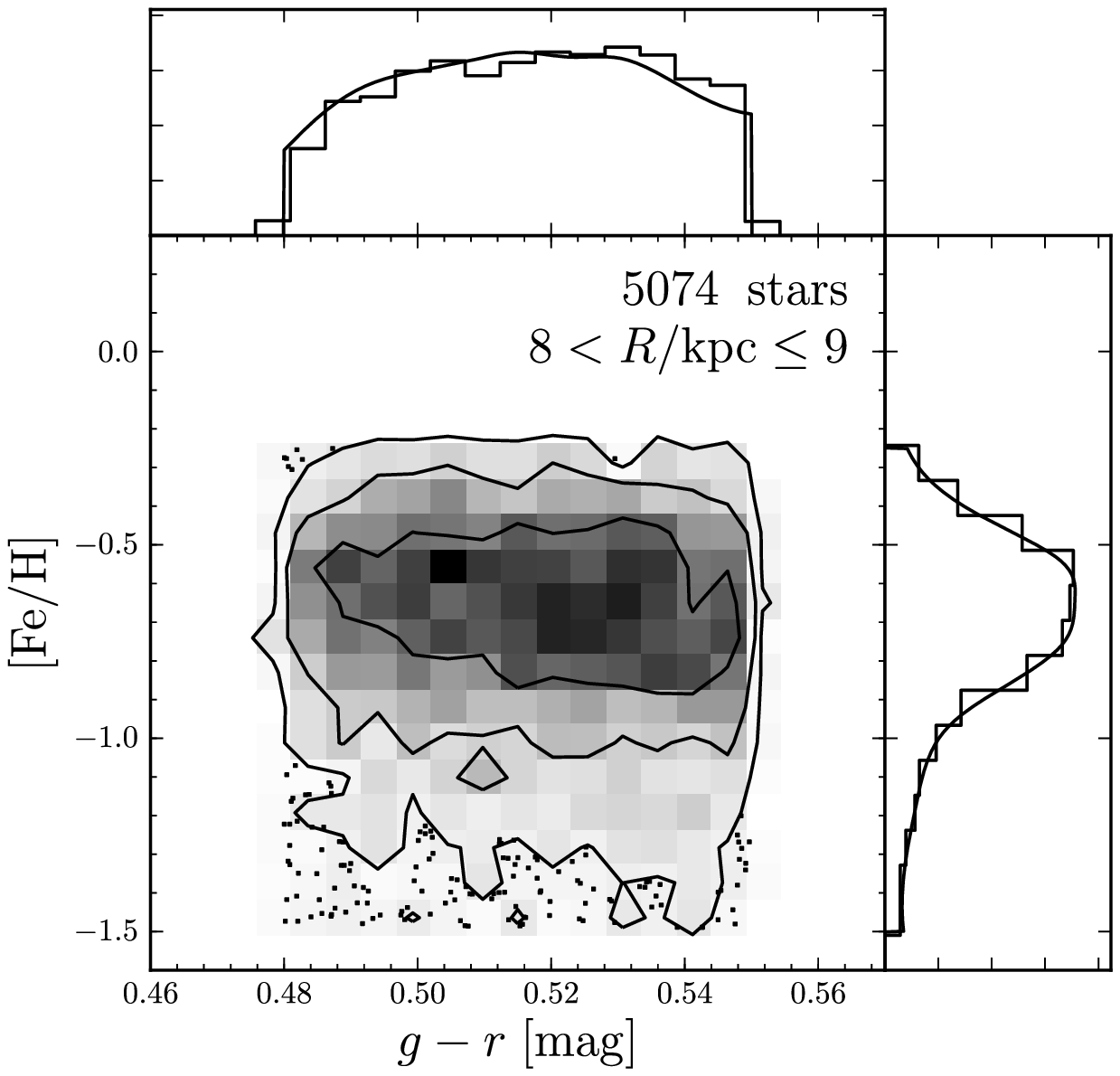}
\includegraphics[width=0.228\textwidth,clip=]{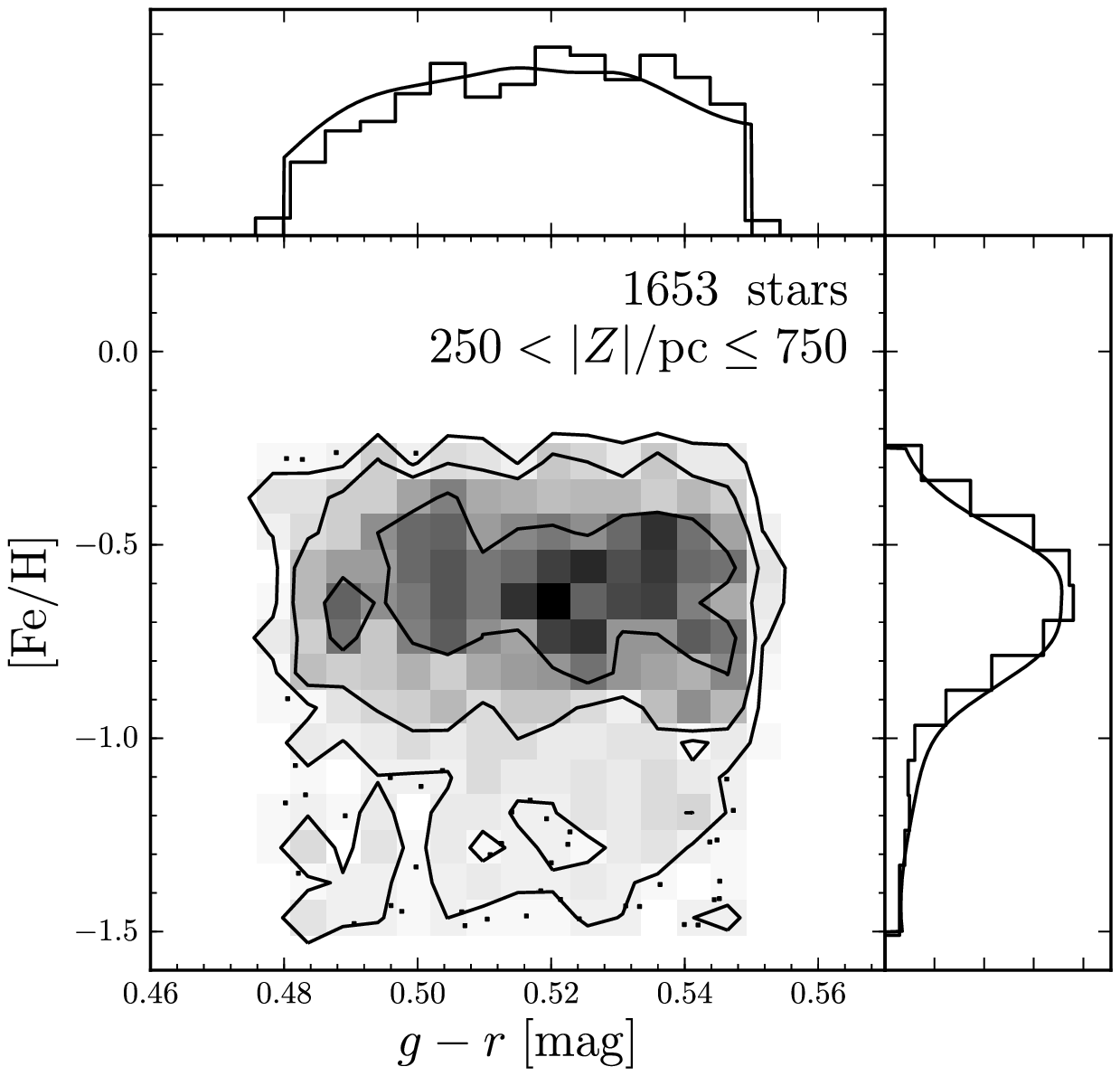}
\includegraphics[width=0.228\textwidth,clip=]{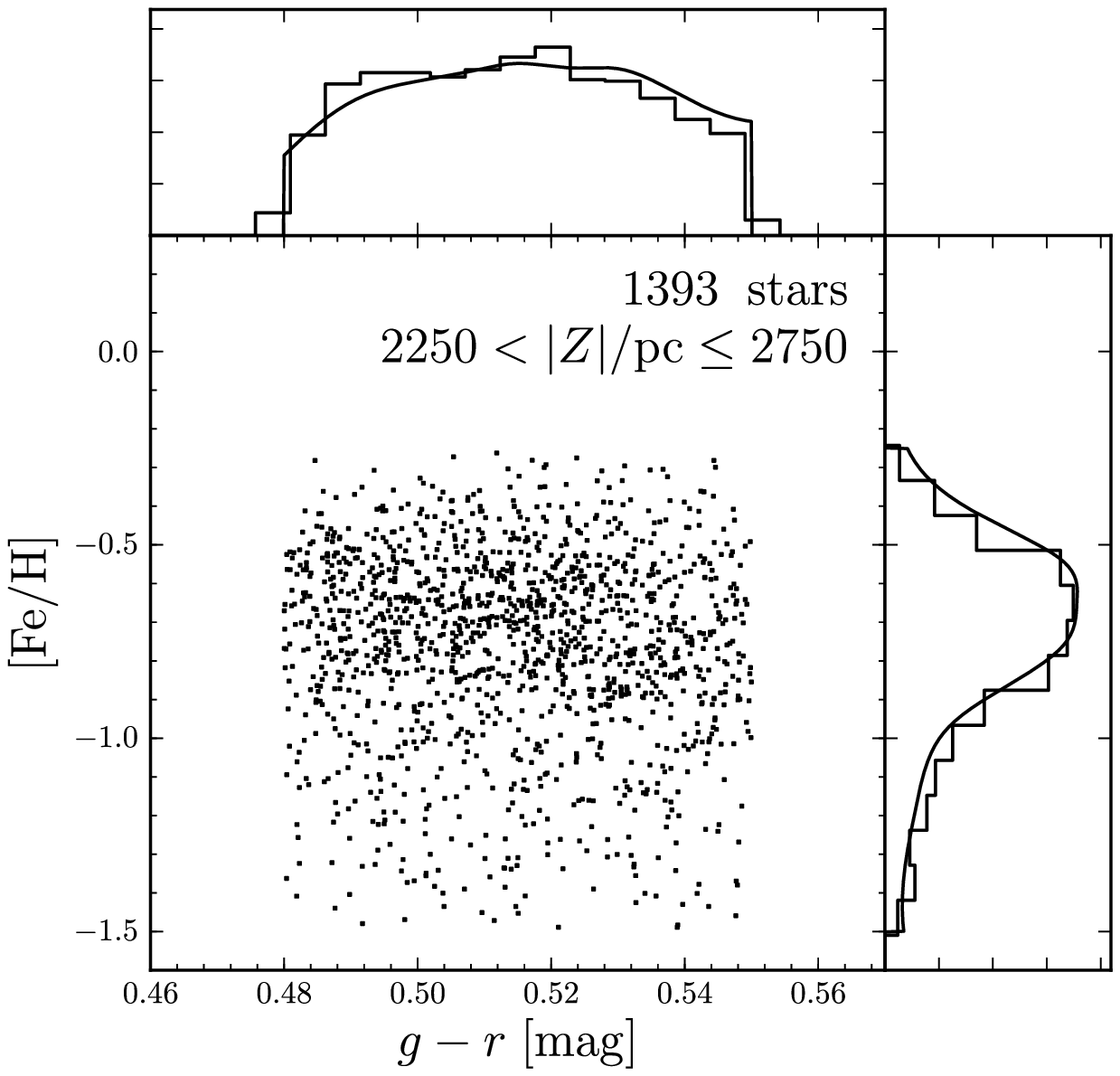}\\
\includegraphics[width=0.26\textwidth,clip=]{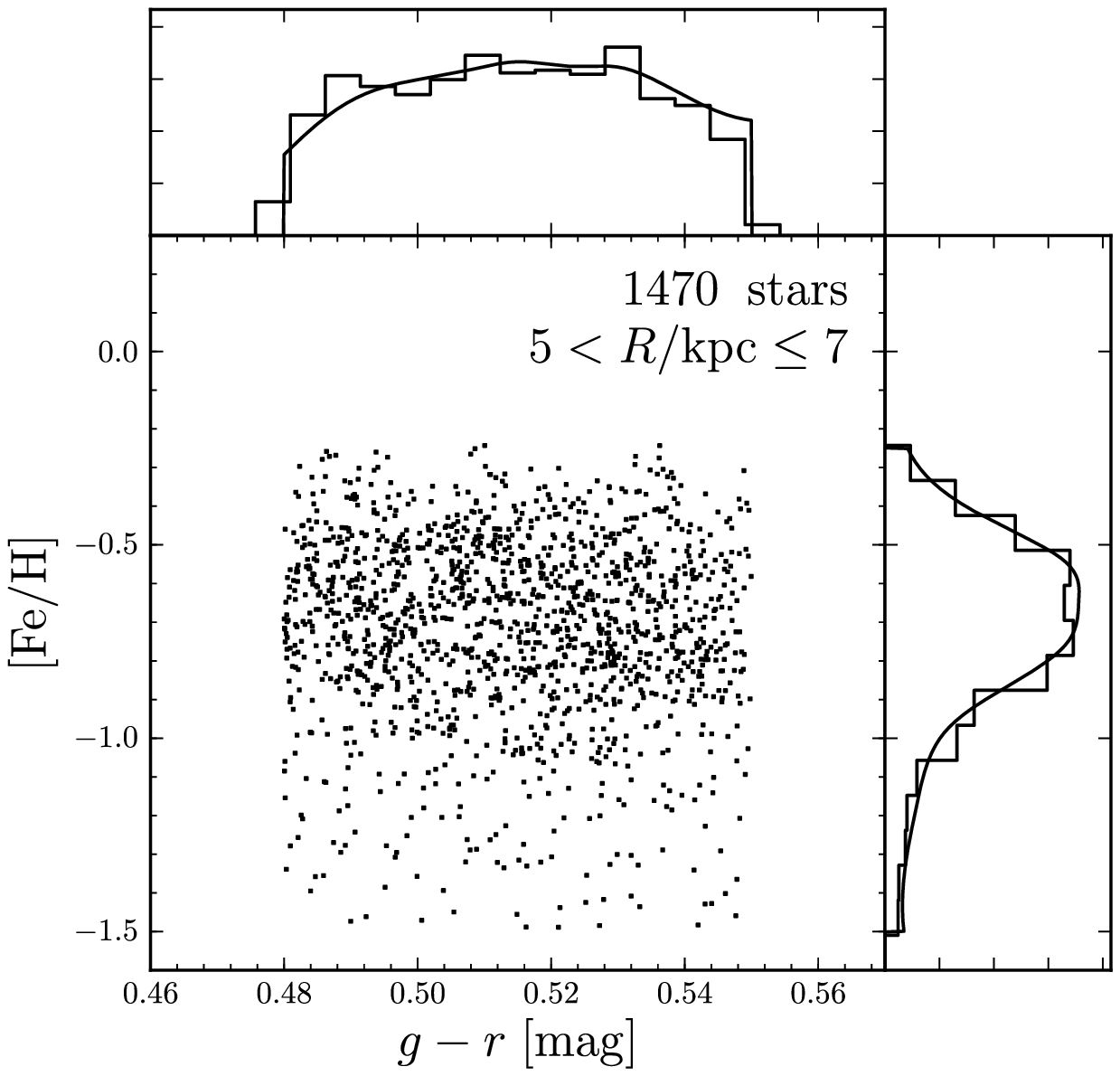}
\includegraphics[width=0.228\textwidth,clip=]{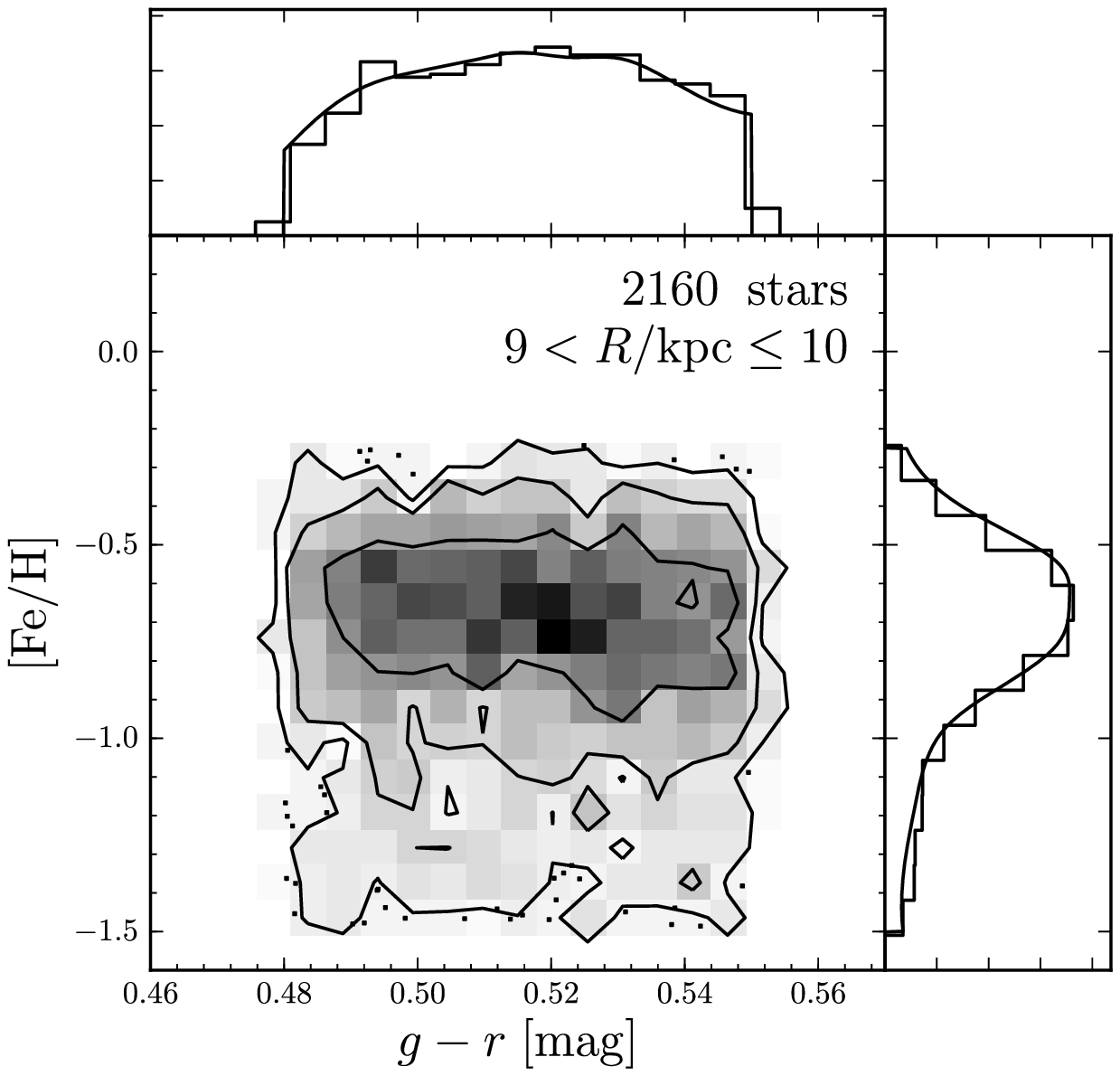}
\includegraphics[width=0.228\textwidth,clip=]{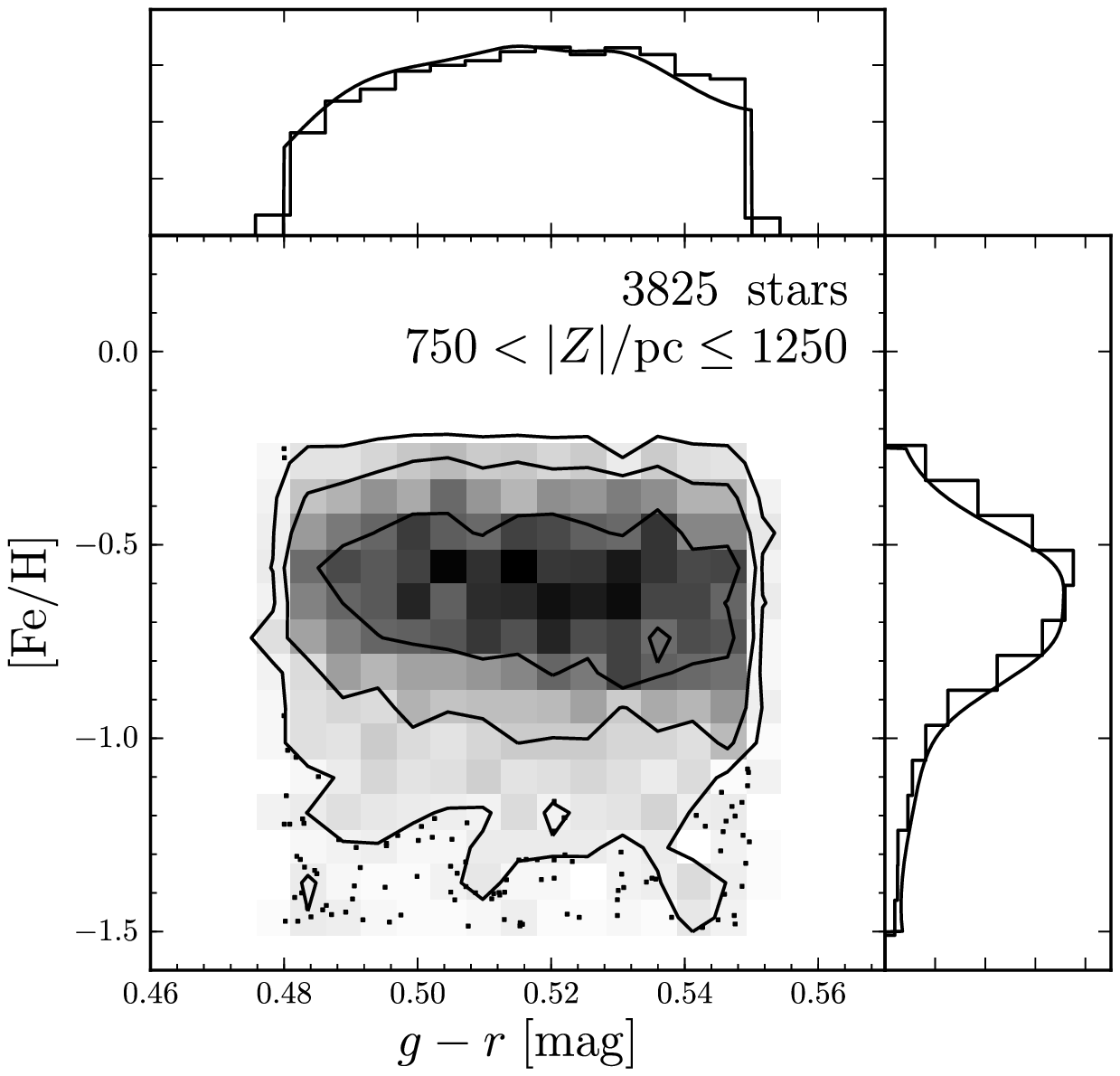}
\includegraphics[width=0.228\textwidth,clip=]{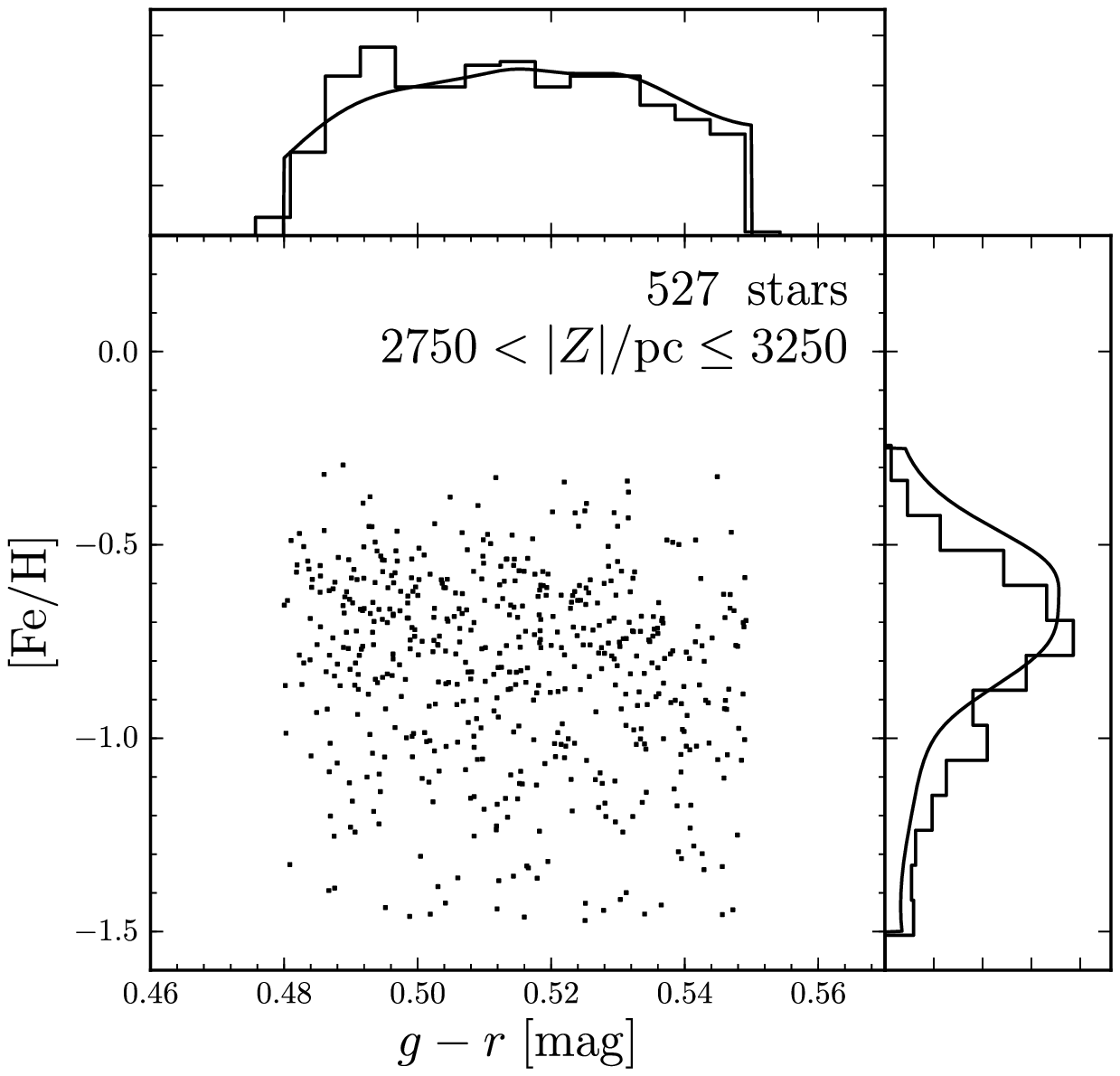}\\
\includegraphics[width=0.26\textwidth,clip=]{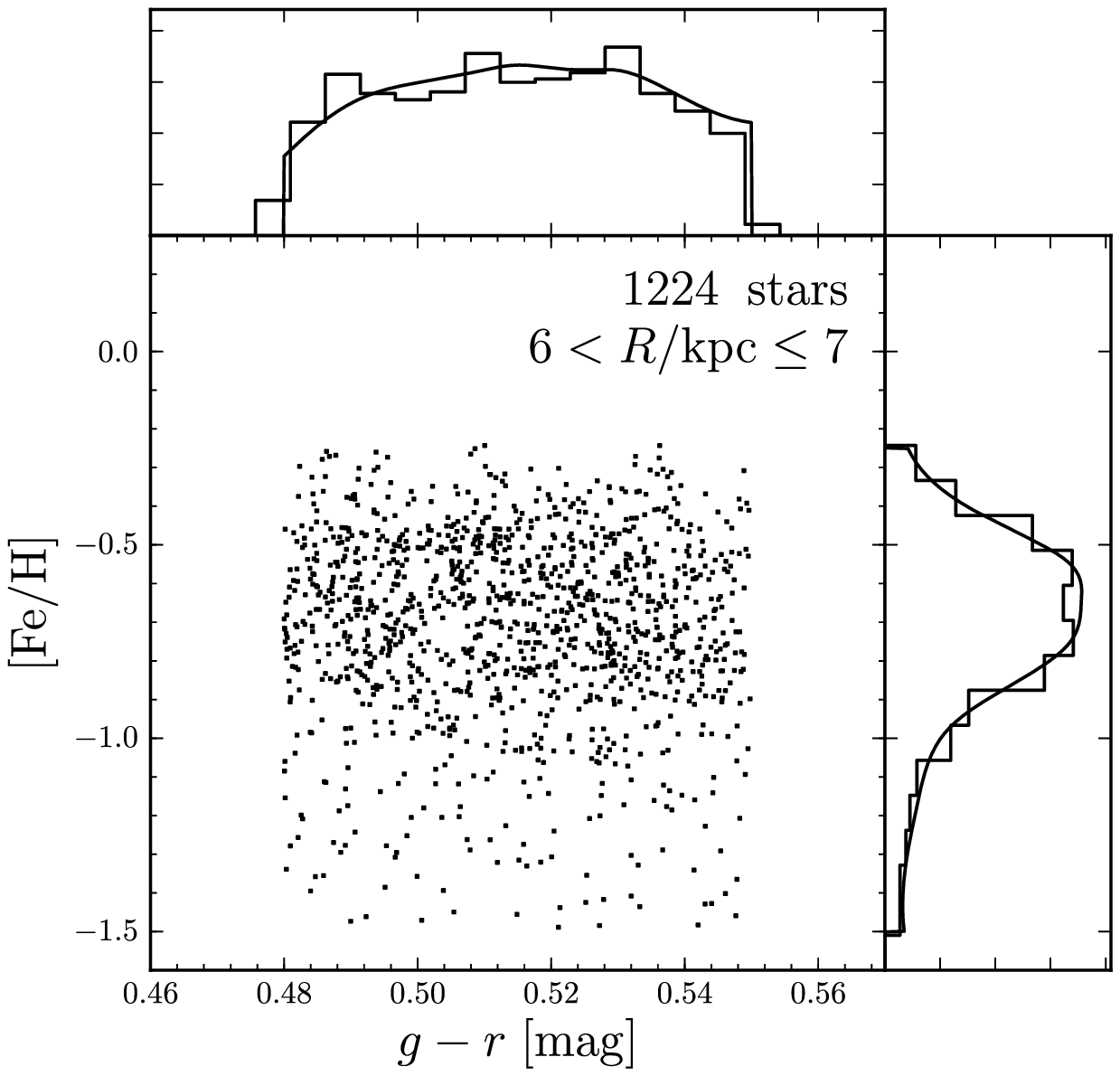}
\includegraphics[width=0.228\textwidth,clip=]{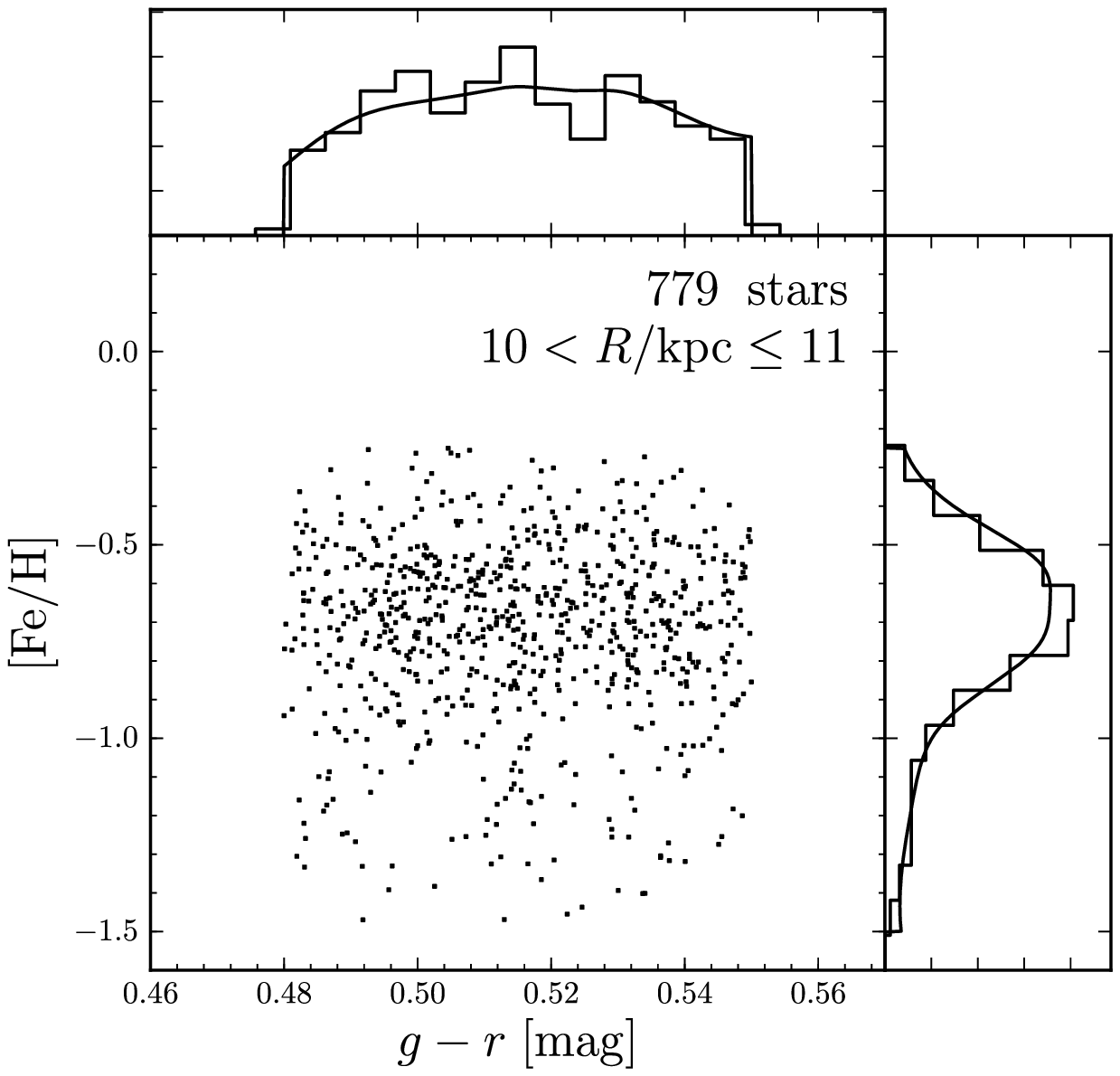}
\includegraphics[width=0.228\textwidth,clip=]{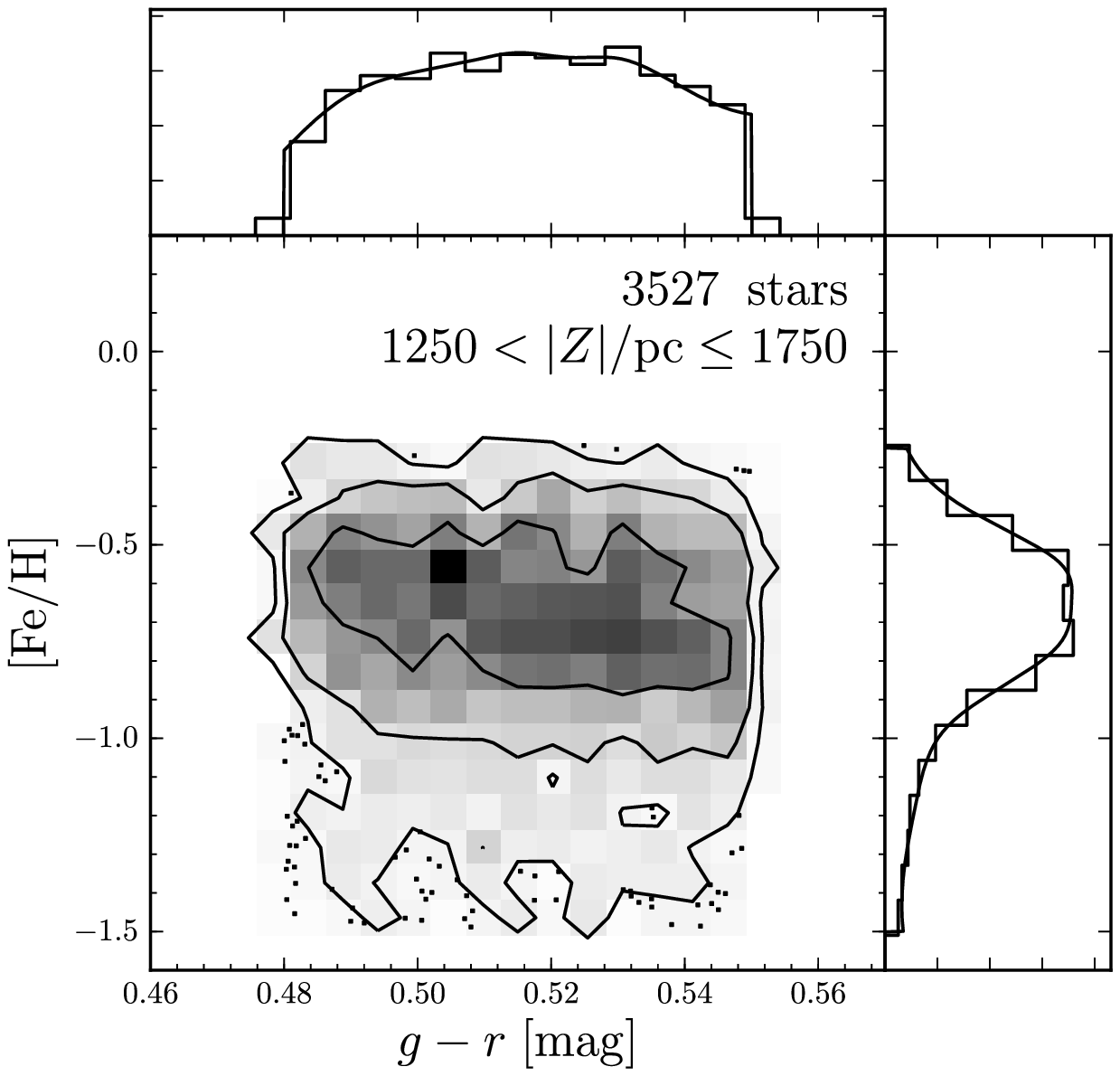}
\includegraphics[width=0.228\textwidth,clip=]{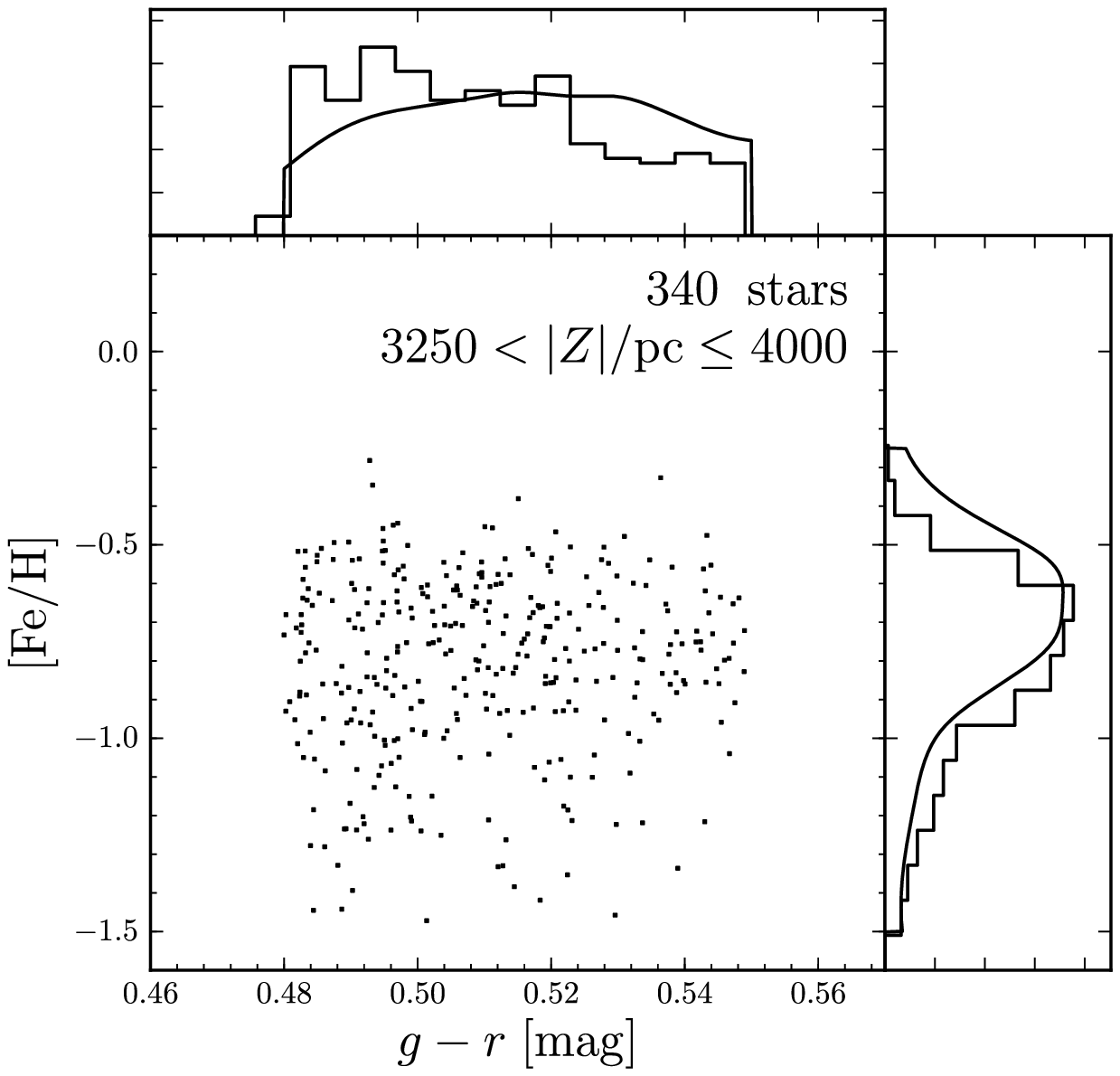}\\
\includegraphics[width=0.26\textwidth,clip=]{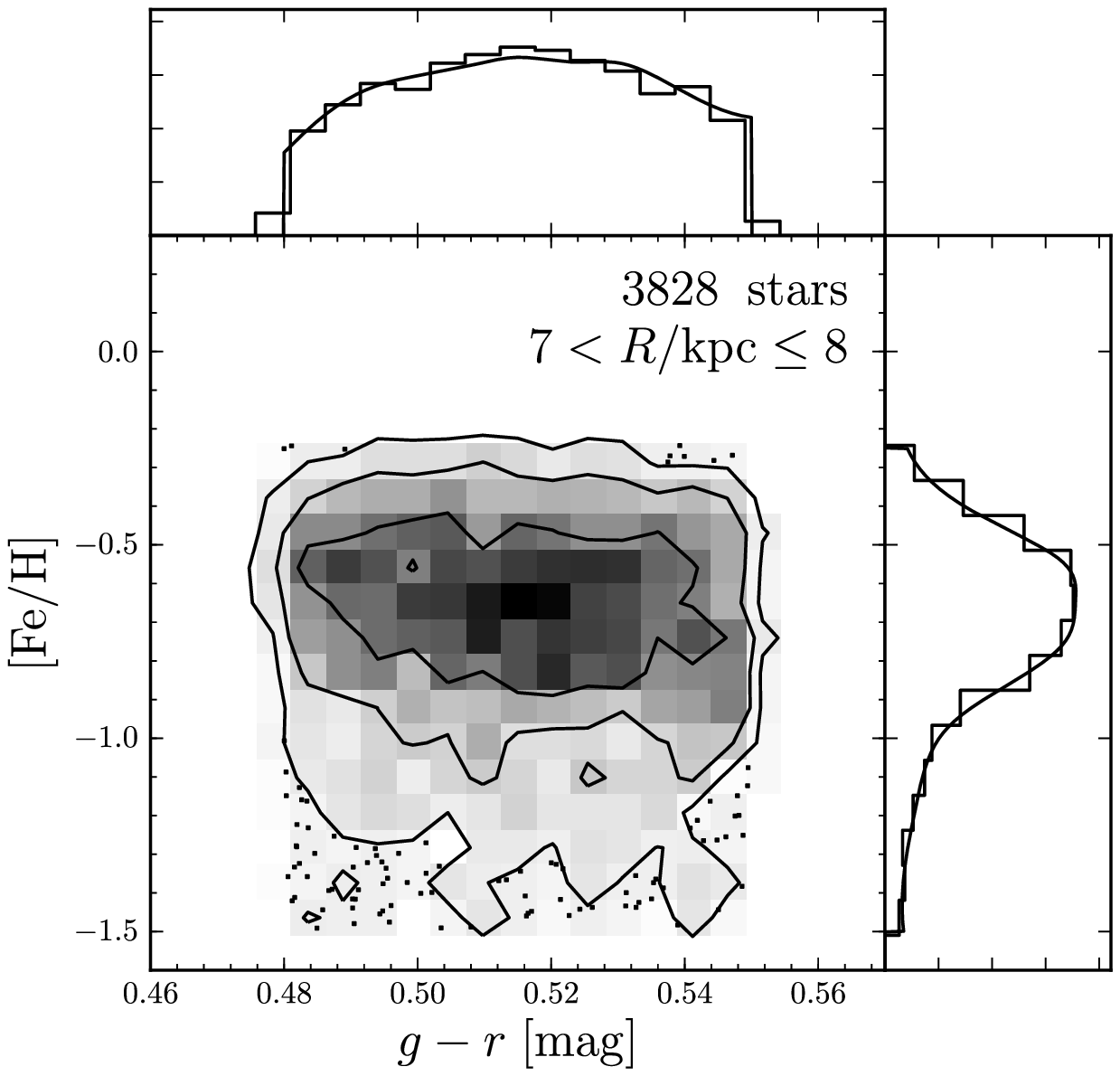}
\includegraphics[width=0.228\textwidth,clip=]{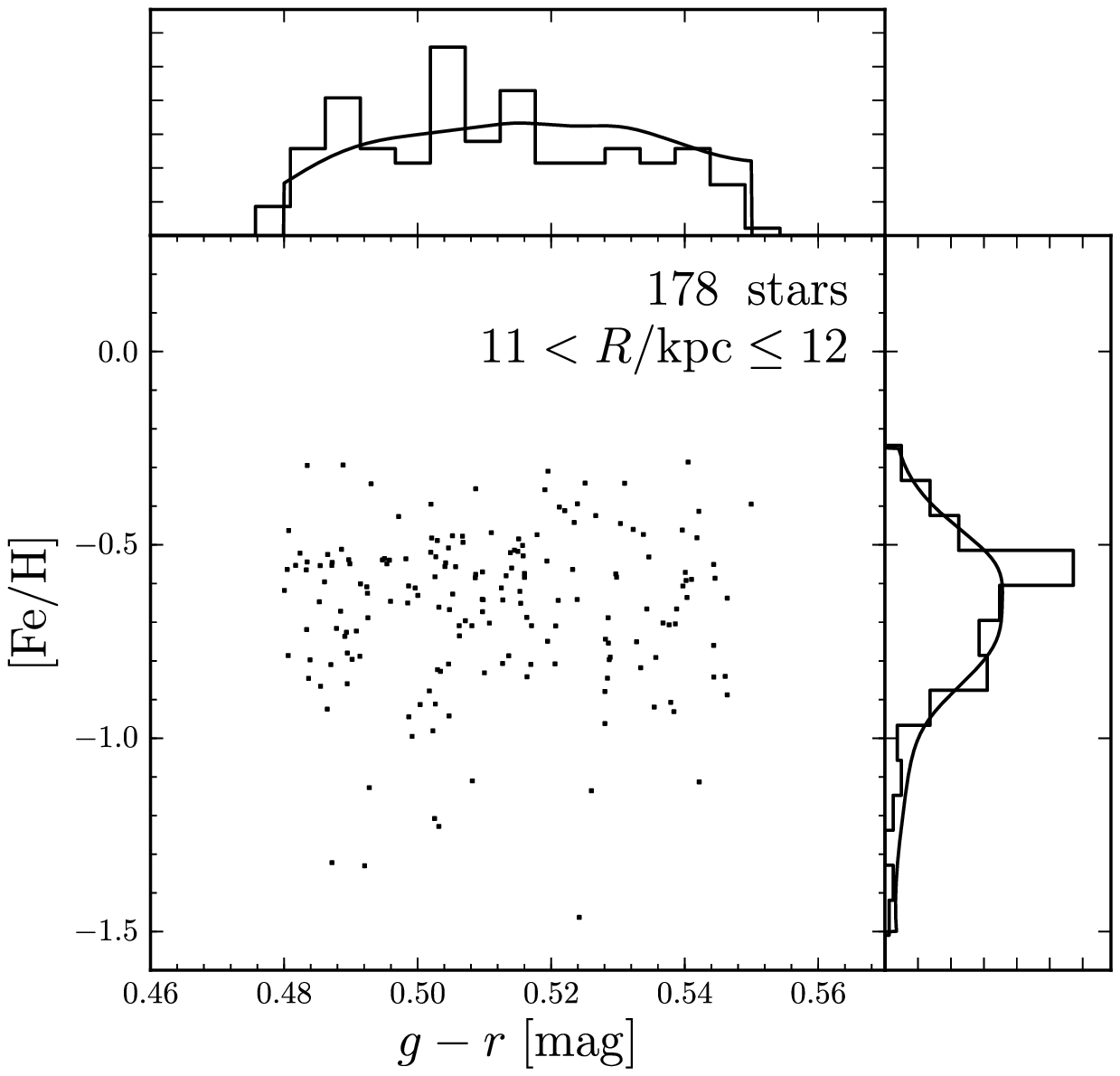}
\includegraphics[width=0.228\textwidth,clip=]{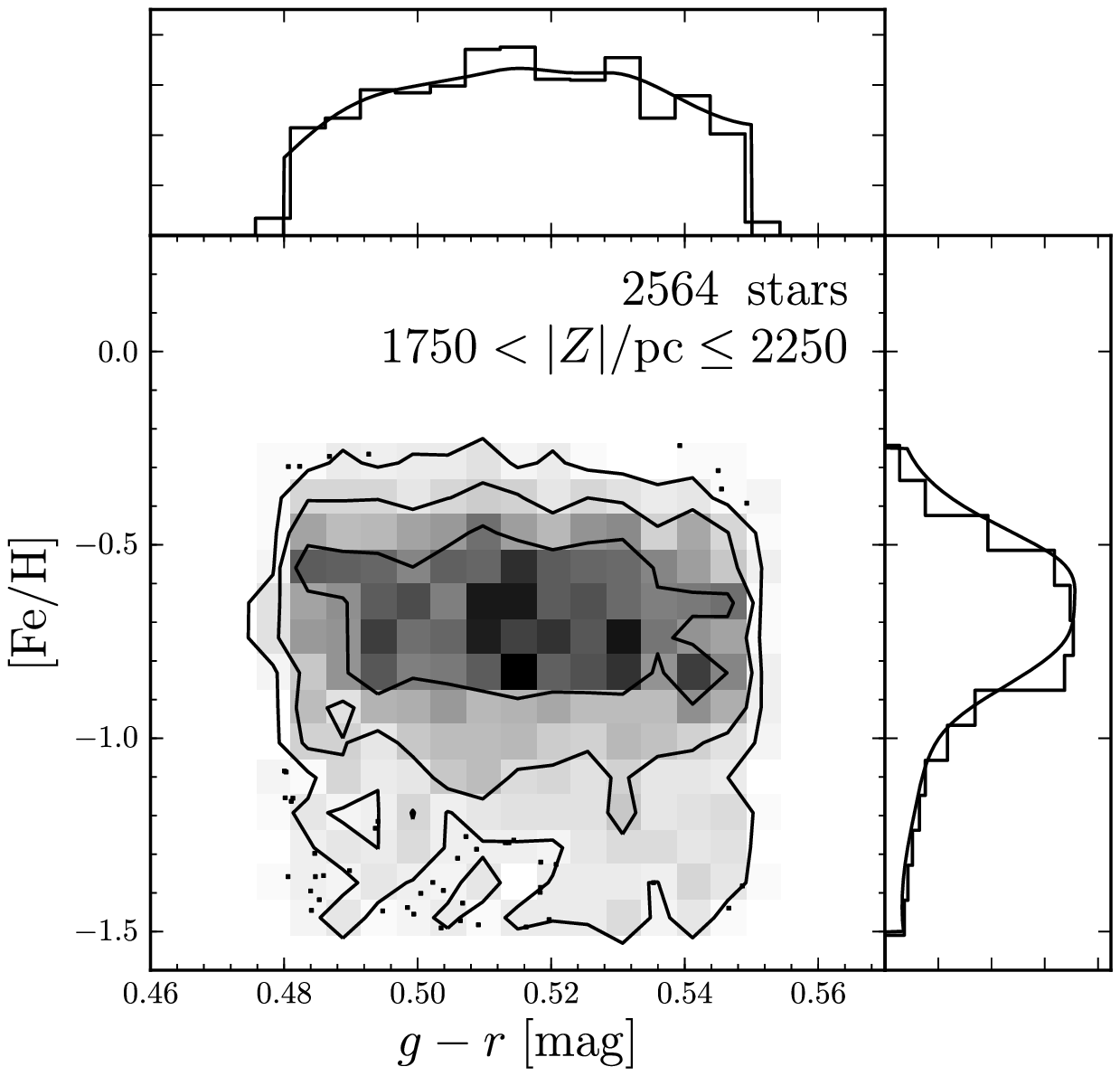}
\includegraphics[width=0.2335\textwidth,clip=]{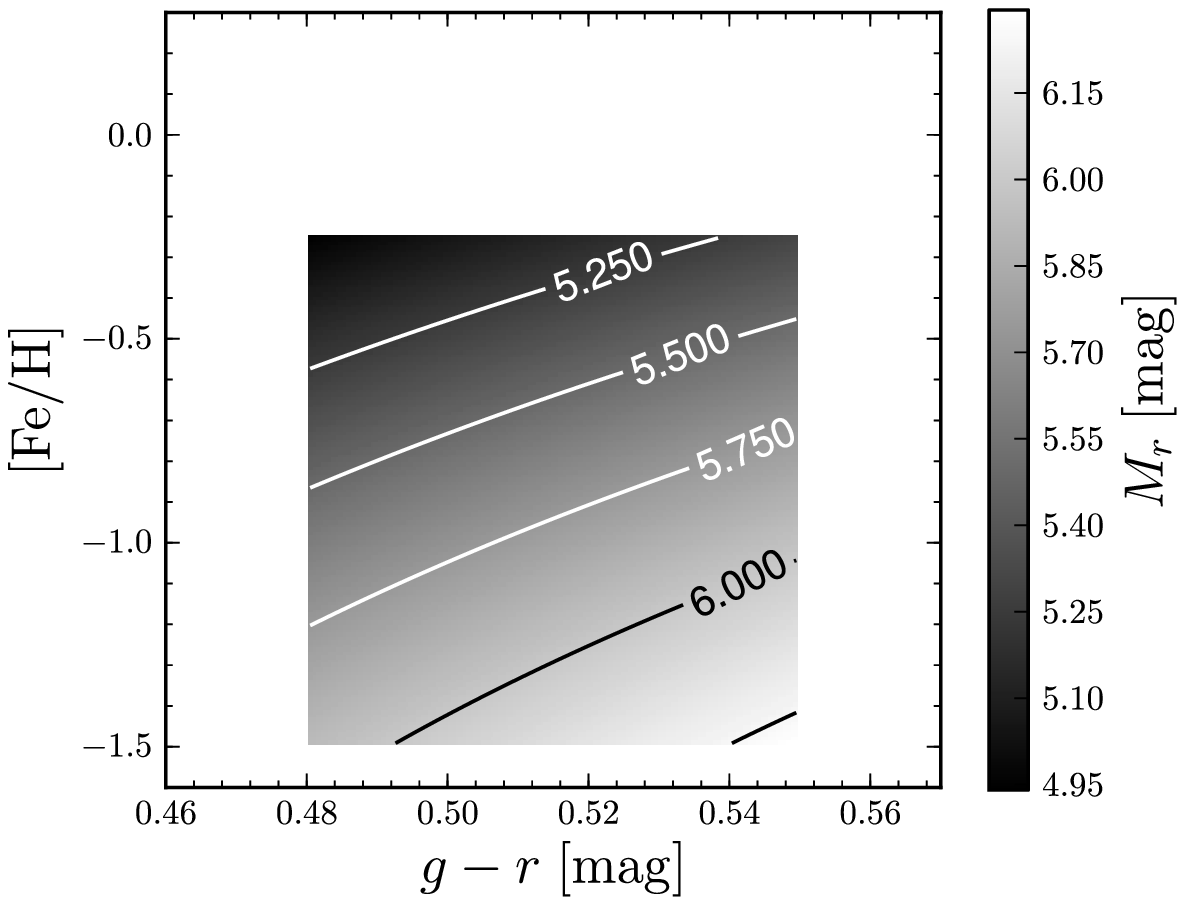}
\caption{Same as \figurename~\ref{fig:FeH_RZ_rich_g}, but for
the \aenhanced\ G-dwarf sample.}\label{fig:FeH_RZ_poor_g}
 \end{figure}

The color--metallicity distribution for the \apoor\ and \aenhanced\
sample is shown in \figurenames~\ref{fig:FeH_RZ_rich_g} and
\ref{fig:FeH_RZ_poor_g}, respectively. The top-left panel shows the distribution for the entire sample; the remaining panels show the
color--metallicity distribution as a function of Galactocentric radius
(including all vertical heights) and as a function of vertical height
(including all Galactocentric radii). For both samples, the
color--metallicity distribution separates into the product of
one-dimensional color and metallicity distributions, thus we
assume that $\rho(g-r,\feh|R,Z)
=\rho^c(g-r|R,Z)\,\rho^{\feh}(\feh|R,Z)$. The $g-r$ distribution is
independent of $R$ and $Z$ for both the \apoor\ and the \aenhanced\
sample; we use a spline interpolation of the color
distribution for the full sample for $\rho^c(g-r|R,Z)$, independent of
$R$ and $Z$. This interpolation is shown in the top histogram in all
panels of \figurenames~\ref{fig:FeH_RZ_rich_g} and
\ref{fig:FeH_RZ_poor_g}. The metallicity distribution of the
\aenhanced\ sample is also mostly independent of $R$ and $Z$, with
only a hint of a trend toward a more metal-poor distribution
at large distances from the plane. The \feh\ distribution of the
\apoor\ sample shows expected trends with $R$ and $Z$: The peak of the
metallicity distribution goes from more metal-rich closer to the
Galactic center and closer to the plane, to more metal-poor at larger
Galactocentric radii and at larger $Z$. These shifts are modest
($\lesssim 0.1$ dex), which is partly due to the fact that farther
from the Solar radius we preferentially see stars at larger distances
from the plane. We stress that these metallicity distributions are the
\emph{observed} distributions uncorrected for selection effects, but
selection effects play a minor role and merely shift the overall
distribution by $\approx$ 0.1 dex (Schlesinger \etal, 2011, in
preparation). We investigate the effect of systematically shifting the
metallicity distribution below.

The effect of metallicity and color on the absolute magnitude using
the \citet{Ivezic08a} color--metallicity--magnitude relation is shown
in the bottom right panel of \figurenames~\ref{fig:FeH_RZ_rich_g} and
\ref{fig:FeH_RZ_poor_g}, for the ranges in color and metallicity
considered for both samples. From the blue and metal-rich to the red
and metal-poor end the shift in absolute magnitude is about 1 mag, or a
factor of about 1.6 in distance.

As the \aenhanced\ metallicity distribution depends only weakly on $R$ and
$Z$, we will assume that it is constant, and use a spline
interpolation of the \feh\ distribution of the full sample as our
model for $\rho^{\feh}(\feh|R,Z)$. We do the same for the \apoor\
sample, even though there are slight trends with $R$ and $Z$. These models are shown in the right histograms of all panels
in \figurenames~\ref{fig:FeH_RZ_rich_g} and
\ref{fig:FeH_RZ_poor_g}. We can then simplify the normalization
integral in \eqnname~(\ref{eq:normint2}) further to
\begin{equation}\label{eq:normint3}
\begin{split}
\int & \dd l\,\dd b\,\dd r\, \dd d\, \dd r\, \dd (g-r) \,\dd \feh \,\lambda(l,b,d,r,g-r,\feh|\theta) \\
&= A_p\,\sum_{\mathrm{plates}\ p}
\int \dd (g-r)\,\rho^c(g-r|R,Z) \int\dd\feh\,\rho^{\feh}(\feh|R,Z) \\
& \,\times 
\int_{d[r_{\mathrm{min}},g-r,\feh]}^{d[r_{\mathrm{max}},g-r,\feh]} \dd d\,
S(p,r[g-r,\feh,d],g-r)\,d^2\,\dens(R,z|l,b,d,\theta)\,.
\end{split}
\end{equation}
If we then determine the overall minimum and maximum heliocentric
distance at which we can observe stars in both samples, we can calculate
the inner integral between these limits, with the understanding that
the selection function is zero outside of the apparent-magnitude range
of the plate in question (since bluer or more metal-rich stars can
only be observed at distances starting at a value that is larger than
the overall minimum distance, and redder and more metal-poor stars can
only be seen out to distances that fall short of the overall maximum
distance, because of the color and metallicity dependence of the
photometric distance method). We can then calculate the integral by
summation on a regular grid as
\begin{equation}\label{eq:normint4}
\begin{split}
\int & \dd l\,\dd b\,\dd r\,\dd d\, \dd (g-r) \,\dd \feh \,\lambda(l,b,d,r,g-r,\feh|\theta) \\
&= A_p\,\sum_{\mathrm{plates}\ p}
\sum_{d}\,d^2\,\dens(R,z|l,b,d,\theta)\,\\
& \qquad \sum_{g-r} \,\sum_{\feh}\,\rho^c(g-r)\,\rho^{\feh}(\feh)\,
S(p,r[g-r,\feh,d],g-r)\,,
\end{split}
\end{equation}
where the distance summation is between the overall minimum and
maximum distance. We dropped integration factors $\Delta d$,
$\Delta(g-r)$, and $\Delta \feh$, as these only contribute terms that
do not depend on the parameters $\theta$ in the log likelihood in
\eqnname~(\ref{eq:densitylike2}) (note that they \emph{do} contribute
when we do not marginalize over the amplitude of the density in
\eqnname~[\ref{eq:densitylike}]). Written in this way, this
normalization integral can be computed efficiently, as all of the
necessary coordinate transformations, selection function evaluations,
and color--metallicity-distribution function calls can be pre-computed
on a dense grid.

\section{Detailed data versus model comparisons}\label{sec:datamodel}

In this \appendixname\ we present detailed comparisons of our best-fit
density models with the observed data, as ultimately the best-fit
density parameters are constrained through the quality of the fit in
the natural coordinates of the spectroscopic data
($l,b,r,g-r,\feh$). We also show that the results we obtain for
different sub-samples of our nominal samples are consistent with the
best fits for the full samples. As we fit density models by forward
modeling the underlying density model, \ie, by taking the spatial
density and running it through the \segue\ selection function and the
photometric distance relation, we cannot show direct maps of the
density in any meaningful way without massaging the data
excessively. Therefore, we compare the observed star counts with the
best-fit model by running the underlying star counts model through the
selection function and photometric magnitude--color--metallicity
relation, and then comparing it with the observed star counts. This
has the added advantage that it shows that the entire framework of (a)
the underlying density, (b) the photometric
magnitude--color--metallicity relation, and (c) our model of the
\segue\ selection function provides a valid description of the
observed data.

\subsection{The \aenhanced\ disk stars}

\begin{figure}[tp]
\includegraphics[width=0.322\textwidth,clip=]{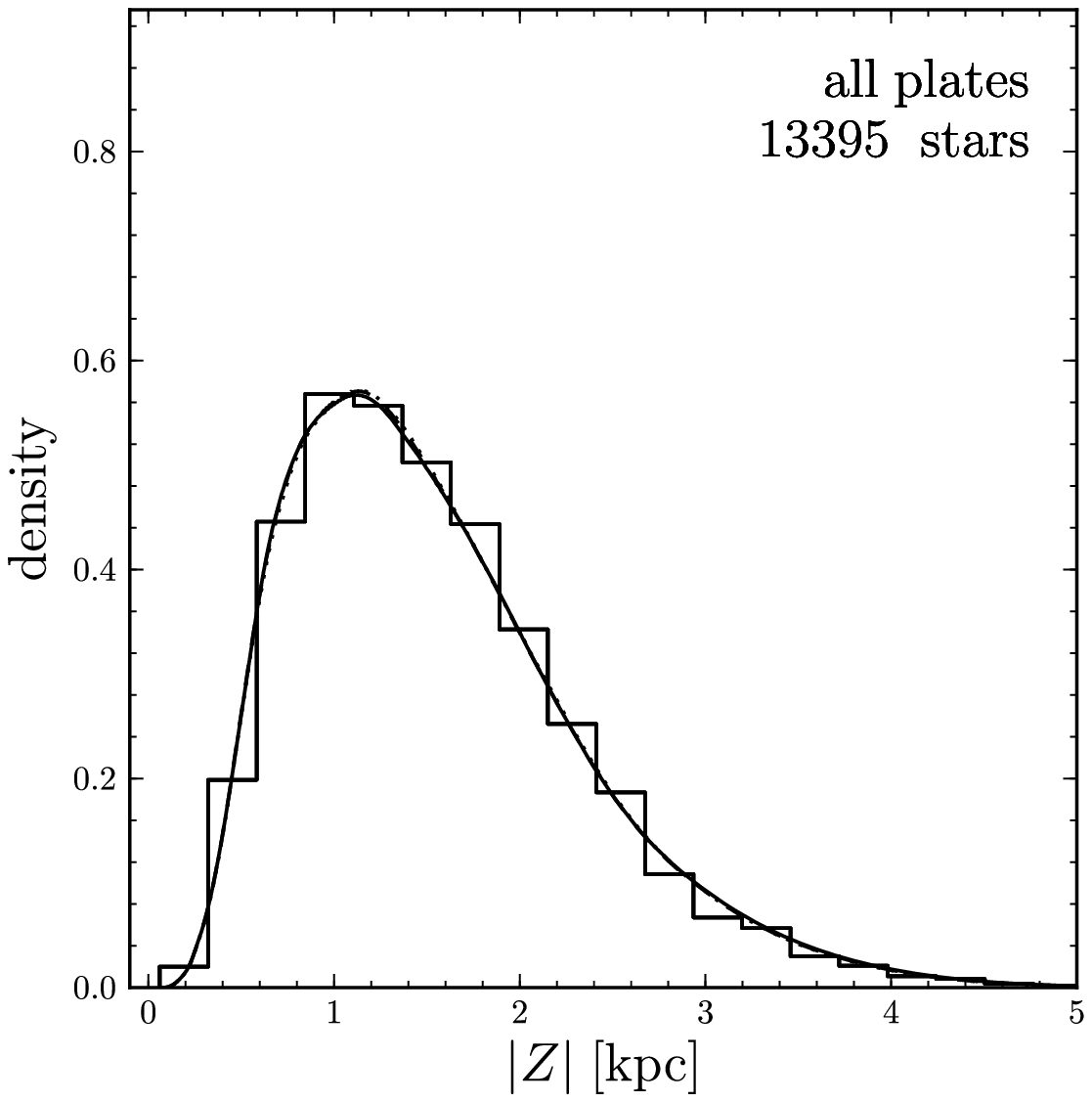}
\includegraphics[width=0.322\textwidth,clip=]{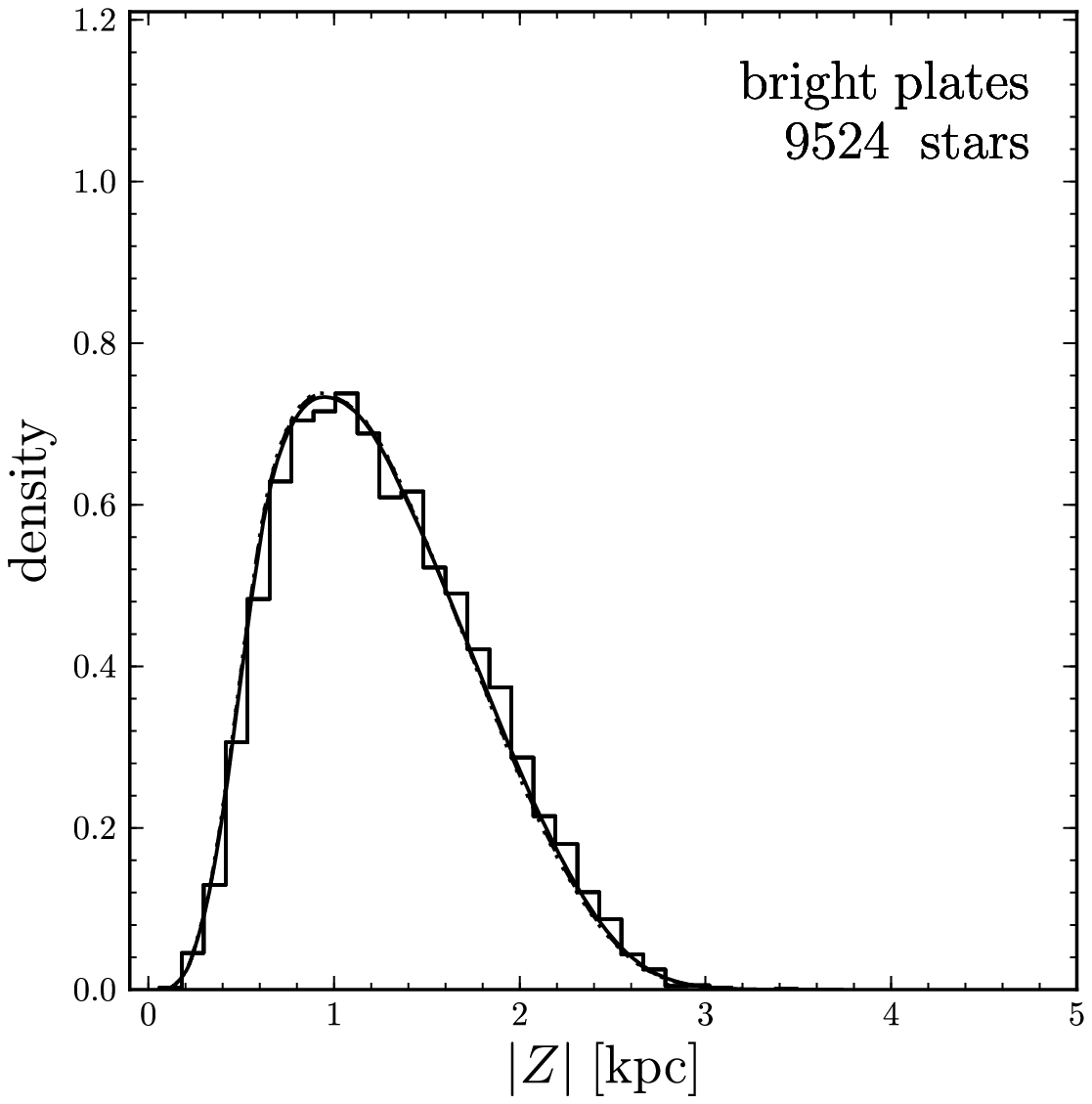}
\includegraphics[width=0.322\textwidth,clip=]{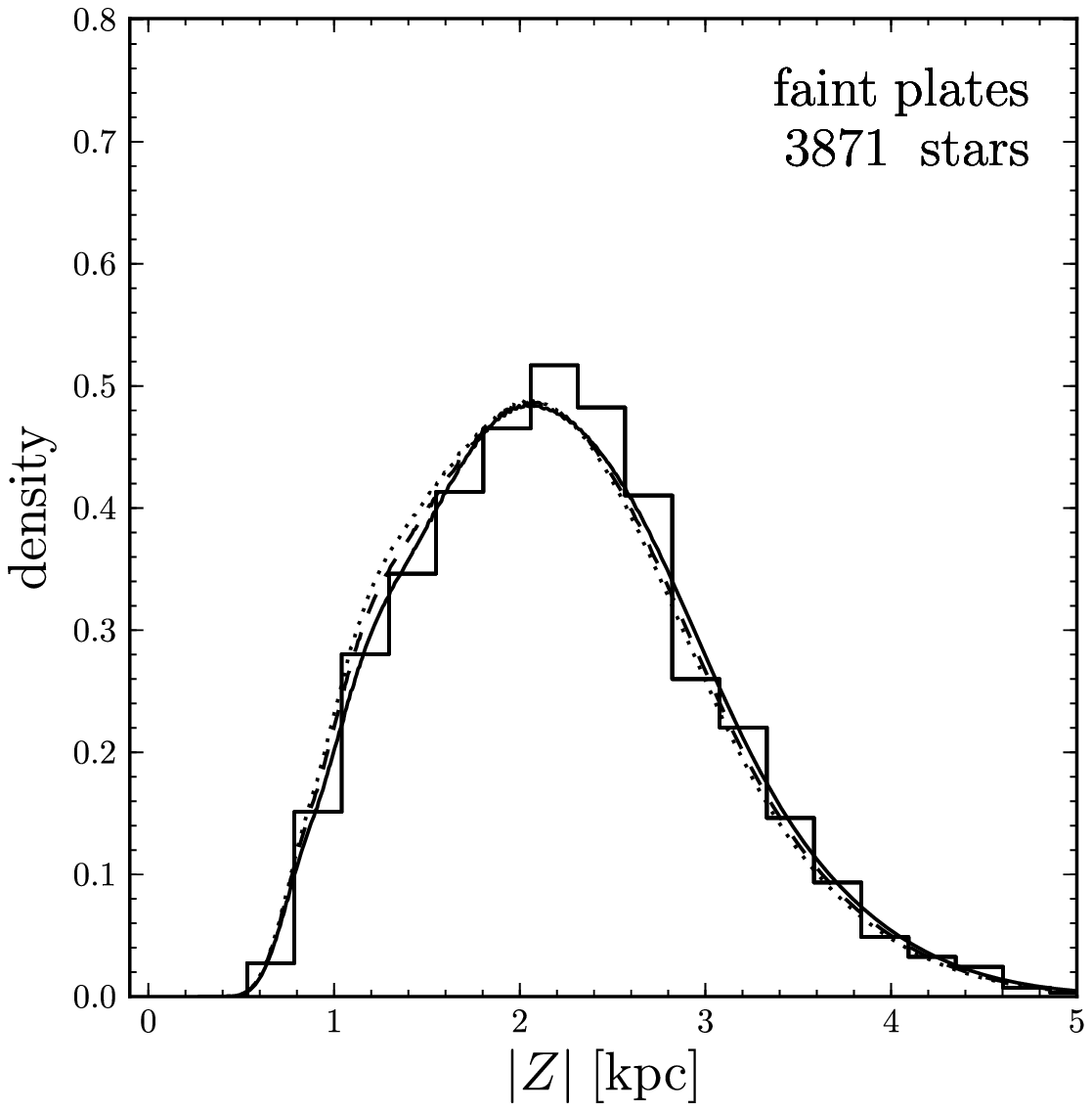}\\
\includegraphics[width=0.322\textwidth,clip=]{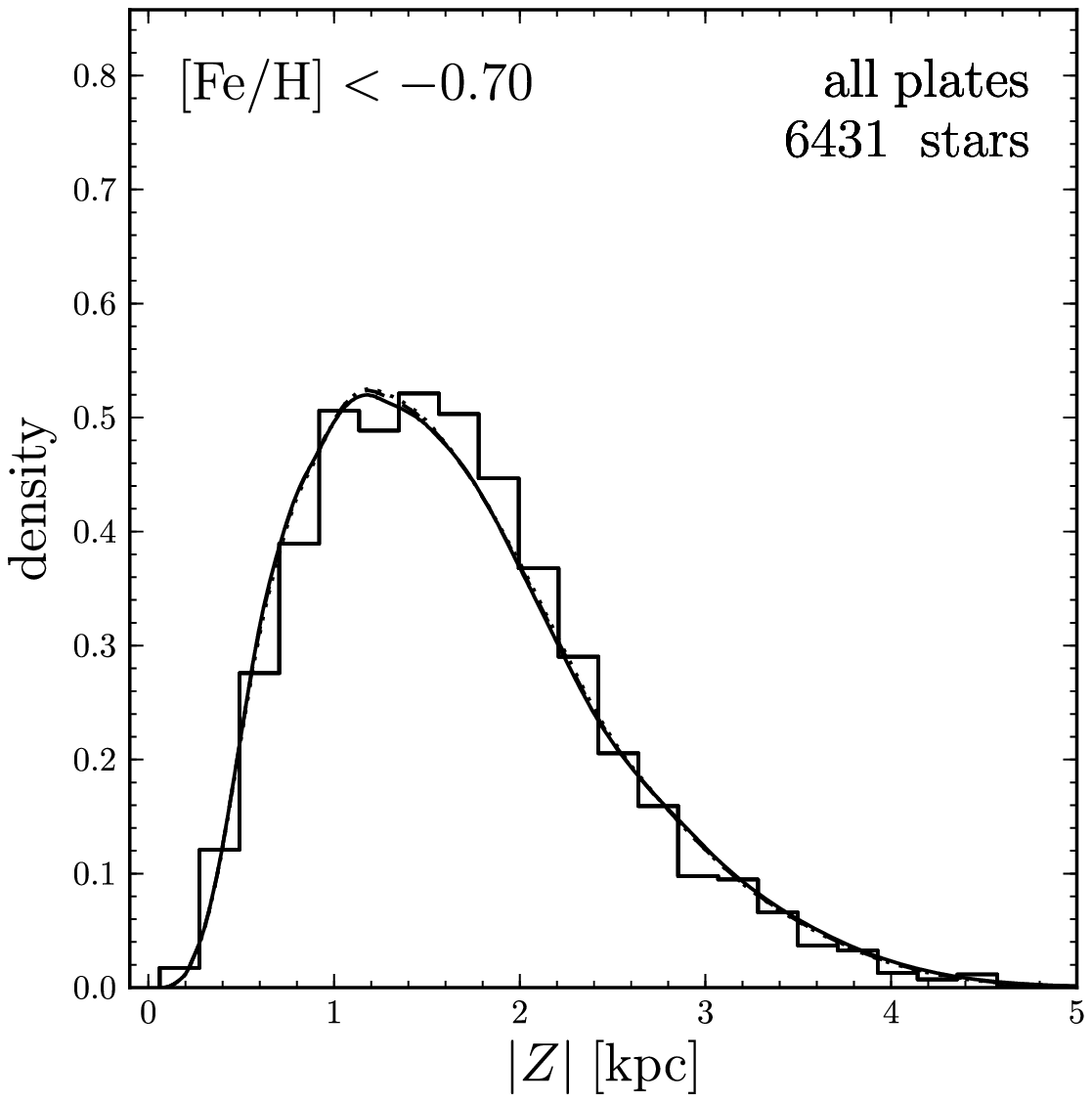}
\includegraphics[width=0.322\textwidth,clip=]{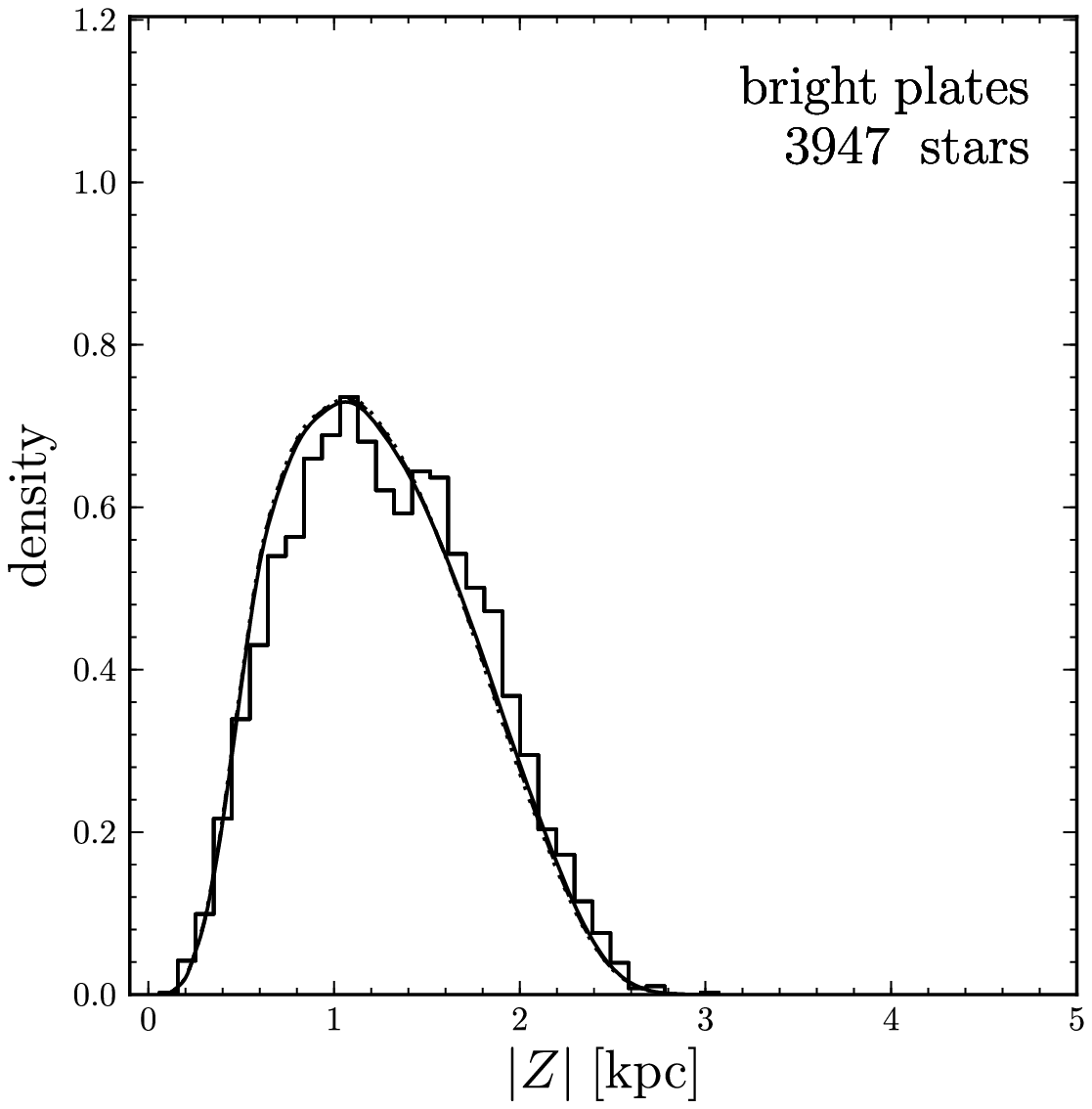}
\includegraphics[width=0.322\textwidth,clip=]{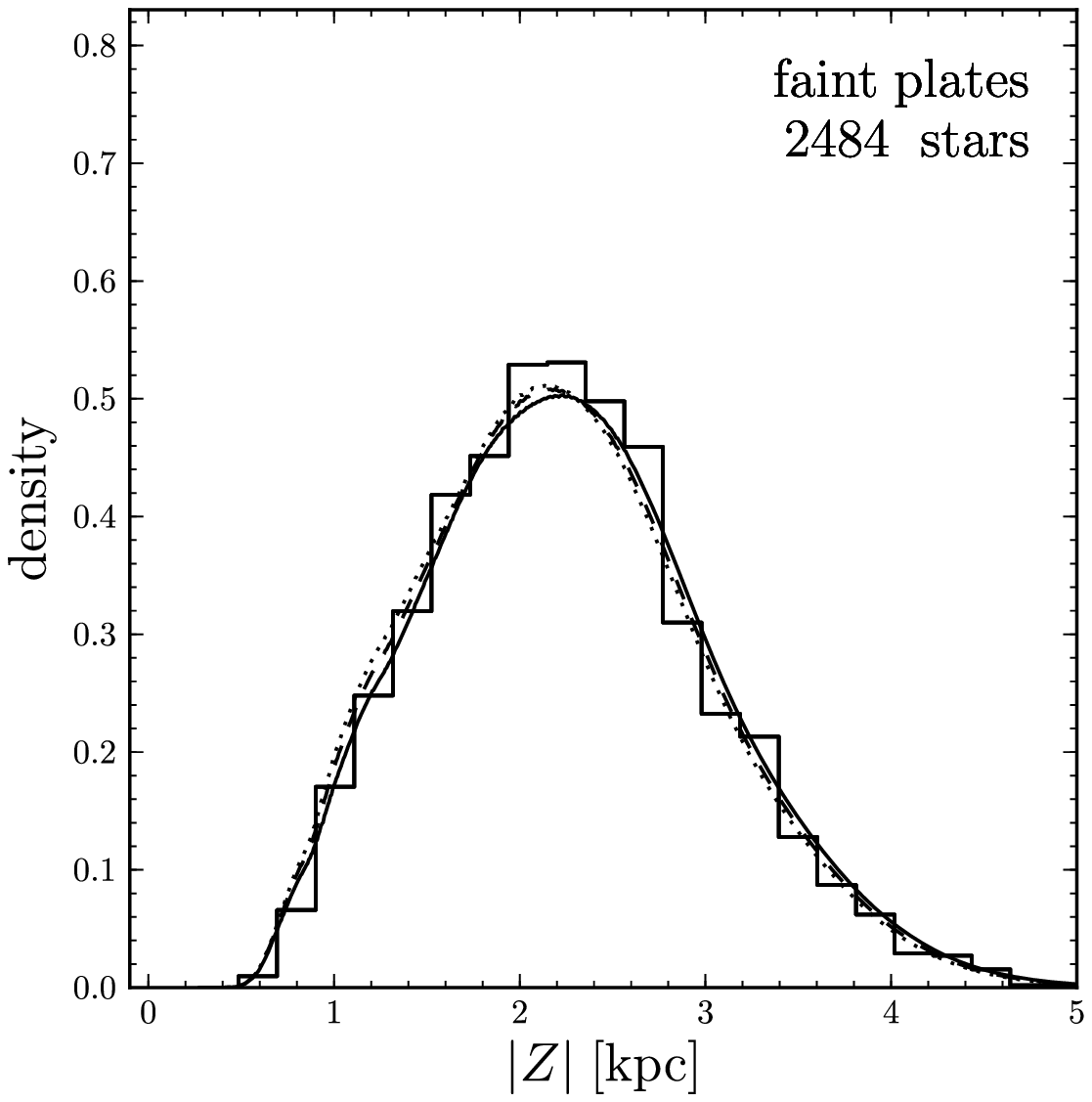}\\
\includegraphics[width=0.322\textwidth,clip=]{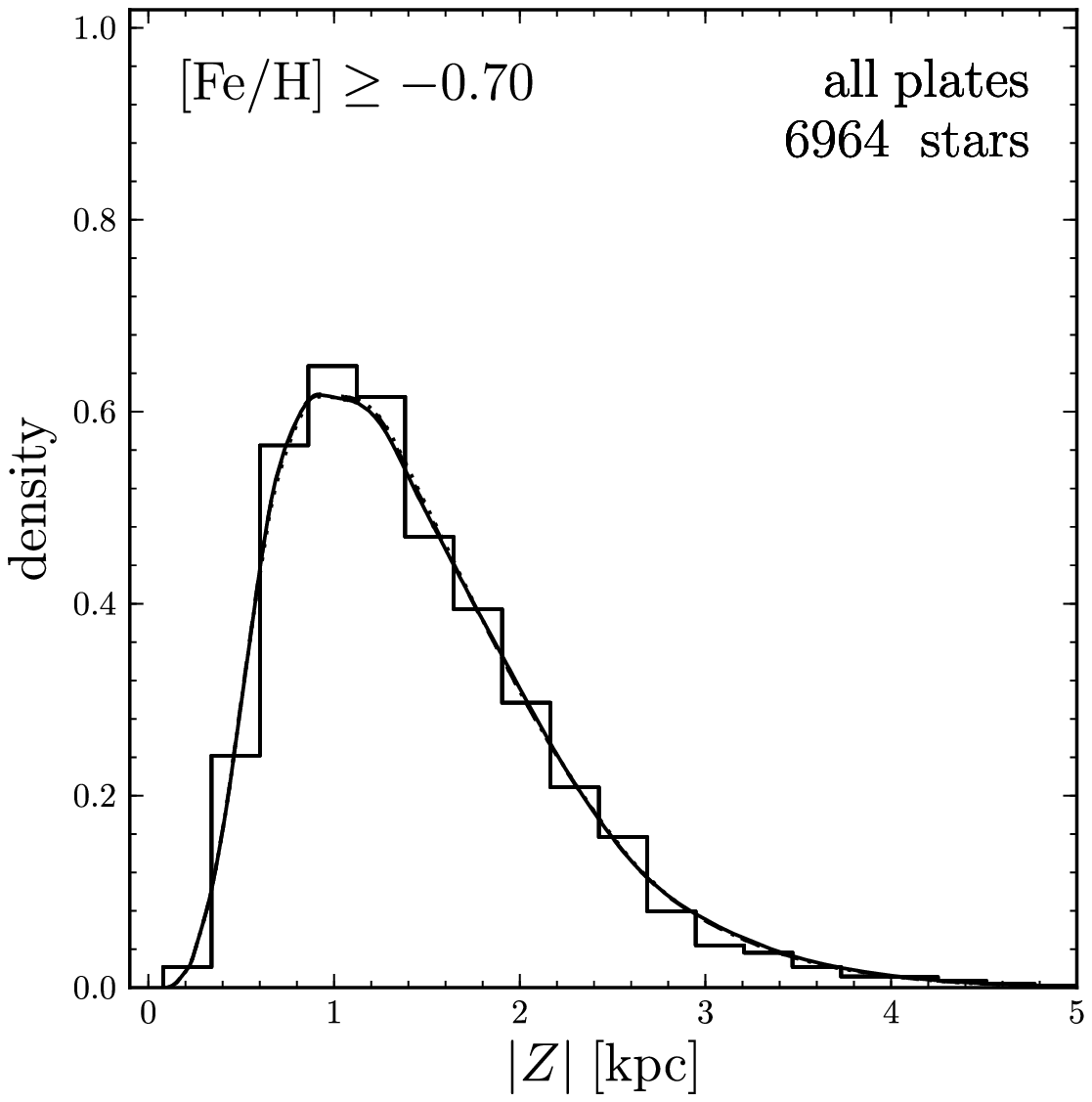}
\includegraphics[width=0.322\textwidth,clip=]{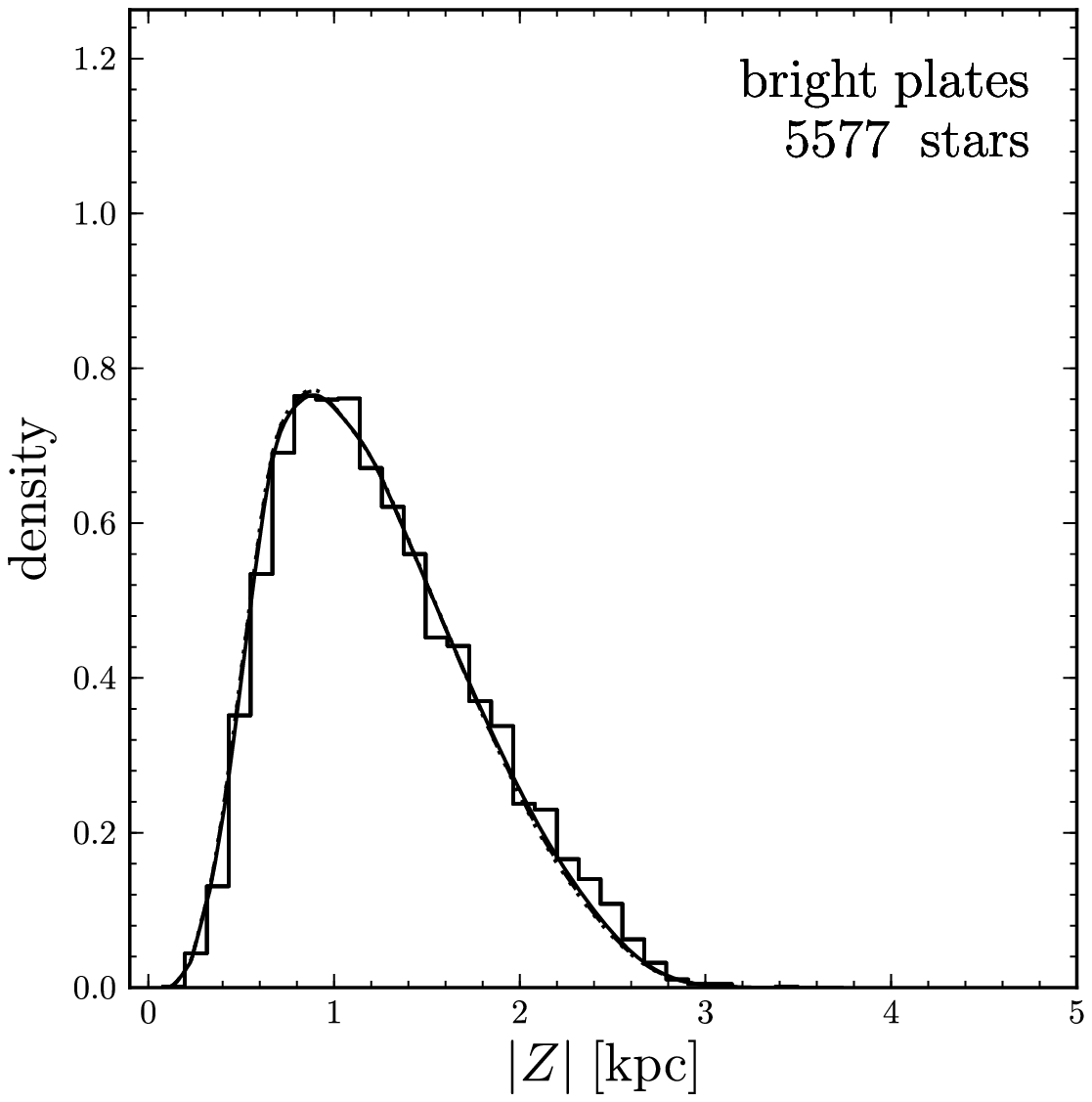}
\includegraphics[width=0.322\textwidth,clip=]{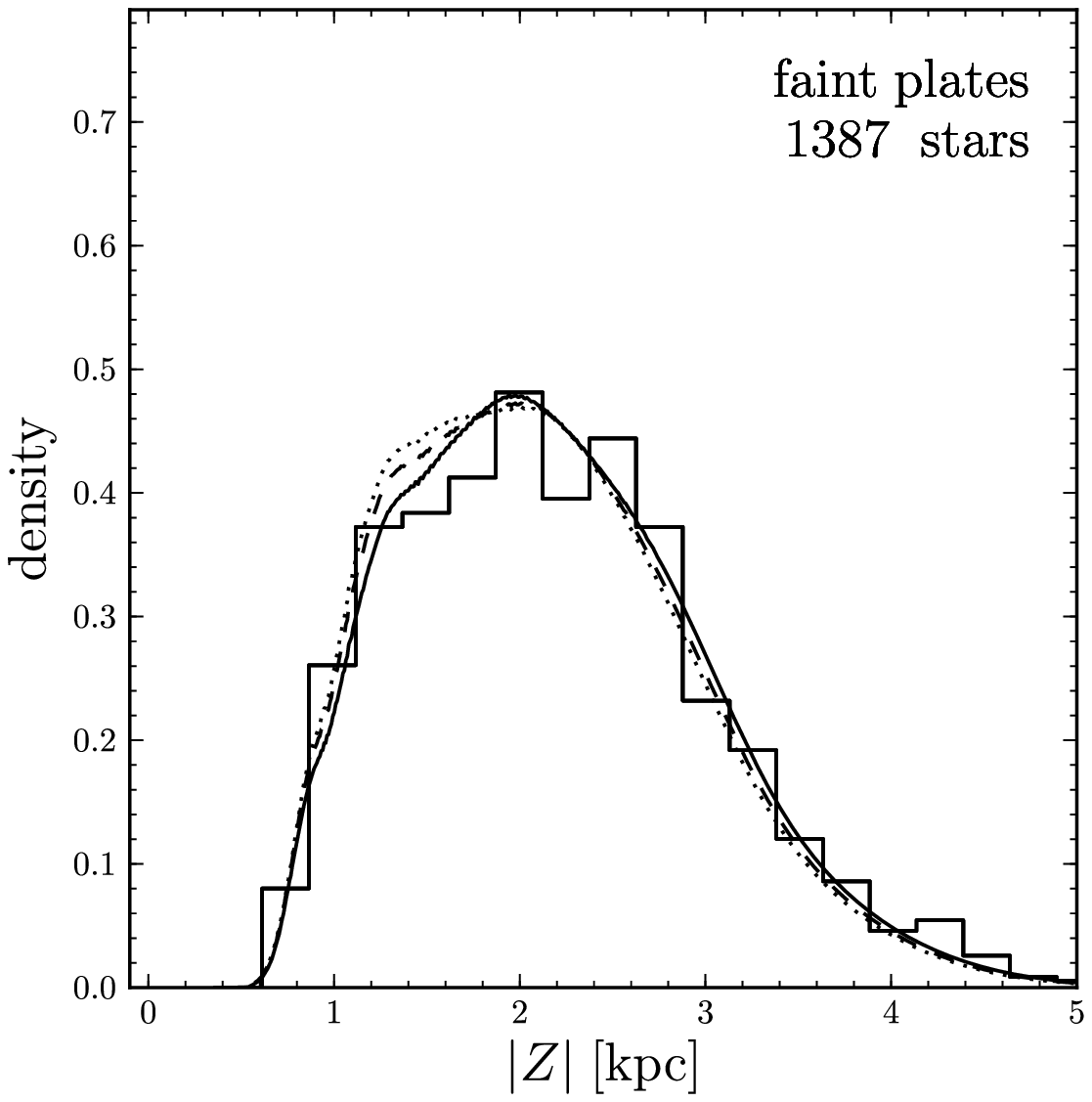}
\caption{Comparison between the observed distribution of vertical
  heights of the \aenhanced\ G-dwarf sample and the distribution predicted by the
  best-fit mixture of two double-exponential disks. The dashed and
  dotted lines show the same model, but with a scale length of 3 and 4
  kpc, respectively. The bottom two rows show the nominal sample used
  in the top row split at \feh\ =
  $-0.7$. See \tablename~\ref{table:poor_results} for the parameters of the
  best-fit models.}\label{fig:model_data_poor_g_zdist}
\end{figure}

\begin{figure}[tp]
\includegraphics[width=0.323\textwidth,clip=]{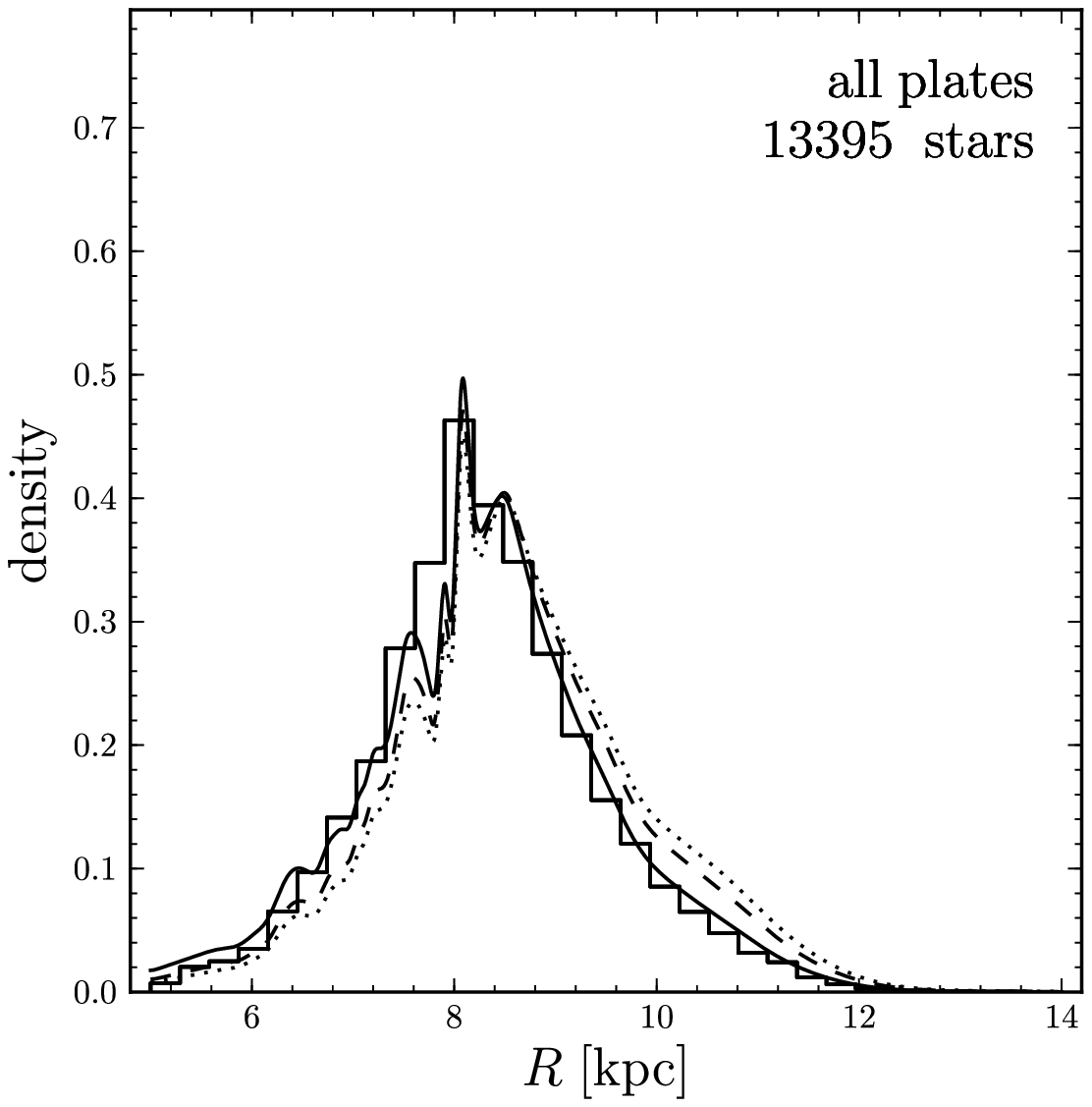}
\includegraphics[width=0.323\textwidth,clip=]{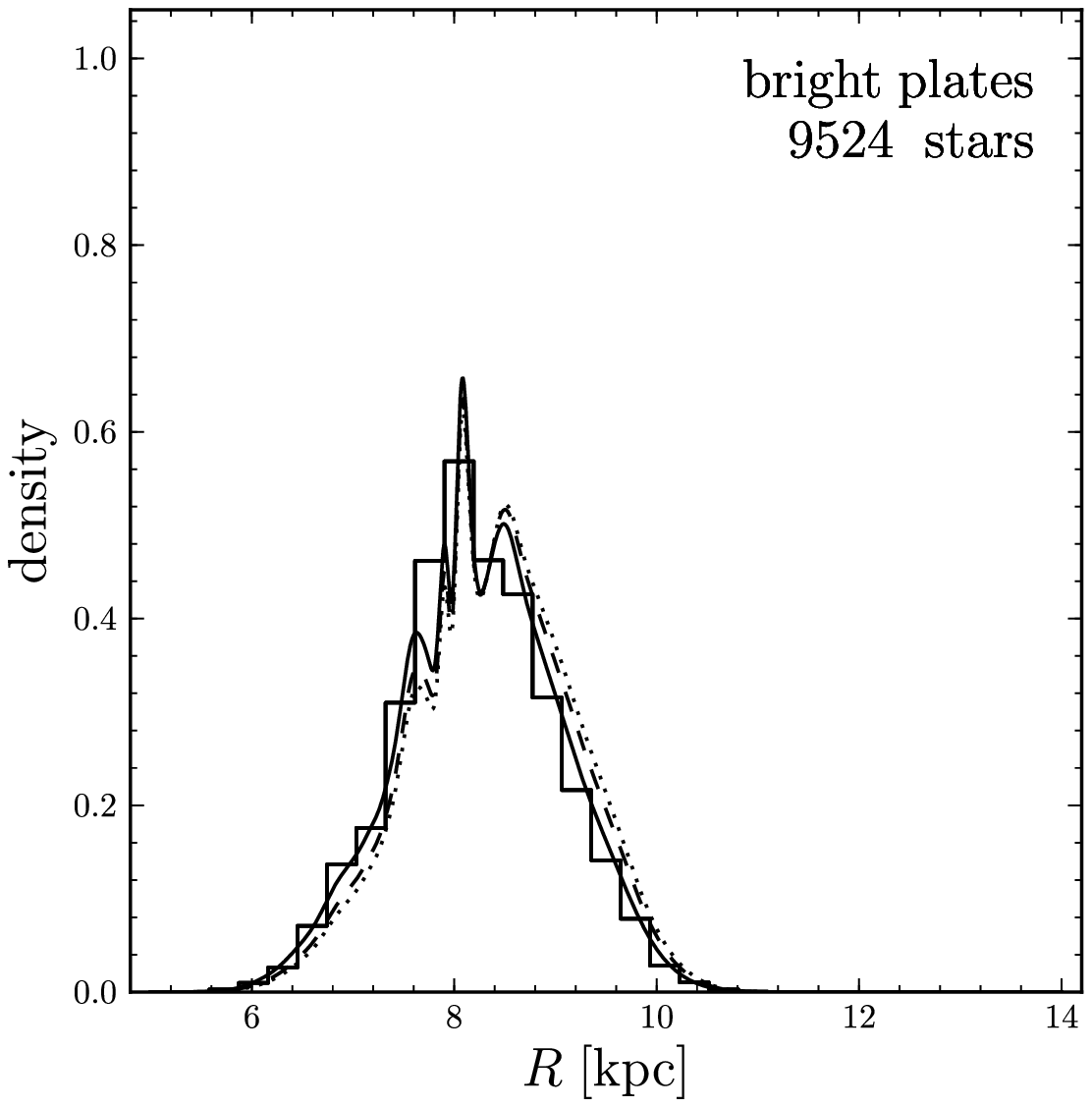}
\includegraphics[width=0.323\textwidth,clip=]{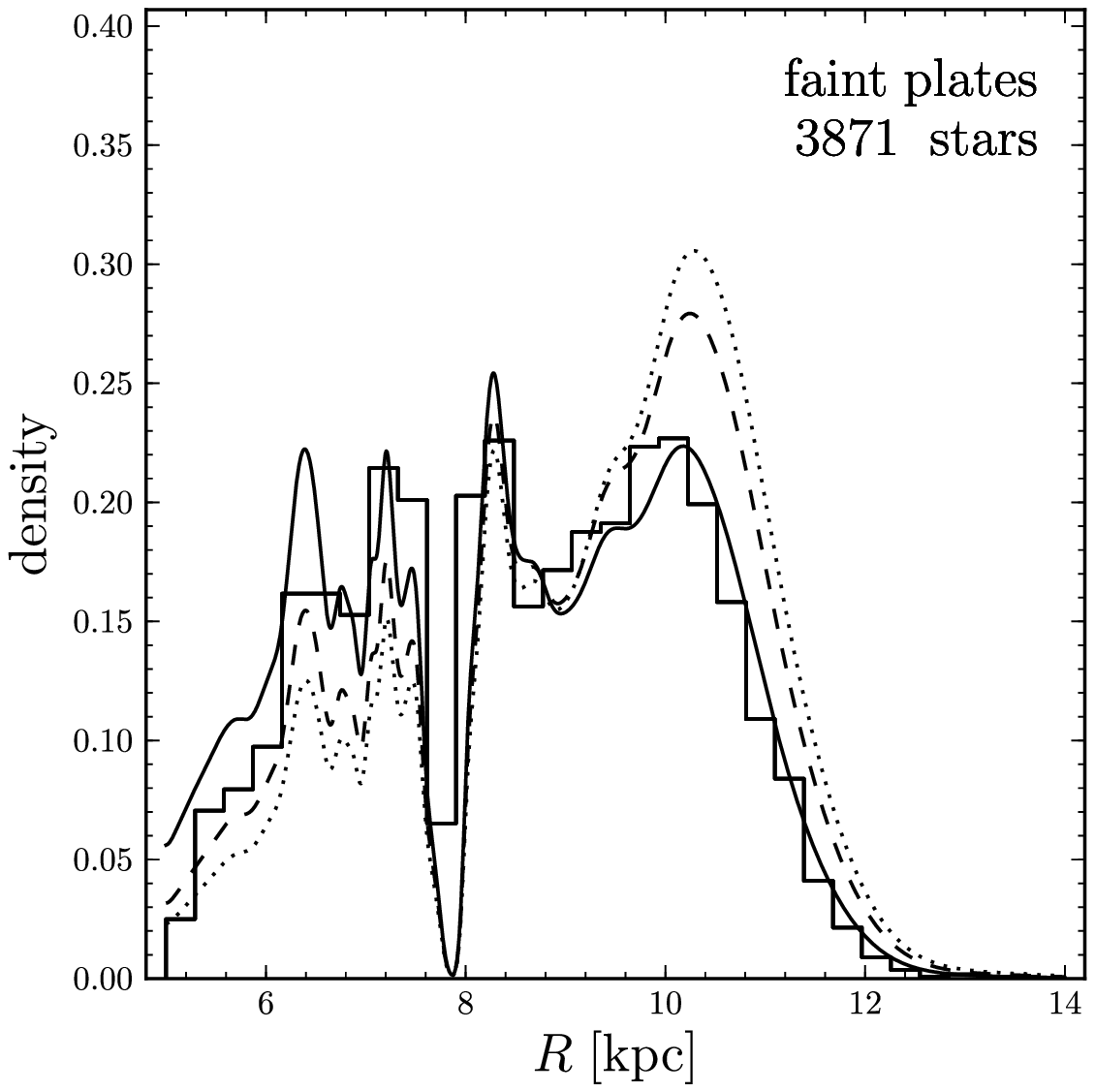}\\
\includegraphics[width=0.323\textwidth,clip=]{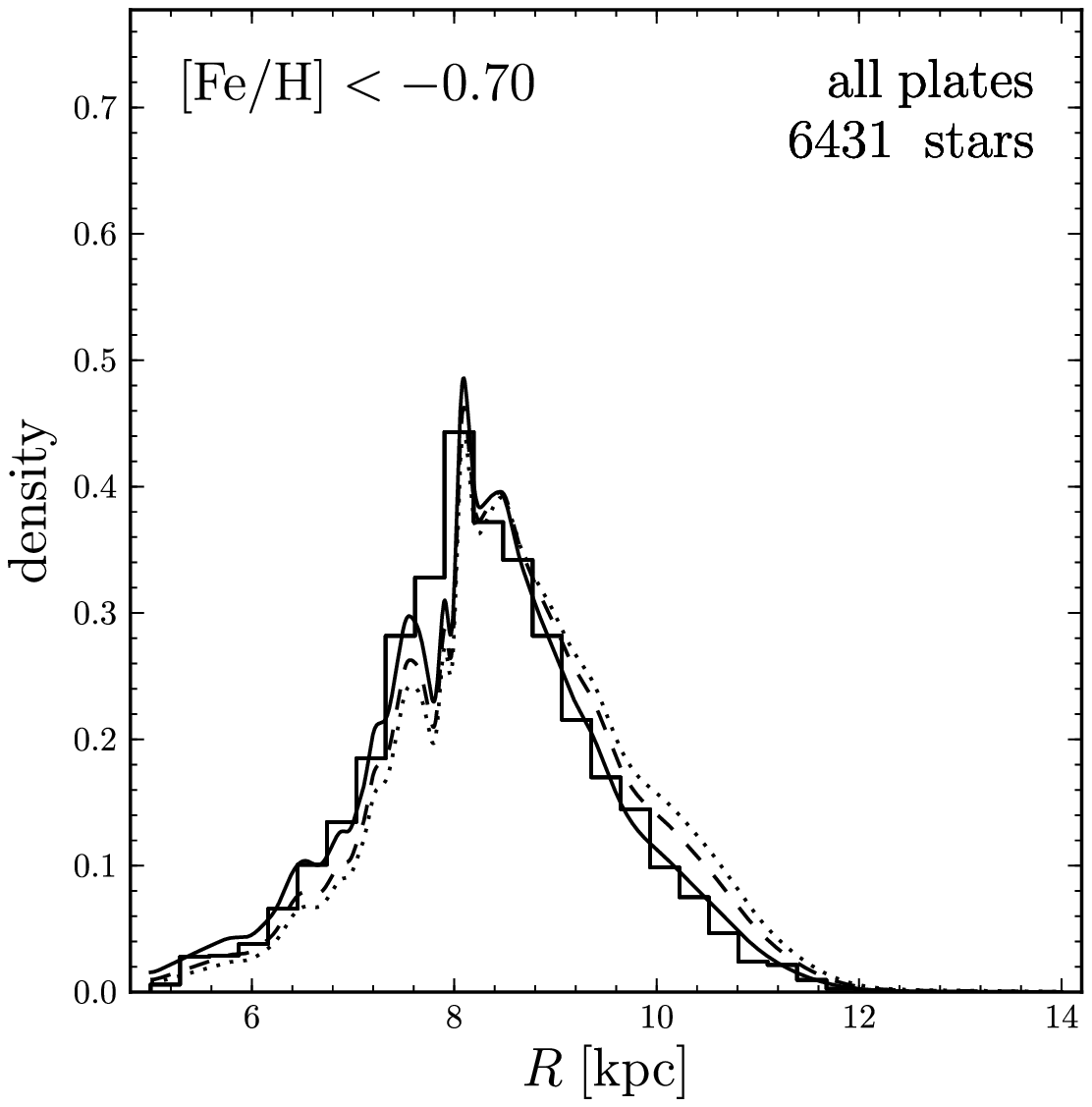}
\includegraphics[width=0.323\textwidth,clip=]{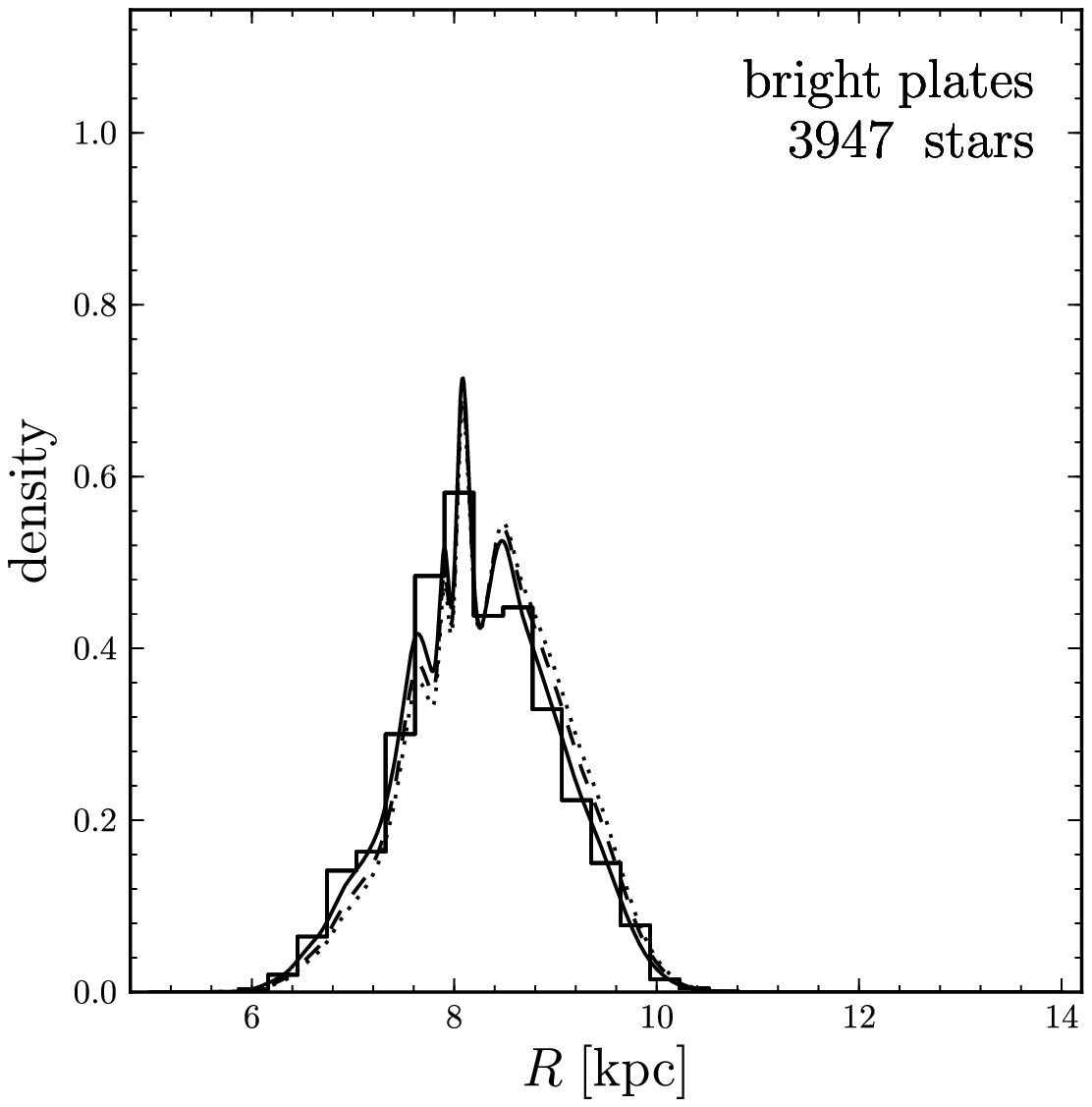}
\includegraphics[width=0.323\textwidth,clip=]{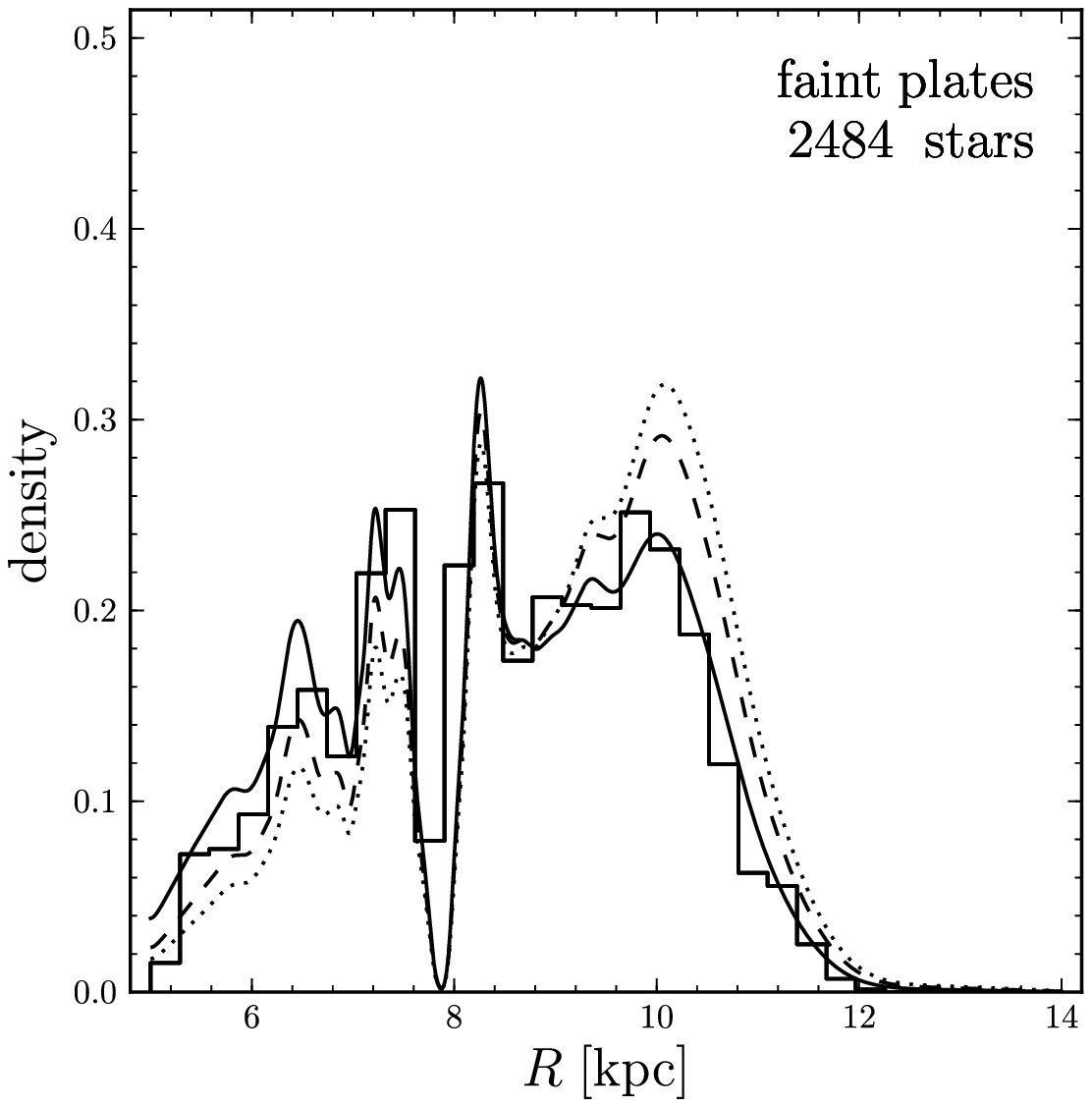}\\
\includegraphics[width=0.323\textwidth,clip=]{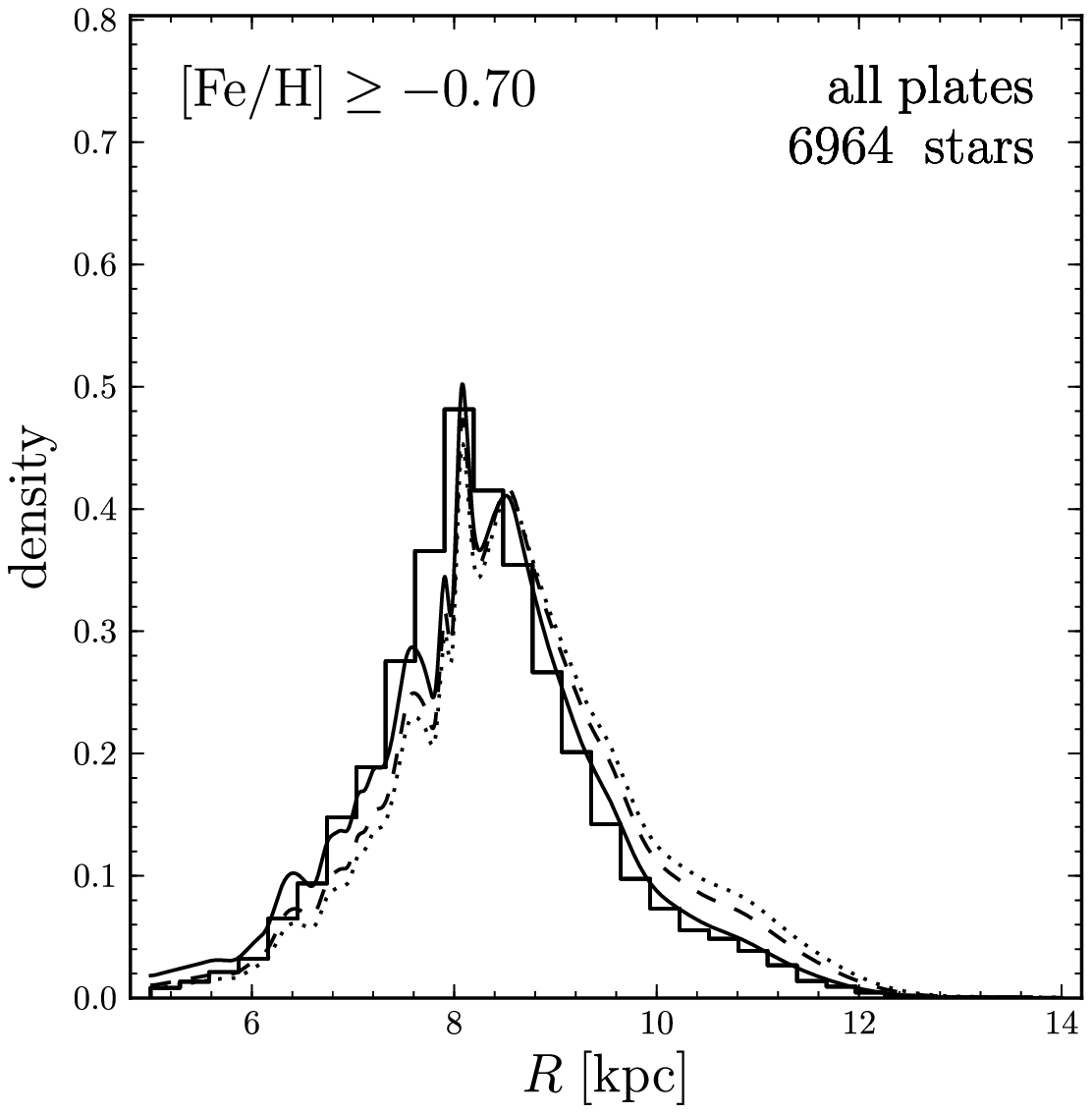}
\includegraphics[width=0.323\textwidth,clip=]{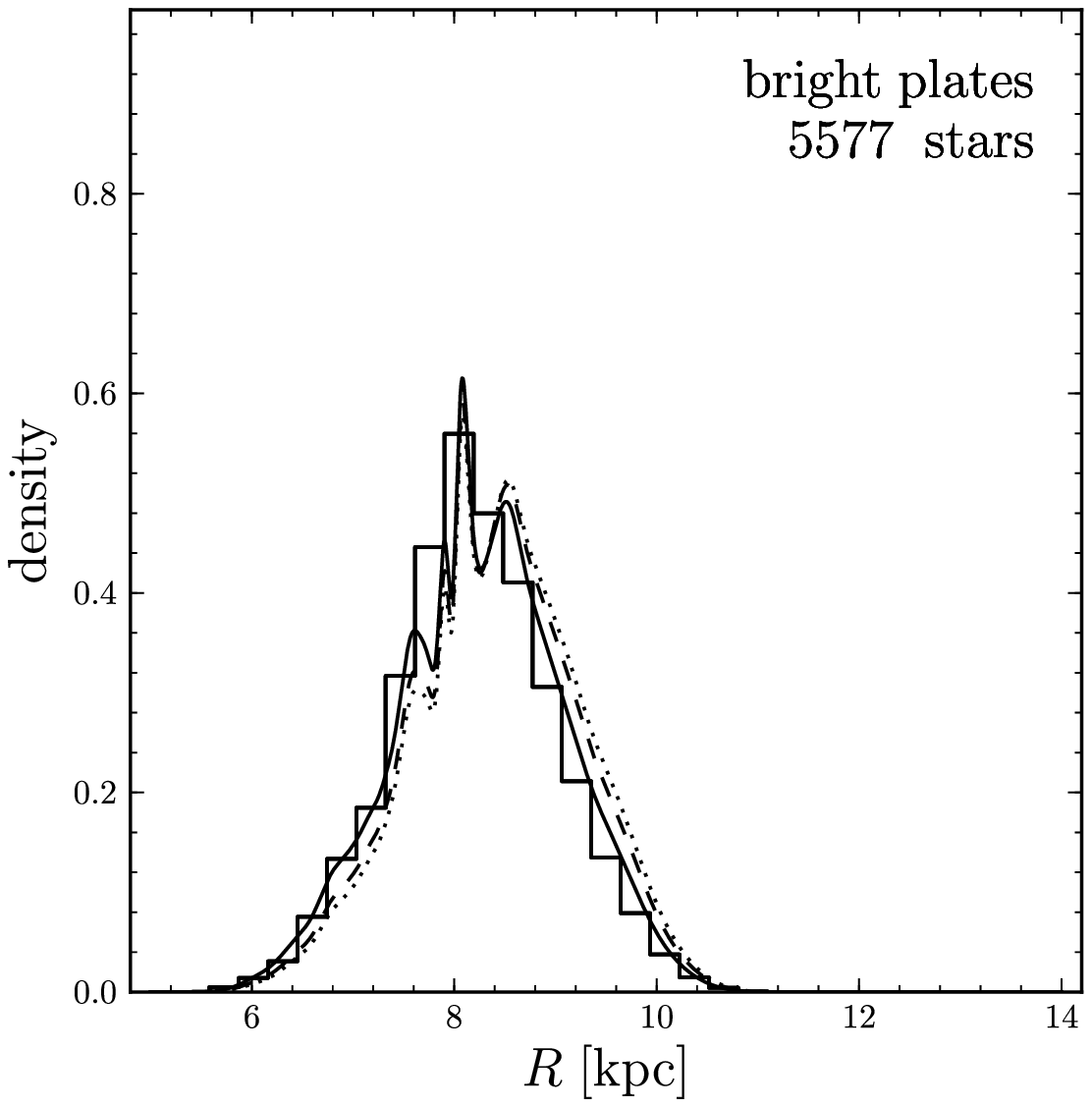}
\includegraphics[width=0.323\textwidth,clip=]{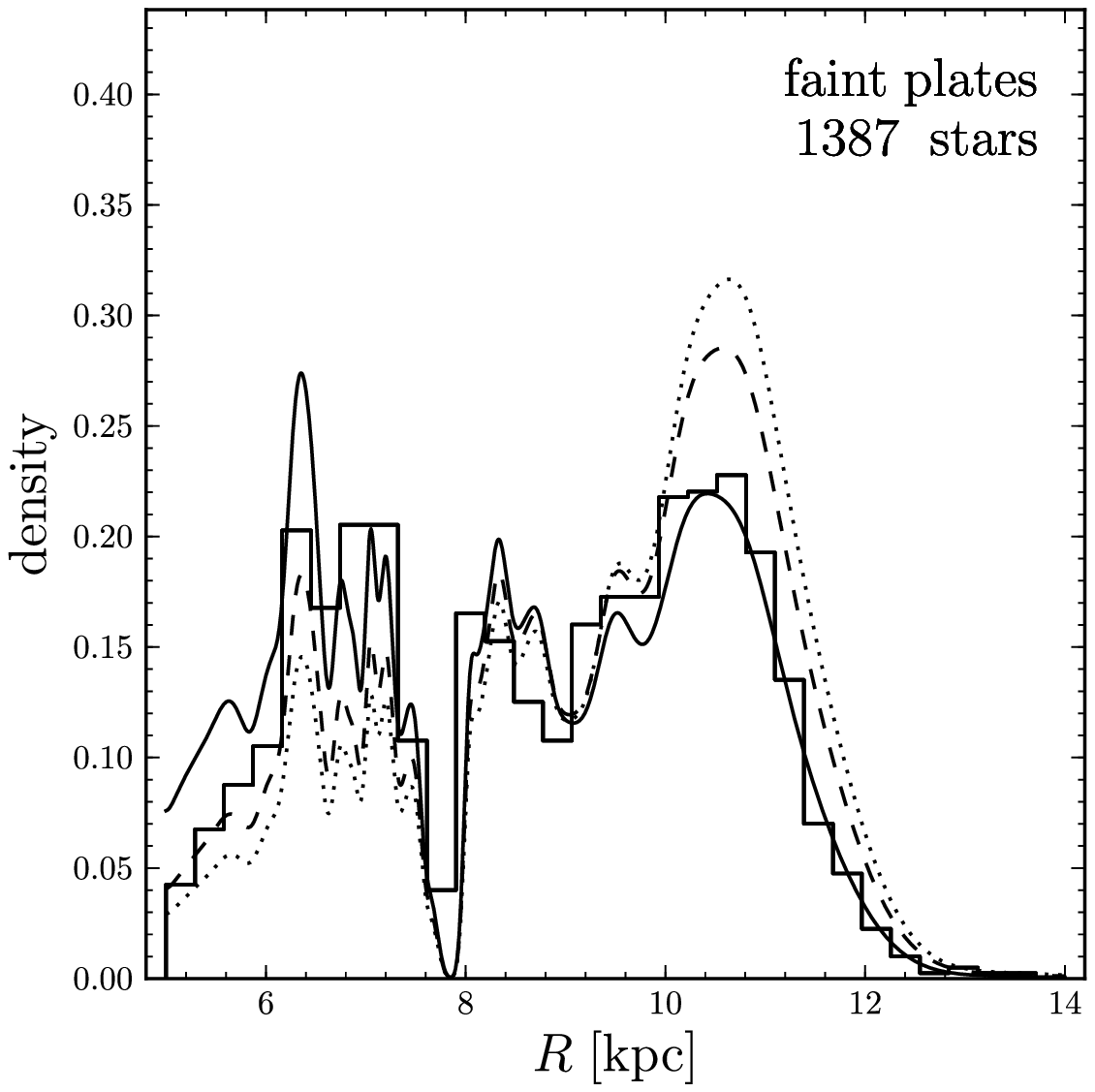}
\caption{Same as \figurename~\ref{fig:model_data_poor_g_zdist}, but
for the distribution of Galactocentric
radii. The dashed and
  dotted lines show the same model but with a scale length of 3 and 4
  kpc, respectively.}\label{fig:model_data_poor_g_Rdist}
\end{figure}

\begin{figure}[tp]
\includegraphics[width=0.30\textwidth,clip=]{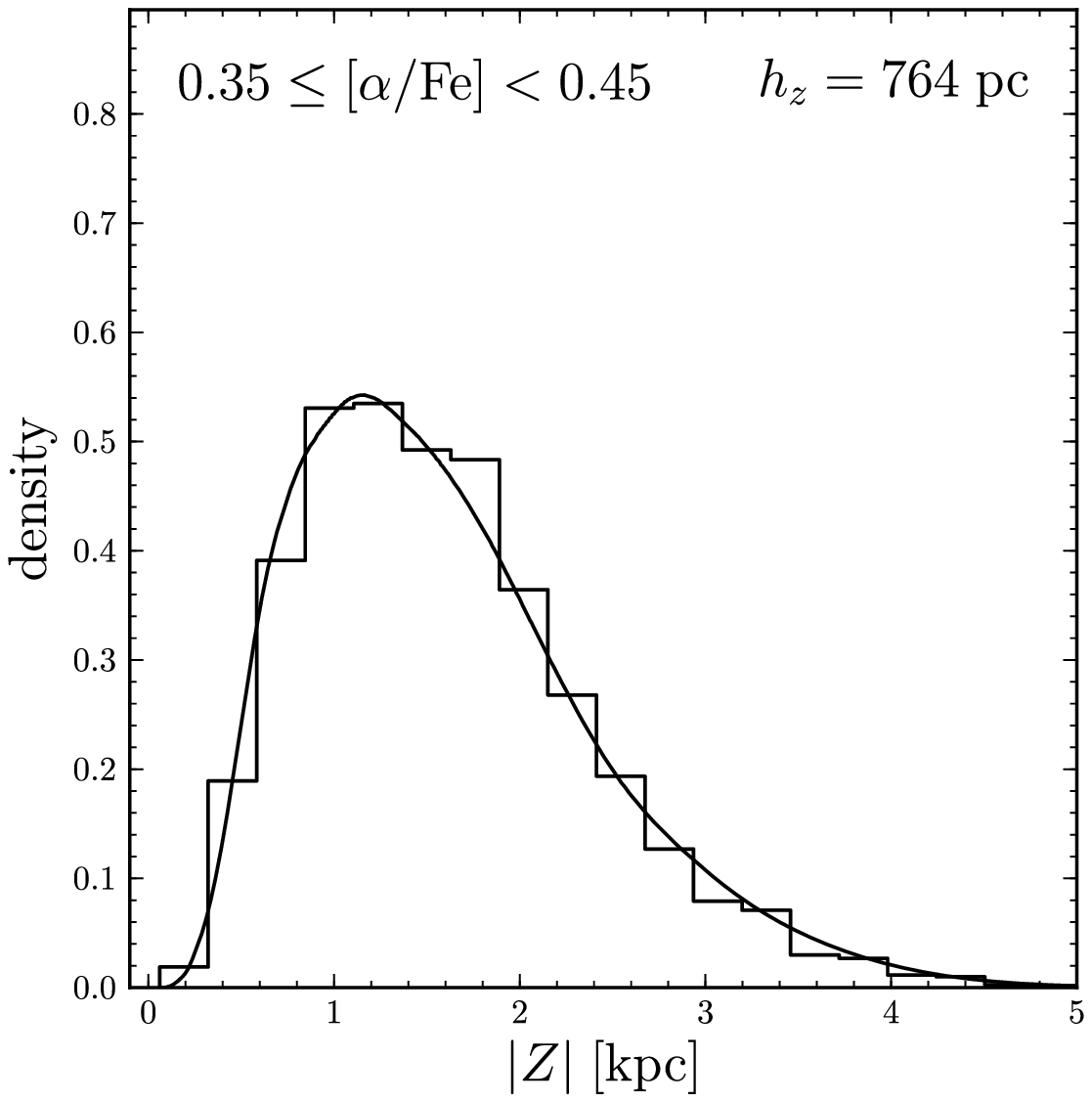}
\includegraphics[width=0.30\textwidth,clip=]{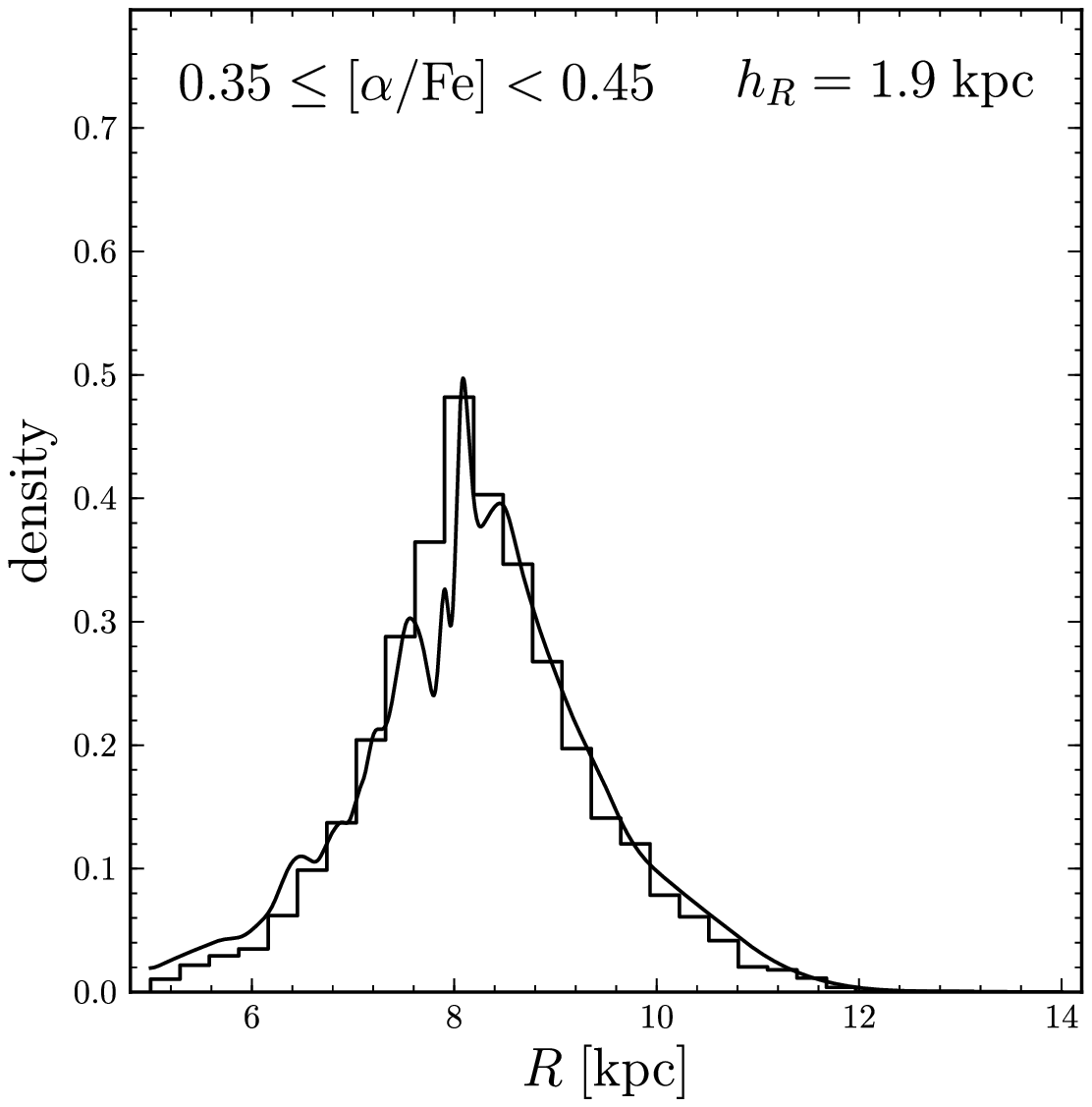}\\
\includegraphics[width=0.30\textwidth,clip=]{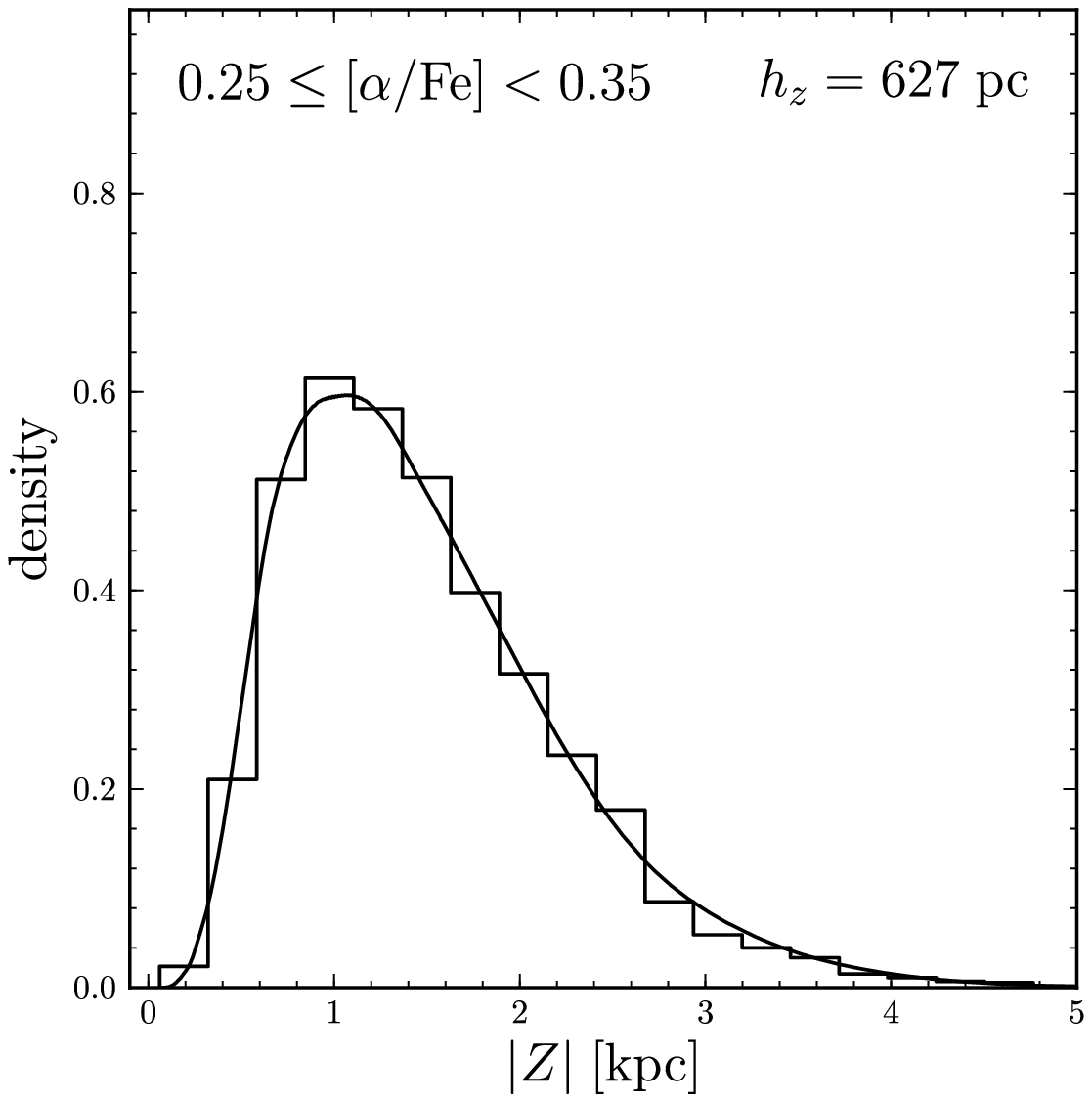}
\includegraphics[width=0.30\textwidth,clip=]{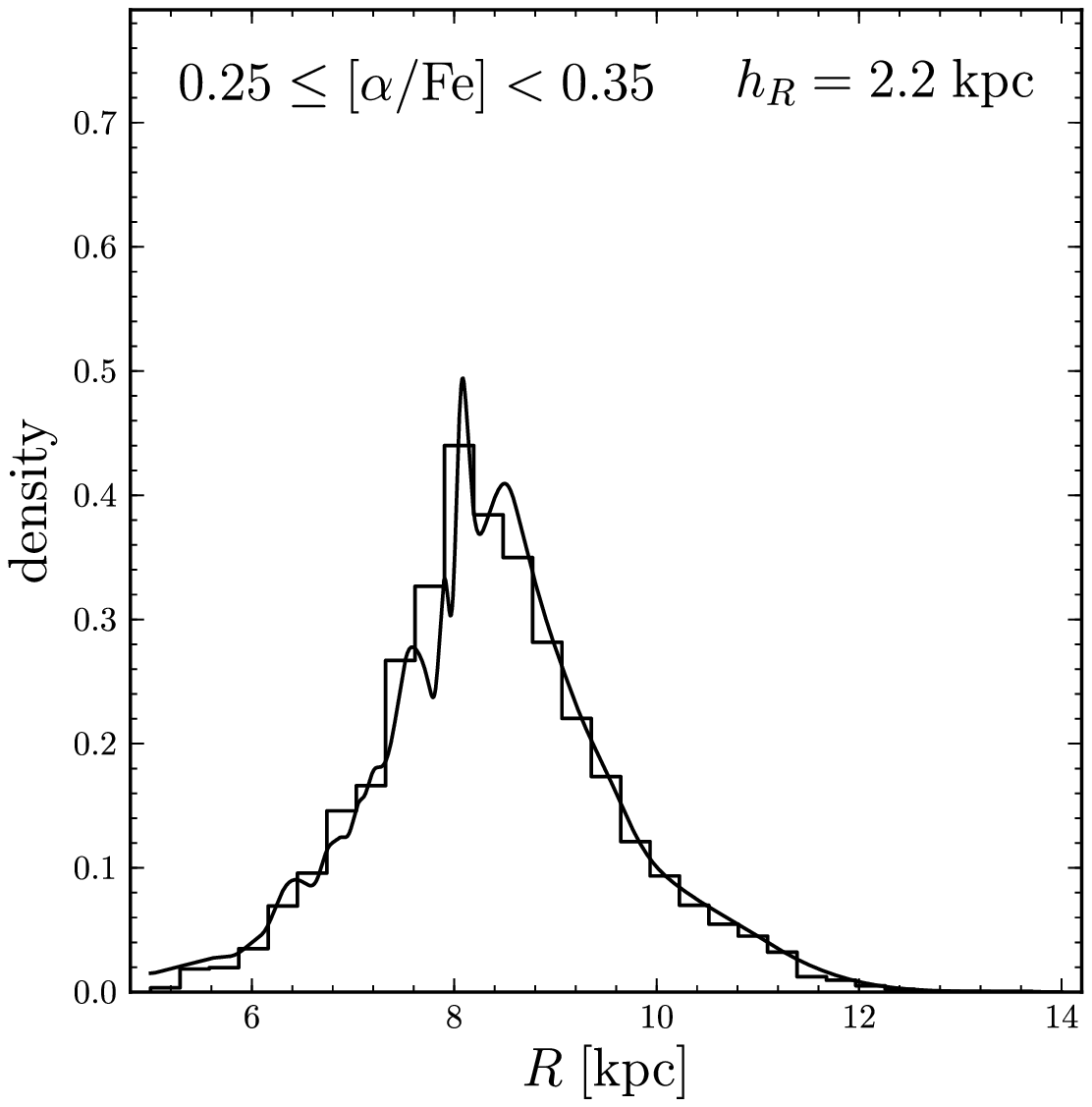}\\
\includegraphics[width=0.30\textwidth,clip=]{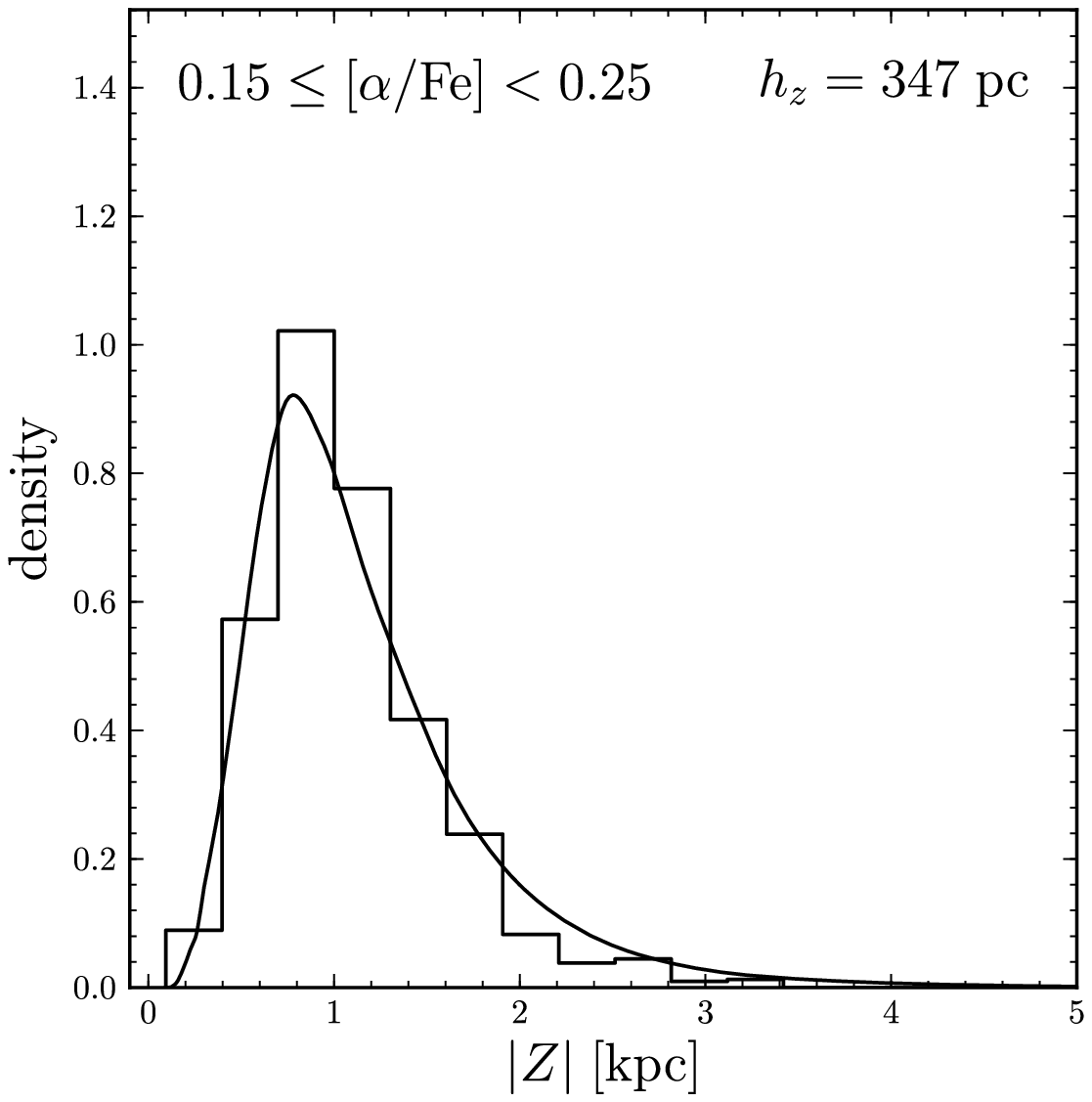}
\includegraphics[width=0.30\textwidth,clip=]{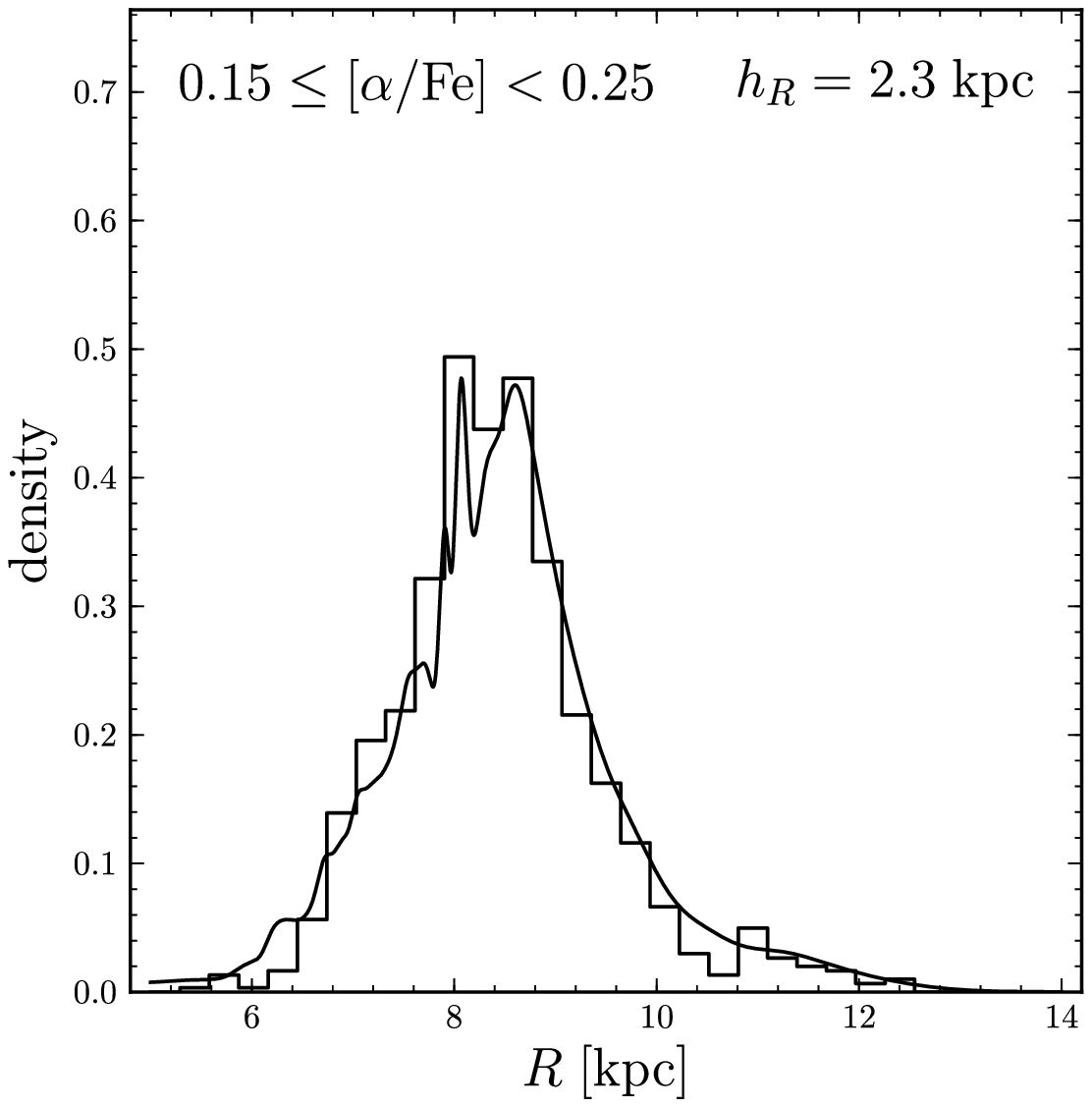}\\
\includegraphics[width=0.30\textwidth,clip=]{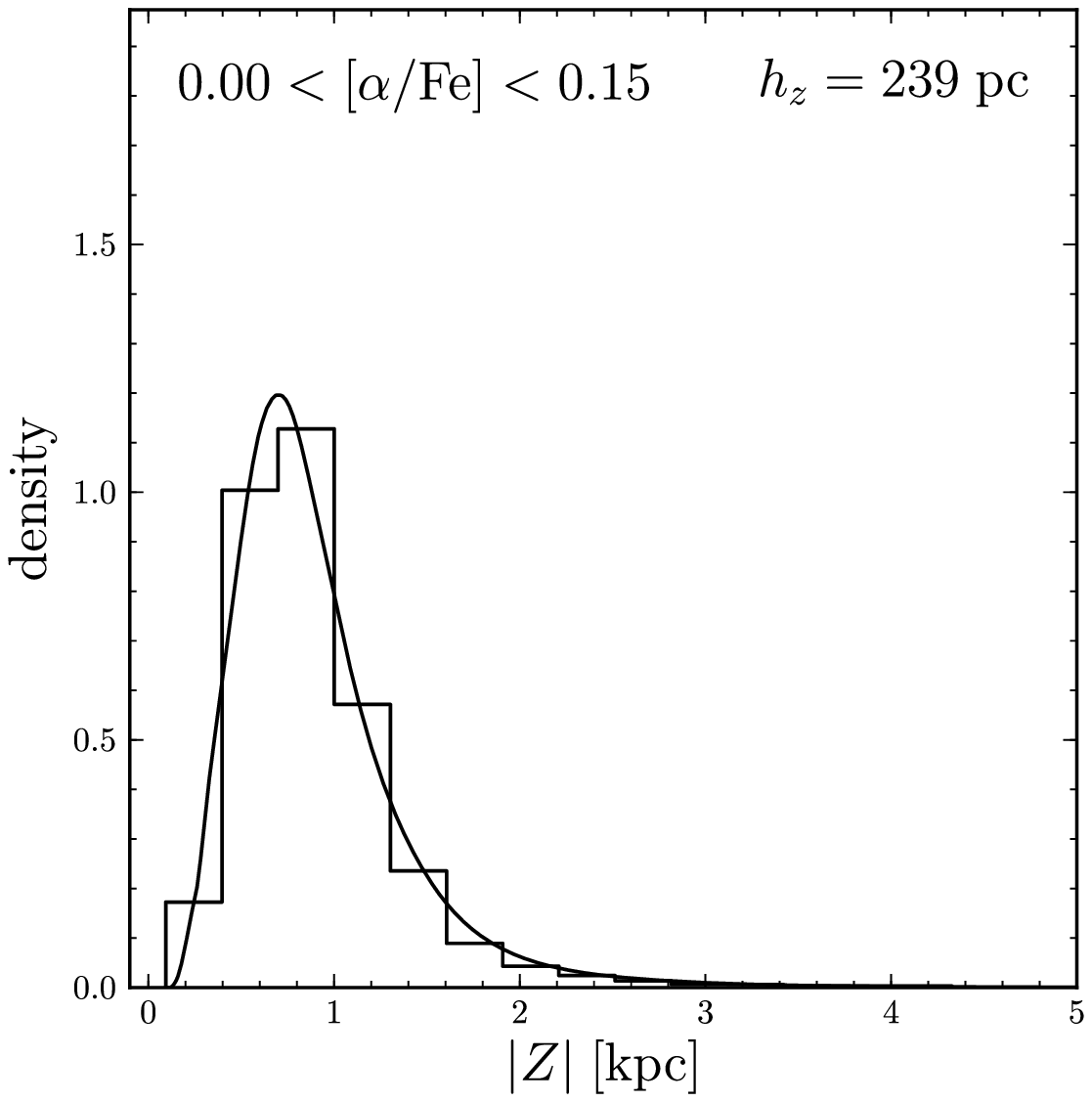}
\includegraphics[width=0.30\textwidth,clip=]{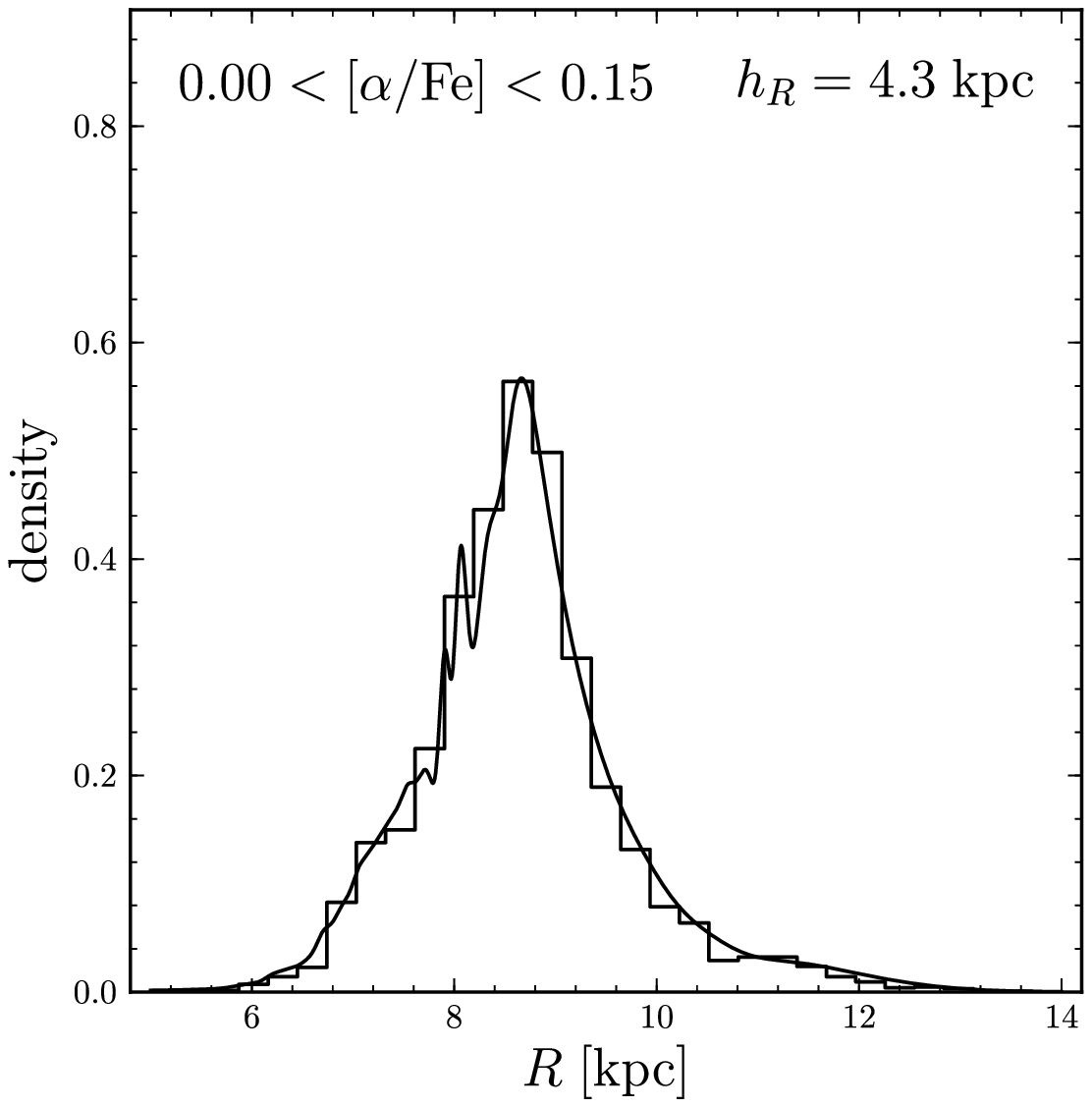}
\caption{Structural parameters of the G-dwarf sample as a function of
\afe. The top two rows show the \aenhanced\ sample split at \afe\ =
0.35; the bottom two rows contain the \apoor\ sample split at \afe\ =
0.15. The left column compares the observed distribution of vertical
heights to the model distribution. The right column does the same for
the distribution of Galactocentric radii. Each sample is fit with a
mixture of two double-exponential disks. The best-fit parameters of
the dominant disk component are shown in the top-left of each panel. In each
case the secondary component only contributes a few percent of the
mass at the solar circle (see \tablename s~\ref{table:poor_results}
and \ref{table:rich_results} for detailed
results).}\label{fig:model_data_afe}
\end{figure}

\begin{figure}[tp]
\includegraphics[width=0.322\textwidth,clip=]{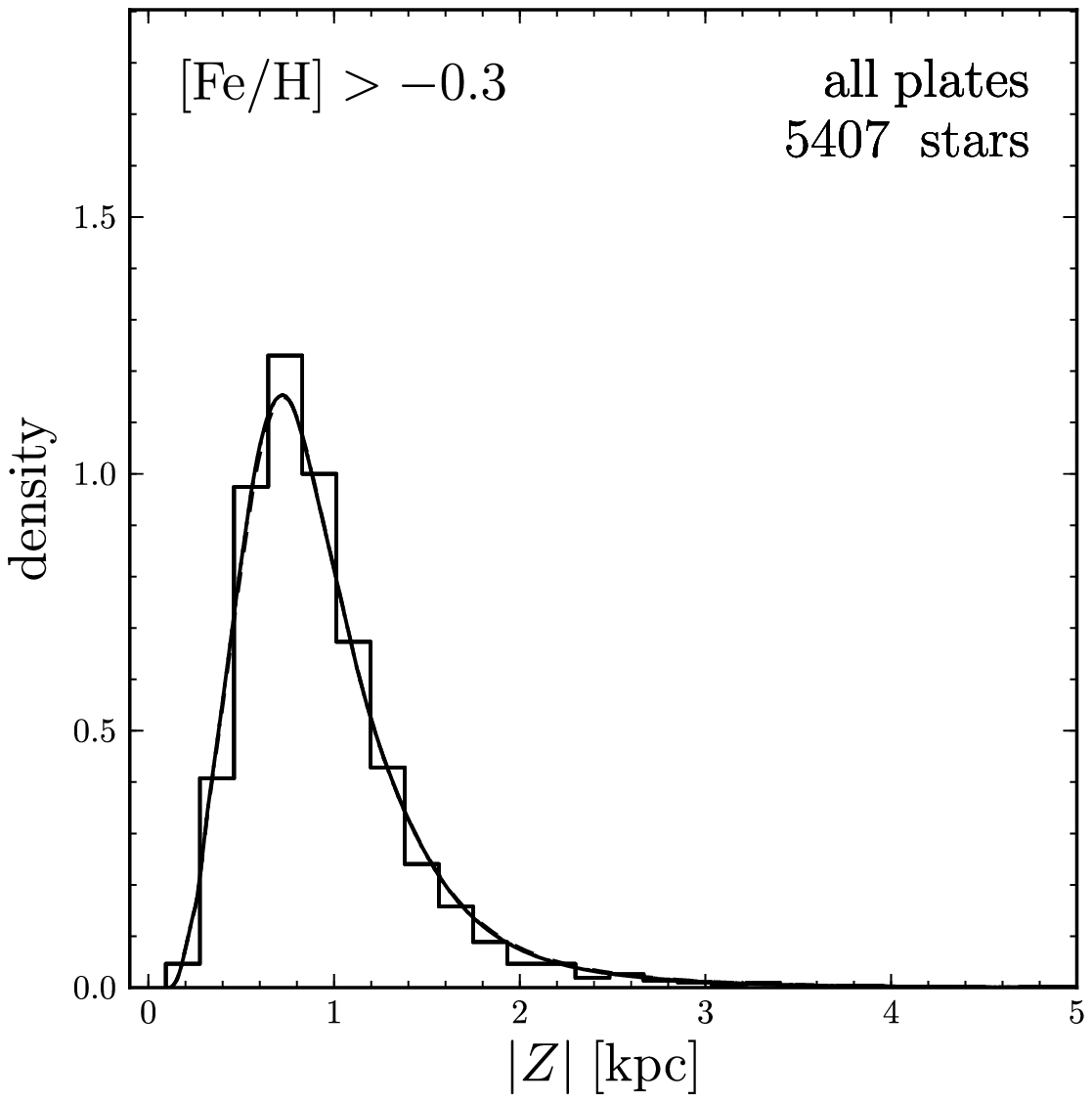}
\includegraphics[width=0.322\textwidth,clip=]{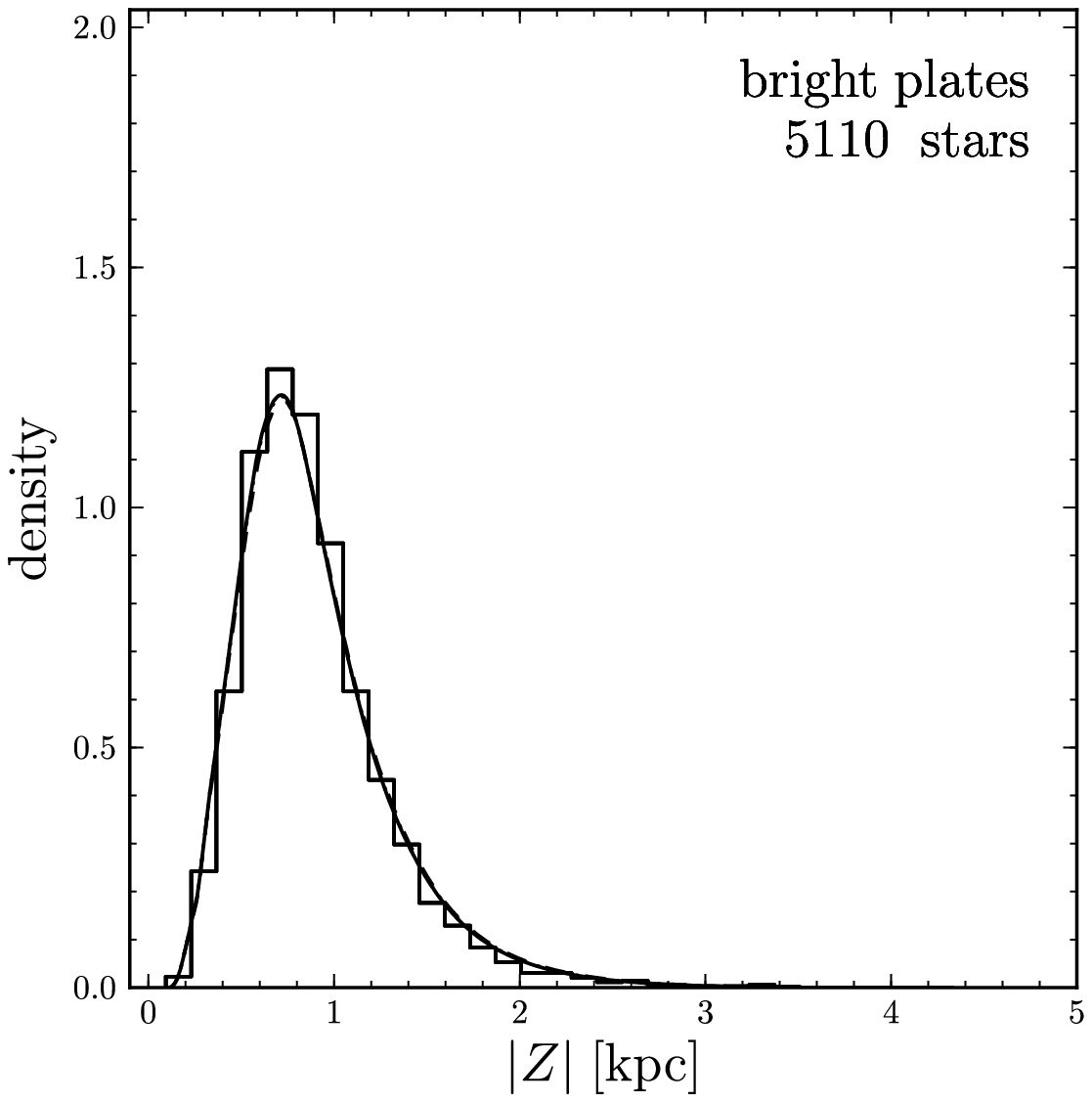}
\includegraphics[width=0.322\textwidth,clip=]{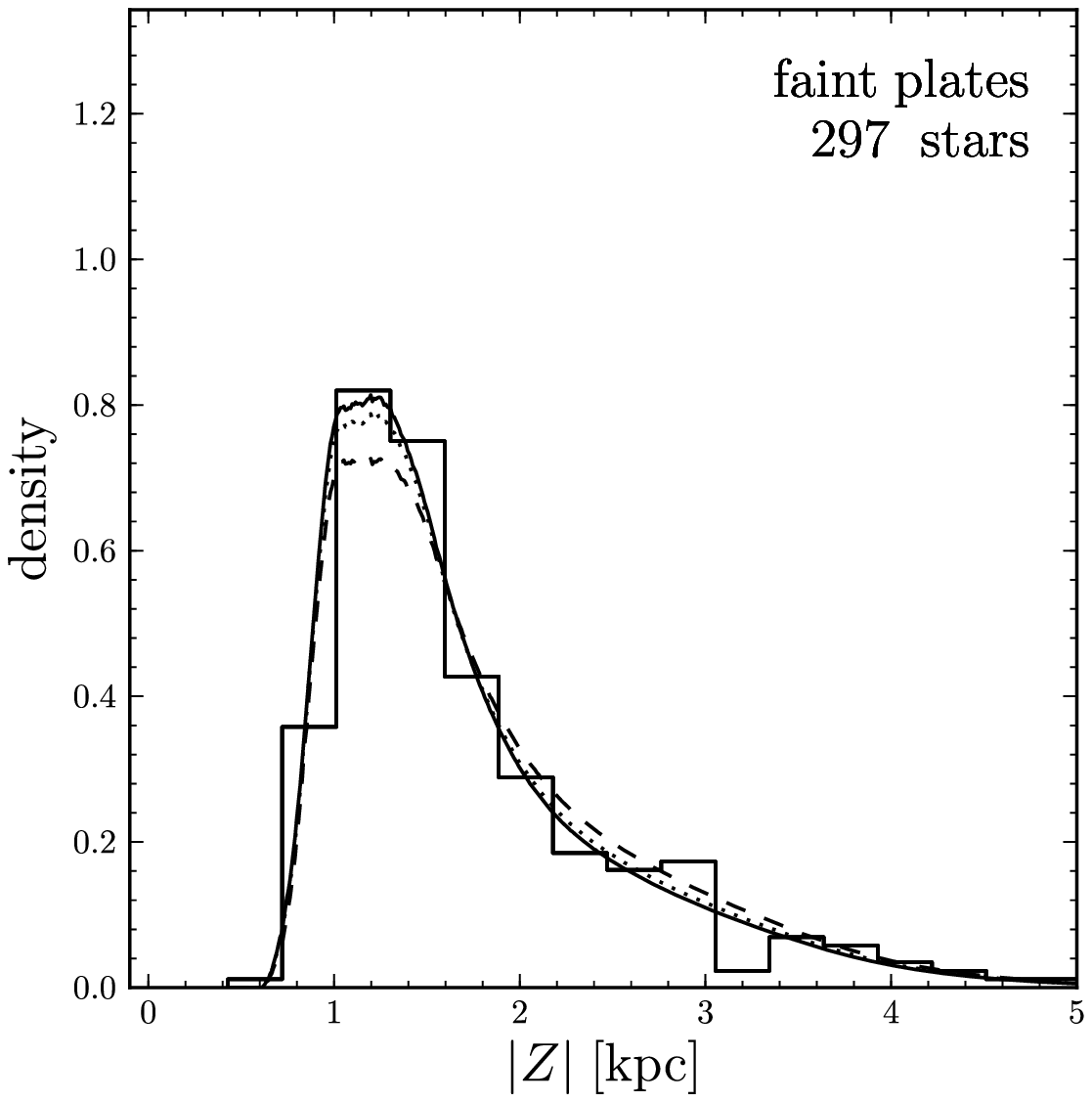}\\
\includegraphics[width=0.322\textwidth,clip=]{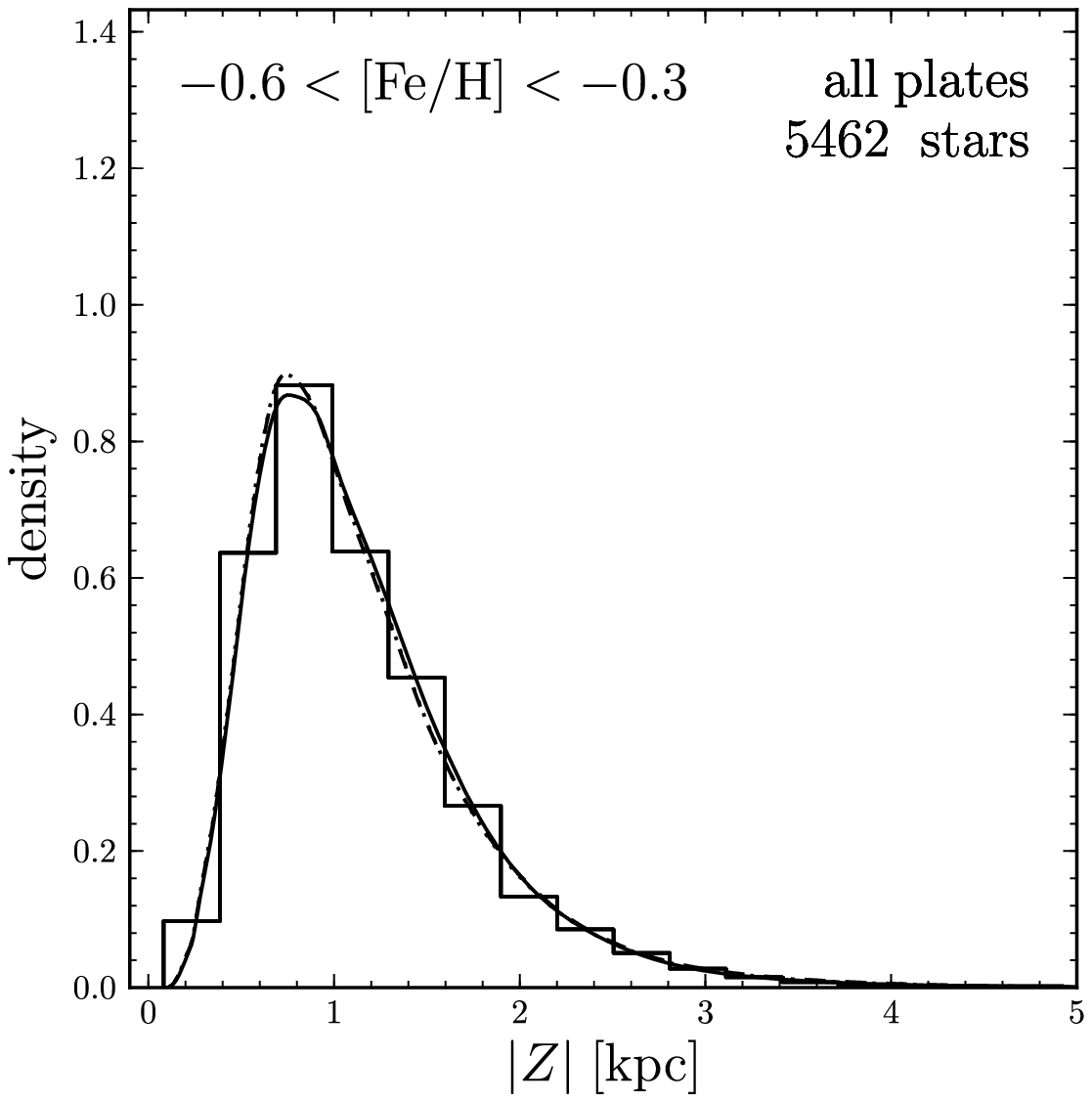}
\includegraphics[width=0.322\textwidth,clip=]{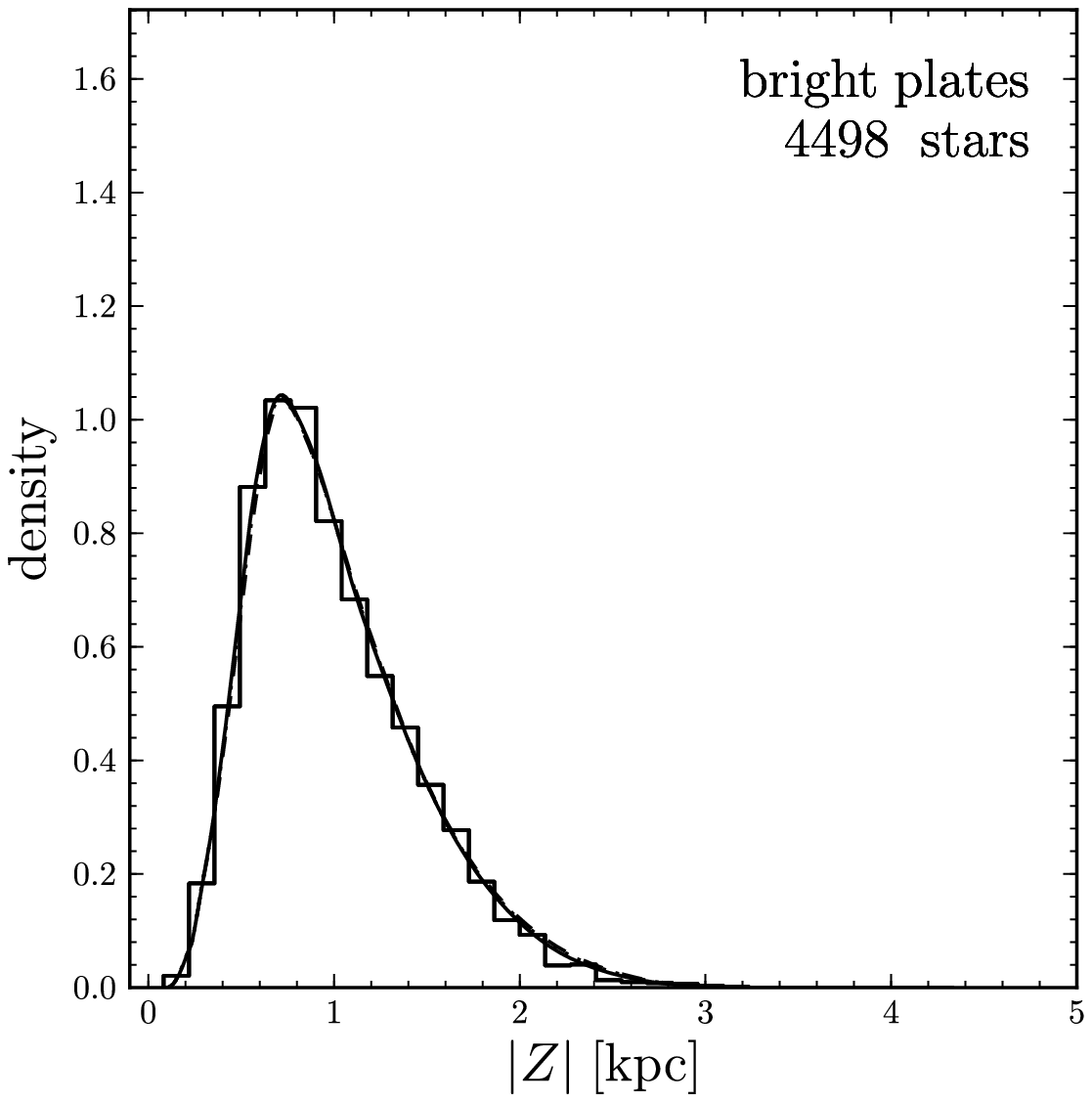}
\includegraphics[width=0.322\textwidth,clip=]{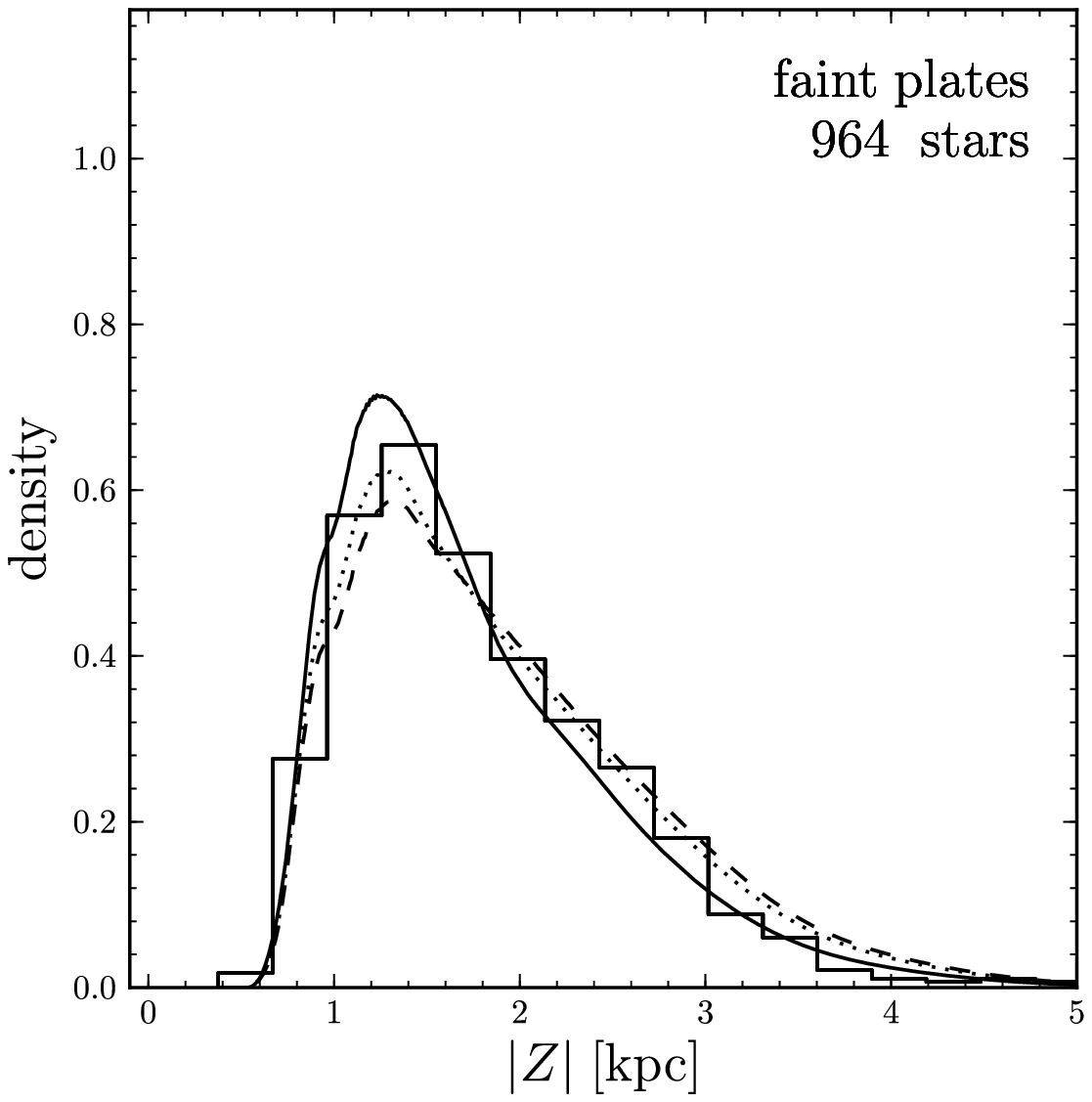}\\
\includegraphics[width=0.322\textwidth,clip=]{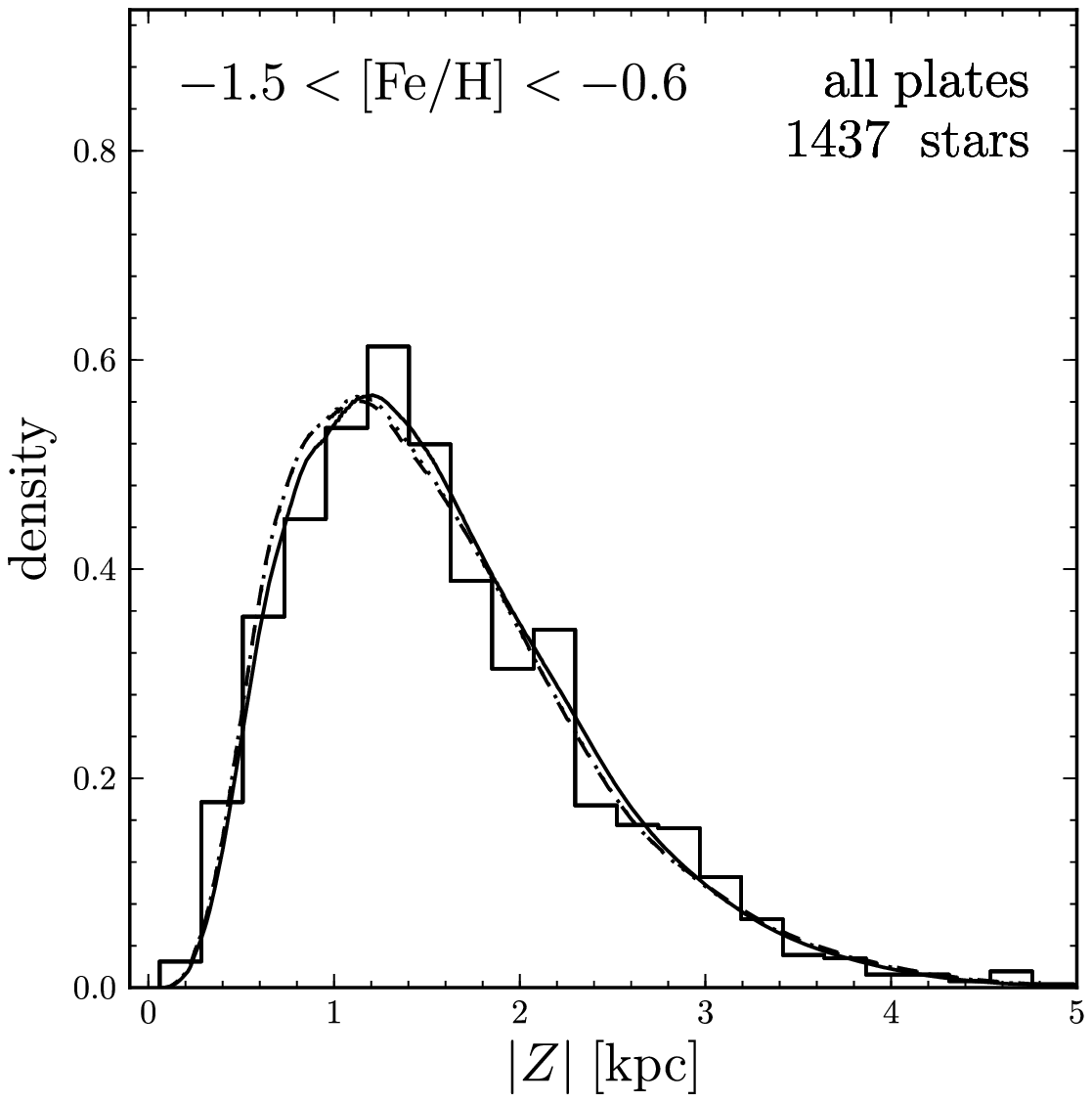}
\includegraphics[width=0.322\textwidth,clip=]{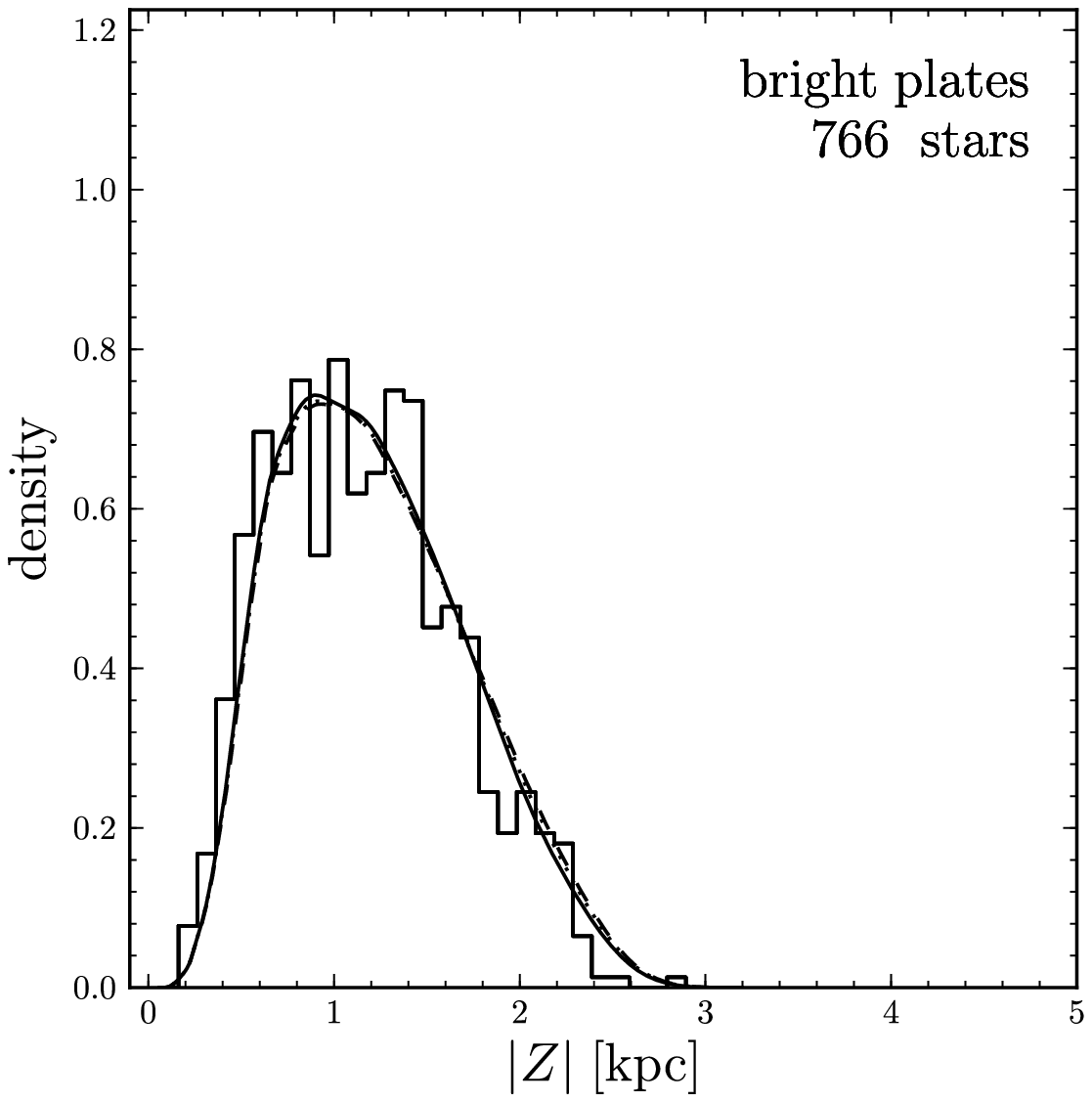}
\includegraphics[width=0.322\textwidth,clip=]{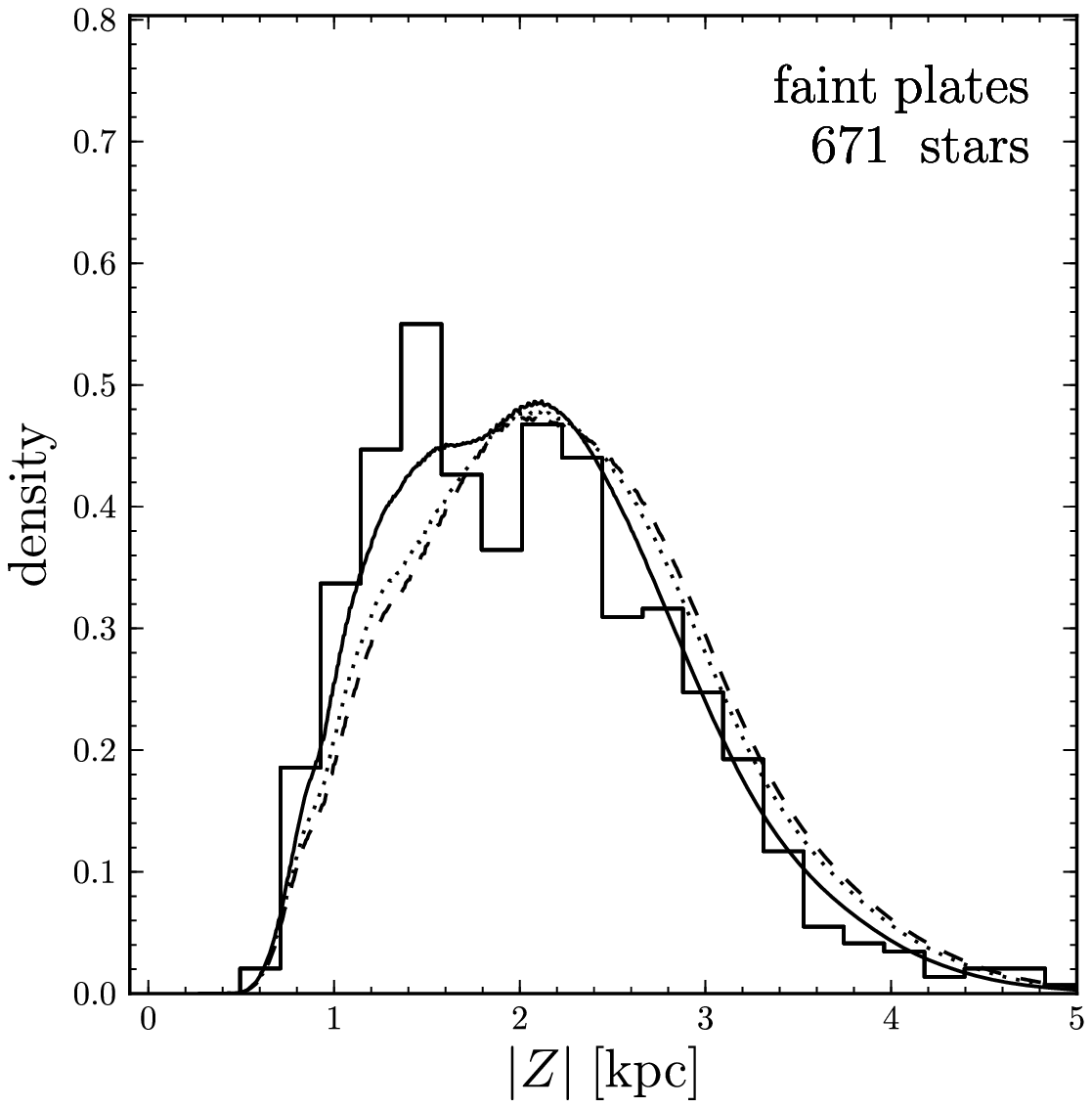}
\caption{Comparison between the observed distribution of vertical
  heights of the \apoor\ G-dwarf sample and the distribution predicted by the
  best-fit mixture of two double-exponential disks model. The dashed and
  dotted lines show the same model, but with a scale length of 2 and 3
  kpc, respectively. The bottom two rows show samples with the same
  \afe\ range of the nominal \apoor\ sample, but with different \feh\
  boundaries. See \tablename~\ref{table:rich_results} for the
  parameters of the best-fit
  models.}\label{fig:model_data_rich_g_zdist}
\end{figure}

\begin{figure}[tp]
\includegraphics[width=0.323\textwidth,clip=]{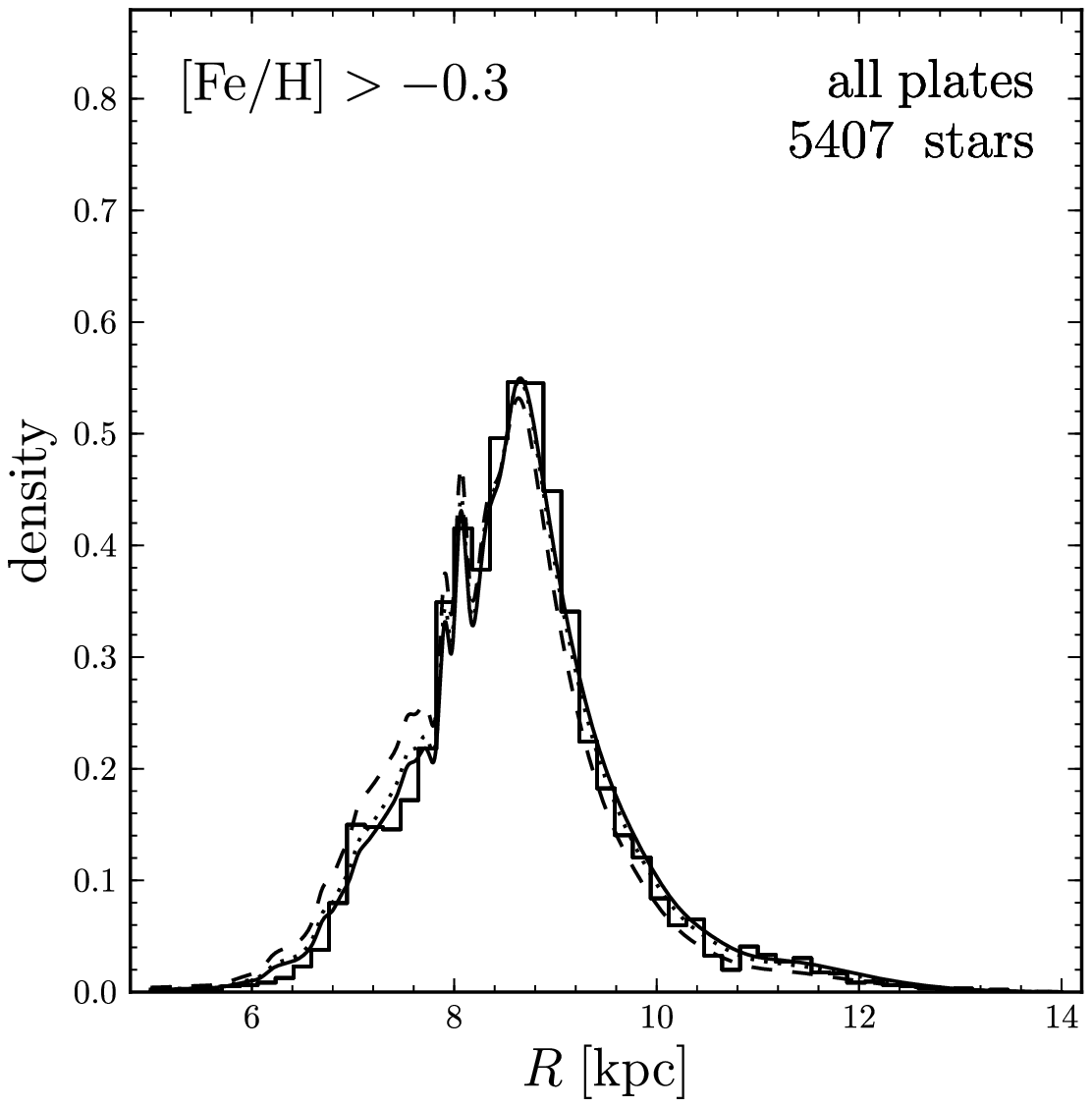}
\includegraphics[width=0.323\textwidth,clip=]{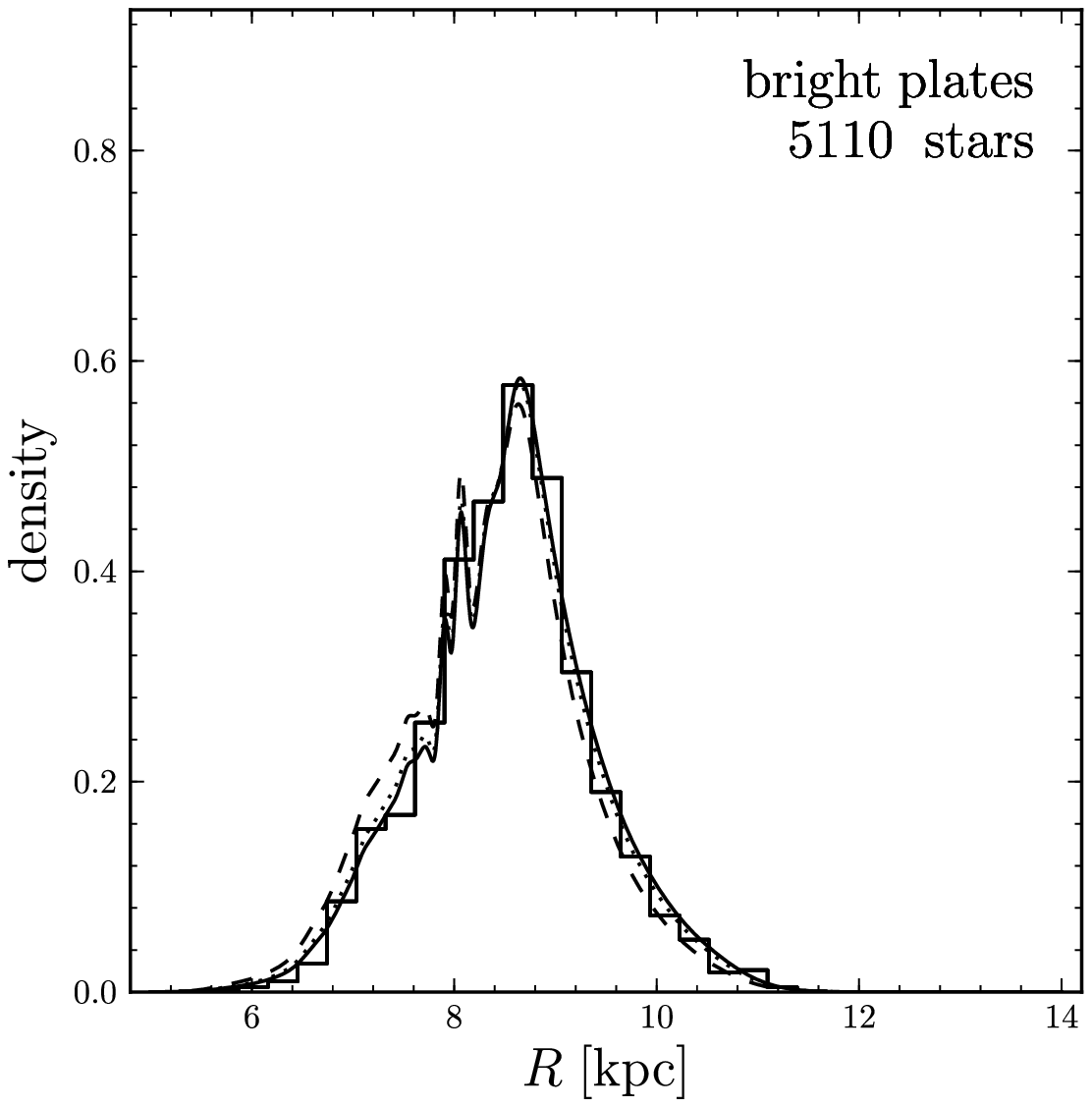}
\includegraphics[width=0.323\textwidth,clip=]{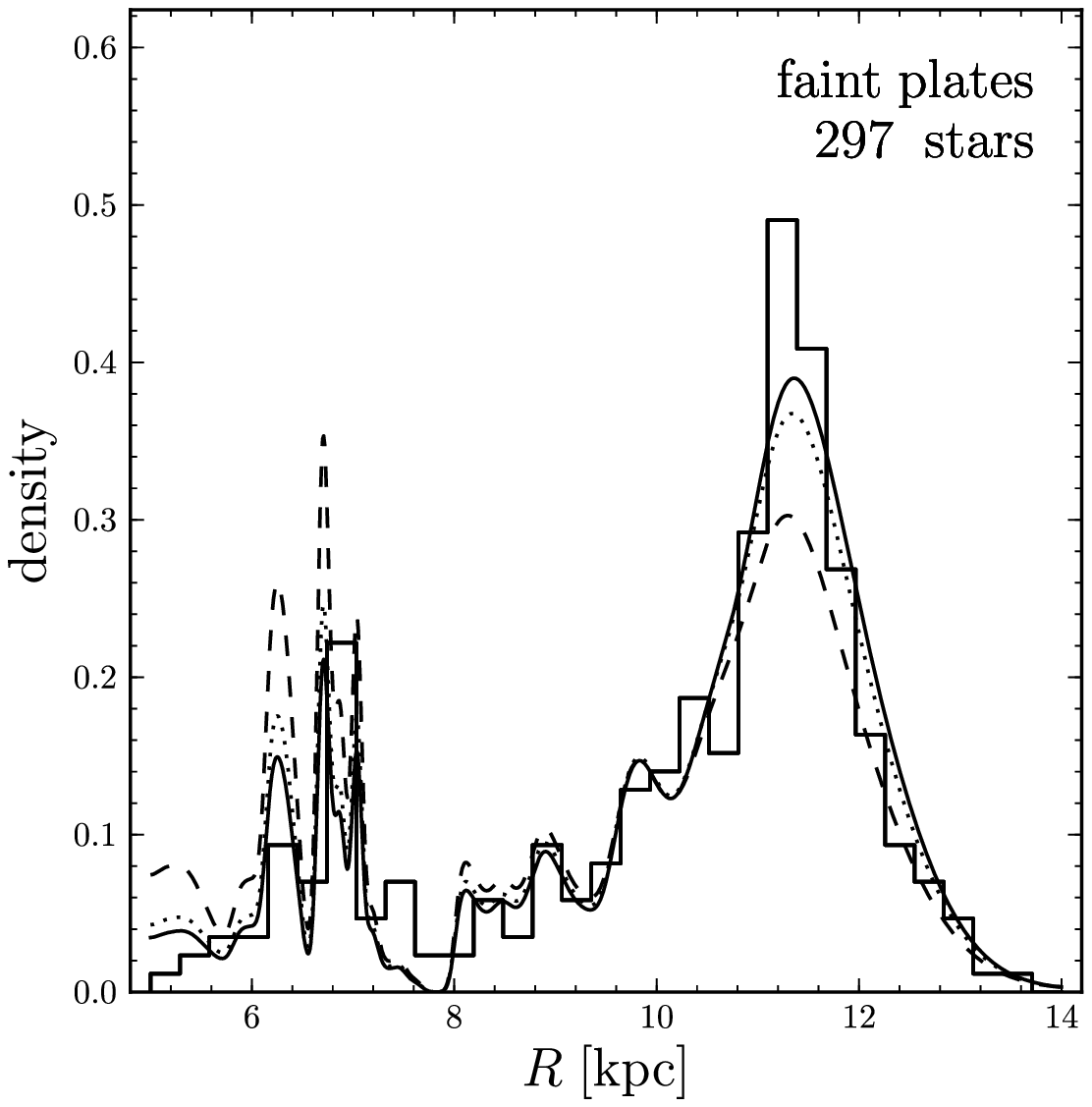}\\
\includegraphics[width=0.323\textwidth,clip=]{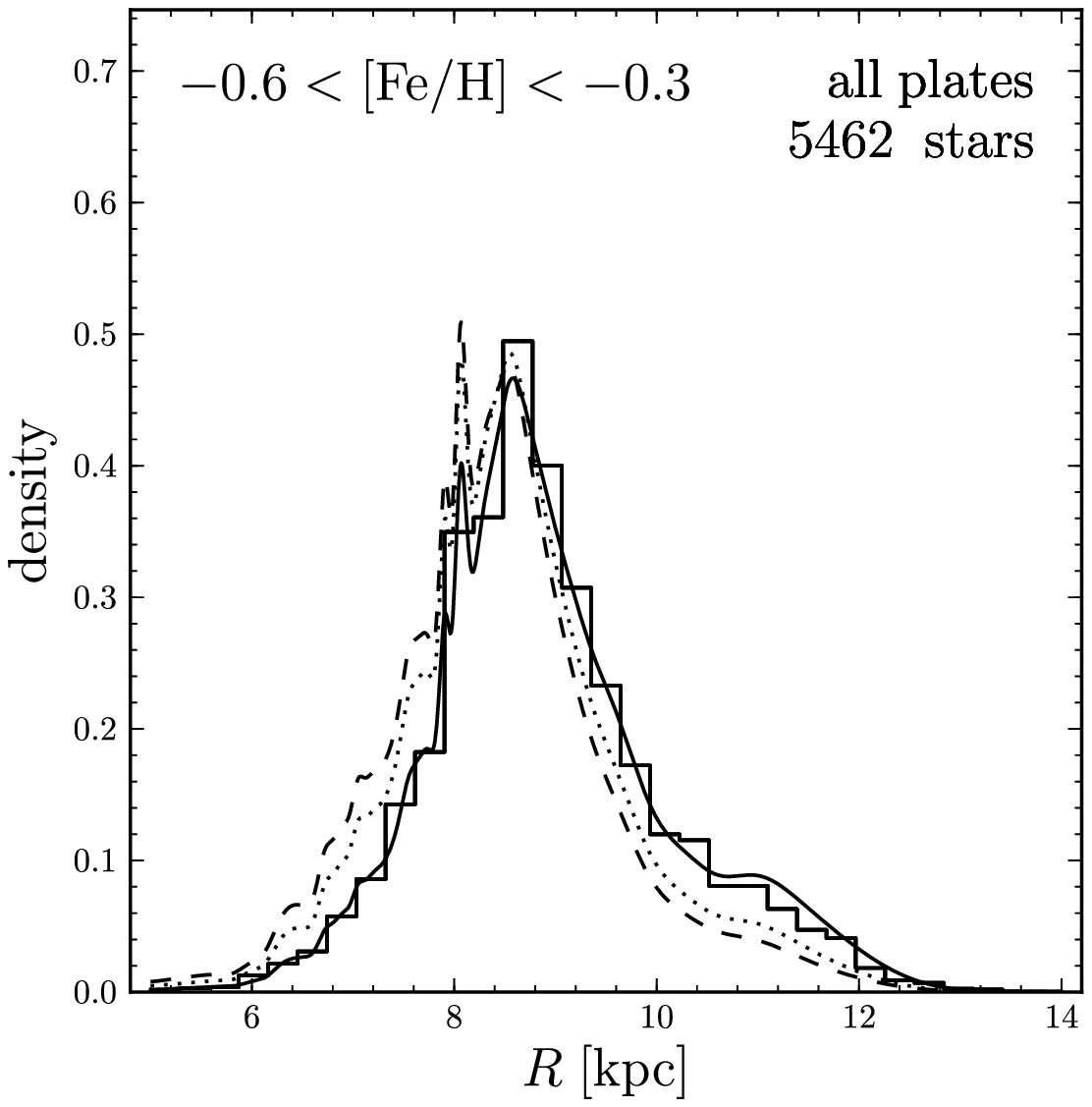}
\includegraphics[width=0.323\textwidth,clip=]{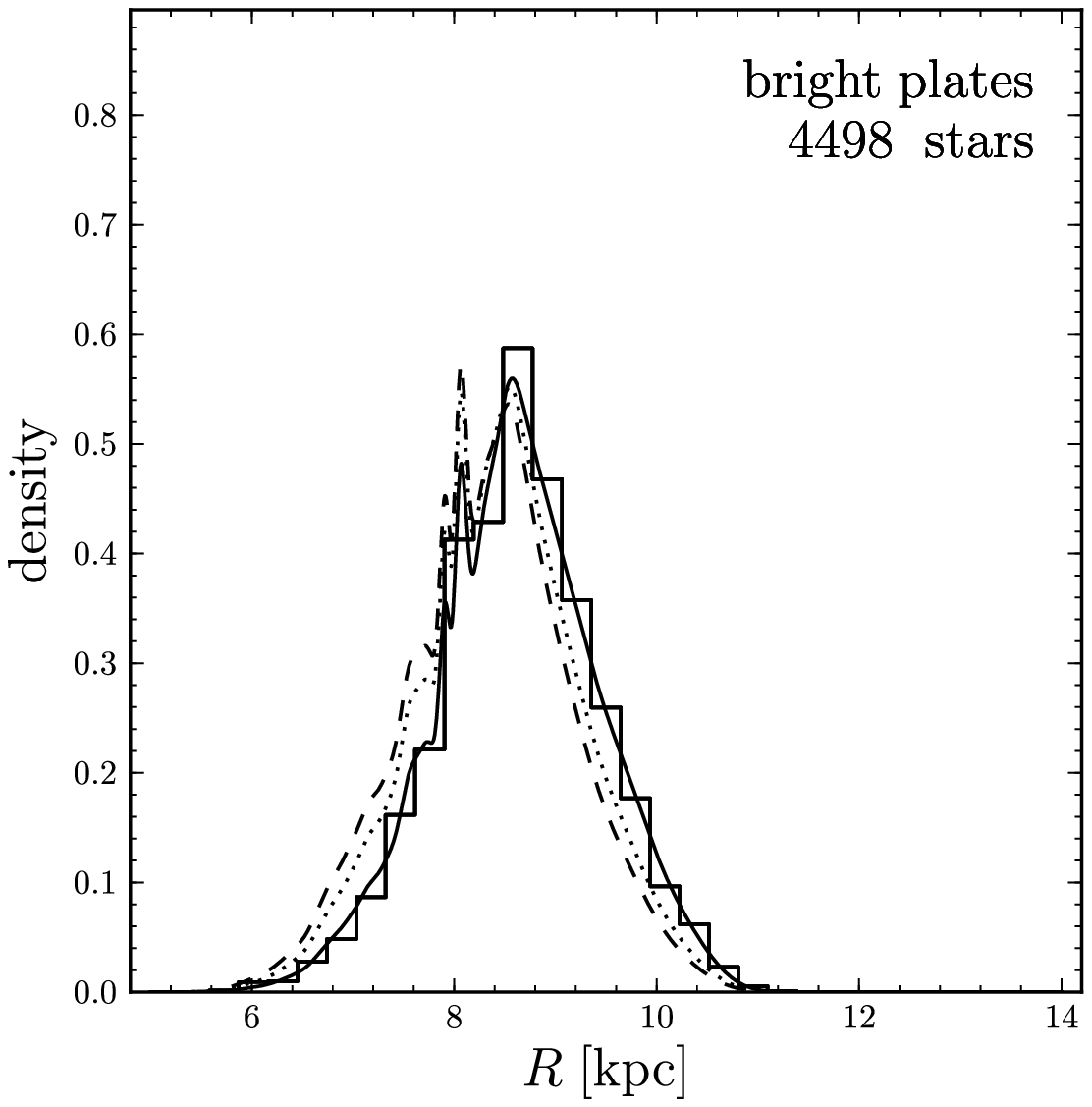}
\includegraphics[width=0.323\textwidth,clip=]{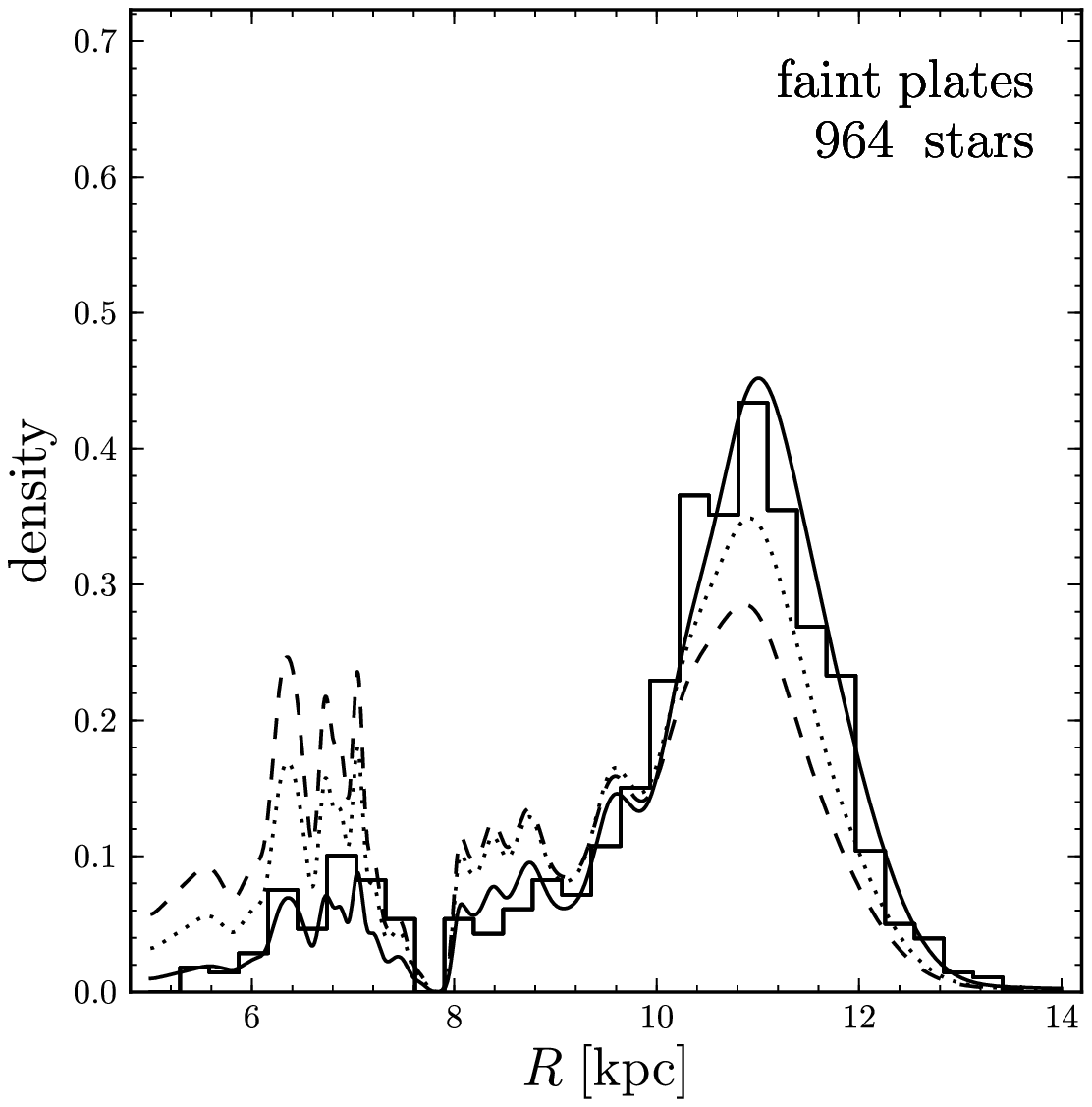}\\
\includegraphics[width=0.323\textwidth,clip=]{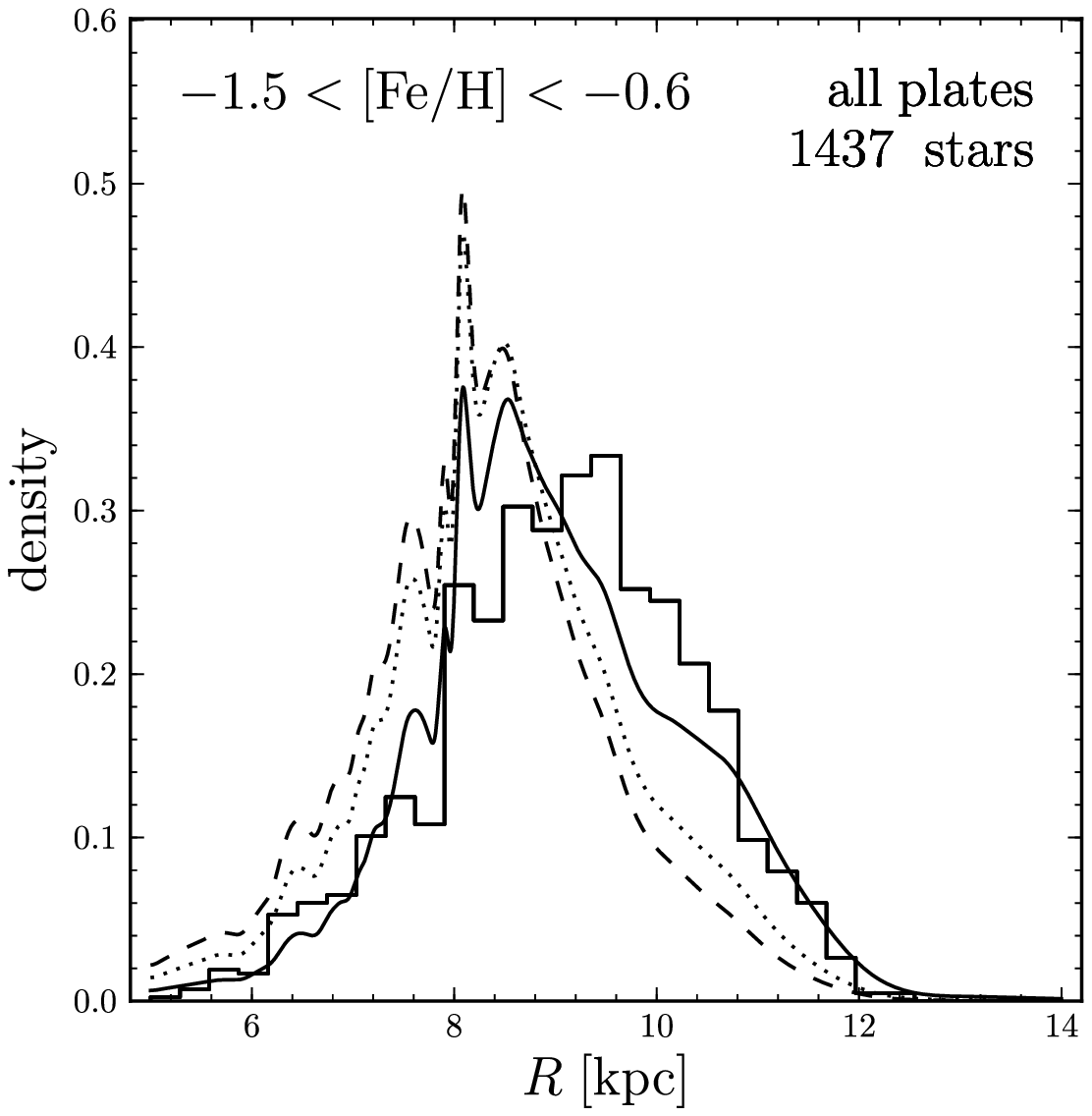}
\includegraphics[width=0.323\textwidth,clip=]{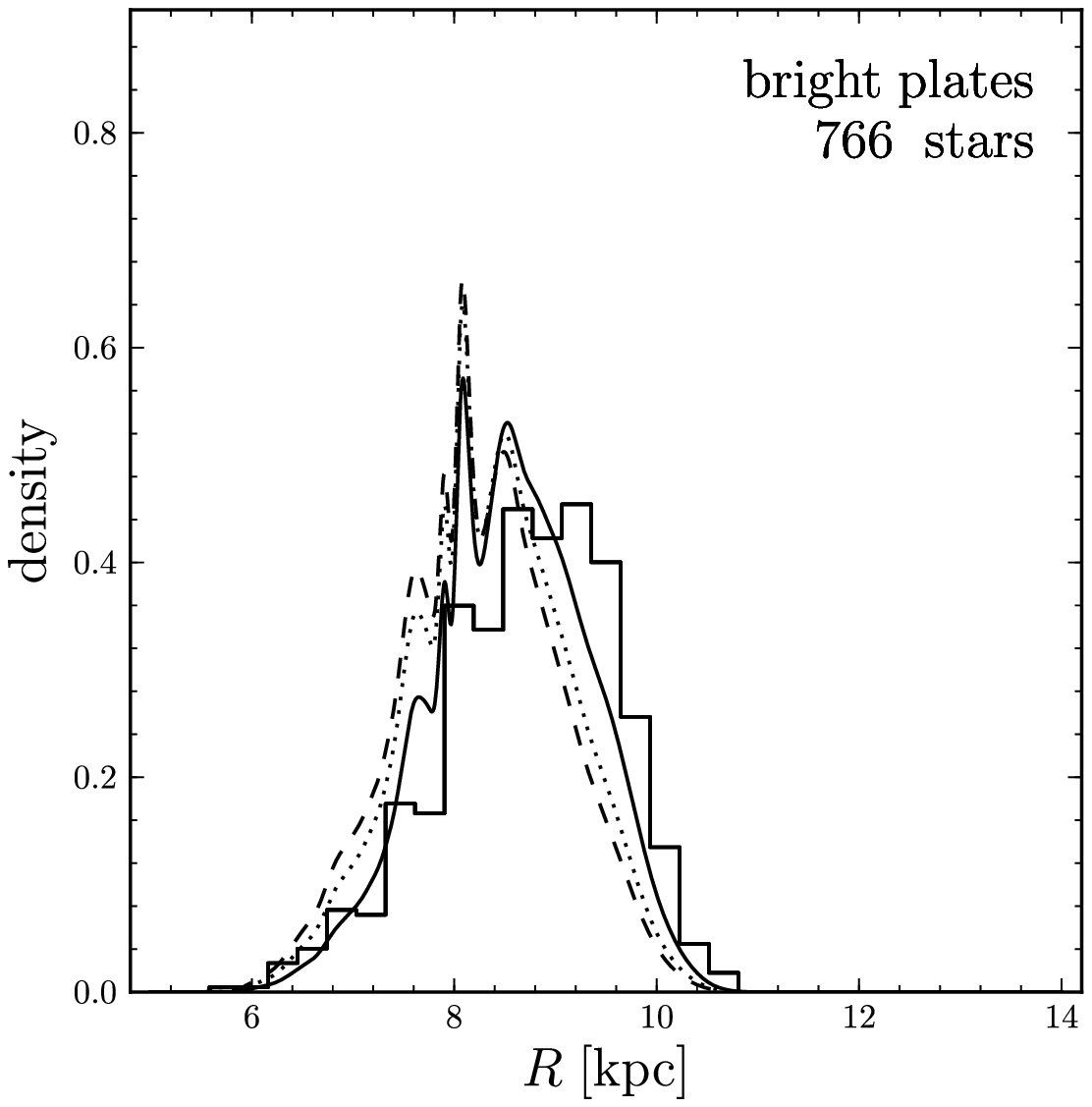}
\includegraphics[width=0.323\textwidth,clip=]{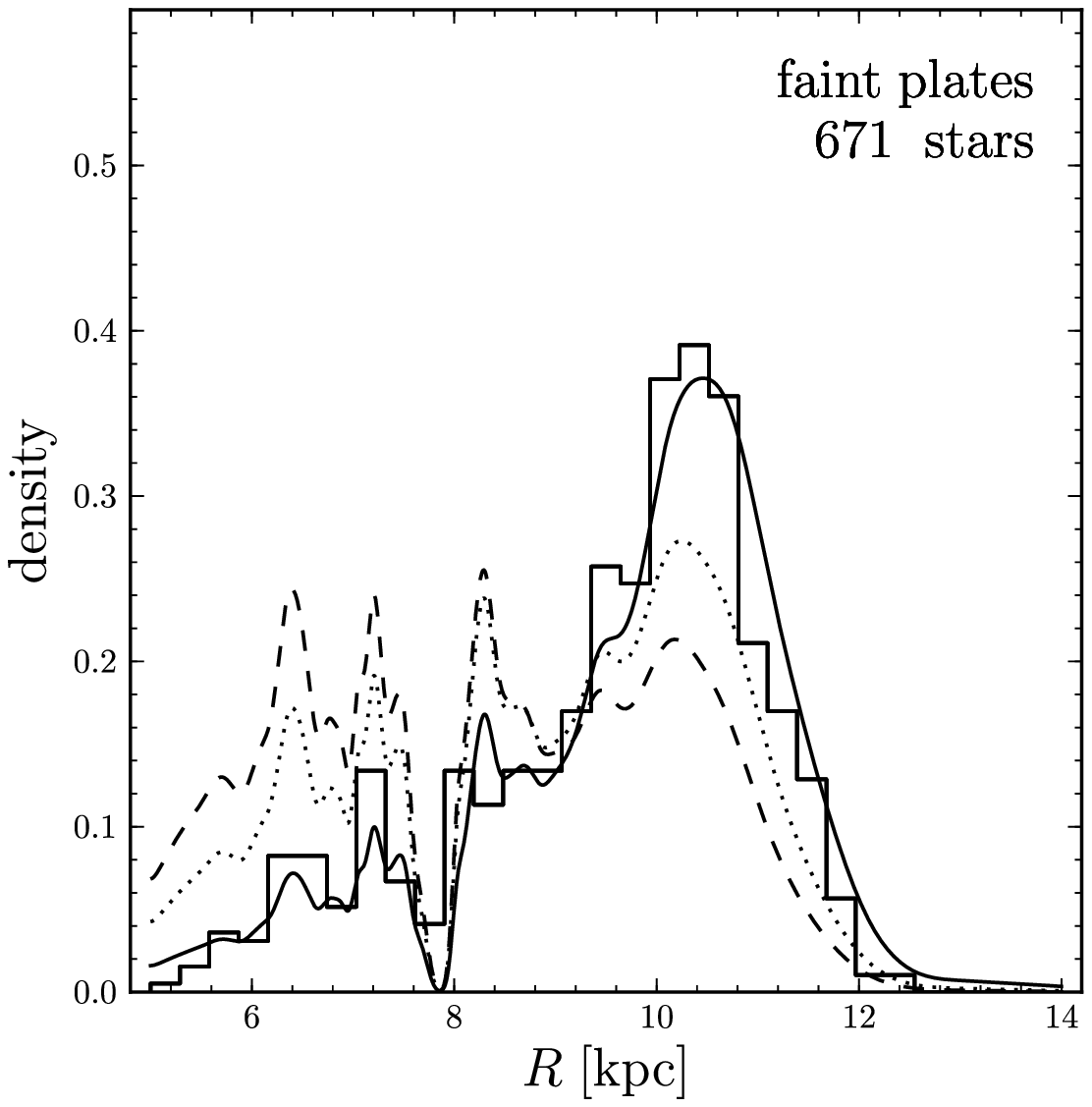}
\caption{Same as \figurename~\ref{fig:model_data_rich_g_zdist}, but for
the distribution of Galactocentric
radii. The dashed and
  dotted lines show the same model, but with a scale length of 2 and 3
  kpc, respectively.}\label{fig:model_data_rich_g_Rdist}
\end{figure}

\figurename~\ref{fig:model_data_poor_g_zdist} compares
the observed distribution of vertical heights $|Z|$ of the \aenhanced\
G-dwarf sample to that predicted by the best-fit model. This
prediction is obtained by running the best-fit density model
integrated over the color--metallicity distribution through our model
for the \segue\ selection function. There are 35 stars with magnitudes that
should put them on faint plates, but that were observed on bright
plates, and 551 stars in the opposite situation are cut from the data
sample to show this comparison. The comparison between the data and
the model is shown for all plates and for the bright and faint plates
separately. The agreement between the model and observed distribution
is excellent for all of these. \figurename~\ref{fig:model_data_poor_g_Rdist} shows a similar comparison for the distribution of Galactocentric radii of the data and in
the model. The model correctly predicts the observed star counts for
most Galactocentric radii, except the smallest around 5 kpc, where the
model slightly overpredicts the number of stars (here, and in further
comparisons below, the model around 8 kpc behaves somewhat erratically,
as this is the boundary between $90^\circ \leq l \leq 270^\circ$ and
$-90^\circ < l < 90^\circ$ plates, and we do not use the finite extent
of the plate in our model distributions). Also shown in this figure
and in \figurename~\ref{fig:model_data_poor_g_zdist} are models that
only differ from the best-fit model in their radial scale length: a
model with a scale length of 3 kpc and one with a scale length of 4
kpc. It is clear that these longer scale lengths are strongly ruled
out by the model, as they strongly overpredict the star counts at large
Galactocentric radii.

\tablename~\ref{table:poor_results} lists the best-fit
parameters for fits that only use (a) bright or faint plates, (b) $b >
0^\circ$ or $b < 0^\circ$ plates, or (c) $|b| > 45^\circ$ or $|b| <
45^\circ$ plates. The results from all of these different samples are
roughly consistent with each other; we note that we can even
measure the radial scale length with high-latitude plates ($|b| >
45^\circ$) alone.

\figurename s~\ref{fig:model_data_poor_g_zdist} and
\ref{fig:model_data_poor_g_Rdist} show comparisons between the
observed star counts and the model, when we split the \aenhanced\
sample into more metal-poor and more metal-rich sub-samples by cutting
the sample at [Fe/H] = $-0.7$. Comparisons for when we split the
\aenhanced\ sample into two bins in \afe, by cutting the sample at
\afe\ = 0.35, are shown in \figurename~\ref{fig:model_data_afe}.

\subsection{The \apoor\ disk sample}

\figurename~\ref{fig:model_data_rich_g_zdist} compares the best-fit
model to the observed star counts as a function of vertical height and
\figurename~\ref{fig:model_data_rich_g_Rdist} shows this comparison as
a function of Galactocentric radius, again removing 4 stars with
magnitudes that should put them on faint plates but that were observed
on bright plates and removing 43 stars in the opposite situation. We
also show models whose parameters are the same as those of the
best-fit model, but with shorter scale lengths of 2 and 3 kpc. The
faint plates, which only contain 6\,percent of the \apoor\ sample,
rule out a short scale length of 2 kpc for the \apoor\ disk. The
best-fit model provides a good fit to the observed star counts.

We again also list the best-fit parameters for fits that only use (a)
bright or faint plates, (b) $b > 0^\circ$ or $b < 0^\circ$ plates, or
(c) $|b| > 45^\circ$ or $|b| < 45^\circ$ plates in
\tablename~\ref{table:rich_results}. The results from all of these
different samples are again roughly consistent, except for the faint
plates fit, which prefer even longer radial scale lengths, but faint
plates only contain 6\,percent of the \apoor\ sample. We have also run
fits for the \apoor\ sample where we (a) employ a
more conservative \sn\ cut of $\sn > 30$, (b) enlarge our sample with
a less conservative \sn\ cut of $\sn > 10$, (c) remove stars on plates
whose K-S probability for the spectroscopic sample to have been drawn
from the underlying photometric sample combined with our model for the
\segue\ selection function (see \figurename~\ref{fig:ks_tanhr}) is
smaller than 0.1, (d) use stars from the \segue\ database that were
explicitly targeted as G-type stars (with all other \logg, \sn, $E(B-V)$
cuts), (e) remove stars with magnitudes that should put them on
\segue\ bright plates, but that were observed as part of a faint plate
and vice versa, and (f) artificially shift the metallicity
distribution 0.1 dex toward the more metal-rich end. The results from these
fits are all consistent with those obtained for our nominal sample
with fiducial cuts.

\figurename s~\ref{fig:model_data_rich_g_zdist} and
\ref{fig:model_data_rich_g_Rdist} show comparisons between the
observed and predicted star counts for the \apoor\ samples that are
more metal-poor than the nominal \apoor\ sample. The fit for the $-0.6
<$ \feh\ $< -0.3$ sample is good, while the fit for the most
metal-poor \apoor\ sub-sample is not entirely
satisfactory. Comparisons between the observed star counts and the
model when we split the \apoor\ sample into two, by cutting at \afe\ =
0.15, are shown in the lower two rows of
\figurename~\ref{fig:model_data_afe}.

\section{Analysis test on mock data samples}\label{sec:fakedata}

\begin{figure}[tp]
\begin{center}
\caption*{Analysis of mock data assuming thin--thick disk dichotomy}
\end{center}
\includegraphics[width=0.5\textwidth,clip=]{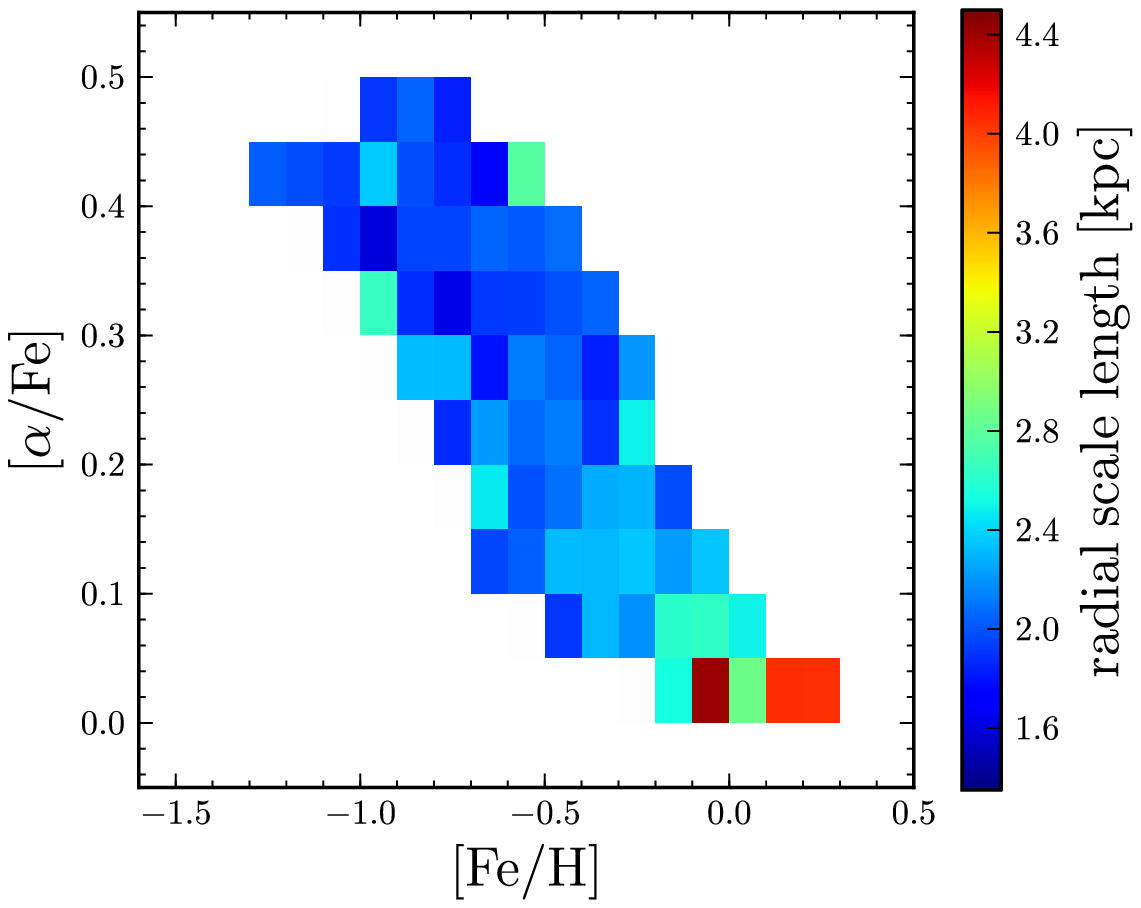}
\includegraphics[width=0.5\textwidth,clip=]{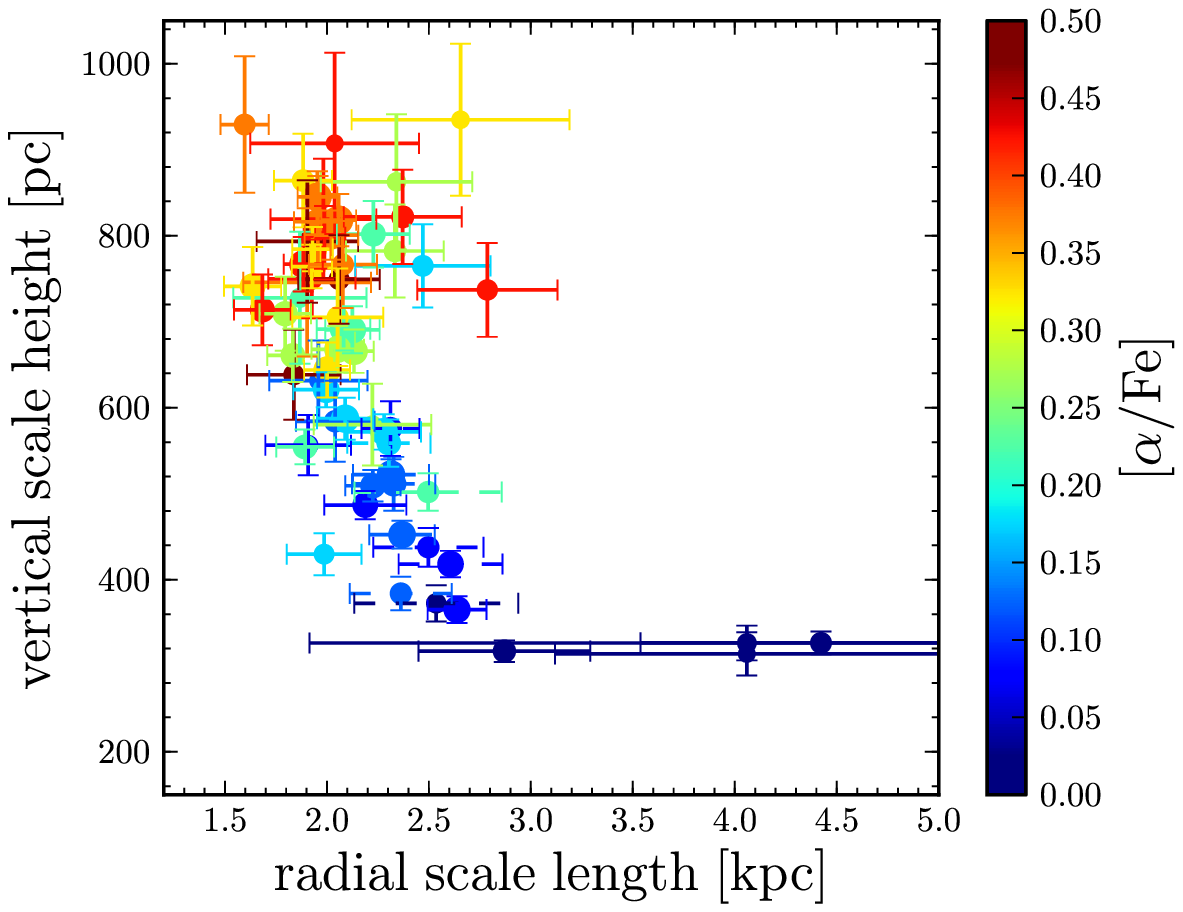}\\
\includegraphics[width=0.5\textwidth,clip=]{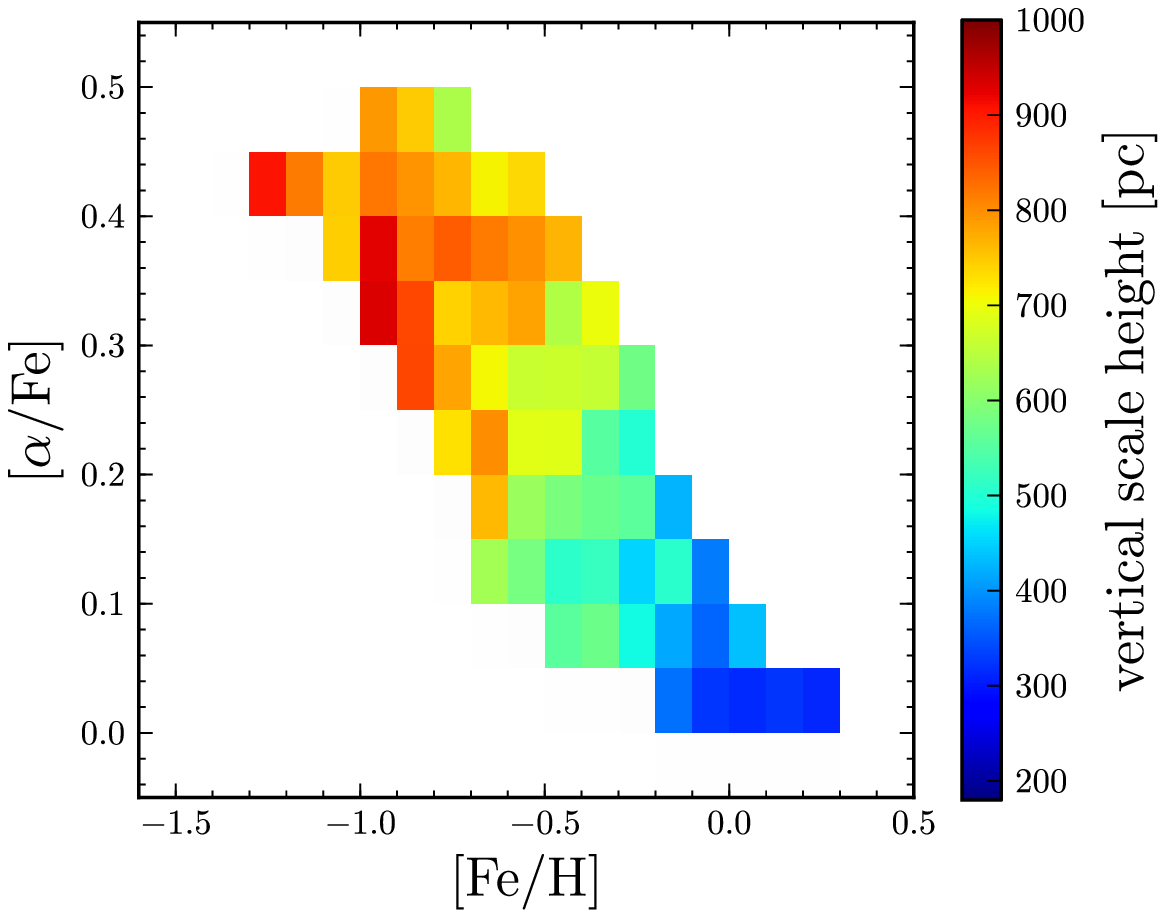}
\includegraphics[width=0.5\textwidth,clip=]{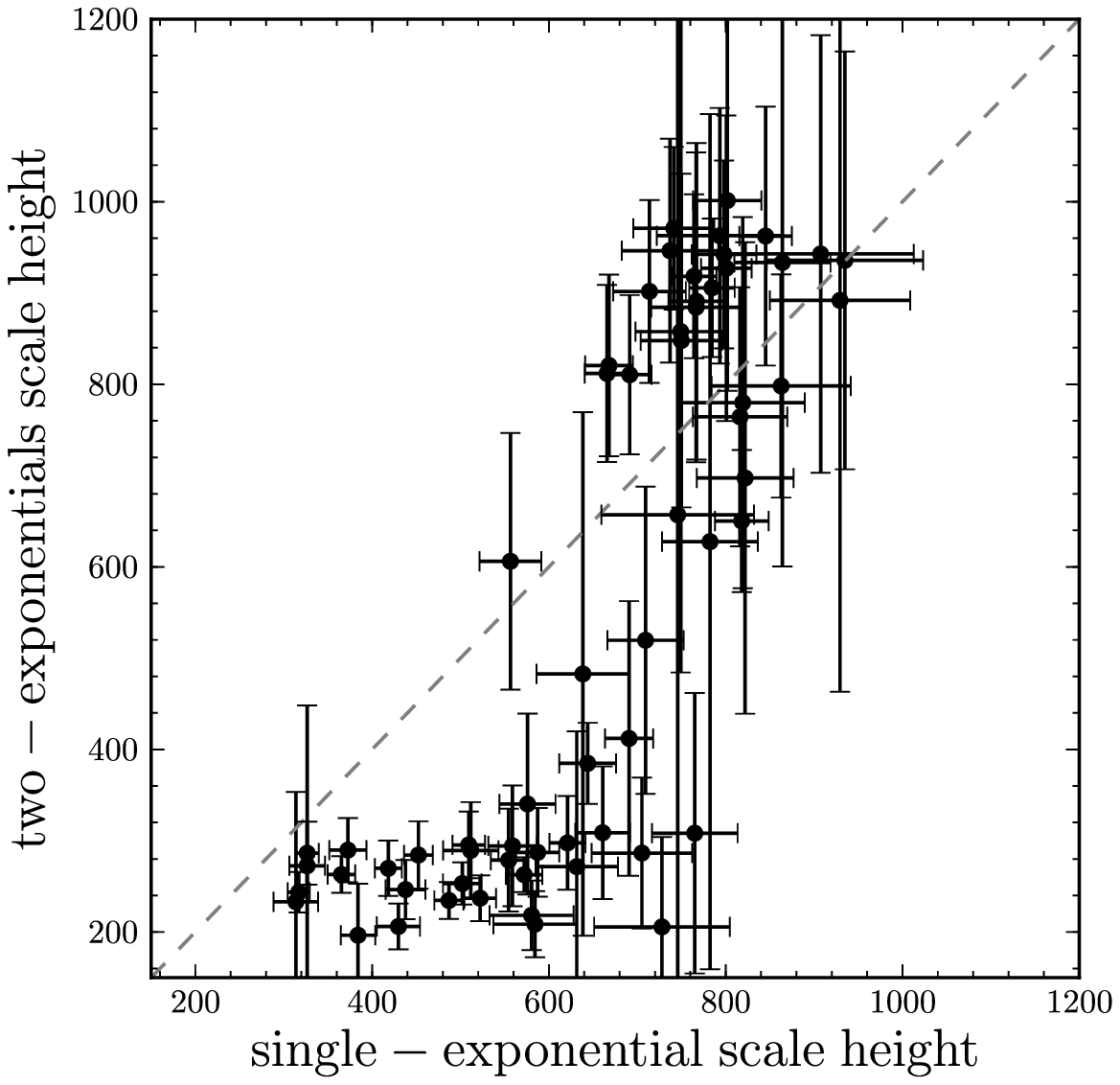}
\caption{Results from applying this paper's analysis to a mock data
sample composed of a two-component thin--thick disk sample. Assuming
abundance errors of 0.2 dex in \feh\ and 0.15 in \afe, we deconvolve
the observed distribution in (\feh,\afe)
(\figurename~\ref{fig:afeh_g}) into two Gaussian components, which are
assumed to correspond to a thin component ($h_z$ = 300 pc, $h_R$ = 3.5
kpc) and a thick component ($h_z$ = 850 pc, $h_R$ = 2 kpc). The panels
on the left show the results of fitting single-exponential models in
each mono-abundance bin, as in \figurename~\ref{fig:pixelFit_g}. The
top panel on the right shows the single-exponential scale height
vs. scale length color-coded by \afe. The bottom right panel shows the
scale height of the dominant component when fitting a mixture of two
expontial components vs. the single-exponential scale height. In
contrast to the fits for the real data sample, this latter plot shows
that the mixture-model is preferred and thus we confidently infer that
in the mock-data sample each bin is made up of two populations (at least for those with single-exponential $h_z \lesssim 800$ pc). Other
differences with the real data are that most mono-abundance bins have
a short single-exponential scale length due to the contamination from
the short, thick disk component and, similarly, that the
single-exponential scale heights quickly reach ``thick'' values of
$h_z \gtrsim 600$ pc when going to more metal-poor and \afe-enhanced
populations. We conclude that abundance errors cannot turn a clearly
thin--thick separated disk sample into the results we observe for
the real data set in \figurename s~\ref{fig:pixelFit_g} to
\ref{fig:one_vs_two}.}\label{fig:fakeBimodalResults}
\end{figure}

In order to test the methodology for fitting the density discussed in
\sectionname~\ref{sec:density}, and as check on the code, we create
mock data samples selected in exactly the same way as the \segue\
G-dwarf sample and fit them using our algorithm. We also use this
framework to test whether the results we obtain can plausibly be the
result of abundance errors smoothing out an underlying two-component
thin--thick disk structure.

We create mock data sampled from a model underlying density by
calculating, for each line of sight, (i) the fraction of stars in the
sample that lies along that line of sight and (ii) the distribution in
$r$-band magnitude as a function of color $g-r$ and metallicity
\feh. For calculating both of these, we take the \segue\ selection
function, described in \appendixname~\ref{sec:selection}, into
account. Thus, we can obtain a sample that is equivalent to what
\segue\ would have observed for a particular density model.

To test the methodology and code we populate each mono-abundance bin
in the (\feh,\afe) plane with a sample drawn from a thin-disk
component with $h_z = 300$ pc and $h_R = 3.5$ kpc, keeping the
abundances and number of stars in each bin the same as in the observed
sample. We then run the same analysis code on this sample as is run to
produce the real data results in \figurename s~\ref{fig:pixelFit_g} to
\figurename s~\ref{fig:one_vs_two}. We find results that are
consistent with the input model within the uncertainties for each
bin. The uncertainties are similar to those found for the real data
near $h_z = 300$ pc and $h_R = 3.5$ kpc. We repeat this for an input
``thick'' disk model with $h_z = 850$ pc and $h_R = 2$ kpc, and again
find results that are consistent with the input model within the
uncertainties.

We use a similar procedure to investigate whether abundance errors can
smooth out an underlying disk model made up of a thin- and thick-disk
component without showing up in our analysis. Assuming \segue\
abundance uncertainties of 0.2 dex in \feh\ and 0.15 dex in \afe, we
first model the underlying abundance distribution using two Gaussian
components, and fit this model to the observed distribution with the
assumed abundance uncertainties using the extreme-deconvolution
technique \citep{Bovy11a}. We use a Gaussian mixture model for the
underlying distribution solely as a convenient of decomposing the
observed distribution for the purpose of this test. The two-Gaussian
mixture model adequately represents the observed distribution after
convolving again with the uncertainties. The best-fit mixture has a
Gaussian centered near solar abundances (40\,percent of the sample;
\feh\ = $-0.3$ dex, \afe\ = 0.1 dex) and one at metal-poor and
$\alpha$-enhanced abundances (60\,percent of the sample; \feh\ =
$-0.7$ dex, \afe\ = 0.35 dex). To reproduce the observed distribution,
these components both need a dispersion of 0.2 dex in \feh\ and 0.07
dex in \afe, with a correlation of $-0.85$ and $-0.6$,
respectively. We then assign stars to these two components with
probabilities computed from their posterior probability of being drawn
from either component, based on their abundances and assumed abundance
uncertainties. We sample new $r$-band magnitudes and coordinates for
these stars based on the component they are assigned to: we draw the
stars assigned to the solar-abundances component from a thin-disk
density with $h_z$ = 300 pc and $h_R = 3.5$ kpc, and stars assigned to
the $\alpha$-enhanced component from a thick-disk distribution with
$h_z = 850$ pc and $h_R$ = 2.5 kpc. We then run the same analysis code
on this sample as is run on the real data.

The results from this test are shown in
\figurename~\ref{fig:fakeBimodalResults}. Although in certain respects
they are similar to the results for the real data, they are different in a few
crucial ways. Most importantly, when fitting a mixture of two
exponential models to each bin we find unambiguous evidence in many
bins for two components. This is shown in the lower right panel of
\figurename~\ref{fig:fakeBimodalResults}, where the scale height of
the dominant component when fitting the mixture is shown vs. the scale
height when fitting a single exponential. For most bins with
single-exponential $h_z \lesssim 800$ pc, the dominant component is
the $h_z = 300$ pc input thin-disk component. Therefore, even though
the abundance pattern of the single-exponential scale height in the
lower left panel of \figurename~\ref{fig:fakeBimodalResults} is smooth
between thin and thick components, most bins are actually resolved
into the two input components. This is a major difference with the
real data, for which the equivalent comparison, shown in
\figurename~\ref{fig:one_vs_two}, shows a striking one-to-one
correlation between the single-exponential and the mixture scale
height, with no evidence for a second component for the vast majority
of the mono-abundance bins.

In addition to the fact that our analysis correctly identifies two
components in each mono-abundance bin in the mock data, the abundance
dependence of the inferred single-exponential scale height and scale
length is also quite different from that of the mock data. The
inferred scale length for the mock data is short for most abundance
bins and only reaches $h_R \gtrsim 3$ kpc for those abundance bins
that are farthest from the center of the metal-poor and
$\alpha$-enhanced abundance component. Thus, the contamination from
the thick-disk component with its short scale length drives the
inferred scale length for most abundance bins to small values. This
behavior is not observed in the real data
(\figurename~\ref{fig:pixelFit_g}). The abundance dependence of the
single-exponential scale height for the mock data is also much steeper
than observed in the real data, with values of $h_z \gtrsim 600$ pc as
metal-rich as \feh\ = -0.3.

From these tests we conclude that abundance errors cannot explain the
single-exponential components we observe in each mono-abundance bin in
the real data or the abundance behavior of the scale height and scale
length. Based on an entirely different argument that uses the observed
isothermality of the vertical kinematics of the same mono-abundance
populations, \citet{Bovy12b} infer that the internal \segue\ abundance
uncertainties are likely somewhat smaller than the values reported by
\segue, with likely uncertainties of 0.15 dex in \feh\ and 0.07 dex in
\afe. Thus, abundance uncertainties do not influence the main
conclusions of this paper.

\end{document}